\documentclass[reprint, amsmath,amssymb, aps, superscriptaddress, floatfix]{revtex4-1}
\pdfoutput=1
\usepackage{graphicx}% Include figure files
\usepackage{dcolumn}% Align table columns on decimal point
\usepackage{bm}% bold math

\usepackage{amsbsy,amssymb,latexsym,amsfonts,amsmath}
\usepackage{graphicx,color}

\allowdisplaybreaks

\newcommand{\Slash}[1]{{\ooalign{\hfil#1\hfil\crcr\raise.167ex\hbox{/}}}}

\begin{document}

\preprint{APS/123-QED}

\title{Global Structure of Conformal Theories in the $SU(3)$ Gauge Theory}

%\thanks{A footnote to the article title}%
\author{K.-I. Ishikawa}
\affiliation{Graduate School of Science, Hiroshima University,Higashi-Hiroshima, Hiroshima 739-8526, Japan}%Lines break automatically or can be forced with \\

\author{Y. Iwasaki}
% \email{CCS, Uniersity of Tsukuba}
\affiliation{Center for Computational Sciences, University of Tsukuba,Tsukuba, Ibaraki 305-8577, Japan}
%\collaboration{MUSO Collaboration}%\noaffiliation

\author{Yu Nakayama}
% \homepage{http://www.Second.institution.edu/~Charlie.Author}
\affiliation{Kavli Institute for the Physics and Mathematics of the Universe (WPI), Todai Institutes for Advanced Study,  Kashiwa, Chiba 277-8583, Japan}

\author{T. Yoshie}
\affiliation{Center for Computational Sciences, University of Tsukuba,Tsukuba, Ibaraki 305-8577, Japan}
%\affiliation{Faculty of Pure and Applied Sciences, University of Tsukuba, Tsukuba, Ibaraki 305-8571, Japan}

\date{\today}% It is always \today, today,
             %  but any date may be explicitly\input{../../../../Volumes/HD-PSU2/file_reserved/iwasaki_scgt}

\begin{abstract}%
We investigate $SU(3)$ gauge theories in four dimensions with $N_f$ fundamental fermions,
on a lattice using the Wilson fermion.
Clarifying the vacuum structure in terms of Polyakov loops in spatial directions and properties of temporal propagators
using a new method ``local analysis'', 
we conjecture that
the ``conformal region'' exists together with the confining region and the deconfining region
in the phase structure parametrized by $\beta$ and $K$,
both in the cases of the large $N_f$  QCD within the conformal window (referred as Conformal QCD) with an
IR cutoff  and small $N_f$ QCD at $T/T_c>1$
with $T_c$ being the chiral transition temperature (referred as High Temperature QCD). 

Our numerical simulation on a lattice of the size $16^3\times 64$ shows the 
following evidence of the conjecture.
In the conformal region we find the vacuum is 
the nontrivial $Z(3)$ twisted vacuum modified by non-perturbative effects
and temporal propagators of meson behave at large $t$ as a power-law corrected Yukawa-type decaying form.
The transition from the conformal region to the deconfining region or the confining region
is  a sharp transition between different vacua and therefore it suggests a first order transition both in Conformal QCD and in High Temperature QCD. 
To confirm the conjecture and distinguish it from the possibility of crossover phenomena,
we need to take the continuum/thermodynamic limit which we do not attempt in this work.

Within our fixed lattice simulation, we find that there is a precise correspondence between
 Conformal QCD and High Temperature QCD in the temporal propagators under the
change of the parameters $N_f$ and $T/T_c$ respectively: the one boundary is
close to meson states and the other is close to free quark states.
In particular, we find the correspondence between Conformal QCD with
$N_f = 7$ and High Temperature QCD with $N_f=2$ at $T\sim 2\, T_c$ being in
close relation to a meson unparticle model.
From this we estimate the anomalous mass dimension $\gamma^* = 1.2 (1)$ for $N_f=7$.
We also show that the asymptotic state in the limit $T/T_c \rightarrow \infty$
 is a free quark state in the $Z(3)$ twisted vacuum.
The approach to a free quark state is very slow; even at $T/T_c \sim 10^5$,
the state is affected by non-perturbative effects.
This is possibly  connected with the slow approach of the free energy to the
Stefan-Boltzmann ideal gas limit.

\end{abstract}

\pacs{Valid PACS appear here}% PACS, the Physics and Astronomy
                             % Classification Scheme.
%\keywords{Suggested keywords}%Use showkeys class option if keyword
                              %display desired
\maketitle

\section{Introduction}
Recently much attention has been paid to conformal theories  in $d=4$ dimension, since conformal theories or nearly conformal theories are attractive candidates for the beyond standard model.
In the evolution of the Universe conformal theories might play  key ingredients
in many aspects, presumably more than we know today.
Every conformal field theories have their own distinct features.
To confront the nature, it is important to understand each conformal theory, and for this purpose,
it is urgent to clarify the global structure of conformal theories\cite{review1}.

One important class of simple conformal field theories in $d=4$ is realized by 
the so-called Banks-Zaks fixed point \cite{Banks1982} in many flavor gauge theories.
The possibility of the existence of a conformal theory in 
$SU(3)$ gauge theory with $N_f$ flavors in the fundamental representation
was first pointed out by W.~Caswell in \cite{Caswell:1974gg}.

From the perturbative computation of the beta function, we believe the upper critical number of flavors $N_f$ for the existence of an infrared (IR) fixed point in $SU(3)$ gauge theory is $16$.
We denote the lower critical number of flavors by $N_f^c$. The region of $N_f  (N_f^{c} \le N_f \le 16)$ that has the IR fixed point is called the conformal window.

In the case of $N_f\simeq 16$, the coupling constant at the IR fixed point is small and therefore the perturbation theory may be applicable. However, in the case of 
$N_f \sim N_f^c$, 
non-perturbative effects will be important and non-perturbative tools are essential.

Lattice gauge theories are systematic and non-perturbative tools to investigate issues such as the lower critical number of flavors $N_f^c$, the anomalous mass dimension and the spectrum.
Many lattice studies were indeed performed~\cite{iwa2004} -- \cite{itou1307}.
Numerical tools such as the step scaling scheme~\cite{step_scaling} and the MCRG method, as well as the calculation of mass spectrum and the analysis of the phase structure
have been used in order to identify the IR fixed point and the conformal window~\cite{review1}.

However the determination of the lower critical number of flavors $N_f^c$ is still 
much controversial. One possible reason for the controversy is pinpointing the IR fixed point for $N_f^c$ is the strong coupling problem and technically it is hard to reach such a region even with the step scaling scheme. 
Another possible reason, when the mass spectrum is used to find the $N_f^c$, is that the suitable mass region for the investigation of conformal properties is limited, as will be clarified in this article. However, in many calculations the mass spectrum outside of this region is used.

In our previous paper~\cite{iwa2004}
we conjectured the lower critical number of flavor $N^c_f=7$.
In Appendix~\ref{our-previous-works} we report a brief summary of the paper  with some update.
In this article we do not assume a particular value of $N_f^{c}$ in general discussions.
The aim of this paper is to establish the properties intrinsic to the conformal window such as the vacuum structure and specific behaviors of temporal propagators. We would like to check these properties for various $N_f$ including $N_f= 7$, and we will verify if our conjecture of $N^c_f= 7$ is consistent with them.

\begin{table*}
\caption{Numerical results for $N_f=16$:
"s" in the second column represents the initial status; the continuation from the lower $K$ (l) or from the higher $K(h)$;
the third column is the number of trajectories for measurement;
the 4th column is the plaquette value;
the $m_q$ in the 5th column is the quark mass defined in Eq.~\ref{eq:quark-mass};
 $m$ in the 6th and 7th columns are the
mass of $PS$ and $V$ channels in the case of the exponential  decay defined in Eq.~\ref{exp};
the $\tilde{m}$  in 8th and 10th columns and $\alpha$ in the 9th and 11th columns are, respectively,  the "mass"  and the exponent of $PS$ and $V$ channels in the case of the Yukawa-type  decay defined in Eq.~\ref{yukawa type}.}
\begin{tabular}{lrrllllllll}
\hline
\hline
 & \multicolumn{4}{c}{$N_f=16$}   & \multicolumn{2}{c}{$\beta=11.5$} &&& \\
\hline
$K$ & s& $N_{tra}$&  plaq & $m_q$ & $m_\pi$ & $m_V$ & $\tilde{m_{\pi}}$ &$\alpha_\pi$ &$\tilde{m_{V}}$ & $\alpha_V$\\
\hline
0.120  & l &1000   &       0.820199(7) &    0.3995(1)   &    0.7930(5)   &    0.7931(5)  &-&-&-&-\\
0.121  & l &  500    &      0.820543(7) &    0.3672(2)  &     0.7451(19)   &    0.7458(20) &-&-&-&-\\
0.122  & l  & 500    &       0.820856(7) &    0.3346(1) &      0.6844(8)  &    0.6847(8)    &-&-&-&-\\
0.123  & l  & 500    &     0.821214(6)  &    0.3039(1)  &     0.6389(9)   &   0.6397(11)    &-&-&-&-\\
0.124  & l  & 500     &        0.821560(7) &   0.2733(1) &       0.5858(8) &     0.5930(18)  &-&-&-&-\\
0.125  & l & 1000    &    0.821918(05)   &     0.2435(1) &        0.5401(27)  &    0.5411(25)  &-&-&-&-\\
0.125  & h & 1000   &       0.821927(06)  &    0.2498(1)    &   -          &     -      &         0.615(5)   &     1.37(5)   &      0.615(6)    &    1.34(6) \\
0.1255 & l &  1000  &       0.822127(06) &    0.2348(1)  &     -        &       -        &       0.599(3) &       1.28(3) &        0.601(3)  &      1.23(3) \\
0.126  & l & 1000    &     0.822324(07)  &    0.2158(1)   &    -         &      -         &      0.504(3)  &      0.99(9)  &       0.502(3)   &     0.99(19) \\ 
0.1262 & h & 1000  &      0.822411(04)  &    0.2122(3)  &     -        &       -        &       0.522(9) &        1.72(7)  &      0.532(7)   &     1.5(4) \\
0.1264 & l &  500    &     0.822497(11)   &    0.2072(1)   &    -         &      -         &      0.548(12)  &     1.45(13)  &     0.545(13) &      1.5(15) \\
0.1266 & h &  1000  &       0.822577(06) &   0.2010(1)   &    -         &      -         &      0.534(3)    &    1.44(8)     &    0.533(4)    &    1.4(5) \\
0.127  & h & 1000   &       0.822745(08)  &    0.1864(1)  &     -        &       -        &       0.451(5)   &     1.29(10)  &      0.446(6)  &      1.34(14) \\
0.130 & h &  1000   &       0.824107(06)  &    0.0998(2)  &     -        &       -        &       0.425(4)   &     1.19(2)    &     0.411(7)    &    1.48(8) \\
0.1315 & h &  1000  &       0.824866(09)  &    0.0552(4) &      -       &        -       &        0.397(4)  &      1.15(2)   &      0.394(7)   &     1.23(9) \\
0.13322 & h &  1000   &     0.825790(08) &    0.0029(4) &      -       &        -       &        0.390(5)  &      1.09(4)   &      0.396(8)   &     1.02(14) \\
\hline
\label{tab:nf16}
\end{tabular}
\end{table*}

\subsection{Strategy and Objectives}
In this article we discuss the following two categories in $SU(3)$ gauge theories with $N_f$ flavors in the fundamental representation which possess an IR fixed point:
\begin{itemize}
\item
Large $N_f  (N_f^{c} \le N_f \le 16)$ QCD within the conformal window  (referred as Conformal QCD) 
\item
small $N_f  (2 \le N_f \le N_f^{c}-1)$ QCD at temperature $T/T_c > 1$ with $T_c$ being the critical temperature (referred as High Temperature QCD) 
\end{itemize}

\begin{table*}
\caption{Numerical results for $N_f=7$: the meanings of the columns are the same as $N_f=16$}
\begin{tabular}{lrrllllllll}
\hline
\hline
 & \multicolumn{4}{c}{$N_f=7$}   & \multicolumn{2}{c}{$\beta=6.0$} &&& \\
\hline
$K$ & s& $N_{tra}$&  plaq & $m_q$ & $m_\pi$ & $m_V$ & $\tilde{m_{\pi}}$ &$\alpha_\pi$ &$\tilde{m_{V}}$ & $\alpha_V$\\
\hline
0.1300 &  l  &    1000 &  0.615510(20) &   0.5500(3)  &    1.2216(10)  &    1.2263(11) &-&-&-&-\\
0.1370 &l    &     1000  & 0.623238(15)  &  0.3081(3)   &    0.8443(25)  &    0.8501(27)&-&-&-&-\\
0.1380& l    &    1000 &  0.624511(16) &   0.2777(2)   &    0.7979(19)  &    0.8032(24) &-&-&-&-\\
0.1390& l    &    1000 &  0.625859(17) &   0.2475(2)  &     0.7358(27)  &    0.7431(31) &-&-&-&-\\
0.1400& l    &    1000 &  0.627285(23) &   0.2181(2) &      0.6824(31)  &    0.6916(33) &-&-&-&-\\
0.1410& l    &     900 &   0.628794(14) &   0.1889(2)  &       0.6304(26) &      0.6394(32) &-&-&-&-\\
0.1412&  l   &    1000  &  0.629129(20) &    0.1833(2)  &     0.6250(21) &     0.6356(23) &-&-&-&-\\
0.1412&  h  &     1000 &  0.629037(12) &   0.1814(2)    &   -                   &   -      &           0.557(5)    &    0.78(5)  &       0.560(5)   &       0.80(7)\\
0.1413&  l    &    500   &   0.629228(12) &   0.1794(3)    &   0.5978(30)  &    0.6022(33) &-&-&-&-\\
0.1413&  h   &    1000  &  0.62927(12)   &   0.1780(4)    &   -                  &	-           &  	  0.512(9)   &     1.39(8)     &    0.514(9)    &      1.36(7)\\
0.1415&¡¡h  &      600  &   0.629559(13)  &   0.1721(2)   &    -            	&   -            &    0.513(3)   &     1.09(7) &        0.516(3) &        1.10(11)\\
0.1420&  h     &  1000    &  0.630328(21) &   0.1587(4)    &   -          		&    -           &     0.525(13)&       1.05(14) &       0.522(15) &      1.21(16)\\
0.1430 & h     &    700   &     0.631951(20) &   0.1309(2)   &   -      		&    -           &     0.523(6)  &      0.39(10)  &      0.529(07)  &     0.51(11)\\
0.1446&  h    &    1000   &     0.634723(22) &   0.0842(5)  &     -   	         &   -            &    0.472(6)   &     0.46(6)     &    0.483(05)    &   0.54(03)\\
0.1452&  h    &     500 &     0.635759(19)   & 0.0614(1)     &      -  	        &     -           &     0.426(12)&       0.80(1)   &      0.426(12)  &     1.03(02)\\
0.1459&  h    &    1000   &     0.637062(17) &   0.0450(2)  &     -   	        &    -            &    0.410(11)   &    0.80(14)   &     0.413(14)   &    1.01(18)\\
0.1464&  h   &      700     &     0.637981(17) &   0.0303(2) &      -  	          &   -           &     0.381(8)    &    0.64(13)   &     0.393(09)   &    0.73(14)\\
0.1472&  h   &     1000     &   0.639496(15)  &  0.0060(2)  &     -   	           & ¡¡-         &      0.405(8)   &     0.75(10)  &      0.406(09)  &     1.06(10)\\
\hline
\end{tabular}
\end{table*}

\begin{table*}
\caption{Numerical results for $N_f=8$ at $\beta=6.0$: the meanings of the columns are the same as $N_f=16$}
\begin{tabular}{lrrllllllll}
\hline
\hline
 & \multicolumn{4}{c}{$N_f=8$}   & \multicolumn{2}{c}{$\beta=6.0$} &&& \\
\hline
$K$ & s& $N_{tra}$&  plaq & $m_q$ & $m_\pi$ & $m_V$ & $\tilde{m_{\pi}}$ &$\alpha_\pi$ &$\tilde{m_{V}}$ & $\alpha_V$\\
\hline
0.1446 & h & 1000 & 0.634723(22) & 0.0738(1)&-&-&      0.422(8) &  0.79(9) & 0.430(9) & 0.84(12)\\
0.1457 & h &1000 & 0.637062(17)  & 0.0342(2) & - & -&   0.386(5) &  0.78(7) & 0.385(7) & 1.05(12)\\
\hline
\end{tabular}
\end{table*}

The existence of an IR fixed point in Conformal QCD is well known as the Banks-Zaks IR fixed point 
\cite{Banks1982}, as mentioned above.
In High Temperature QCD the existence of an IR fixed point has been recently pointed out  in Ref.~\cite{coll2}. 
We will clarify the precise relation between the IR fixed point and the ``conformal symmetry" in detail below since this is not literally true at first sight (e.g. non-vanishing trace anomaly for High Temperature QCD). See also Appendix B for a brief review of our argument in ~\cite{coll2}. 

Let us further consider the case where there is an IR cutoff in the theory which possesses an IR fixed point.
In the case of Conformal QCD in the continuum limit, the compact space and/or time gives an IR cutoff.
In the case of High Temperature QCD, the temperature $T$ plays a role of an IR cutoff together with a cutoff due to possible compact space, depending on how to take the continuum limit.
We note any lattice calculation  is performed on a finite lattice. Thus any calculation on a lattice possesses an IR cutoff.

In the case there is an IR cutoff,
we introduce a new concept ``conformal theories with an IR cutoff'': In the ``conformal region'' where the quark mass is smaller than the critical value,
 temporal propagators $G_H(t)$ of meson behave at large $t$ as
a power-law corrected Yukawa-type decaying form instead of the exponential decaying form observed in the ``confining region" and ``deconfining region''. We note the exponential decay form in the deconfining region is approximate due to the finiteness of the $t$ region. This point will be discussed in some detail below.

One of the objectives of this article is to verify the existence of the conformal region and a power-law corrected Yukawa-type decaying form instead of the exponential decaying form of meson propagators in the conformal region, on a finite lattice with fixed size $16^3\times 64$. 

We stress that QCD in compact space and/or time 
is a conformal theory with an IR cutoff for $\beta \ge \beta^c$, as will be discussed below.
Here the $\beta^c$ is the critical bare coupling constant ($\beta = {6}/{g_0^2}$)
 at which a chiral transition occurs for the massless quark.
In the case of the compact space, the temperature may be defined by $1/N_t \, a$ as usual. 

On the other hand, one of our final goals is the verification of the conjecture of the conformal theory with an IR cutoff 
for the case of the thermodynamical limit of High Temperature QCD in the flat space,
in addition to the continuum limit of the Conformal QCD in the flat space at zero temperature. Since we define the conformal region from the properties of the temporal propagators, we need a lattice with large $N_t$ in order to verify the idea of the conformal theories with an IR cutoff. Therefore we take the same lattice size $16^3\times 64$ for the simulation of High Temperature QCD.

We understand that our quantitative predictions for thermodynamic properties for High Temperature QCD will be affected by the small spatial lattice size.
However, since our theoretical argument only relies on the vanishing beta function and  the existence of an IR cutoff (either by temporal one or spatial one),
this lattice size does not spoil our objective to investigate qualitatively the behavior of propagators. 
If we could confirm our concepts of conformal field theories with an IR cutoff on this size lattice,
we would be able to naturally conjecture that our proposal  will be realized on a larger spatial lattice such as $256^3\times 64$.
While testing our conjecture on a larger lattice to take the thermodynamic limit is important if we would like to compare our results with the experiment, it is beyond the scope of this article. We will make a small comment in section XII.

After verification of the existence of the conformal region and a power-law corrected Yukawa-type decaying form on the lattice with size $16^3\times 64$, 
we would like to reveal the properties of the conformal region and the temporal propagators
in all cases of Conformal QCD and High Temperature QCD as a whole. In particular we would like to clarify the underlying physics leading to the behavior of the  characteristic form of the propagators and extract the physical properties of each theory.

We utilize two tools to investigate the issues: One is a new method of analyzing the propagators  of mesons which we call the local-analysis of propagators. The other is the analysis of the vacuum in terms of
the Polyakov loops in spatial directions.

We find that the vacuum corresponding to the conformal region is the nontrivial $Z(3)$ twisted vacuum modified by non-perturbative effects.
Clarifying the relation between the vacuum structure and properties of temporal propagators in each vacuum,
we show the transition from the power-law corrected Yukawa-type decaying form to the exponential decay
is  a transition between different vacua and therefore it is a first order transition
both in Conformal QCD and in High Temperature QCD.
 
Finally, we argue from our theoretical analysis based on the renormalization group (RG) flow and our
numerical simulations that there is a precise correspondence between
the Conformal QCD and High Temperature QCD within the conformal region.
The correspondence between the two sets of conformal theories with an IR cutoff 
is realized between 
a continuous parameter $T/T_c$  and a discrete parameter $N_f$: the one boundary is
close to meson states and the other is close to free quark states.

In particular, we find the correspondence between Conformal QCD with
$N_f = 7$ and High Temperature QCD with $N_f=2$ at $T\sim 2\, T_c$ being in
close relation to a meson unparticle model.
From this we estimate the anomalous mass dimension $\gamma^* = 1.2 (1)$ for $N_f=7$.
We also show that the asymptotic state in the limit $T/T_c \rightarrow \infty$
 is a free quark state in the $Z(3)$ twisted vacuum.
The approach to a free quark state is very slow; even at $T/T_c \sim 10^5$,
the state is affected by non-perturbative effects.
We believe that this is related with the slow approach of the free energy to the
Stefan-Boltzmann ideal gas limit. To conclude the precise relation we need to perform similar analysis in the thermodynamical limit.

The fact above is consistent with  our conjecture that the lower critical flavor number $N_f^{c}=7$
\cite{iwa2012}.

\subsection{Outline of the Paper}
The rest of the paper is organized as follows.
In section II, we provide theoretical and numerical background to study an IR fixed point in lattice QCD. In section III, we introduce the new concept of ``conformal theories with an IR cutoff" and discuss its implications. In section IV, we study the structure of propagators to confirm the existence of conformal regions. In section V, we study the vacuum structure of lattice QCD. In section VI, we analyze the propagators in each vacuum. In section VII, we discuss the detailed relation between the vacuum structure and the existence of conformal region. In section VIII, we analyze the effects of boundary conditions on the structure of vacuum. In section IX, we introduce the concept of unparticle models as an effective description in the conformal region. In section X, we propose the correspondence between Conformal QCD and High Temperature QCD. In section XI, the correspondence is further studied to predict the mass anomalous dimensions. In section XII, we conclude the paper with further discussions.

We have six appendices. In Appendix A, we review the history of our previous works. In Appendix B, we review the (finite temperature) beta function and its relation to trace anomaly. In Appendix C, we derive the unparticle propagators. In Appendix D, we report the computation of one-loop vacuum energy in lattice QCD with various boundary conditions. In Appendix E, we review our viewpoint on the chiral phase transition in $N_f =2$ QCD. We collect the figures not listed in the main text in Appendix F.

\section{Background}
\subsection{Action and Observables}
We define continuous gauge theories as the continuum limit of lattice gauge theories,
defined on the Euclidean lattice of the size $N_x=N_y=N_z=N$ and $N_t$. To obtain the thermodynamic interpretation, we in general impose an anti-periodic boundary condition in the time direction for fermion fields and
periodic boundary conditions otherwise.
We also discuss the case when anti-periodic boundary conditions in spatial directions for fermion fields are imposed.

Our general argument that follows can be applied to any gauge theories with (vector-like) fermions in arbitrary representations, but to be specific, we focus on $SU(3)$ gauge theories with $N_f$ fundamental fermions (``quarks").
We employ the Wilson quark action and the standard one-plaquette gauge action.
The theory is defined by two parameters; the bare coupling constant $g_0$ and the bare degenerate quark mass $m_0$ at ultraviolet (UV) cutoff.
We also use, instead of $g_0$ and $m_0$, 
$\beta={6}/{g_0^2}$
and the hopping parameter
$K= 1/2(m_0a+4)$. 

We measure together with the plaquette and the Polyakov loop in each space-time direction,
the quark mass $m_q$ defined 
through Ward-Takahashi identities
\cite{Bocc, ItohNP}
\begin{equation}
m_q 
=  \frac{\langle 0 | \nabla_4 A_4 | {\rm PS} \rangle}
        {2\langle  0 | P | {\rm PS} \rangle}, 
\label{eq:quark-mass}
\end{equation}
where $P$ is the pseudo-scalar density and $A_4$ the fourth component of the
local axial vector current, with renormalization constants being suppressed.
The quark mass $m_q$ thus defined 
does only depend on $\beta$ and $K$ and does not depend on whether it is in the confining region or
the deconfining region up to order $a$ corrections\cite{iwa_chiral}.

In addition to them 
we investigate in detail  the $t$ dependence of the propagator of the local meson operator in the $H$ channel:
\begin{equation}
G_H(t) = \sum_{x} \langle \bar{\psi}\gamma_H \psi(x,t) \bar{\psi} \gamma_H \psi(0) \rangle \ ,
\label{propagator}
\end{equation}
where the summation is over the spatial lattice points.
In this paper, we mostly focus on the pseudo-scalar (PS) channel $H=PS$, but¡¡
we also measure other channels and use the vector channel to see the chiral symmetry.

\subsection{Simulations}
We make numerical simulations for $N_f=7, 8, 12, 16$  as candidates of Conformal QCD
and for $N_f=2$ as High Temperature QCD
on a lattice with fixed size $16^3\times 64$.

The algorithms we employ are the blocked HMC algorithm \cite{Hayakawa:2010gm} in the case $N_f=2\, \mathbb{N}$ and the RHMC algorithm \cite{Clark:2006fx} for $N_f=1$ in the case $N_f=2\, \mathbb{N} +1$.

We specify the coupling constant  $\beta=11.5$ for $N_f=16,$
taking account of the fact that the IR fixed point for $N_f=16$ is $\beta=11.48$ in two-loop approximation~ (see subsection:~\ref{sec:Banks-Zaks}),
while $\beta=6.0$ for $N_f=7, 8, 12, 16$,
varying the hopping parameter $K$ so that the quark mass takes the value from 0.40 to 0.0,
except for a few cases in the deconfining region or the confining region for comparison.
We further perform simulations for $N_f=12$ at $\beta=8.0$.

For High Temperature QCD,
identifying the chiral transition around $K=0.151$ at $\beta=6.0$ on a lattice $16^3\times 64$
by the ''on-Kc method'' in Ref.~\cite{iwa96}, we choose the following values of $\beta's$\,:
$\beta=6.5, 7.0, 8.0. 10.0$ and  $15.0$. 

 If we formally estimate the temperatures based on $\Delta \beta \sim 0.5$ for the scale change of a factor $2$ from the beta function in the one-loop approximation with $N_f=2$, we obtain 
$T/T_c  \sim 2,\, T/T_c  \sim 4,\, T/T_c  \sim 16,\, T/T_c \sim 100,$ and $T/T_c \sim 10^5$, respectively.
We take several values of $K$ for each $\beta$ in such a way that the quark masses $m_q's$ take values $0.00 \le m_q \le 0.30,$ except for a few cases. 

We show the parameters for simulations and the numerical results in Tables $(\mathrm{I} - \mathrm{VII})$.
All results for masses of mesons and the quarks are expressed in units of the inverse of the lattice spacing $a^{-1}$ in the text and the tables.

We choose the run-parameters in such a way that the acceptance of the global metropolis test is about $70\%.$
The statistics are 1,000 MD trajectories for thermalization and 1,000 MD trajectories for the measurement, or 
500 MD trajectories for thermalization and $500\sim900$ MD trajectories for the measurement.
We estimate the errors by the jack-knife method
with a bin size corresponding to 100 HMC trajectories.

\subsection{Continuum limit}\label{contlimit}
The continuum limit of a lattice theory is defined by taking the lattice space $a \rightarrow 0$ with $N \rightarrow \infty$ and $N_t \rightarrow \infty$, keeping $L =N \, a$  and $ L_t =N_t \, a$ fixed.

When $N_f \leq 16$, the point $g_0=0$ and $m_0=0$ 
in the two parameter space $(g_0, m_0)$
is an UV fixed point. Therefore a theory governed by this fixed point is an asymptotically free theory. We restrict ourselves to the theory defined by this UV fixed point in this article.

There are four cases in the continuum limit:
\begin{enumerate}
\item
$L$ and $L_t$ are finite:\\ the space is the three-torus $\mathbb{T}^3$, finite temperature $T=1/L_t$
\item
$L$ is finite and $L_t$ is $\infty$:\\the space is $\mathbb{T}^3$, zero temperature
\item
$L$ is $\infty$ and $L_t$ is finite:\\ the space is the Euclidean plane $\mathbb{R}^3$, finite temperature $T=1/L_t$
\item
$L$ and $L_t$ are $\infty$: \\the space is $\mathbb{R}^3$, zero temperature 
\end{enumerate}

When  $L$ and/or $L_t$ are finite, the system is bounded by an IR cutoff $\Lambda_{\mathrm{IR}} $.
Practically in numerical simulations, to achieve the limit corresponding to case 3, we may first take the limit $N_t $ infinity (thermodynamical limit) and then we take the limit $N$ infinity.
On the other hand, to achieve the limit corresponding to case 4, we may put $N=r\, N_t$ with $r$ an aspect ratio,
and finally we take the limit $N$ infinity simultaneously keeping $r$ fixed. 

When we take the continuum limit, we have to fix the physical scale. In confining QCD, the natural choice is to fix the mass of the hadron by demanding
\begin{align}
L \cdot m_{PS} > c
\end{align}
and take $L \to \infty$ limit,
where the value $c\sim 5$ is typically used in the literature. For the conformal QCD, this choice turns out to be 
subtle because there should be no scale in conformal field theories
 after taking the continuum limit.
As we will argue, the choice of $c$ leads to different phases in the continuum limit. We will also evaluate the upper limit value of $c$ to retain the conformal behavior in the continuum limit.

\subsection{Banks-Zaks fixed point}
\label{sec:Banks-Zaks}
Within the perturbation theory, the zero temperature beta function for the $SU(3)$ gauge coupling constant can be computed as
\begin{align}
\mathcal{B}(g) = -\frac{(33-2N_f)}{48\pi^2} g^3 -\frac{ \left(102 - \frac{38}{3}N_f\right) }{(16\pi^2)^2}g^5 + \mathcal{O}(g^7) \ . \label{betaff}
\end{align}
The fixed point $\mathcal{B} (g^*) = 0$ exists for $8.05 < N_f <16.5$ with the two-loop approximation. 
Of course, the two-loop result is not trustful for lower values of $N_f$, so the lower bound of the conformal window cannot be obtained from the perturbation theory.

As we review in Appendix B, the beta function is related to the trace anomaly. The trace of the energy-momentum tensor in massless QCD is given by
\begin{align}
T^{\mu}_{\ \mu} = \mathcal{B}(g) \mathrm{Tr} F_{\mu\nu}^2 \ 
\end{align}
as an operator identity.
It vanishes when the theory is at the IR fixed point $g=g^*$ and vanishing of the trace anomaly means it is conformal invariant.

At the conformal fixed point, one may compute the anomalous mass dimension. The perturbation theory predicts
\begin{align}
\gamma_m = \frac{1}{2\pi^2}g^2 \ , 
\end{align}
which should be compared with the lattice simulation after establishing how to read the anomalous mass dimension from the temporal propagators as we show later.
For reference, for $N_f=16$, the fixed point value from \eqref{betaff} is $\beta_0 = 11.48$ and the mass anomalous dimension is $\gamma^* = 0.026$. 
We further note that the conserved current operator is not renormalized. Therefore the anomalous dimension vanishes. 

\subsection{Phase structure} 
\label{sec:phase} 

\begin{figure*}[thb]
 \includegraphics [width=7.5cm]{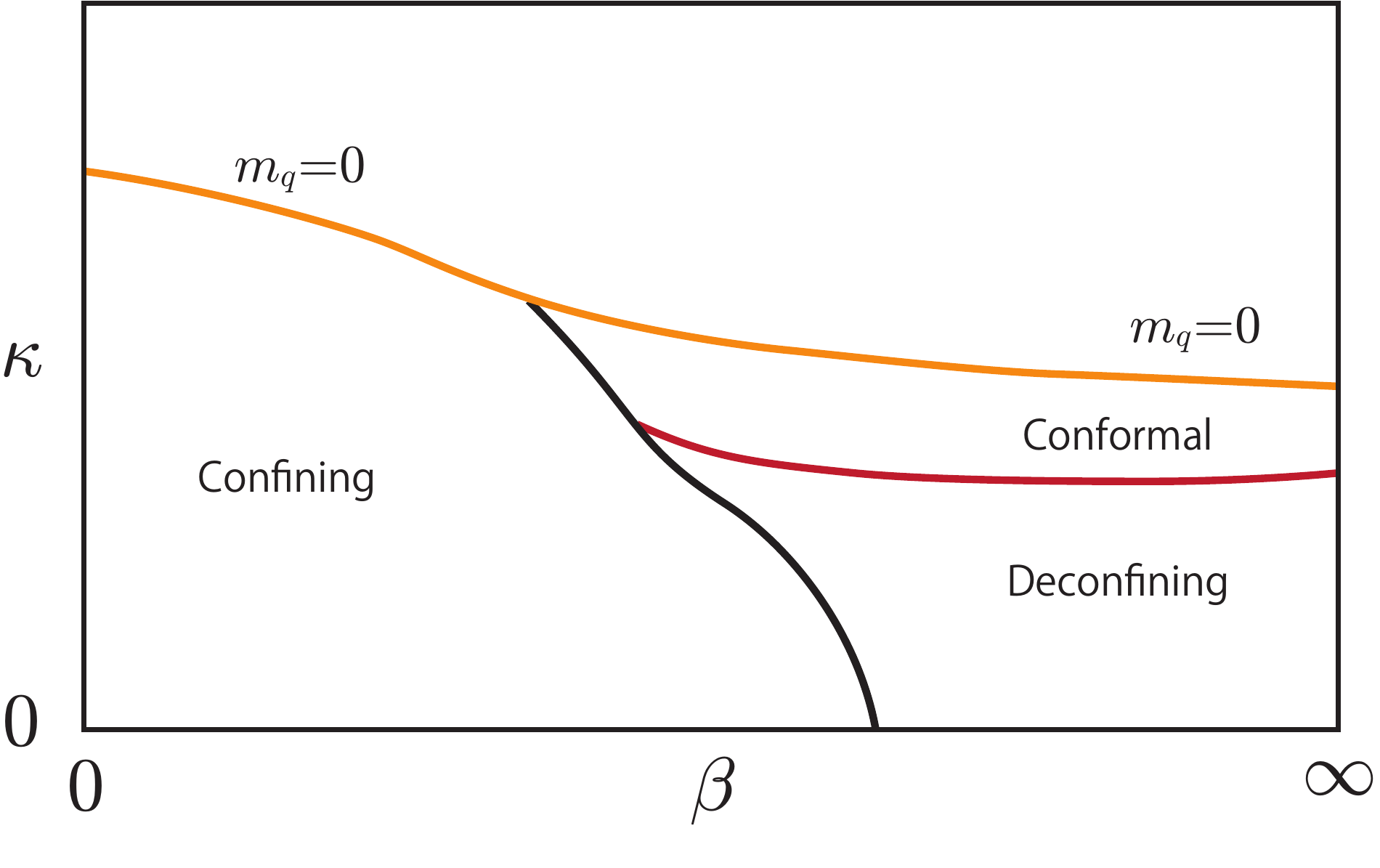}
 \hspace{1cm}
 \includegraphics [width=7.5cm]{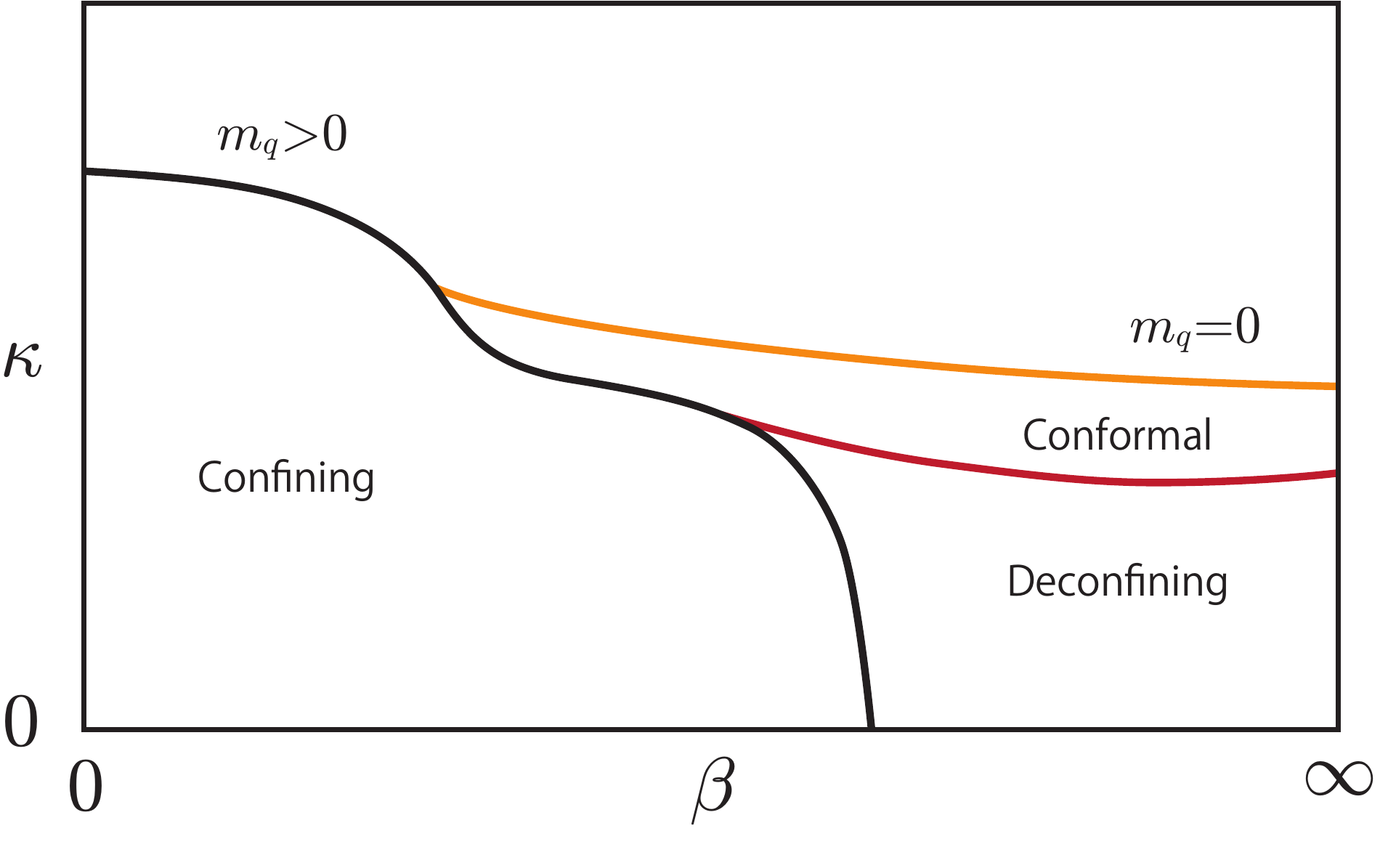}
\caption{(color online) Phase diagram on a finite lattice : (left) $1 \le N_f \le N_f^c -1$ ; (right) $N_f^c \le N_f \le 16$;
 In the case $N_f^c \le N_f \le 16$ the massless quark line originating from the UV fixed point hits the bulk transition point at finite $\beta$ and no massless line exists in the confining region.  The region above the bulk transition corresponds to the doublers of the Wilson fermion. On the other hand, in the case  $1 \le N_f \le N_f^c -1$
 on the massless quark line there is a chiral phase transition point. Below the critical point the massless line is in the confining region.}
\label{phase diagram finite lattice}
\end{figure*}

In order to investigate properties of the theory in the
continuum limit,
it is vital to clarify the phase structure of lattice 
QCD.

Although we would like to extract the phase diagram 
in the continuum limit,
we have to perform simulations at finite $N$.
Therefore the phase diagram is a three dimensional space parameterized by $g_0$,  $m_0$ and 
$N$.  
Thus, first of all, one has to make clear what kind of phases there are in this three dimensional space.

We claim that we  are able to classify the phase space into three regions:
\begin{enumerate}
\item 
confining region
\item 
deconfining region
\item 
conformal region
\end{enumerate}

Let us first recall that there are no order parameters which distinguish the ``deconfining phase'' from the ``confining phase'', except for the two limits;  in the limit $m_q\rightarrow\infty$ the Polyakov loop for the $t$ direction can be used at the deconfining phase transition, and in the limit $m_q\rightarrow 0 $, the chiral scalar density can be used at the chiral transition.
Therefore it is not possible in principle to state which phase is realized at the intermediate quark mass $m_q$. 
However,  in the $3 \le N_f \le 6$ case there is a first order transition line from the chiral transition point toward heavier quark mass.
In such a case, one may state either the chiral symmetry is restored or 
the chiral symmetry is broken, depending on the region it belongs to.

The deconfining transition is a first order phase transition, and similarly to the above there is a first order transition line from the deconfining transition point to lighter quark mass.
In this case one may say either the quark confinement or deconfinement, depending on the region.
However, in the intermediate quark mass region the first order transition becomes weak and probably
disappears. Thus the confining region and the deconfining region are connected, therefore in strict 
meaning both regions belong to one phase.

We use the terminology ``confining region" instead of the confining phase, since as mentioned above there is no order parameter in general.

When there is a Banks-Zaks IR fixed point \cite{Banks1982}
 at finite coupling constant $g$ on the massless line which starts from the UV fixed point, the long distance behavior is determined by the IR fixed point. This defines a conformal theory. The meaning of the conformal theory will be discussed later.
We recall the region of $N_f$ for the existence of the IR fixed point
$N_f^{c} \le N_f \le 16$ is called the conformal window.

On the other hand, when $1 \leq N_f \leq N_f^{c}-1$, the beta function does not possess an IR fixed point in the continuum limit with infinite space-time, which implies the quark confinement.
When the lattice size is finite, the vicinity of the point $g_0=0$ along $m_q=0$ is the chiral symmetric phase
 and there is a chiral phase transition on the quark massless line $m_q=0$.

The scenario above is a common lore, 
although there is an alternative possibility that in-between the confinement (chiral symmetry broken) region and
the chiral symmetric conformal region at zero temperature, 
a new phase like the magnetic phase may exist.

\begin{figure*}[thb]
\includegraphics[width=7.5cm]{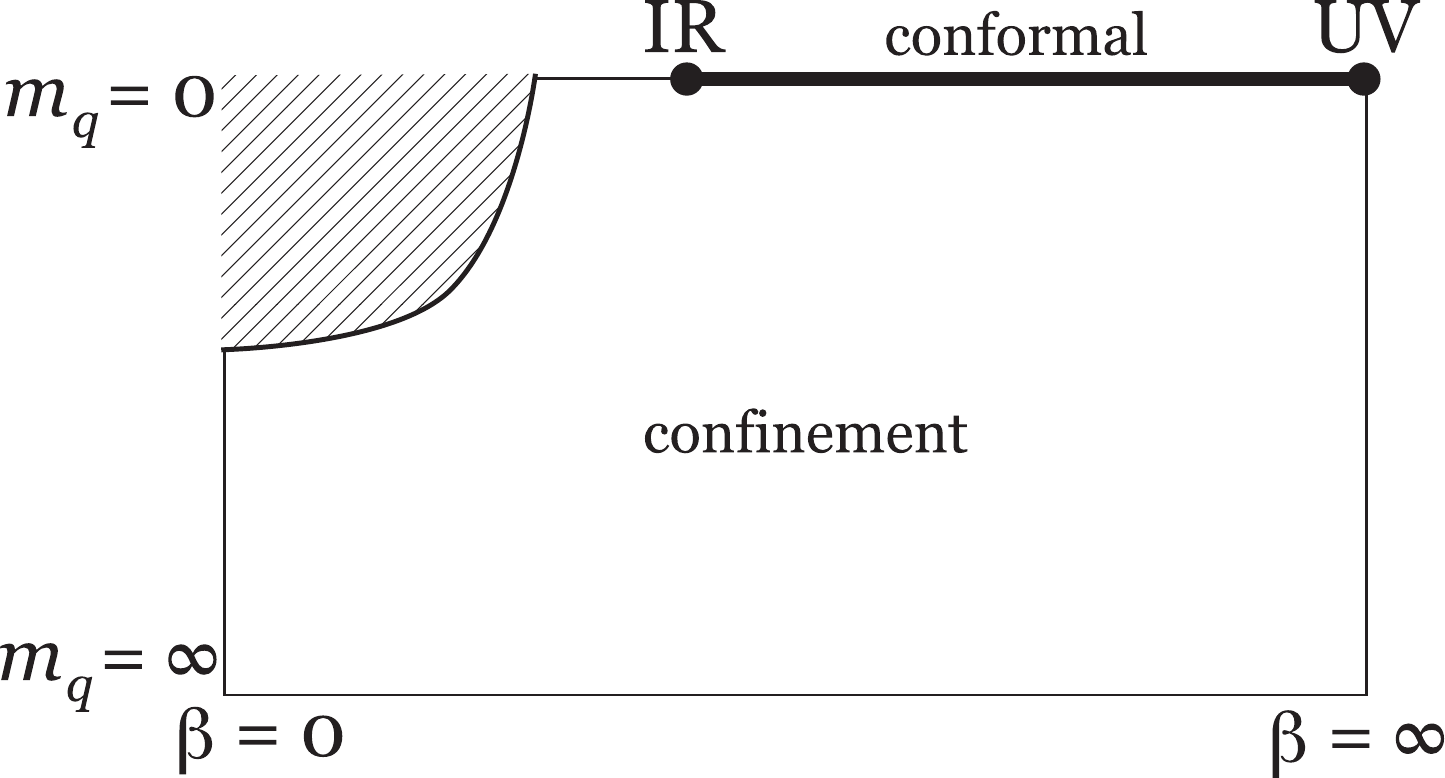}
\hspace{1.5cm}
\includegraphics[width=7.5cm]{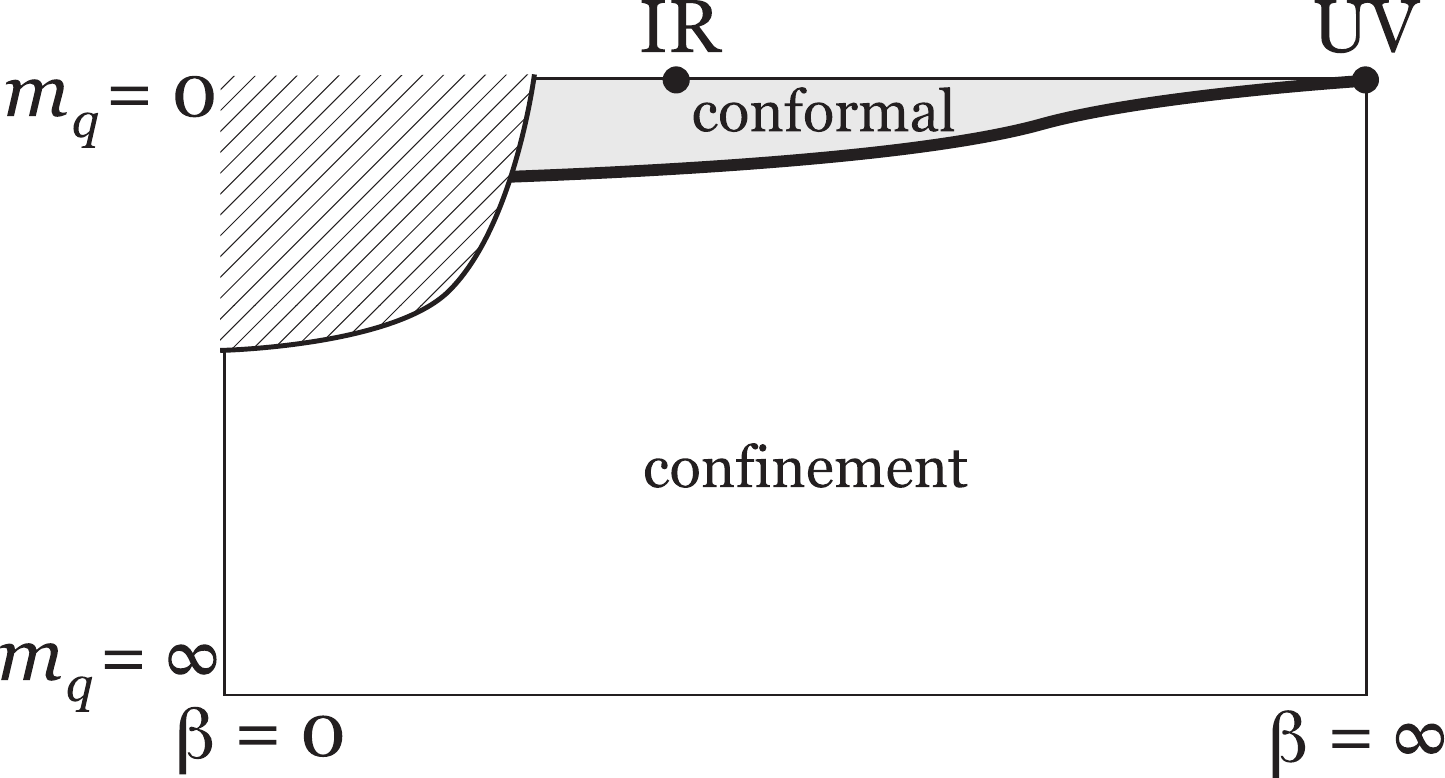}
\caption{ %\label{fig:epsart} 
The phase diagram for the case $N_f^c \le N_f \le 16$ predicted from the RG argument: (left: for $\Lambda_{\mathrm{IR}}=0$) and (right: for $\Lambda_{\mathrm{IR}}=$ finite).
The shaded strong coupling region for small quark masses does not exist in the $\beta - m_q$ plane {\cite{iwa2012}},
because the region corresponds to Wilson doublers  when mapped in terms of $\beta - K$.}
\label{phase diagram infinity lattice}
\end{figure*}

\begin{figure}[bht]
\includegraphics[width=7.5cm]{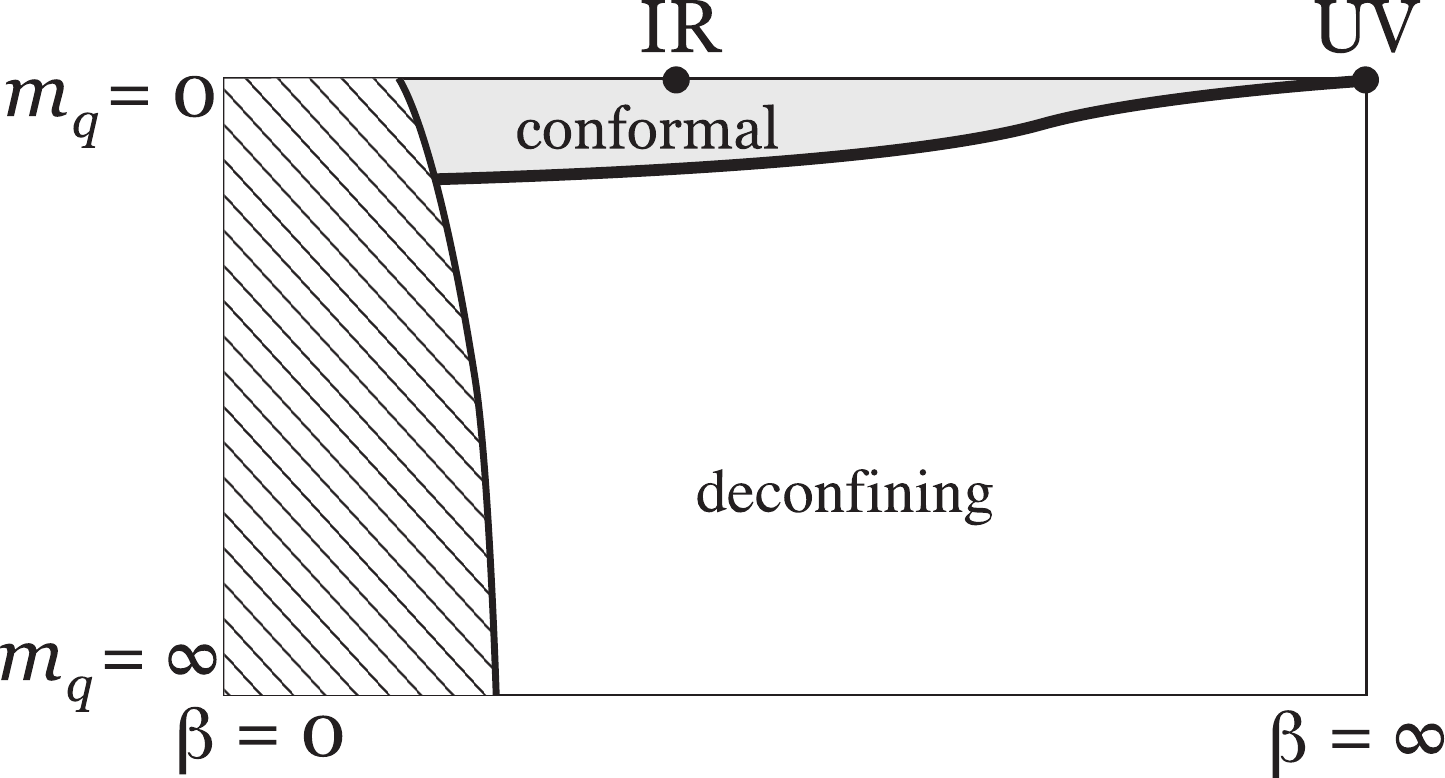}
\caption{%\label{fig:epsart} 
The phase diagram predicted from the RG argument for fixed temperature $T$ with $T/T_c > 1$ in the case $1 \le N_f \le N_f^c -1$ : 
The shaded strong coupling region does not correspond to the temperature $T/T_c >1$.}
\label{phase diagram infinity lattice deconfine}
\end{figure}

In this article we propose a new concept of ``conformal region'', which will be discussed in detail in later sections,
in particular, in Sec.~\ref{sec:conformal_region}.
Analyzing the vacuum structure in terms of the Polyakov loops in spatial directions and the specific behavior of the temporal propagators of meson,
we will show that there exists the conformal region in addition to the confining region and the deconfining region,
as shown in Fig.~\ref{phase diagram finite lattice}.
On a finite lattice, both in the cases $1\le N_f\le N_f^c -1$ and $N_f^c \le N_f \le 16$, when the bare coupling constant
$g_0$ is small enough, and when the quark mass $m_q$ is larger than the critical mass, the system is in the deconfining region. On the other hand, when $g_0$ is larger, it is in the confining region.

In the continuum limit, we will argue that in the case $N_f^c \le N_f \le 16$ only the confining region outside of the
conformal  region remains as shown in Fig,\ref{phase diagram infinity lattice}: (right panel) when the infrared cutoff 
$\Lambda_{\mathrm{IR}}$ is finite and (left panel) when $\Lambda_{\mathrm{IR}}=0$.
In the case $1\le N_f\le N_f^c -1$, only the deconfining region outside of the conformal region remains,
as shown in Fig. ~\ref{phase diagram infinity lattice deconfine}.

\begin{figure*}[thb]
\includegraphics [width=5.0cm]{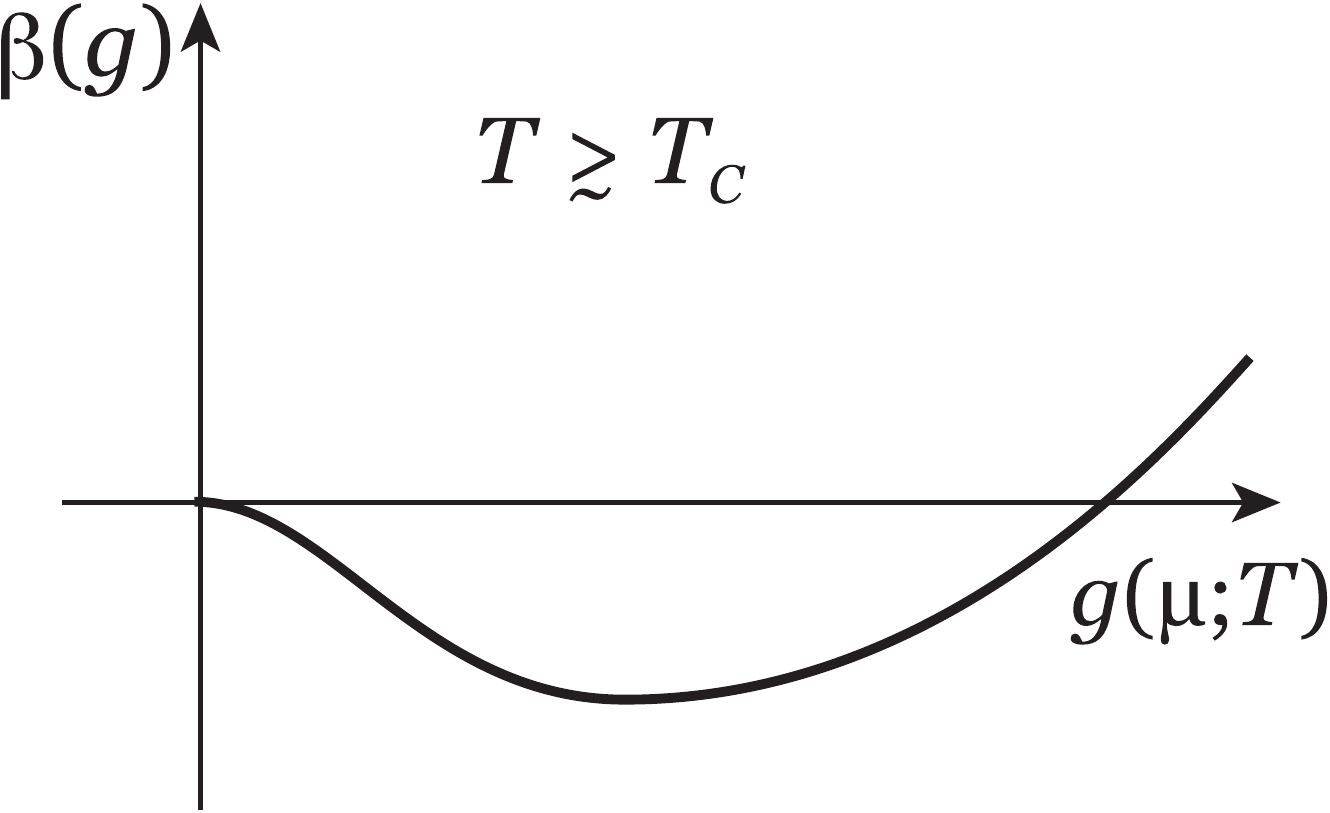}
\hspace{3.0cm}
%\includegraphics[width=5.0cm]{fig/beta_gt-2s.pdf}
%\hspace{1.0cm}
\includegraphics[width=5.0cm]{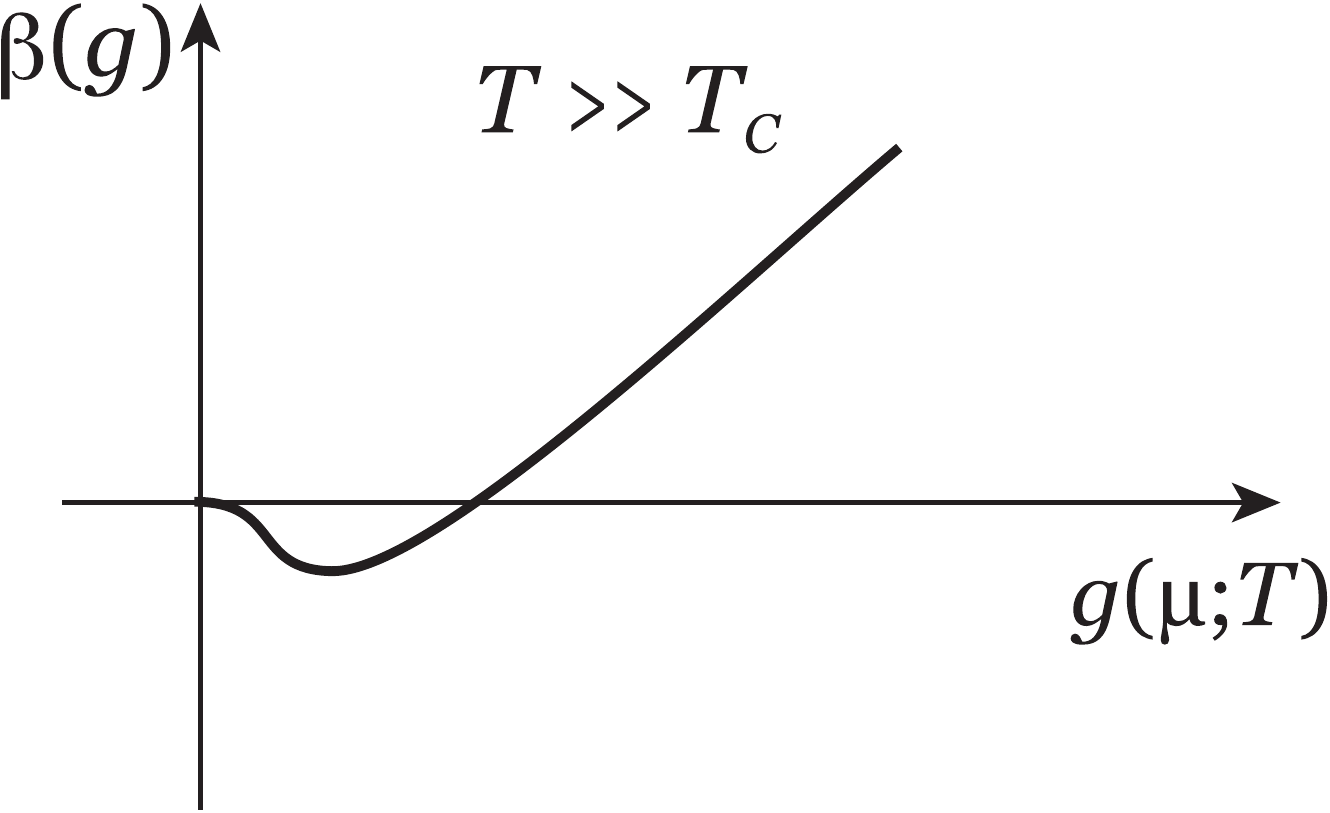}
\caption{%\label{fig:epsart} 
The beta function $\beta(g(\mu; T))$ at finite temperature $T$.
}
\label{beta_gT}
\end{figure*}

\subsection{``Phase transition" on a finite lattice}
As mentioned above, our final goal is to investigate conformal theories in the continuum limit.
However, we have to restrict ourselves to the calculations on a lattice with fixed size in this article.

The phase transition occurs only in the system with infinite degrees of freedom.
On a finite lattice, when physical quantities exhibit a ``discontinuous gap" (or more precisely a sharp 
transition) at some point and 
when theoretical argument supports the existence of the phase transition,
we identify the transition as a first order transition in the continuum limit.
Strictly speaking, all the ``discontinuities" we discuss only exist in the continuum limit,
and our numerical simulation on a fixed lattice gives only an indication of the discontinuity
as a sharp transition. With this respect, without taking the continuum limit, we cannot exclude the possibility that 
the ``phase transition" we propose may result in the crossover.

For a second order transition we have to carefully apply a scaling law in order to judge the existence of the phase transition.
The same remark for the discontinuity applies here as well.

\section{Conformal theories with an IR cutoff}
We have recently investigated field theories which posses an IR fixed point with an IR cutoff,
and introduced the nomenclature ``conformal field theories with an IR cutoff"
reported in short reports (Refs.\cite{coll1} and \cite{coll2}).

\subsection{Definition and Examples}
We first define the conformal field theories with an IR cutoff. The first assumption is that the beta function (either zero-temperature or finite temperature) vanishes. Of course, if there were no other dimensionful quantities, this would imply that the theory is scale invariant and all the correlation functions show a strict power behavior. In (perturbative) QCD at zero temperature, they will be further conformal invariant due to vanishing of trace anomaly (see Appendix B for more details).

Our new observation is that when such theories have a finite cutoff, then they will show the universal behavior that we call ``conformal field theories with an IR cutoff". In particular, we claim that within the suitable parameter region that we call ``conformal region", the temporal propagators show power-law corrected Yukawa-type decaying form. 
In the examples we study in this paper, the conformal field theories with an IR cutoff are realized as

\subsubsection{ Conformal QCD}
When the flavor number $N_f$ is within the conformal window $N_f^{c} \le N_f \le 16$,
the beta function possesses the Banks-Zaks IR fixed point.
The continuum limit  1, or 2, or 3. defines a theory with an IR cutoff.

\subsubsection{High Temperature QCD}
When the flavor number $N_f$ is exclusive with the conformal window,  $1\le N_f \le N_f^{c}-1$, and
$T \ge T_c$ with $T_c$ the chiral phase transition point,
the beta function of a running coupling constant $g(\mu; T)$  at temperature $T$ 
possesses an IR fixed point as shown in Ref~\cite{coll2}. (It is recapitulated in Appendix B.)
The temperature $T$ plays a role of the IR cutoff, together with a possible cutoff due to compact space.

As long as  $T < T_c,$ the beta function is negative all through $g$.
As the temperature is increased further, the form of the beta function will change
as in Fig.\ref{beta_gT}:  (left) When $T > T_c$ but $T \sim T_c$, the beta function changes the sign from negative to positive at large $g$;  As temperature increases the fixed point moves toward smaller $g$; (right) When $T \gg T_c$ it changes the sign at small $g$.

\subsubsection{Numerical simulations on a finite lattice}
All numerical simulations are performed on a finite lattice, which introduces an IR cutoff.
Therefore any lattice Conformal QCD 
($N_f$ is within the conformal window $N_f^{c} \le N_f \le 16$) and lattice High Temperature QCD
($N_f$ is $1\le N_f \le N_f^{c}-1$, and $T \ge T_c$ ), are
conformal field theories with an IR cutoff.

\subsection{Phase Transition from RG argument}

We have shown two examples of conformal field theories with an IR-cutoff: Conformal QCD and High temperature QCD. 
One of the main claims of the paper is that within the conformal region, there is a correspondence between a set of theories in each class. We now argue that when we increase the quark mass in each theories with the other parameters fixed (e.g. $N$, $\beta_0$ and $N_f$), they show the first order phase transition and leave the conformal region. Depending on the parameters, the region outside of the conformal region can be either ``confining region" or ``deconfining region" for a finite lattice simulation, as shown in Fig.~\ref{phase diagram finite lattice}.

\subsubsection{From conformal to confining}

Let us discuss the mechanism of the phase transition from the conformal region to the confining region from the argument based on the RG flow. For this purpose we quickly remind ourselves of the properties of the RG flow of the quark mass and the gauge coupling constant when the beta function posses an IR fixed point.

Suppose the IR cutoff is zero. When quarks have tiny masses,  the RG trajectory would stay close to the critical line, approaching the IR fixed point and finally would pass away from the IR fixed point to infinity. Therefore the IR behavior is governed by the ``confining region''. Only on the massless quark line the scale invariance is realized at the IR fixed point. See the left panel of Fig.~\ref{phase diagram infinity lattice}.

When the cutoff $\Lambda_{\mathrm{IR}}$ is finite, the RG flow from UV to IR does stop evolving at the scale $\Lambda_{\mathrm{IR}}$.
When the typical mass scale (e.g. that of a meson) $m_H$ is smaller than $\Lambda_{\mathrm{IR}}$, it is in the ``conformal region''.
On the other hand, when $m_H$ is larger than $\Lambda_{\mathrm{IR}}$, the flow passes  away from the IR fixed point to infinity  with relevant variables integrated out, thus being in the  ``confining region''. See the right panel of Fig.~\ref{phase diagram infinity lattice}.

This scenario implies that
when physical quantities in the IR limit (e.g. hadron masses) are mapped into a diagram in terms of physical parameters at UV (e.g. the bare coupling constant and the bare quark mass), there will be gaps in the physical quantities along the boundary between the two phases.
Thus the phase transition will be a first order transition.

\begin{table*}
\caption{Numerical results for $N_f=12$ at $\beta=6.0$: the meanings of the columns are the same as $N_f=16$}
\begin{tabular}{lrrllllllll}
\hline
\hline
 & \multicolumn{4}{c}{$N_f=12$}   & \multicolumn{2}{c}{$\beta=6.0$} &&& \\
\hline
$K$ & s& $N_{tra}$&  plaq & $m_q$ & $m_\pi$ & $m_V$ & $\tilde{m_{\pi}}$ &$\alpha_\pi$ &$\tilde{m_{V}}$ & $\alpha_V$\\
\hline
0.120  & l  &     500 &   0.616908(25)   &      0.9515(73)  &       1.632(12) & 0.1635(12) &-&-&-&-\\
0.125  & l   &    900 &   0.622451(14)   &      0.7035(71)  &       1.388(19) &  0.1390(21)&-&-&-&-\\
0.130 &  l   &    500 &   0.629336(94)   &      0.4944(98)  &       1.123((19) &  1.125(18)  &-&-&-&-\\
0.135 &   l  &     500 &   0.637354(48)  &       0.3184(27) &        0.8534(26) &  0.8574(20)&-&-&-&-\\
0.136 &   l  &     1000 &   0.639298(10)  &       0.2854(3)  &       0.7960(41) & 0.8004(45) &-&-&-&-\\
0.136 &   h &      1000 &   0.639307(11)  &       0.2850(1) &   -      &         -      &         0.781(4)   &    0.72(5)  &    0.769(3) & 0.75(5)\\
0.137 &   h  &     500 &  0.641257(21)  &       0.2521(47)  & -        &       -        &       0.687(10)     &  1.11(12)    &    0.688(10) & 1.16(11)\\
0.140 &   h  &     500 &  0.647566(24)  &       0.1576(17)  &  -       &        -       &        0.550(5)      &    0.83(7)    &     0.546(5)  & 1.00(11)\\
0.1425 &  h &      500 &  0.653517(15) &        0.0781(2)   &  -       &        -       &        0.371(10)     &   1.32(14)  &      0.364(0)  &  1.49(19)\\
0.144 &  h   &    500   &  0.657296(17) &       0.0304(22)   &  -      &         -      &         0.406(5)      &    0.62(11)  &      0.406(6) &  0.99(5)\\
\hline
\end{tabular}
\end{table*}

\begin{table*}
\caption{Numerical results for $N_f=12$ at $\beta=8.0$: the meanings of the columns are the same as $N_f=16$}
\begin{tabular}{lrrllllllll}
\hline
\hline
 & \multicolumn{4}{c}{$N_f=12$}   & \multicolumn{2}{c}{$\beta=8.0$} &&& \\
\hline
$K$ & s& $N_{tra}$&  plaq & $m_q$ & $m_\pi$ & $m_V$ & $\tilde{m_{\pi}}$ &$\alpha_\pi$ &$\tilde{m_{V}}$ & $\alpha_V$\\
\hline
0.120 &  l   &     600  & 0.730676(9)   &      0.5685(3)   &      1.0882(11)  &      1.0881(11) &-&-&-&-\\
0.125 &  l   &     600 &   0.733366(10)  &       0.3940(1)  &       0.8293(7) &       0.8297(6) &-&-&-&-\\
0.128 &  l   &     500 &   0.735300(9)  &       0.3002(1)  &       0.6680(5) &       0.6690(4) &-&-&-&-\\
0.129 & l    &    500 &   0.736007(10)   &      0.2705(1)   &      0.6186(22)  &      0.6207(23) &-&-&-&-\\
0.129 &  h  &      500 &  0.736011(16)  &       0.2784(2)  &       -                  & -        &   0.685(5)  &  1.27(7) &  0.686(6) & 1.27(8)\\
0.130 &  h   &     700 &  0.736759(9)  &       0.2485(2)    &      -                    & -      &     0.668(4) &   0.86(5) &  0.670(4) &  0.85(6)\\
0.133 &  h   &    1000 & 0.739240(14)  &       0.1598(2)    &      -                 	&  -       &    0.411(12) &  0.80(1)  &  0.413(11) &  1.01(18)\\
0.135  & h    &    800 &  0.741071(7)   &        0.1008(3)   &      -                    & -      &     0.389(24)  &  1.52(22) &   0.378(35) &  1.73(43)\\
0.138  & h &   600 &  0.741081(9)  &       0.01261(1)  &       -       &   -  &        0.381(6) & 1.06(8)&  0.370(15) & 1.33(22) \\
\hline
\end{tabular}
\end{table*}

\subsubsection{From conformal to deconfining}
A similar argument for the RG flow applies for the phase transition into the deconfining region.
Let us consider the case where the bare coupling constant $g_0$ is sufficiently small on a finite lattice
(See Fig.~\ref{phase diagram finite lattice}.).
When the typical mass scale $m_H$ is smaller than $\Lambda_{\mathrm{IR}}$, it is in the ``conformal region'' as before.
However, when $m_H$ is larger than $\Lambda_{\mathrm{IR}}$,
the RG flow passes  away from the IR fixed point to a point in the deconfining region.

The transition from the conformal region to the deconfining region is a first order 
when the lattice size is finite,
as in the case of the conformal region to the confining region.

Since deconfining region and conformal region are supposed to be in the same universality class in the infinite volume limit (see e.g. \cite{Alho:2012mh}), the ``phase transition" between the two will become weaker in the same limit.

\subsubsection{Confining or Deconfining in conformal QCD?}
With the finite lattice size, whether the theory is confining or deconfining by increasing the quark mass depends on the bare coupling $\beta_0$ and the lattice size as well as the number of flavor $N_f$.

In later sections \ref{sec:structure of vacuum}-\ref{sec:conformal_region} after presenting our numerical simulations, we argue that these  two different possibilities realized in finite lattice simulations may be a potential source of the controversy of the ``conformal behavior" of the intermediate ranges of the conformal window such as $N_f=12$, when one tried to study the  mass spectrum by including the mass values outside of the conformal region.

\begin{table*}
\caption{Numerical results for $N_f=2$:
The symbol a, b and c in the second column at $\beta=100.0$ and $K=0.1258$
means the state is close to $(1/3, 1/3, 1/3)$, $(0, 1/3, 1/3, 0)$ and $(0, 0, 1/3)$, respectively (see  subsection V-A for the definition of state).
Otherwise, the meanings of the columns are the same as $N_f=16$}
\begin{tabular}{lrrllllllll}
\hline
\hline
 & \multicolumn{4}{c}{$N_f=2$}   &&&&&& \\
\hline
$K$ & s& $N_{tra}$&  plaq & $m_q$ & $m_\pi$ & $m_V$ & $\tilde{m_{\pi}}$ &$\alpha_\pi$ &$\tilde{m_{V}}$ & $\alpha_V$\\
\hline
&&&&&&$\beta=5.9$&&&&\\
\hline
0.152 & l & 1000 & 0.602192(18) &   0.0332(1) &      0.3280(60) &     0.4492(59) &-&-&-&-\\
\hline
&&&&&&$\beta=6.5$&&&&\\
\hline
0.110 & l &  500 &   0.596366(17)  & 1.6932(23)  &    2.1372(15)  &    2.1398(15)  &-&-&-&-\\
0.145 & l   & 1000 &  0.648107(13)  &   0.0587(2)   &    0.4249(45)  &    0.4414(49) &-&-&-&-\\
0.1455 & l  & 600 &  0.648321(13) &   0.0465(3)  &     0.4112(56)  &    0.4194(70)  &-&-&-&-\\
0.146 & l & 1000 & 0.648546(14) &   0.0337(3)  &     -         &      -       &                0.371(9) &       0.71(8)     &    0.371(12)   &    0.98(14)\\
0.1465 & l & 1000 &  0.648799(14) &   0.0213(4) &      -       &        -      &                 0.286(19) &      0.73(19) &       0.279(14) &      1.08(26)\\
0.147 & l & 1000  &  0.649046(14)  &   0.0083(2) &      -       &        -      &                 0.295(16)  &     1.00(16) &       0.286(6)   &     1.41(20)\\
\hline
&&&&&&$\beta=7.0$&&&&\\
\hline
0.142 &  l &  700 &  0.678445(09)  & 0.0592(3)     &  -   &            -     &                  0.386(13)   &    0.74(15)    &     0.402(11)  &   0.66(10)\\
0.143  & l  &  500 &   0.678788(10) &  0.0333(4) &      - &              -   &                 0.360(13)  &     0.69(22)   &      0.356(10) &     0.94(23)\\
0.144 &  l  &  600 &  0.679108(16)  & 0.0074(2)    &   -    &           -      &                 0.354(14)     &    1.02(14)   &  0.320(14)  &   1.87(18)\\
\hline
&&&&&&$\beta=8.0$&&&&\\
\hline
0.139 & l &  700 & 0.725022(14)  &  0.0345(2) &      - &              -   &                    0.318(12)  &     0.97(14)   &     0.299(12)   &    1.41(21)\\
0.140 & l &  800 &  0.725140(91)  &  0.0084(1) &       - &              -   &                    0.376(7)  &     1.02(7)  &     0.403(6)  &     0.67(9)\\
\hline
&&&&&&$\beta=10.0$&&&&\\
\hline
0.110 &  l &  600 &  0.783954(05) &  0.8644(2)  &     1.3959(5)   &    1.3953(5) &-&-&-&-\\
0.125 & l &  600  &  0.784657(10) &  0.3046(1)  &     0.6518(16) &     0.6520(16) &-&-&-&-\\
0.130 &  l &  700 &   0.785016(08) &   0.1626(1) &       0.3887(5) &       0.3907(7) &-&-&-&-\\
0.130 & h &  900 &   0.785036(11) &  0.1676(1)  &     -        &       -       &                0.495(11)   &    1.40(11)  &       0.498(10)  &      1.32(11)\\
0.135 &  l & 1000 &   0.785549(08) &  0.0280(2) &      -       &        -      &                 0.372(69)  &     1.11(6)  &       0.373(3)  &      1.14(10)\\
\hline
&&&&&&$\beta=15.0$&&&&\\
\hline
0.130 & l & 1000 &  0.860880(03) &  0.0455(1)  &     -   &            -    &                   0.385(55)  &     1.21(4)  &       0.3972(6)  &     1.00(9)\\
\hline
&&&&&&$\beta=100.0$&&&&\\
\hline
0.100 & l  & 1000 &   0.979878(01)  & 1.0054(1)   &    -   &            -   &                    1.466(1)  &      1.14(1)  &       1.467(1)   &     1.11(1)\\
0.120 &  l &  600 &  0.979884(01)  & 0.1860(1)    &   -     &          -      &                 0.519(1)    &    1.35(1)    &     0.510(4)     &   1.49(7)\\
0.122 &  l &  800 &  0.979885(01) &  0.1227(1)     &  -     &          -      &                 0.454(2)    &    1.25(1) &        0.447(4)    &    1.36(7)\\
0.125 &  l &  900 &  0.979888(01)  &  0.2741(1)    &   -    &           -     &                  0.389(10)   &  1.29(12)  &     0.415(5)  &     0.825(65)\\
0.1258 & a & 1000 &  0.979889(01) &   0.0016(1)  &     -  &             -   &                    0.373(7) &       1.39(6)  &       0.346(16) &      1.87(22)\\
0.1258 & b & 1000 &  0.979889(01)  &  0.0014(1)  &     -   &            -   &                    0.318(3)  &      0.79(2)  &       0.326(14) &      0.65(23)\\
0.1258  & c & 1000 &  0.979889(01) &   0.0012(1) &    - &              -   &                    0.224(3)  &      0.45(2)  &       0.238(10) &      0.24(16)\\
\hline
&&&&&&$\beta=1000.0$&&&&\\
\hline
0.125 & l &  800 &  0.9979990(01) & 0.0031(1) &      -  &             -     &                  0.396(4)  &      1.19(4) &        0.441(6)  &      0.37(4)\\
\hline
\end{tabular}
\end{table*}

\subsubsection{Continuum limit}
When the lattice size is finite,  the large $t$ behavior of the meson propagator $G(t)$ is the exponential type
both in the confining region and the deconfining region. More strictly, in the deconfining region the exponential 
decay form is an approximate form due to the finiteness of the $t$ range.

In the continuum limit,
in the case of the confining region with $T=0$, the asymptotic behavior is exactly exponential with the mass of the ground state, while 
in the case of the deconfining region with $T$ finite, 
the spectral decomposition of $G(t)$ is necessary to accommodate the exact asymptotic behavior.

When the continuum limit is taken in the case where the $N_f$ is within conformal region $N_f^c \le N_f \le 16$, the confining region for larger quark masses enlarges and the deconfining region finally disappears with $N \rightarrow \infty$ (See Fig.2).

In the case $1\le N_f \le N_f^c -1$, when $g_0$ is larger than the critical coupling constant, there is no conformal
region and  only the confining region dominates in the continuum limit.
On the other hand, when $g_0$ is smaller (that is, $T/T_c$ kept larger than unity), only the deconfining region remains for larger quark masses (See Fig.3).

In our discussions, we have  considered the simplified RG flow in which only the gauge coupling constant and the quark mass are the relevant parameters.
As long as we are in the perturbative regime, this is completely justified. 
In the non-perturbative regime, there is a theoretical possibility that perturbatively irrelevant operators become relevant in the IR,
changing the renormalization group flow. If this were the case, the lattice simulations of conformal window would become 
 much harder because we have to tune these extra parameters (e.g. $O(a)$ lattice action) to reach the fixed point. 
 As far as our numerical analysis with the fixed lattice size suggests, this does not seem to happen. 
 See also the functional RG group analysis of the extended RG flow at zero temperature and at finite temperature in \cite{Braun:2005uj}\cite{Braun:2006jd}.
 Up to non-universal scheme dependence, our RG argument in relation to how the confinement/deconfinement
 occurs does not contradict with their analysis.

\begin{figure*}[htb]%
\includegraphics[width=7.8cm]{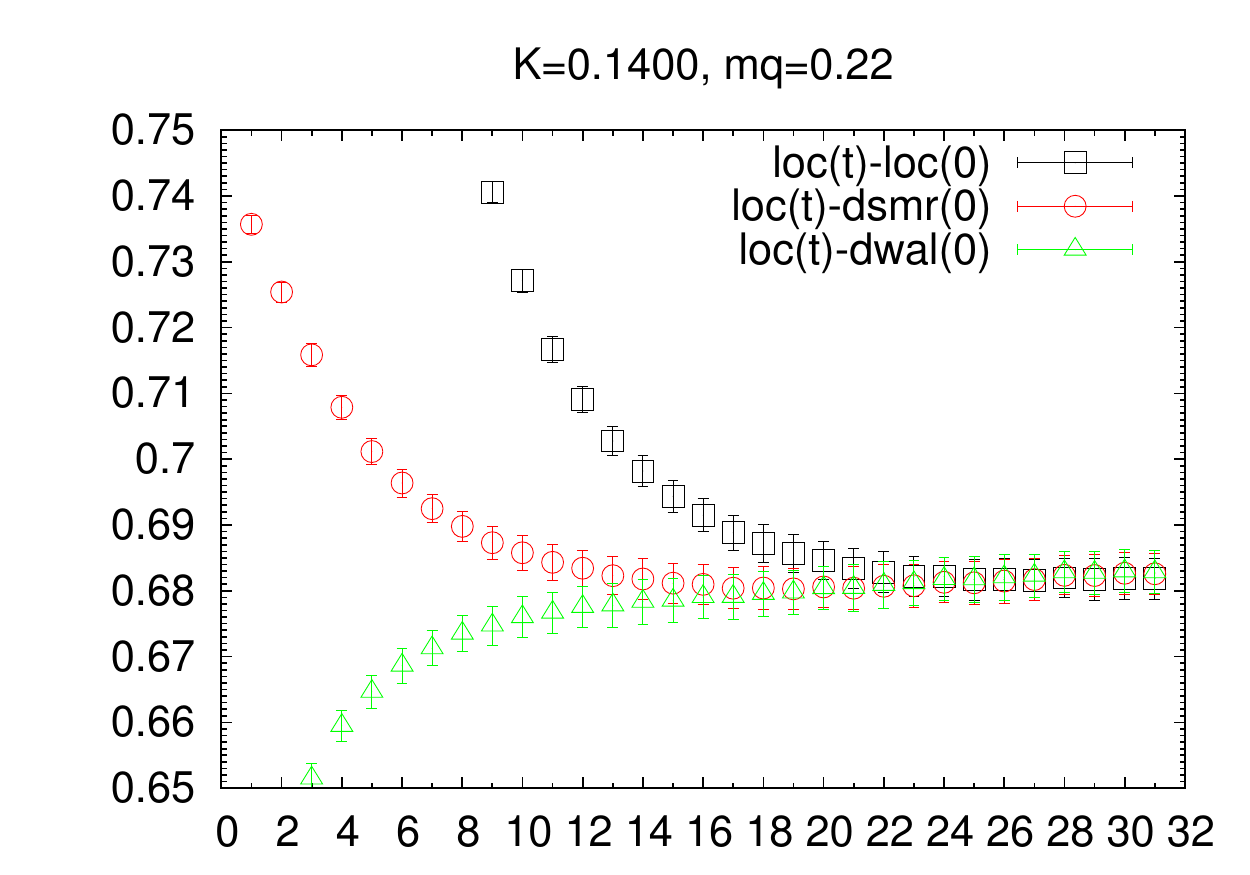}%\figsubcap{a}
\hspace{1.5cm}
\includegraphics [width=7.5cm]{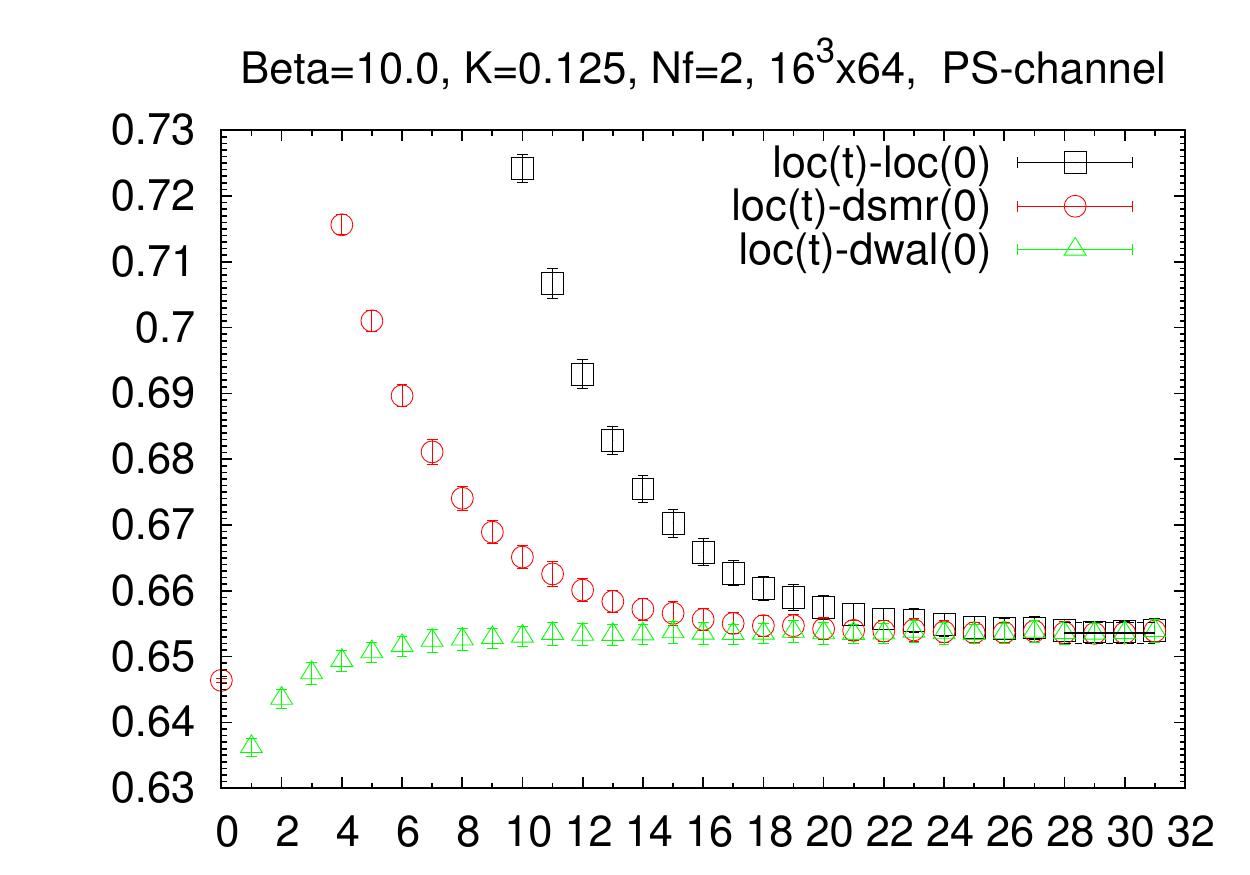}
  \caption{(color online)  The effective mass: (left) $N_f=7$ at $\beta=6.0$ and $K=0.1400$  and (right) $N_f=2$ at $\beta = 10.0$ and $K=0.125$;
See the text for the three types of sources.}
  \label{exp-decay}
\end{figure*}

\begin{figure*}[bht]%
  \includegraphics[width=7.8cm]{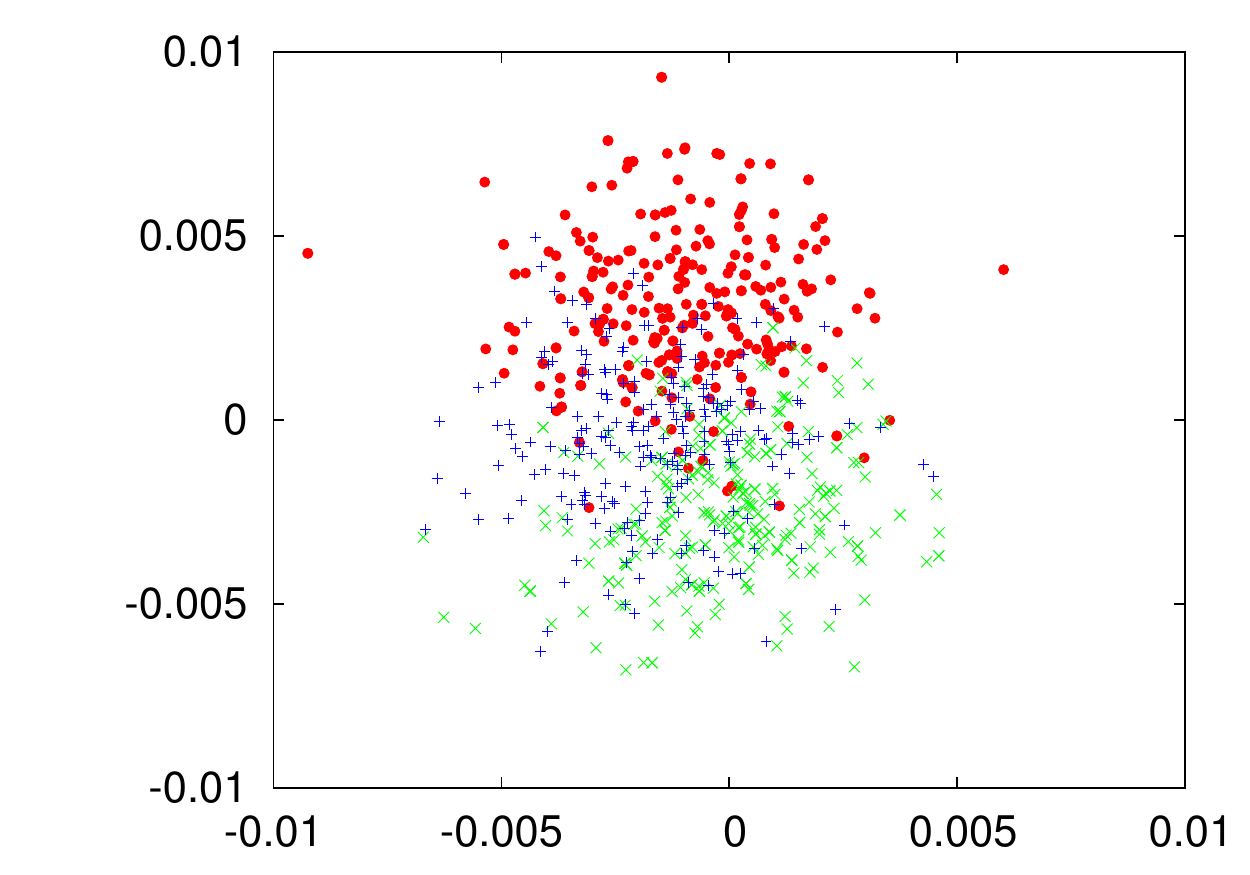}
  \hspace{1.5cm}
    \includegraphics[width=7.8cm]{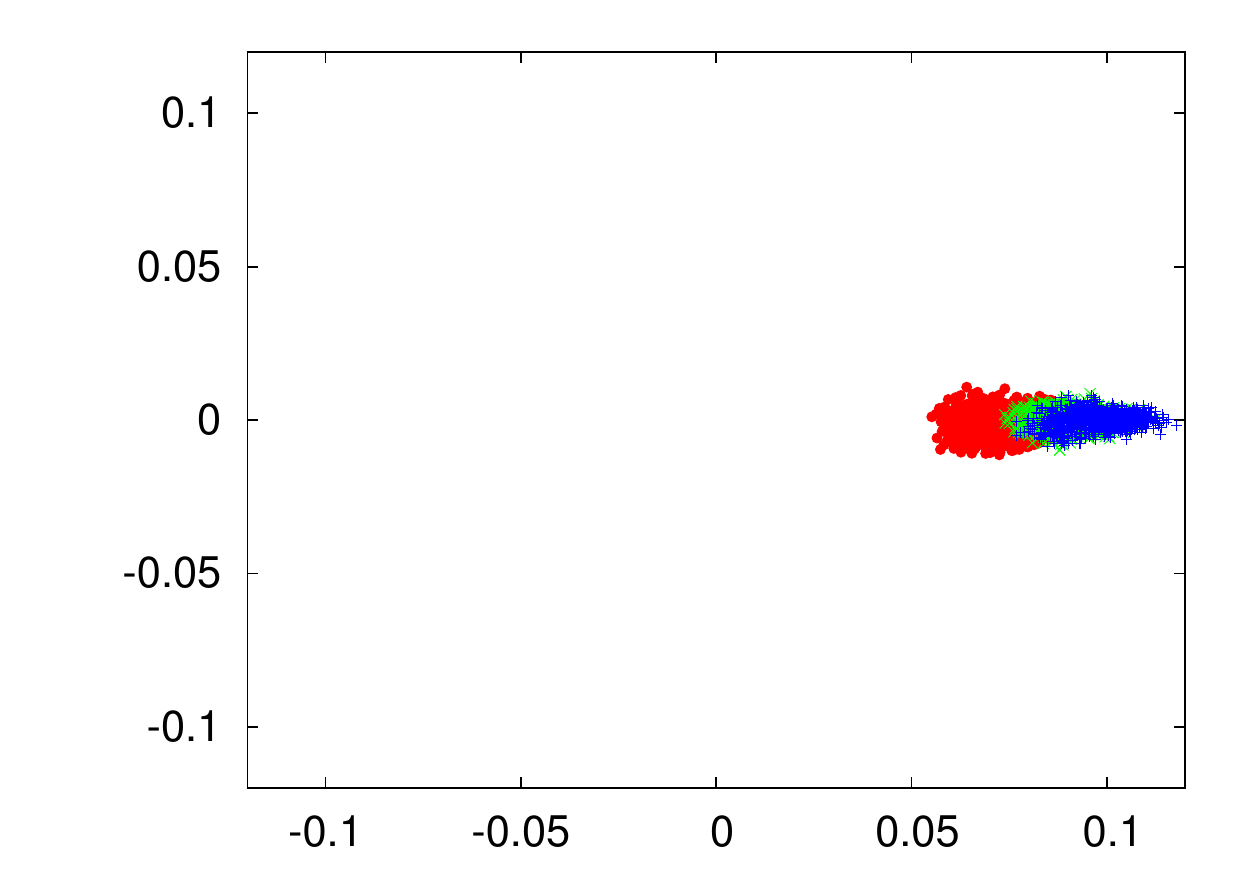}
        \label{expnf7nf16}
        \caption{(color online)   
        The scattered plots of Polyakov loops in the $x$, $y$ and $z$ directions overlaid;  (left) $N_f=7$ at $\beta=6.0$ and $K=0.1400$  and (right) $N_f=2$ at $\beta = 10.0$ and $K=0.125$.}
\label{polyakov-1}
\end{figure*}

\subsection{In relation to mass deformed CFT}

As we have already emphasized, in the continuum limit without an IR cut-off, the conformal region only exists
on the strict massless line $m_q= 0$. Once we have any non-zero quark mass, the theory is in the confining phase. 
The first order phase transition line we have proposed between confining region and conformal region
becomes coincident with the massless line. The phase transition clearly occurs at $m_q = 0$ for the 
conformal QCD in the continuum limit without an IR cut-off.
While the free energy becomes continuous there because the energy gap behaves as $1/L$, some other physical observables may become discontinuous. 

In this continuum limit, the confining region with the tiny quark mass has been known as the mass
deformed conformal field theory and has been studied intensively in the literature (see e.g. \cite{DelDebbio:2010ze}
\cite{DelDebbio:2010jy}). 
As long as the renormalization group flow stays for a
 sufficiently long time close to the fixed point, which requires that mass is significantly smaller compared
  with the UV cut-off or any other energy scale, various physical observables 
in the mass deformed conformal field theory will show the scaling behavior. 
Here the quark mass serves as the effective IR cut-off in the fermion sector \cite{Fodor:2012ty}\cite{Fodor:2012et}, 
and the competition with the intrinsic IR cut-off from the finite lattice size is our main focus.
Approaching the massless line makes the correlation length divergent, and the critical exponent is determined from the properties of the conformal fixed point.
However it is not obvious 
  if this criterion has been really achieved in the finite lattice simulations. 

The fate of the mass deformed conformal field theory under the presence of a finite cut-off is two-fold. It could be either our conformal
 region (when the mass is small enough), or confining region (when the mass is larger). 
In the literature, guided by the expected scaling behavior without the IR cut-off
in the mass deformed conformal field theory, the simulations have
 been mainly aimed at the confining region.
In contrast, our main focus in the following is the conformal region which is directly connected to the 
conformal field theory in the continuum limit on the massless line. 
Note again that the conformal region with non-zero mass exists whenever the IR cut-off is non-zero.
As we will discuss in the following sections,
 this will enable us to continuously connect the propagators, in principle, to the massless and continuum limit 
without encountering the phase transition. Indeed, we will see the remnant of the power-law
behavior in the propagators in the finite lattice simulations in the conformal region, which is not visible in the confining region.

One important remark is in order.
In the above paragraphs, we have started with the mass deformed conformal field theory defined in the continuum limit with no IR cut-off. With the finite IR cut-off, we have already discussed in section II and III, there exists yet another possible phase, the deconfining region. The properties of this region are remotely distinguished from the mass deformed conformal field theory, and we should not be able to test the prediction of the mass deformed conformal field theory in the deconfining region. 
The region remains in the continuum limit if we keep the temperature finite, but as we have already mentioned, it should go away in the conformal QCD at zero temperature. 

We will see that in the finite size lattice simulations, it will depend on the details of the simulation parameters i.e. coupling constant or lattice size whether the confining region or deconfining region will appear above the phase transition line from the conformal region. This will be elaborated in section \ref{12} in the example of $N_f =12$.
A further remark on the finite size scaling will appear in section \ref{finite}.

\section{Analysis of Propagators}
\subsection{Long distance behavior of Propagators}
\label{long_distance}
Based on the above RG argument, we conjecture the long distance behavior 
for the propagator of the local meson operator 
$$
G_H(t) = \sum_{x} \langle \bar{\psi}\gamma_H \psi(x,t) \bar{\psi} \gamma_H \psi(0) \rangle \ 
$$
qualitatively differs depending  on whether the quark mass is smaller than the critical mass or not.

When the theory is in the relatively heavy quark region, it decays exponentially at large $t$ as
\begin{equation} G_H(t) = c_H \, \exp(-m_H t)\label{exp} \end{equation}.

 \begin{figure*}[htb]
\includegraphics[width=7.8cm]{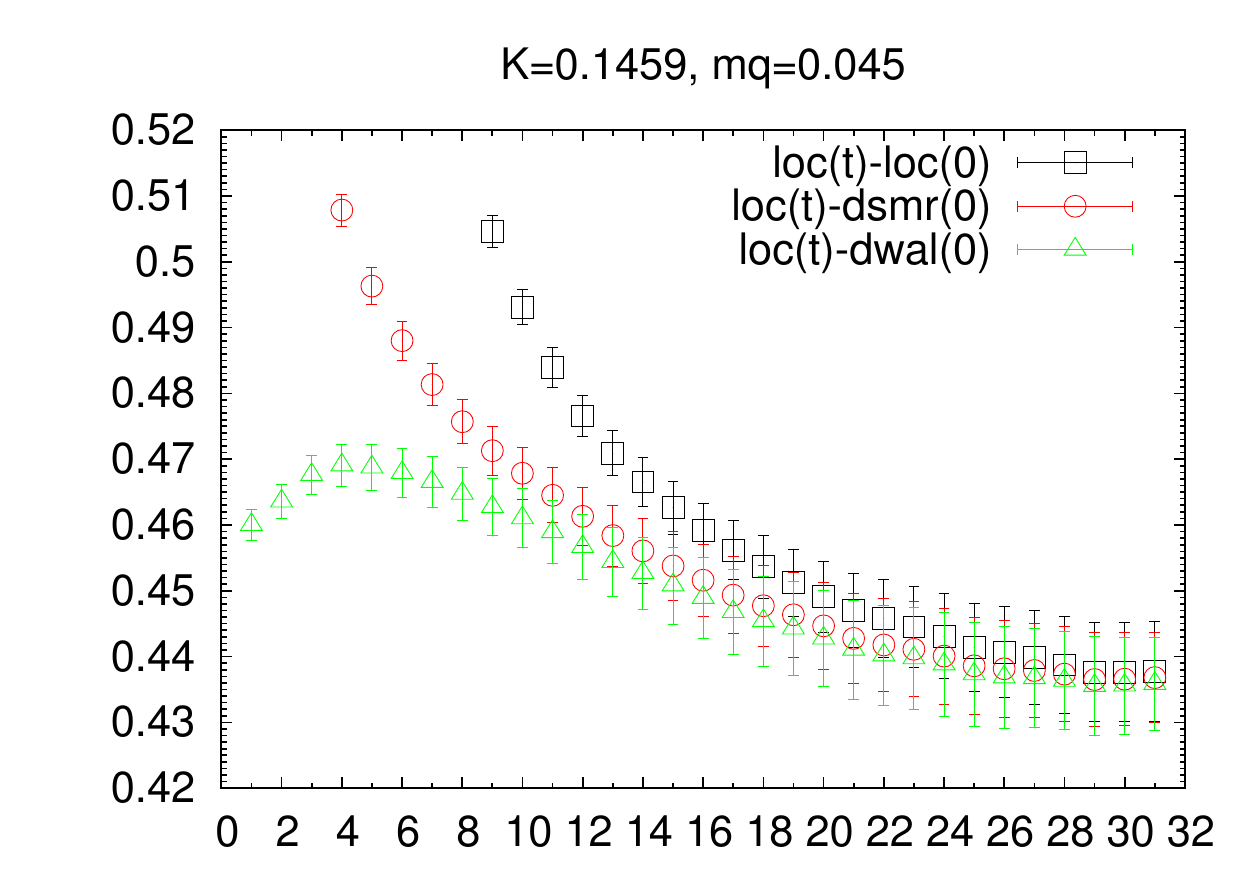}
\hspace{1cm}
\includegraphics [width=7.5cm]{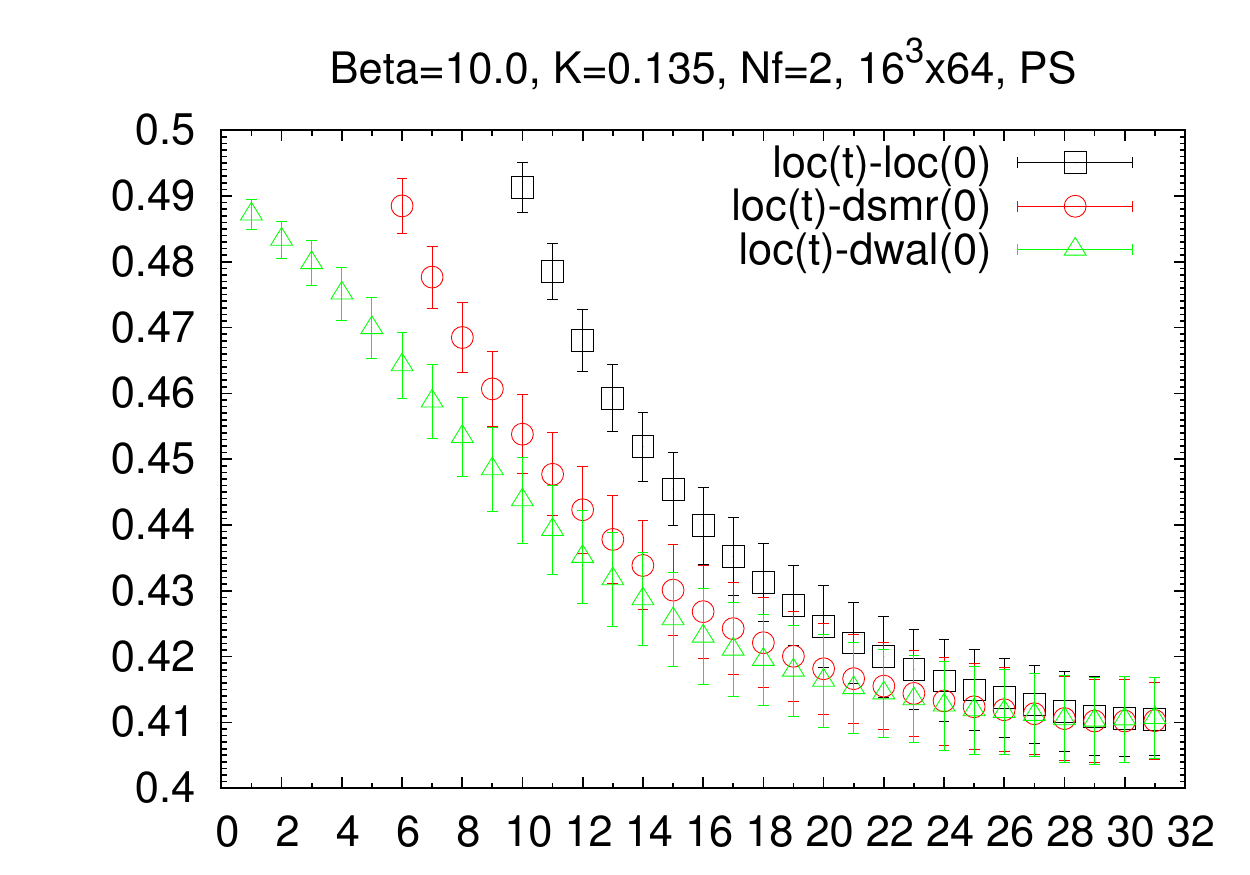}
\caption{(color online)  %\label{fig:eps art} 
The effective mass: (left) $N_f=7$ at $\beta=6.0$ and $K=0.1459$  and (right) $N_f=2$ at $\beta = 10.0$ and $K=0.135$;
See the text for the three types of sources.}
\label{yukawa-decay-1}
\end{figure*}
 
 \begin{figure*}[bht]
\includegraphics[width=7.8cm]{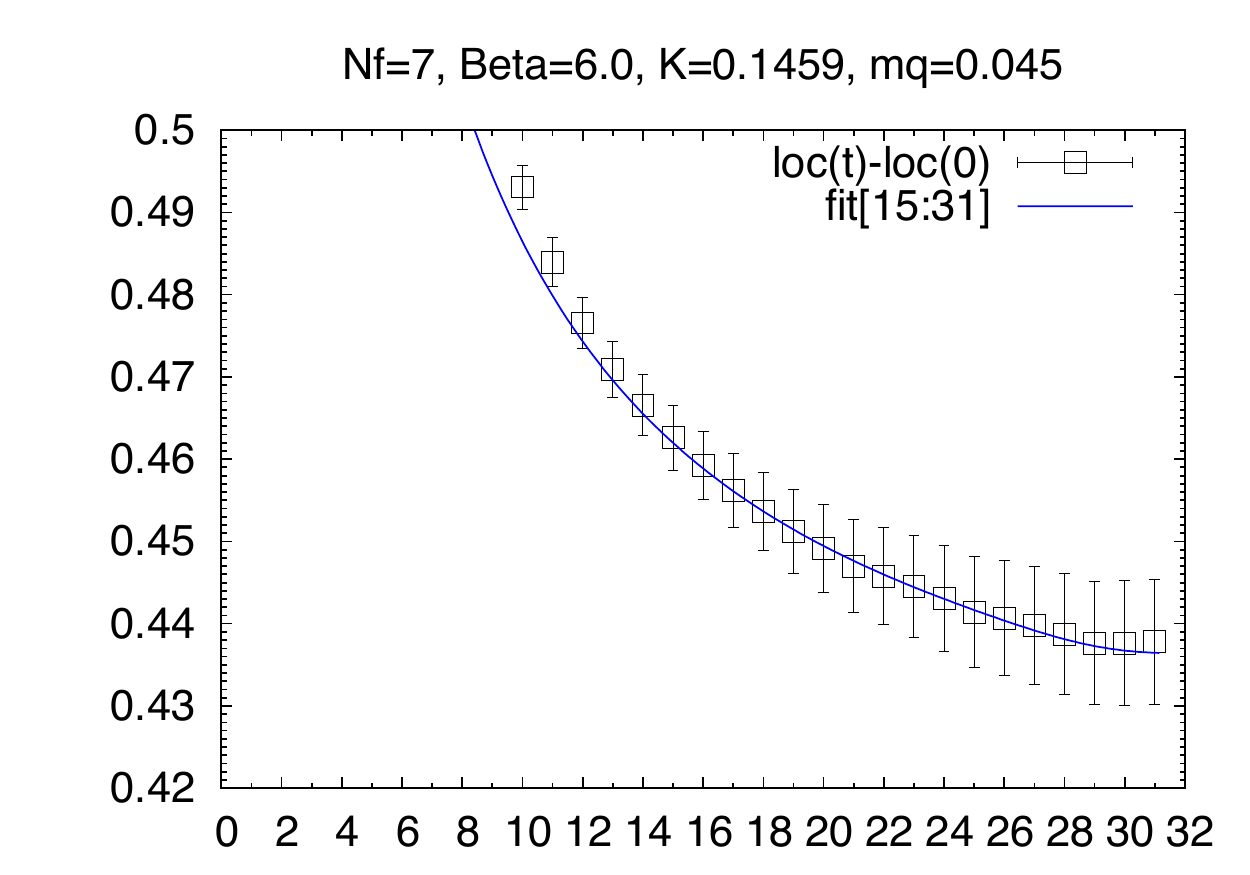}
\hspace{1cm}
\includegraphics [width=7.5cm]{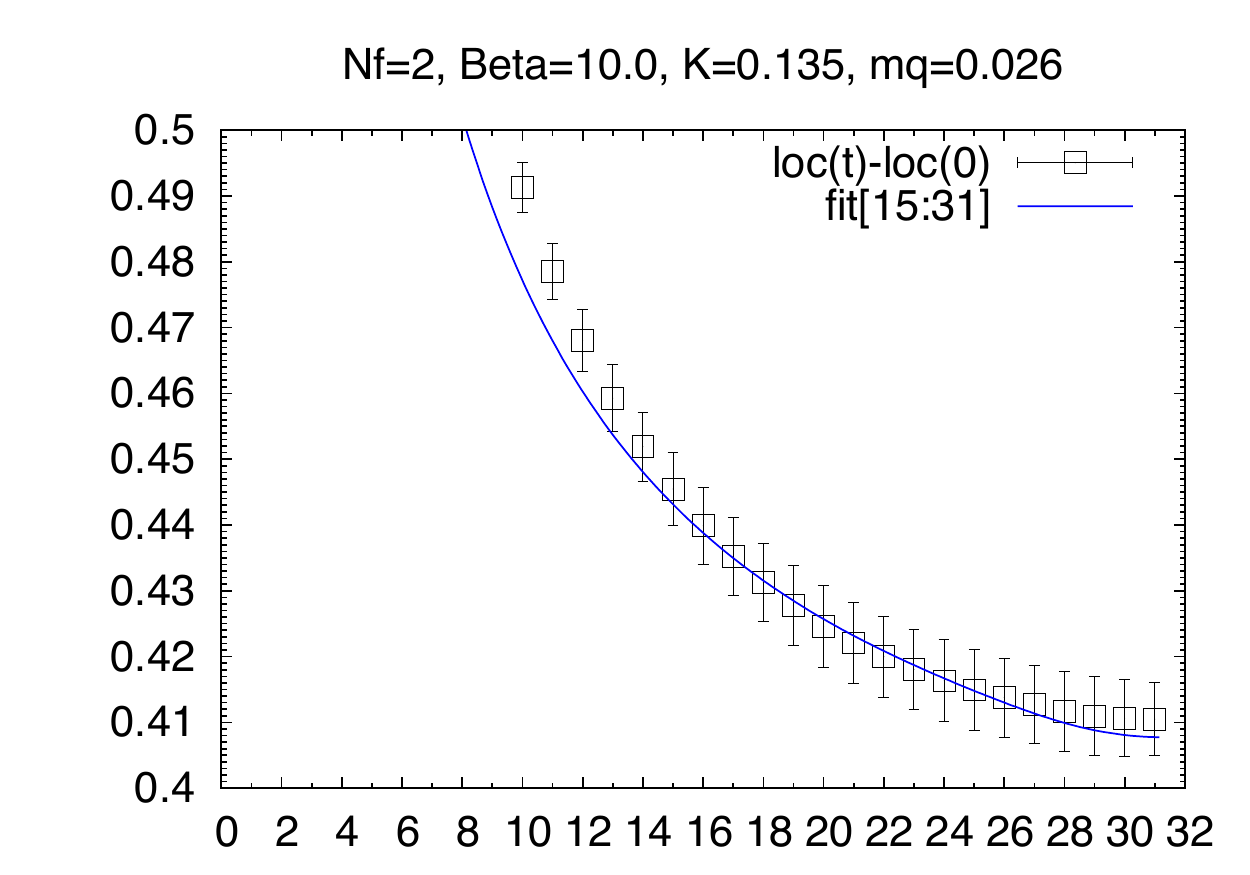}
\caption{(color online)  %\label{fig:eps art} 
The effective mass plots for local-sink local-source case and fits by power-law corrected Yukawa type decay;
(left) $N_f=7$ at $\beta=6.0$ and $K=0.1459$  and (right) $N_f=2$ at $\beta = 10.0$ and $K=0.135$.}
\label{yukawa-fit}
\end{figure*}

\begin{figure*}[bht]%
  \includegraphics[width=7.8cm]{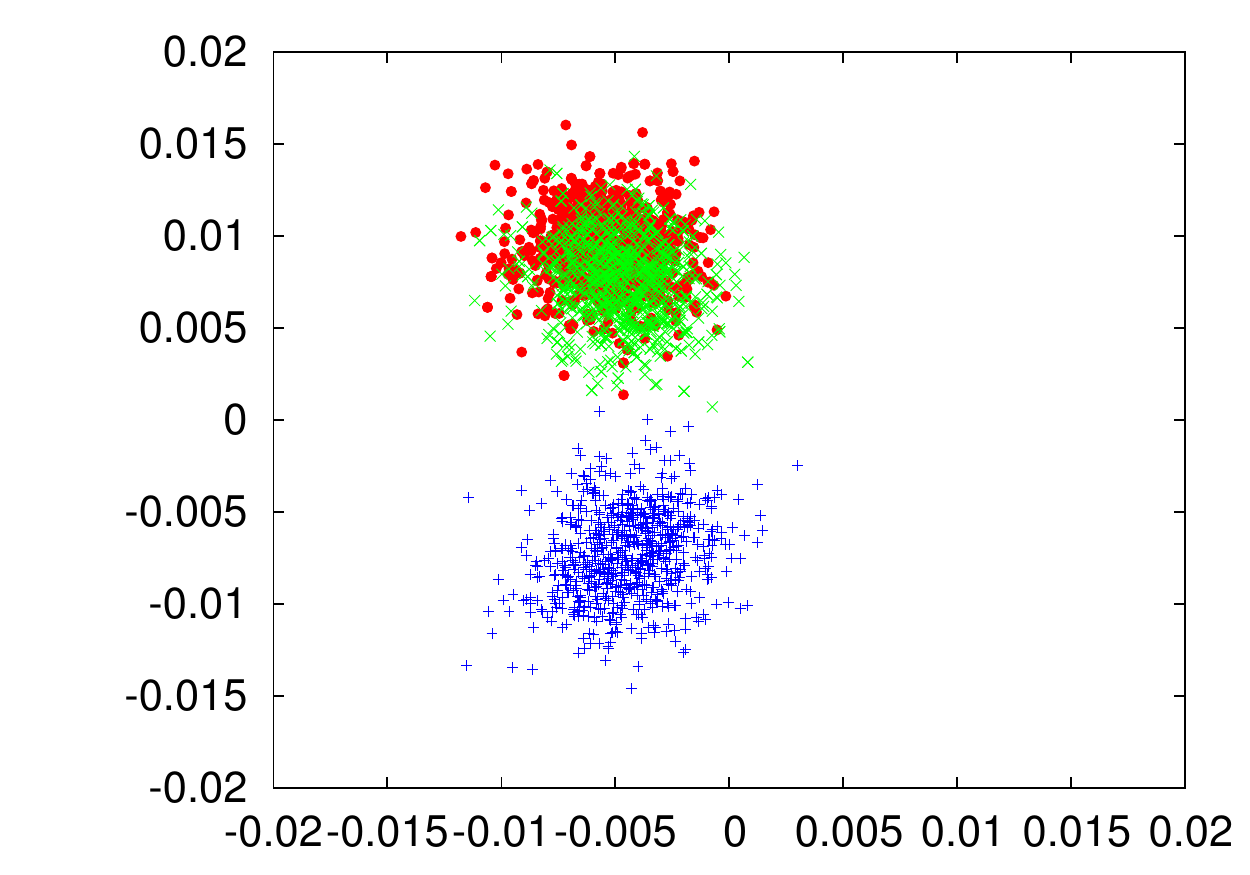}
    \hspace{1.5cm}
    \includegraphics[width=7.8cm]{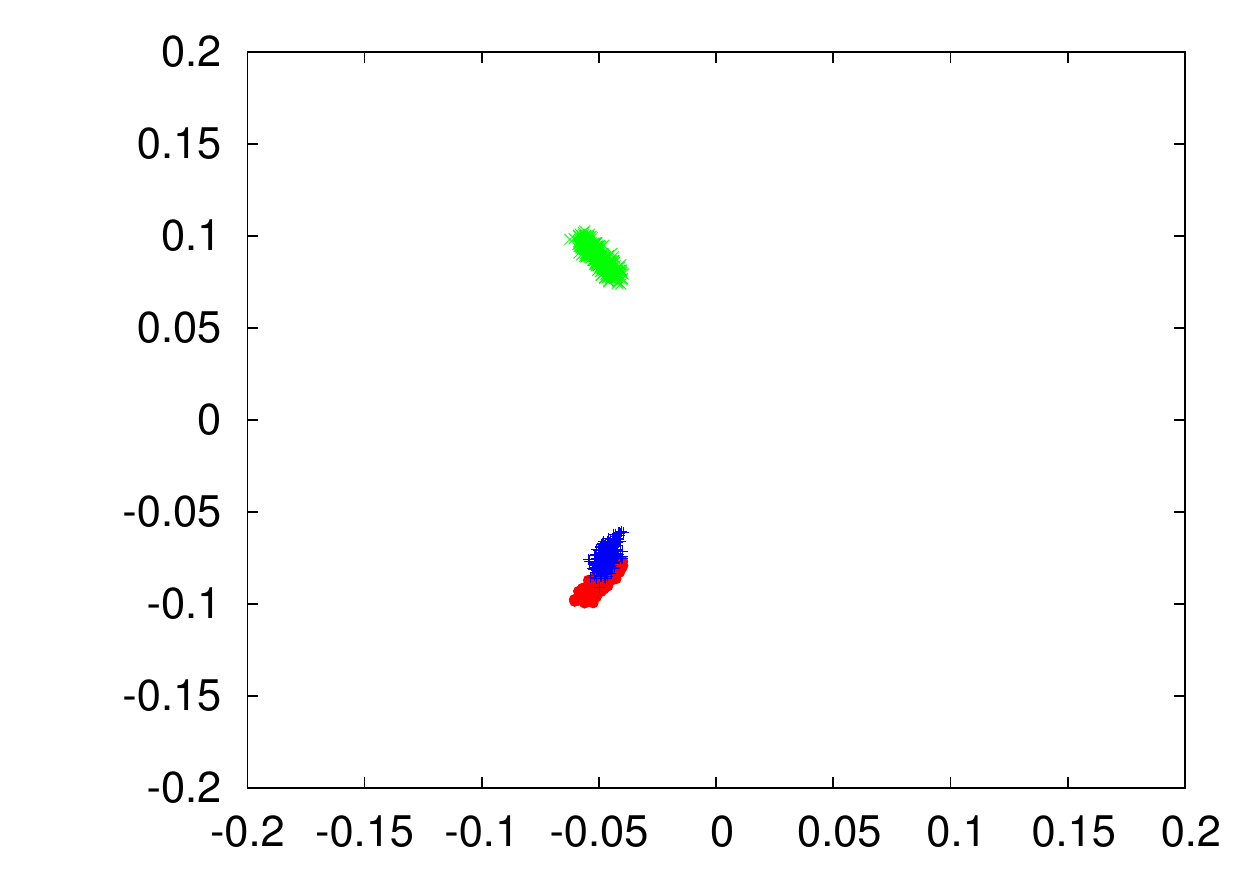}
                \label{complexnf7nf16}
        \caption{(color online)   
         The scattered plots of Polyakov loops in the $x$, $y$ and $z$ directions overlaid; 
          (left) $N_f=7$ at $\beta=6.0$ and $K=0.1459$  and (right) $N_f=2$ at $\beta = 10.0$ and $K=0.135$.}
         \label{polyakov-2}
\end{figure*}

\begin{figure*}[htb]
\includegraphics [width=7.0cm]{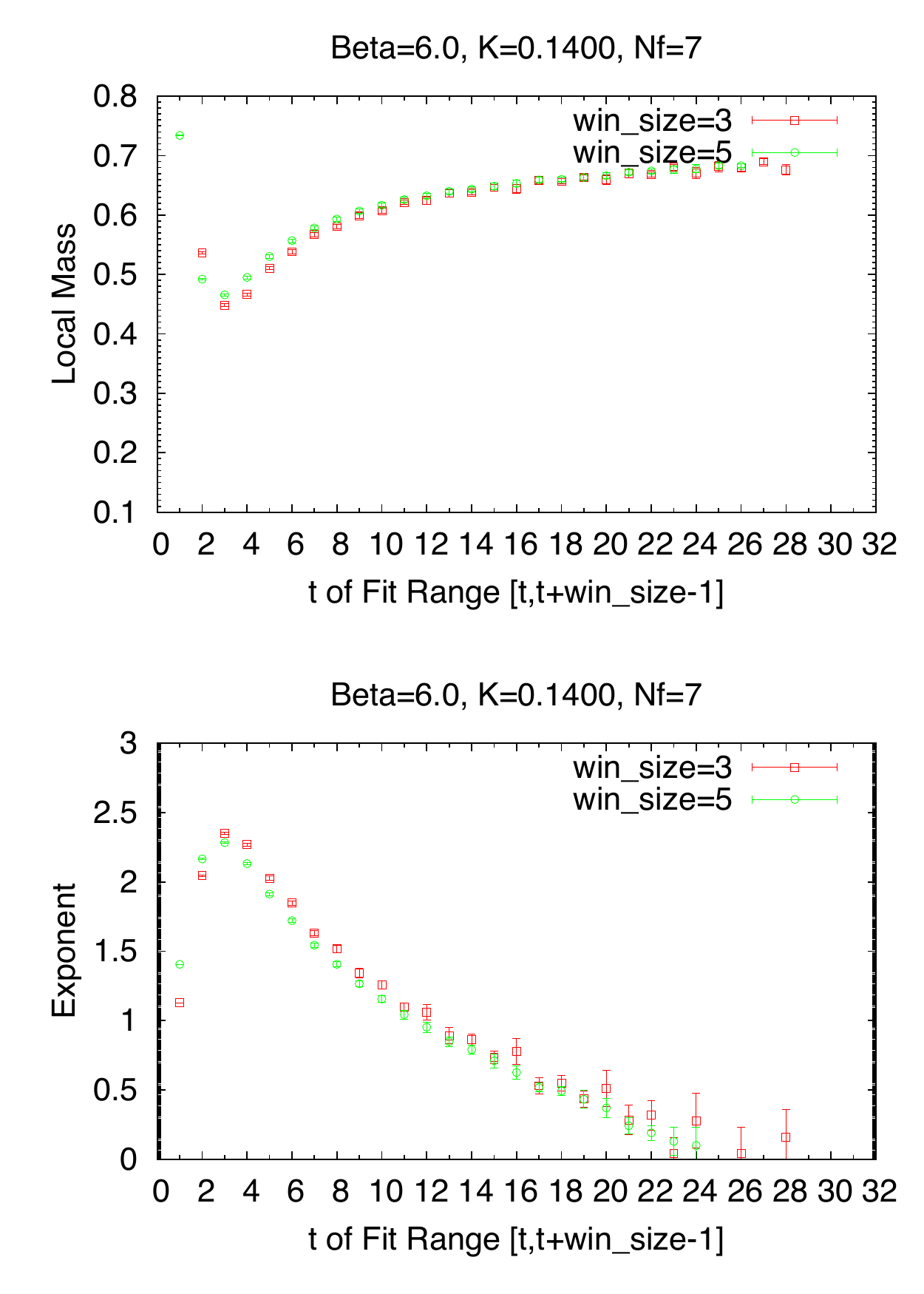}
%\caption{(color online)  The local mass $m(t)$ and local exponent $\alpha(t)$ for $N_f=7$ at $\beta=6.0$ and $K=0.1400$.}
%\label{nf7k1400}
%\end{figure*}
%
\hspace{1.5cm}
%\begin{figure*}[htb]
\includegraphics [width=7.0cm]{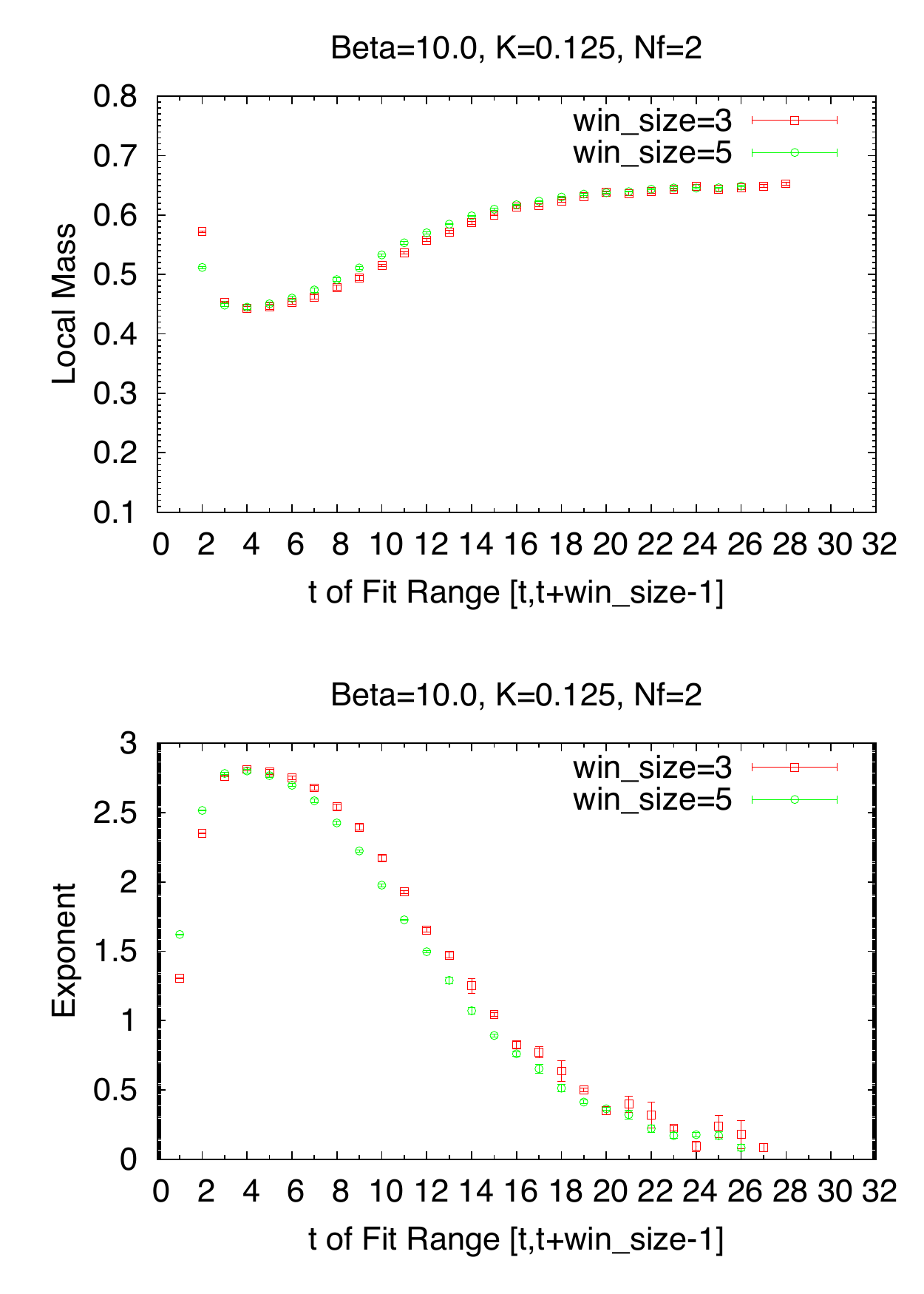}
\caption{(color online)  The local mass $m(t)$ and local exponent $\alpha(t)$:
(left) $N_f=7$ at $\beta=6.0$ and $K=0.1400$; (right) $N_f=2$ at $\beta=10.0$ and $K=0.125$.}
\label{nf2b10.0k125}
\end{figure*}

In contrast, in the ``conformal region" 
 defined by
\begin{equation} 
m_H   \leq c \,  \Lambda_{\mathrm{IR}},
\label{critical mass}
\end{equation}
where $c$ is a constant of order 1, the propagator $G(t)$ behaves at large $t$  as
 \begin{equation}
G_H(t) = \tilde{c}_H\\ \frac {\exp(-\tilde{m}_Ht)}{t^{\alpha_H}},
\label{yukawa type}
 \end{equation}
which is a power-law corrected Yukawa-type decaying form instead of the exponential decaying form (Eq.(\ref{exp})). 

In the continuum limit we have to discuss Conformal QCD and High Temperature QCD separately.

When the theory is in the ``confining region" in Conformal QCD,  $m_H$ in Eq.~\ref{exp} is the mass of the ground state hadron in the channel $H$. 
In the continuum limit  with $L= \infty$ (i.e. $\Lambda_{\mathrm{IR}} = 0$), the propagator on the massless quark line takes the form
\begin{equation} G_H(t) = \tilde{c} \, \frac {1}{t^{\alpha_H}}.\label{massless}\end{equation}
%consistent with $\tilde{m}_H=0$ limit of Eq. (\ref{yukawa type}).
%since   there is no physical quantities with physical dimensions.
If we take the coupling constant $g_0=g^{*}$ at the UV cutoff, $\alpha_H$ takes a constant value, and the RG equation demands 
\begin{equation} \alpha_H=3 - 2 \gamma^{*},\label{anoma}\end{equation}
for the PS channel
with $\gamma^{*}$ being the anomalous mass dimension $\gamma$ at $g=g^{*}$.
The theory is scale invariant and shown to be conformal
invariant within perturbation theory \cite{Polchinski:1987dy}.
See e.g. \cite{Nakayama:2010zz}\cite{Nakayama:2013is} and references therein from AdS/CFT approach. The distinction between scale invariance and conformal invariance in lattice QCD was addressed in \cite{DelDebbio:2013qta}.
When  $0 \le g_0 <  g^*$, $\alpha_H$ depends slowly on $t$ as a solution of the  RG equation. In the IR limit $t\to \infty$, we must retain $\alpha_H(t) \to 3-2\gamma^{*}$.

\begin{figure*}[htb]
\includegraphics [width=7.5cm]{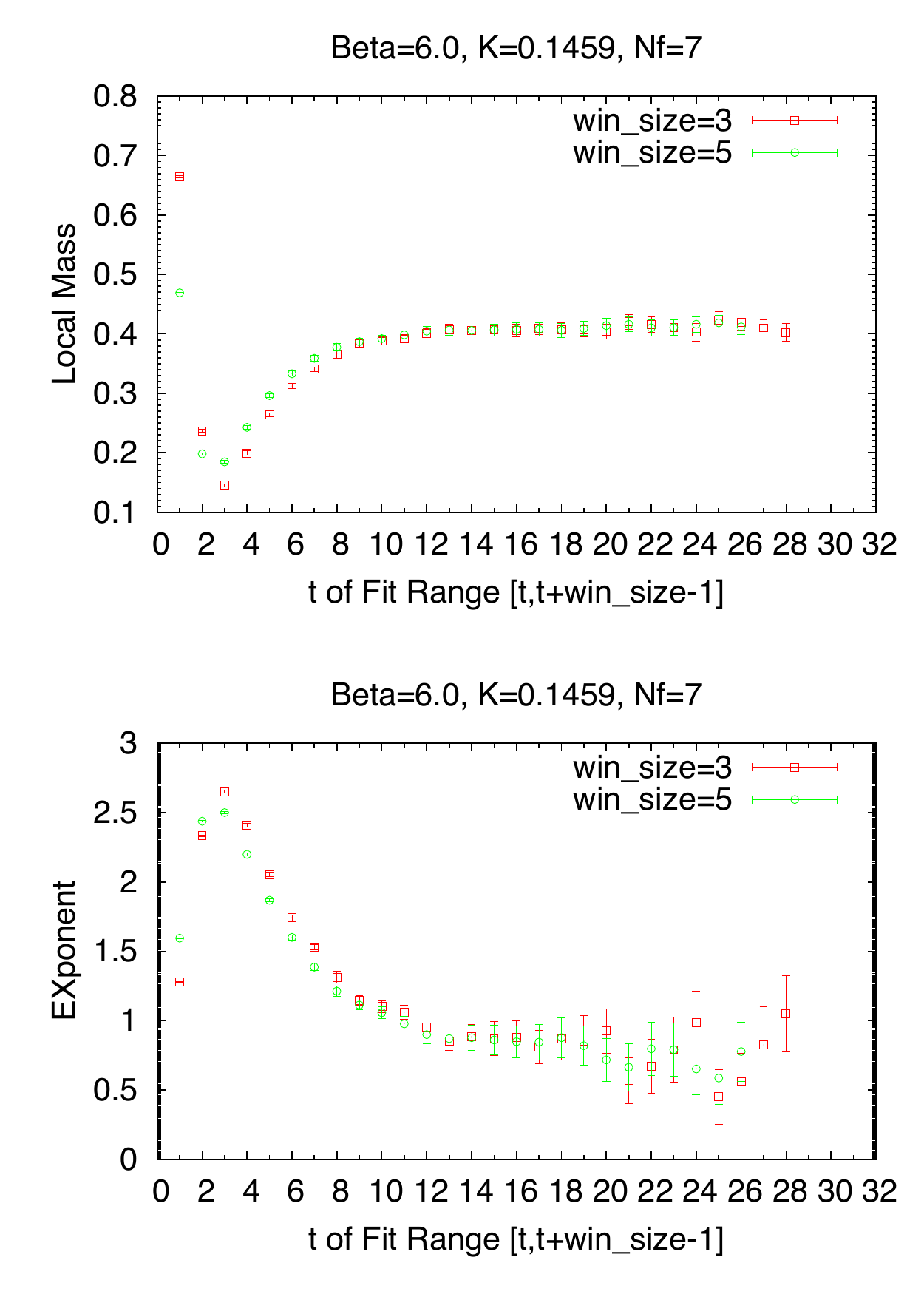}
%\caption{(color online) The local mass $m(t)$ and local exponent $\alpha(t)$ for $N_f=7$ at $\beta=6.0$ and $K=0.1459$.}
%\label{nf7k1459}
%\end{figure*}
%
\vspace{1.5cm}
%\begin{figure*}[bht]
\includegraphics [width=7.5cm]{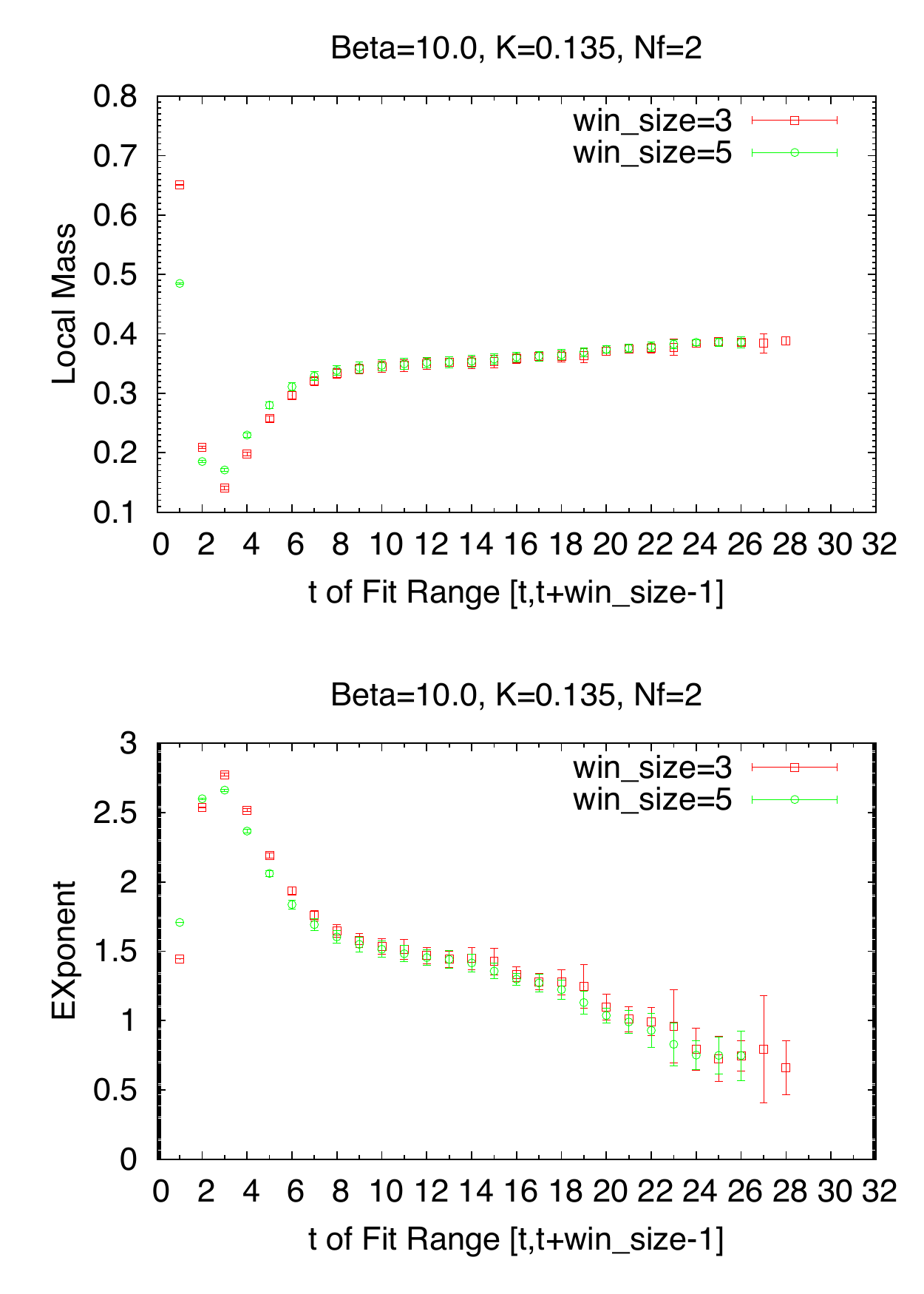}
\caption{(color online)  The local mass $m(t)$ and local exponent $\alpha(t)$:
(left) $N_f=7$ at $\beta=6.0$ and $K=0.1459$; (right) $N_f=2$ at $\beta=10.0$ and $K=0.135$.}
\label{nf2b10.0k135}
\end{figure*}

In the thermodynamical limit at finite temperature, eqs.~ (\ref{exp}) and (\ref{yukawa type}) are valid only approximately due to the finiteness of the $t$ range.
A more rigorous way to  obtain physical implication 
would be to make the spectral decomposition of $G_H(t)$.

However, our objective in this article is to verify the existence of the conformal region and the  power-law corrected Yukawa-type decaying form on a finite lattice in the case of conformal theories with an IR cutoff.
It is beyond the scope of our objective to obtain the thermodynamical physical quantities.

 The conjecture should be satisfied in
(1) large  $N_f$ QCD within the conformal window with an IR cutoff
and
(2)  small $N_f$ QCD at high temperature $T/T_c>1$ with $T_c$ being the chiral transition temperature.

In order to investigate the large $t$ behavior of a propagator,
we define the effective mass $m_H(t)$  by 
$$\frac{\cosh(m_H(t)(t-N_t/2))}{\cosh(m_H(t)(t+1-N_t/2))}=\frac{G_H(t)}{G_H(t+1)}.$$
In the case of exponential-type decay the effective mass approaches a constant in the large $t$ region which is called a plateau.

We show the $t$ dependence of the effective mass
in the PS channel with three types of sources
for the four examples (light quark mass and relatively heavy quark mass cases in each categories:
\begin{enumerate}
\item
\begin{itemize}
\item
$N_f=7, \beta =6.0, K=0.1400$ ($m_q=0.25$). 
\item
$N_f=7, \beta=6.0, K=0.1459$ ($m_q=0.045$).
\end{itemize}
\item
\begin{itemize}
\item
$N_f=2, \beta =10.0, K=0.125$ ($m_q=0.30$).
\item
$N_f=2, \beta=10.0, K=0.135$ ($m_q=0.028$).
\end{itemize}
\end{enumerate}

Three types of symbols represent three types of source-sink;  the local-sink local-source (squares)
local-sink (quark-anti-quark) doubly exponentially-smeared-source of a radius 5 lattice units  (circles) and 
local-sink doubly wall-source (triangles).

We show in Fig.~\ref{exp-decay},
typical examples of exponential decay
in two cases of  relatively heavy quark mass cases:
$N_f=7, \beta=6.0, K=0.1400$ ($m_q=0.25$) and $N_f=2, \beta=10.0, K=0.125$ ($m_q=0.30$).

 We see the clear plateau of the effective mass at $t=24\sim 31$ in both cases.
 
Next we show the scattered plot of the Polyakov loop in the complex plane in Fig.~\ref{polyakov-1} in these cases.
(It should be noted the difference for the scales.)
Apparently the former is a disordered state which implies the confining region, while the latter is an ordered state which implies the deconfining region. These results can be better understood when
we consider the phase structure. We will discuss this point in Sec.~\ref{sec:conformal_region}.

Now we show in Fig.~\ref{yukawa-decay-1},
typical examples of Yukawa-type decay
in two cases of the very light quark masses:
$N_f=7, \beta=6.0, K=0.1459$ ($m_q=0.045$) and $N_f=2, \beta=10.0, K=0.135$ ($m_q=0.028$).

We see in both cases the effective mass is slowly decreasing without  plateau up to $t=31$,
  suggesting the power-law correction.
We show the power-law corrected fit for the local-local data with the fitting range $t=[15:31]$ 
in Fig.~\ref{yukawa-fit}.
The fits with $\alpha_H=0.8(1)$ and $\alpha_H=1.1(1)$ reproduce the data very well.

The $\chi^2/{\rm dof}= 0.2166\times 10^{-2}( \pm 0.3352\times10^{-2})/14$ and  $0.1375\times 10^{-1}( \pm 0.7847\times10^{-2})/14$ are very small. This does not mean the fits are excellent, but it reflects  that the correlation  in the $t$ direction is not taken account.
It is well known that it requires data in high statistics to take into account the correlation. 
Furthermore it is a notorious problem to fit data with power terms.
Therefore it is hard to estimate the error including the correlation.
We have estimated the errors by a jack knife method.

We have confirmed in all cases with $m_q \le 0.4$ that
the propagator of a meson $G_H(t)$ behaves  at large $t$ as a power-law corrected Yukawa-type decaying form $G_H(t) = \tilde{c}_H\, \exp{(-\tilde{m}_H t)}/t^{\alpha_H}$  instead of the exponentially decaying form $c_H\exp{(-m_H t)}$.

We show in Fig.~\ref{polyakov-2} the scattered plot of the Polyakov loop in spatial directions.
The patterns are apparently different from those in the confining region and the deconfining region.
They exhibit the characteristics in the conformal region.

We defer the detailed discussion on the Polyakov loop and the boundary of the conformal region 
 after the discussion of the structure of the vacuum in Sec.~\ref{sec:conformal_region}.
We will show that the boundary between the ``conformal region'' and the ``confining region'' is a first order transition in the $N_f=7, 12$ and $16$ cases.

 \subsection{Local analysis of propagators}
We are able to extract the properties of a quark and anti-quark system in the IR region 
from the long distance behavior of temporal propagators.
However, the propagators contain more information for the properties of a quark and anti-quark system.
For example from the short distance behavior we may extract the properties in the UV region.

In order to investigate the dynamics of the theory,
we make a detailed analysis of temporal propagators 
which we call the ``local-analysis" of propagators~\cite{coll2}.
We restrict ourselves to the case of the local-sink local-source  for the local analysis.
We parametrize the propagator $G(t)$ as
\begin{equation} 
G(t) = c\,  \frac {\exp(-m(t)\, t)}{t^{\, \alpha(t)}},
\label{local}
\end{equation}

It is possible to determine $c, m(t_0), \alpha(t_0)$ locally, using three point data
$G(t_0 ), G(t_0+1), G(t_0 +2).$ This is not a fit.
One important point is  $m(t)$ and $\alpha(t)$ are smooth functions in $t$. 
In spirit this is analogous to the Callan-Symanzik RG approach where we interpret $m(t)$ as the scale dependent mass and $\alpha(t)$ as the wave-function renormalization factor.
We have also
made fits to the form Eq.(\ref{local}) using 5 points data. The fit gives generally similar results with
the 3-point determination. It implies the 3-point determination well represent the dynamics of the system.

The  $m(t)$ and $\alpha(t)$ at short distance are governed by the UV fixed point and take
the value of a free quark and an
 anti-quark: $m(t)=2\, m_q$ and $\alpha(t)=3.0$ in the limit $t \rightarrow 0$ of the continuum theory.
While in the limit $t \rightarrow \infty$, 
$m(t)$ and $\alpha(t)$ are governed by the IR fixed point.
First of all, the $m(t)$ should approach a hadronic mass $m_H$.
The exponent $\alpha_H$ in $t\to \infty$ with $t \,\tilde{m}_H\ \ll 1$ takes the universal formula $3-2\gamma^{*},  $ while with $ t \,\tilde{m}_H \gg 1$ it takes a value depending on the dynamics.

The  $m(t)$ and $\alpha(t)$ evolve with $t$ from UV to IR and thereby contain useful information on the dynamics, which we discuss below.

We show the $m(t)$ and $\alpha(t)$ for the four examples discussed above:
\begin{itemize}
\item \ Fig.~\ref{nf2b10.0k125}; exponential decay
\begin{itemize}
\item
$N_f=7, \beta=6.0, K=0.1400$ ($m_q=0.25$) 
\item
$N_f=2, \beta=10.0, K=0.125$ ($m_q=0.30$)
\end{itemize}
\item \ Fig.~\ref{nf2b10.0k135}; power corrected Yukawa-type decay
\begin{itemize}
\item
 $N_f=7, \beta=6.0, K=0.1459$ ($m_q=0.045$)
\item
 $N_f=2, \beta=10.0, K=0.135$ ($m_q=0.028$)
\end{itemize}
\end{itemize}
%\end{enumerate}

In Fig.~\ref{nf2b10.0k125} where the propagators decay exponentially, 
the exponent $\alpha(t)$ take values close to $3.0$ at $t=3$ (we disregard the data at $t=1$ and $2$, as they are
affected by the boundary) and decrease monotonously down to $0.0$.
In the confining region (left) it decays without particular pattern, while in the deconfining region (right) it stays around $3.0$
which means a free quark and anti-quark pair, at $t  = 4 \sim 8$.
The $m(t)$ take values close to $2 m_q$ at $t=3$ and increase  to the values of a meson state $m_H$, 
which are around $0.6\sim 0.7$.

In Fig.~\ref{nf2b10.0k135}, both of the exponents $\alpha(t)$ exhibit characteristic $t-$dependence.
However, they are quite different from each other.
On the left panel it shows a plateau at $t= 14\sim 31$.
On the other hand, it shows a shoulder $t=10\sim 16$ on the right panel of the Figure.
The difference arises from the difference of the dynamics.

The four examples exhibit the usefulness of the local-analysis of propagators.
We are able to learn not only the phase structure of the theories but also they reveal the detailed dynamics.
We will fully utilize the technique in the following.

\begin{figure*}[htb]
\includegraphics [width=7.5cm]{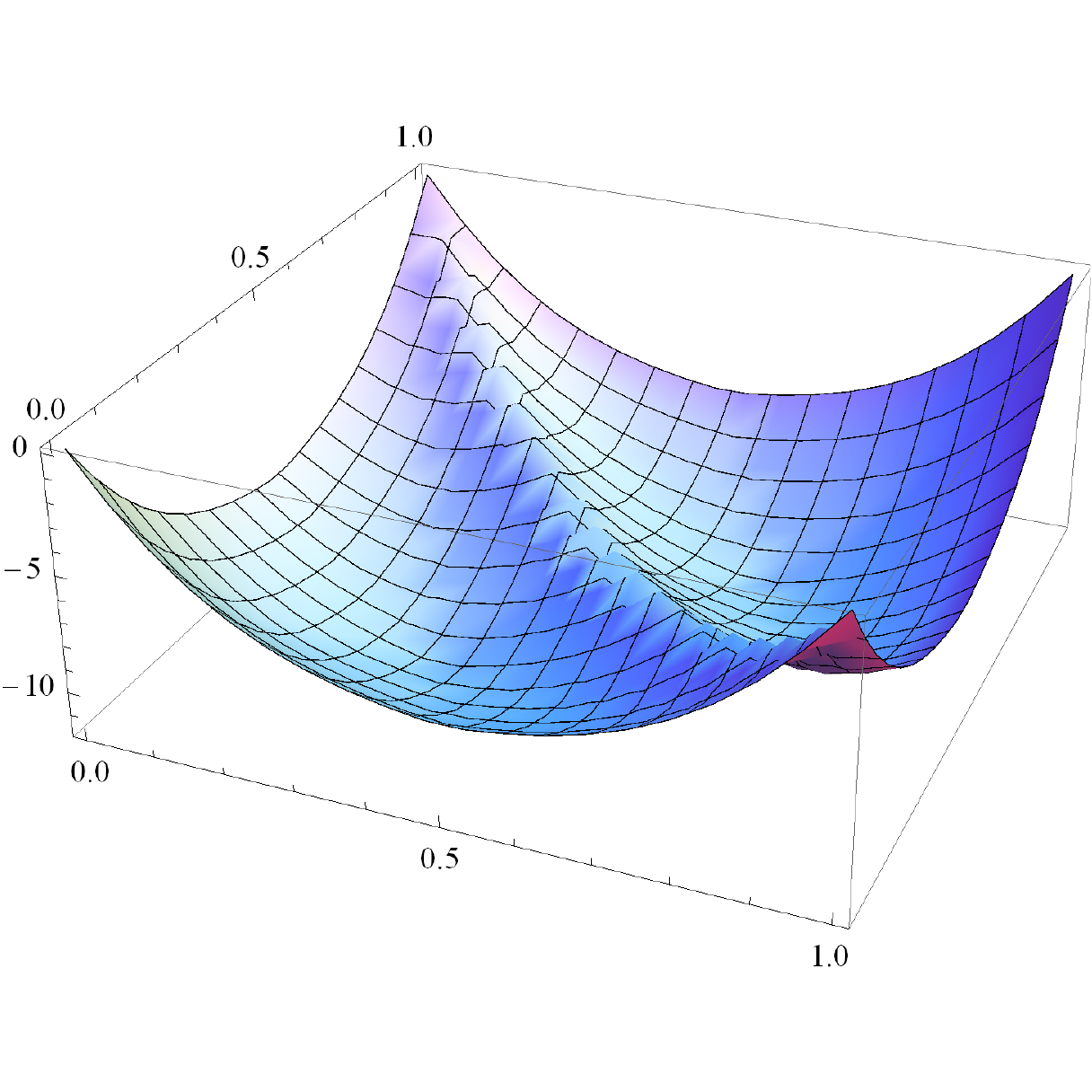}
 \hspace{1cm}
\includegraphics [width=7.5cm]{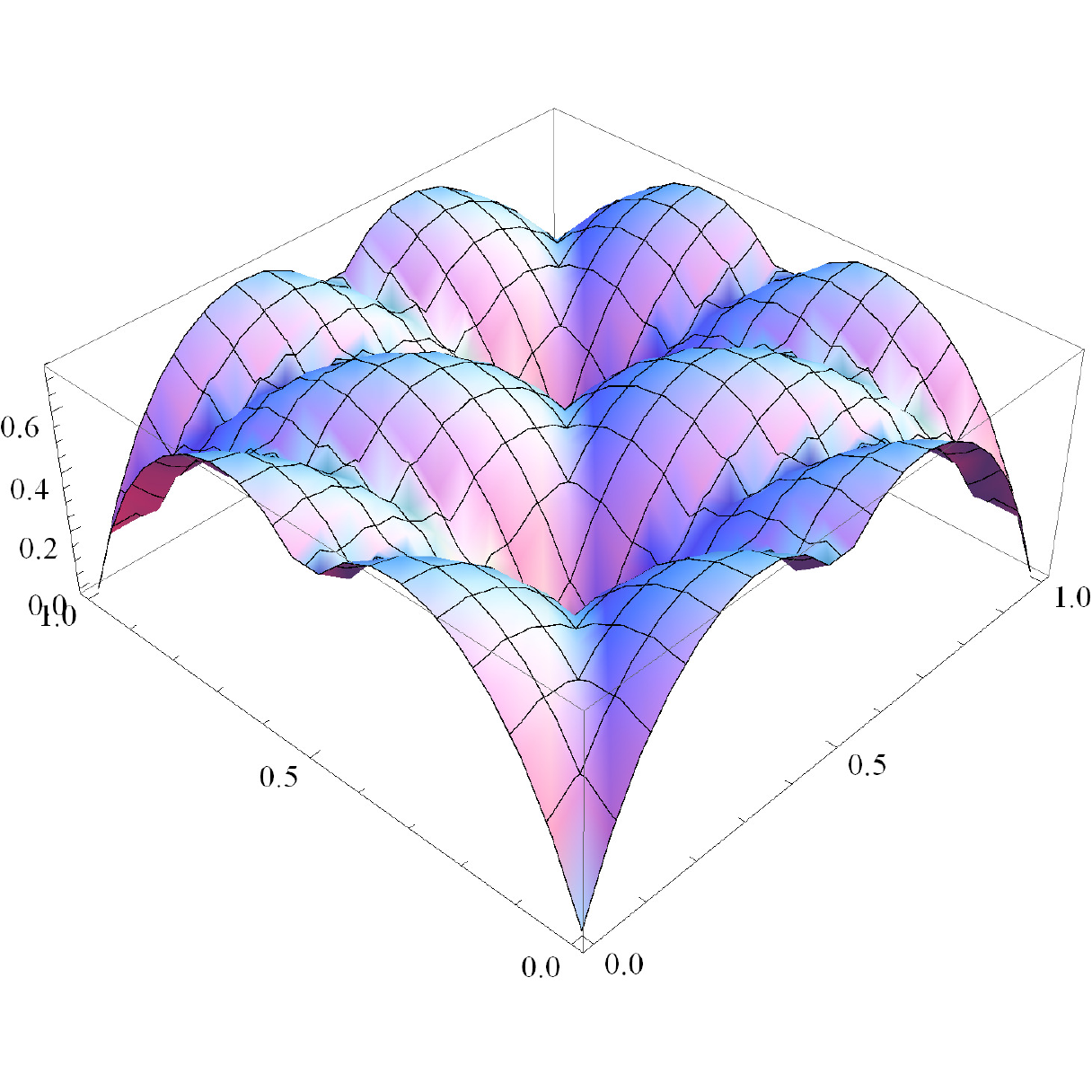}
\caption{(color online)  The effective potential $V_{\mathrm{eff}}(a, b)$ in terms of $a$ and $b$: $m=0.0$ (left) and $m=1.0$ (right).}
\label{effective potential}
\end{figure*}

\begin{figure*}[htb]
\includegraphics [width=7.5cm]{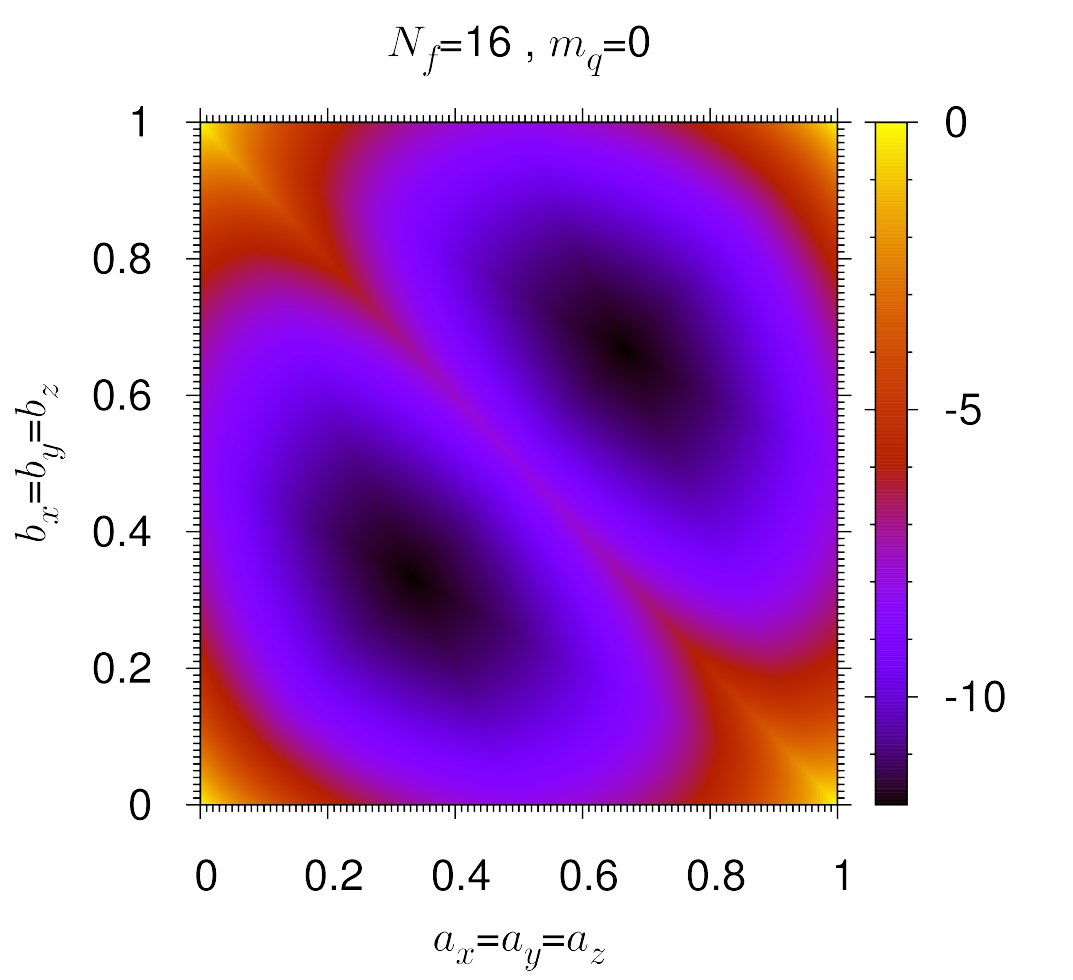}
 \hspace{1cm}
\includegraphics [width=7.5cm]{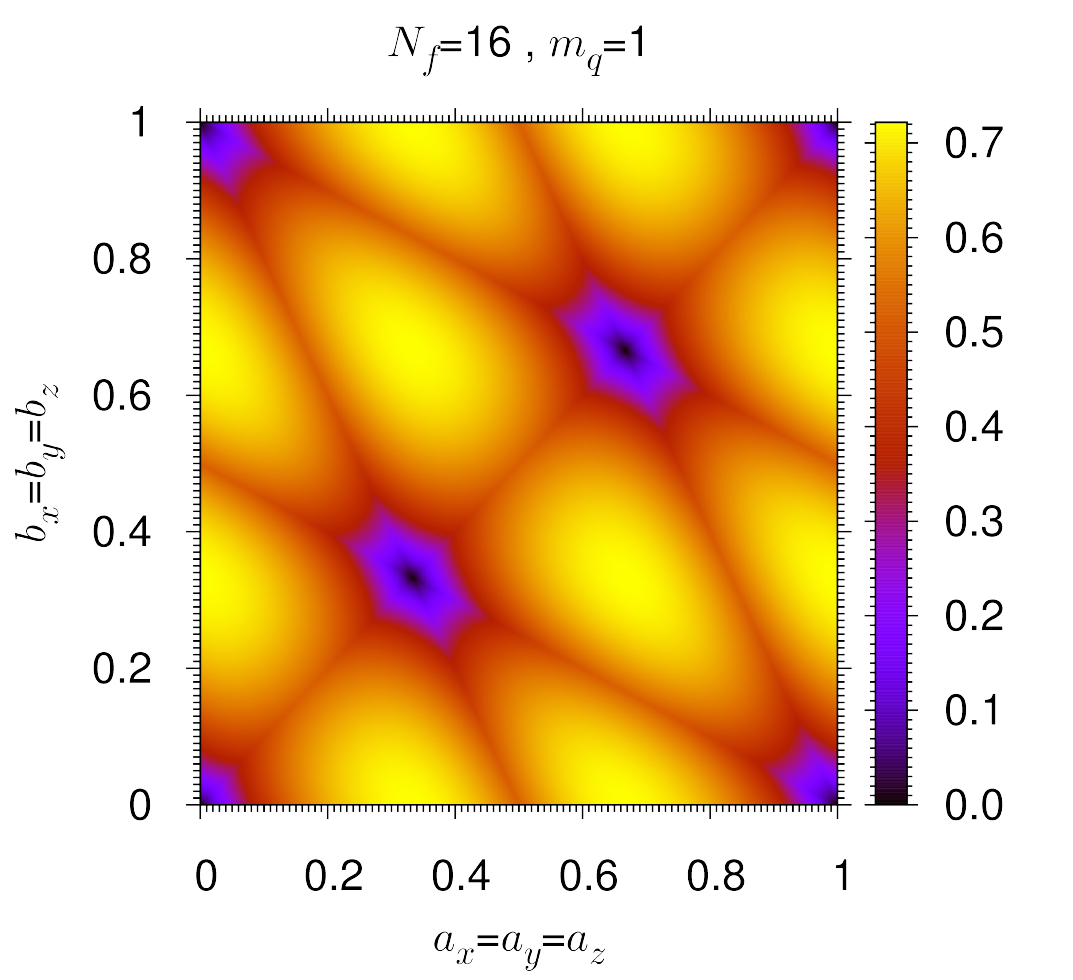}
\caption{(color online)  The contour of the effective potential $V_{\mathrm{eff}}(a, b)$  in terms of $a$ and $b$:: $m=0.0$ (left) and $m=1.0$ (right).}
\label{contour}
\end{figure*}

\section{Structure of The Vacuum and Polyakov loops}
\label{sec:structure of vacuum}
\subsection{The $Z(3)$ twisted vacuum}
To understand the phase structure in relation to the expectation values of the Polyakov loops, we would like to discuss the vacuum structure of the perturbative QCD on the  lattice 
in the one-loop approximation by computing the zero temperature vacuum energy.

In the perturbative QCD in the finite volume, the classical vacua are characterized by the flat connection. In the case of our torus lattice, the flat connection is given by the Polyakov loop in each $x,y,z$ directions (in fundamental representation of $SU(3)$) 
\begin{align}
 U_x &= \exp( i\int A_x dx) =  \mathrm{diag} (e^{i2\pi a_x}, e^{ i2\pi b_x}, e^{i 2\pi c_x}) \cr
 U_y &= \exp( i\int A_y dy)  = \mathrm{diag} (e^{i2\pi a_y}, e^{ i2\pi b_y}, e^{i 2\pi c_y}) \cr
 U_z &= \exp( i\int A_z dz)  = \mathrm{diag} (e^{i2\pi a_z}, e^{ i2\pi b_z}, e^{i 2\pi c_z})  
\end{align}
with $a_i + b_i + c_i \in \mathbb{Z} $ for $(i=x,y,z)$ from the unitary condition.
Note that $a_i = b_i = c_i = \frac{1}{3}, \frac{2}{3}$ gives a non-trivial center of the gauge group.

\begin{figure*}[thb]
\includegraphics [width=6.7cm]{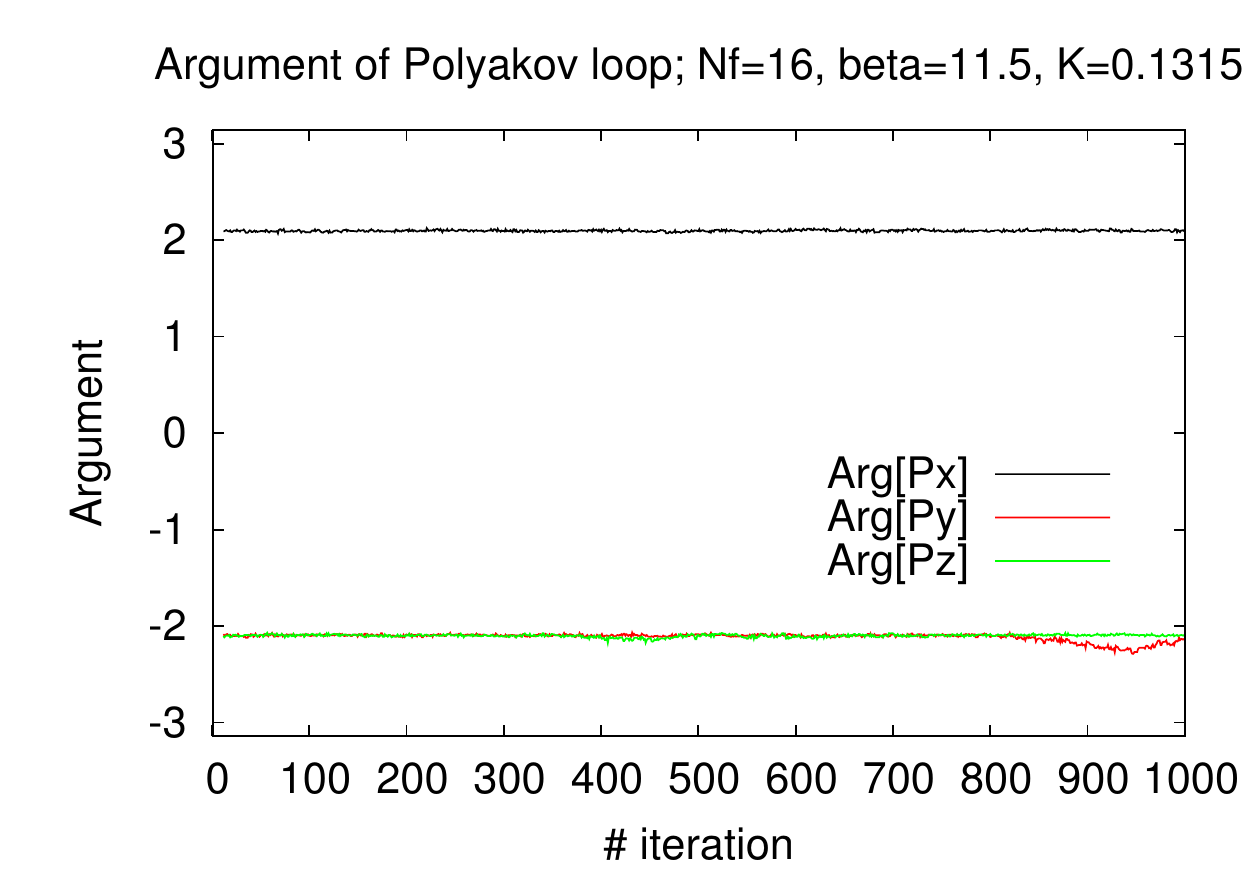}
\hspace{1cm}
\includegraphics [width=6.7cm]{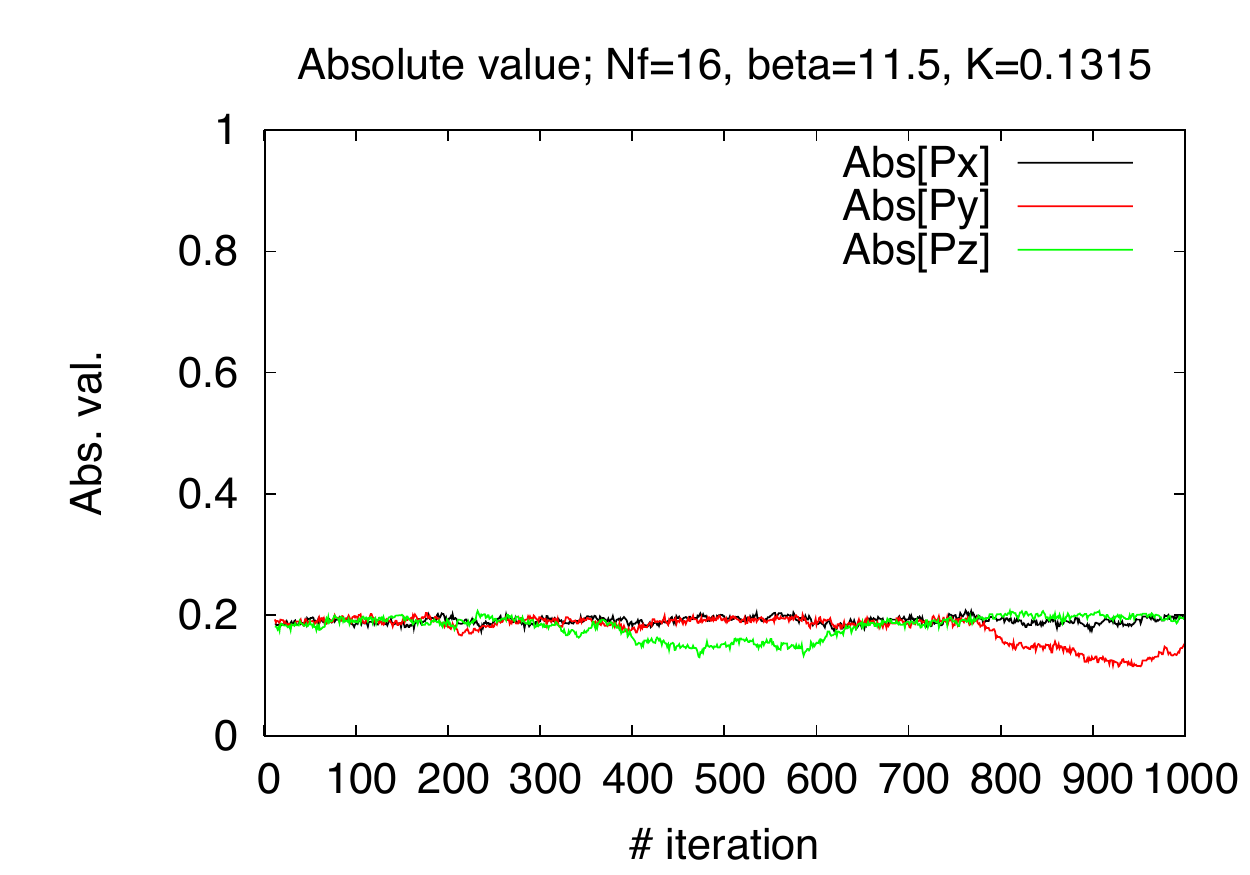}
%\vspace{1cm}
\caption{(color online)  The time history of the argument and the absolute value of Polyakov loops for $N_f=16$ at $\beta=11.5$ and $K=0.1315$.}
\label{nf16_poly}
\end{figure*}

When the space is compact, we expect that a non-trivial potential for the flat direction is quantum mechanically generated similarly to the Hosotani mechanism~\cite{hosotani83}
The one-loop effective energy including both fermion loops and gauge field loops
for the $N_f=16$ case with $m_q=0.0$ on a $16^3$ lattice are calculated at the zero temperature in the 6 parameter space;
$a_i$, $b_i$ in the $x$, $y$ and $z$ directions.
The details of calculation are given in Appendix D.
The effective potential  and the contour map in terms of two parameters; $a$, $b$ in one direction among 6 parameters are shown in Fig.~\ref{effective potential} and Fig.~\ref{contour}.

\begin{figure*}[thb]
\includegraphics [width=6.7cm]{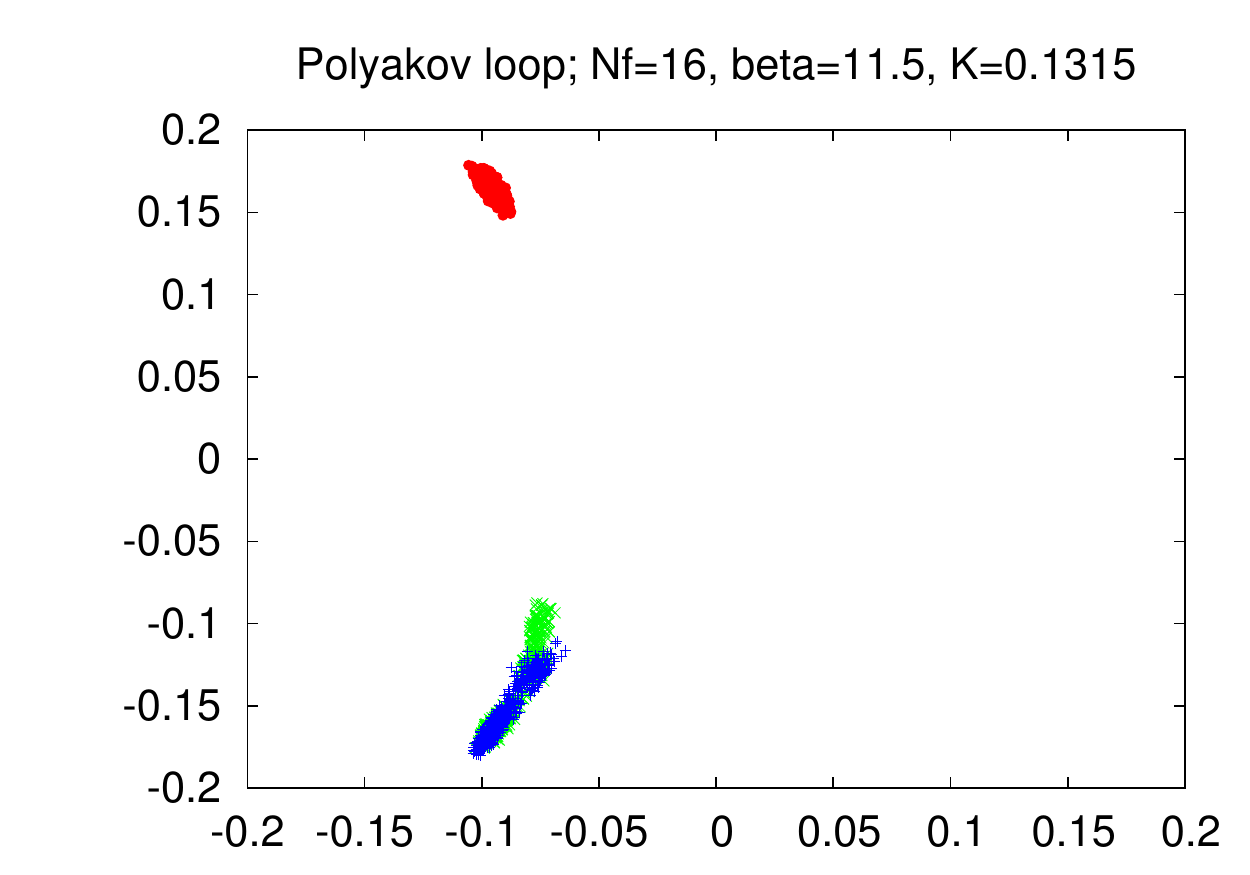}
\hspace{1cm}
\includegraphics [width=6.7cm]{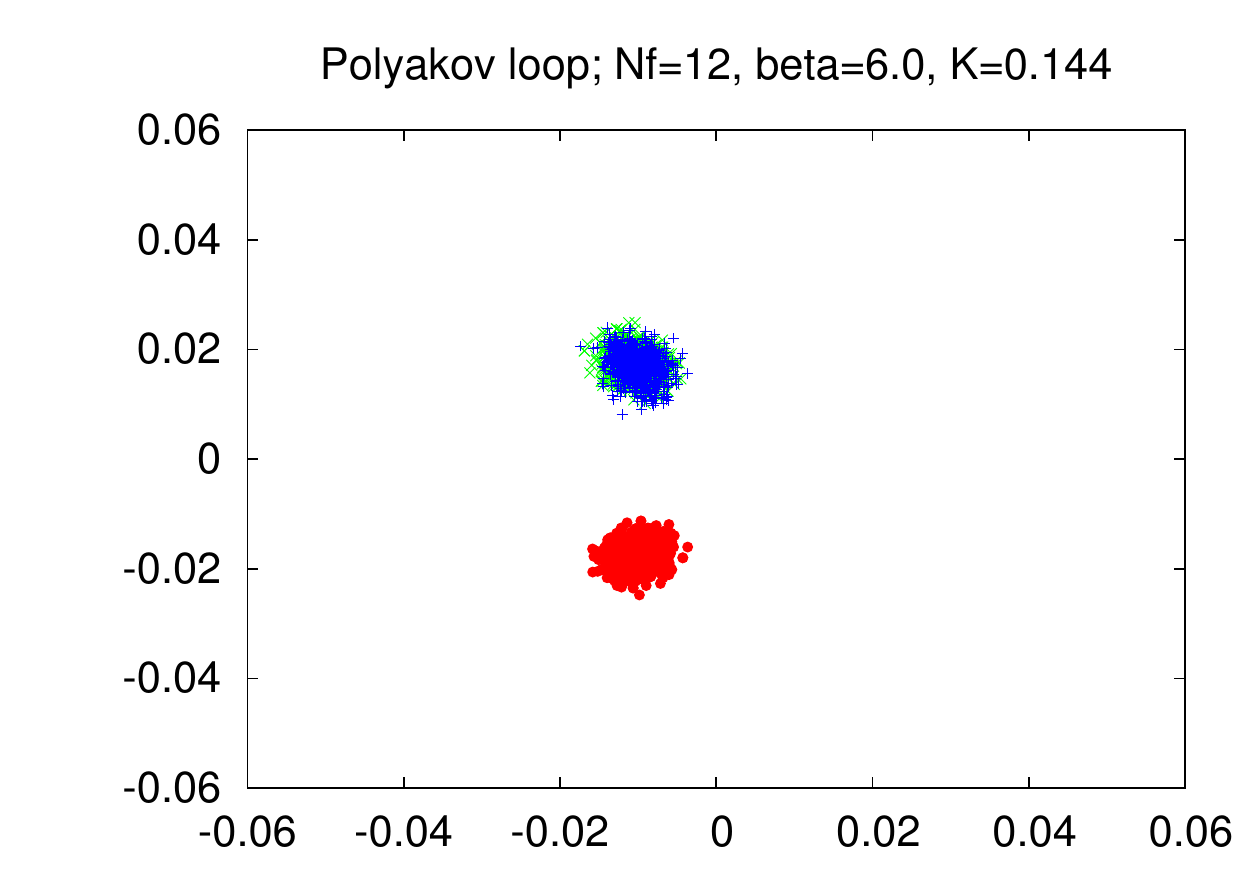}
\includegraphics [width=6.7cm]{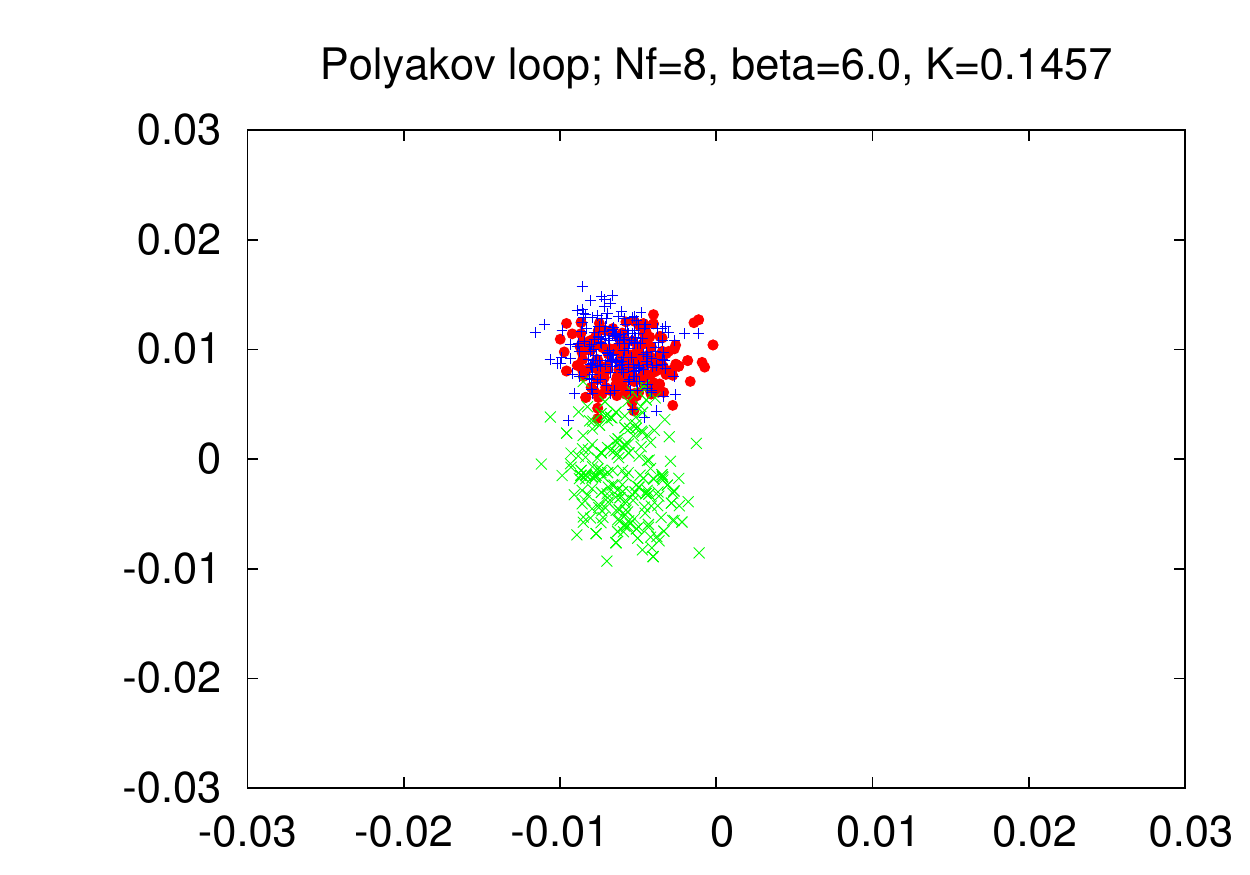}
\hspace{1cm}
\includegraphics [width=6.7cm]{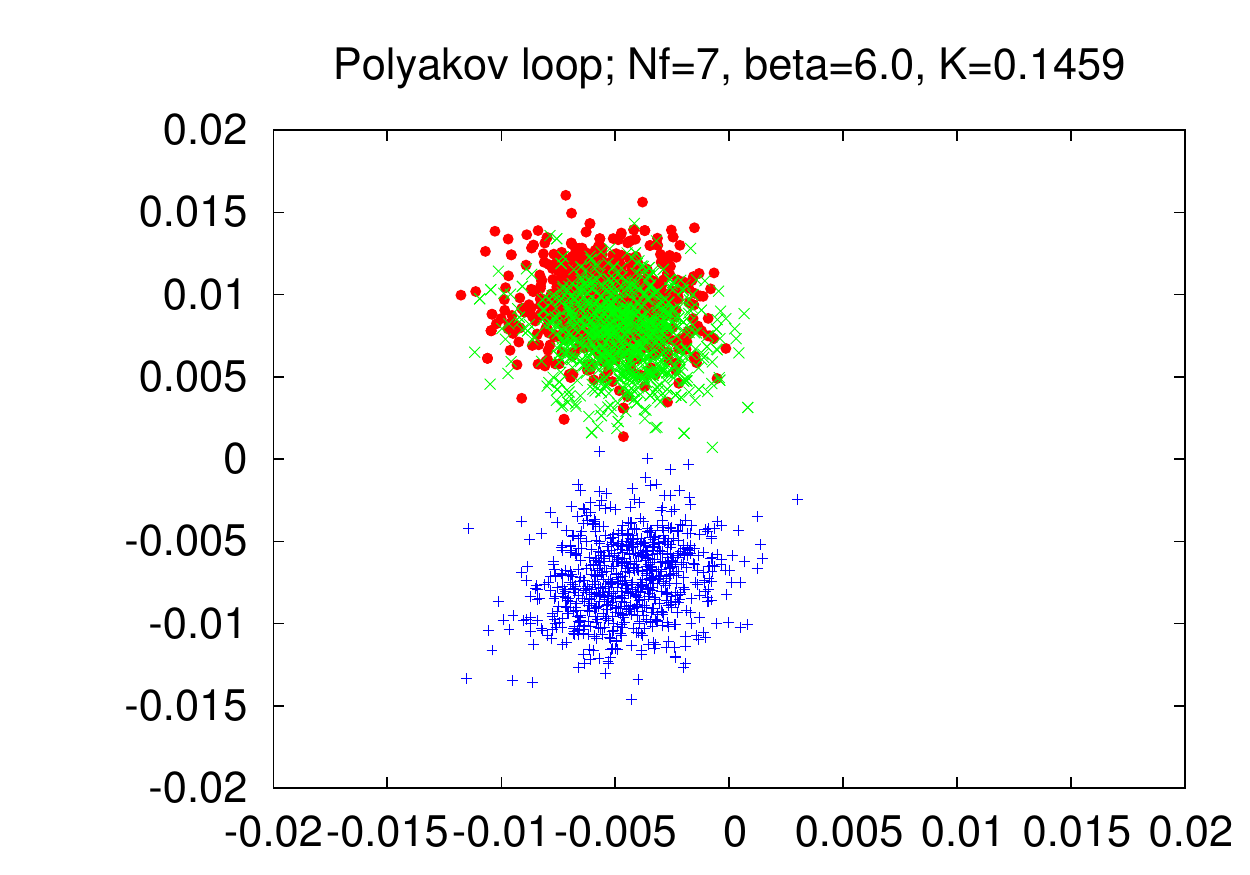}
\caption{(color online)  The scattered plots of Polyakov loops in the $x$, $y$ and $z$ directions overlaid; 
         $N_f=16$ at $\beta=11.5$ and $K=0.1315$, $N_f=12$ at $\beta = 6.0$ and $K=0.144$, $N_f=8$ at $\beta = 6.0$ and $K=0.1457$ and $N_f=7$ at $\beta = 6.0$ and $K=0.1459$.}
\label{Polyakov bf-7-16}
\end{figure*}

\begin{figure*}[thb]
\includegraphics [width=6.7cm]{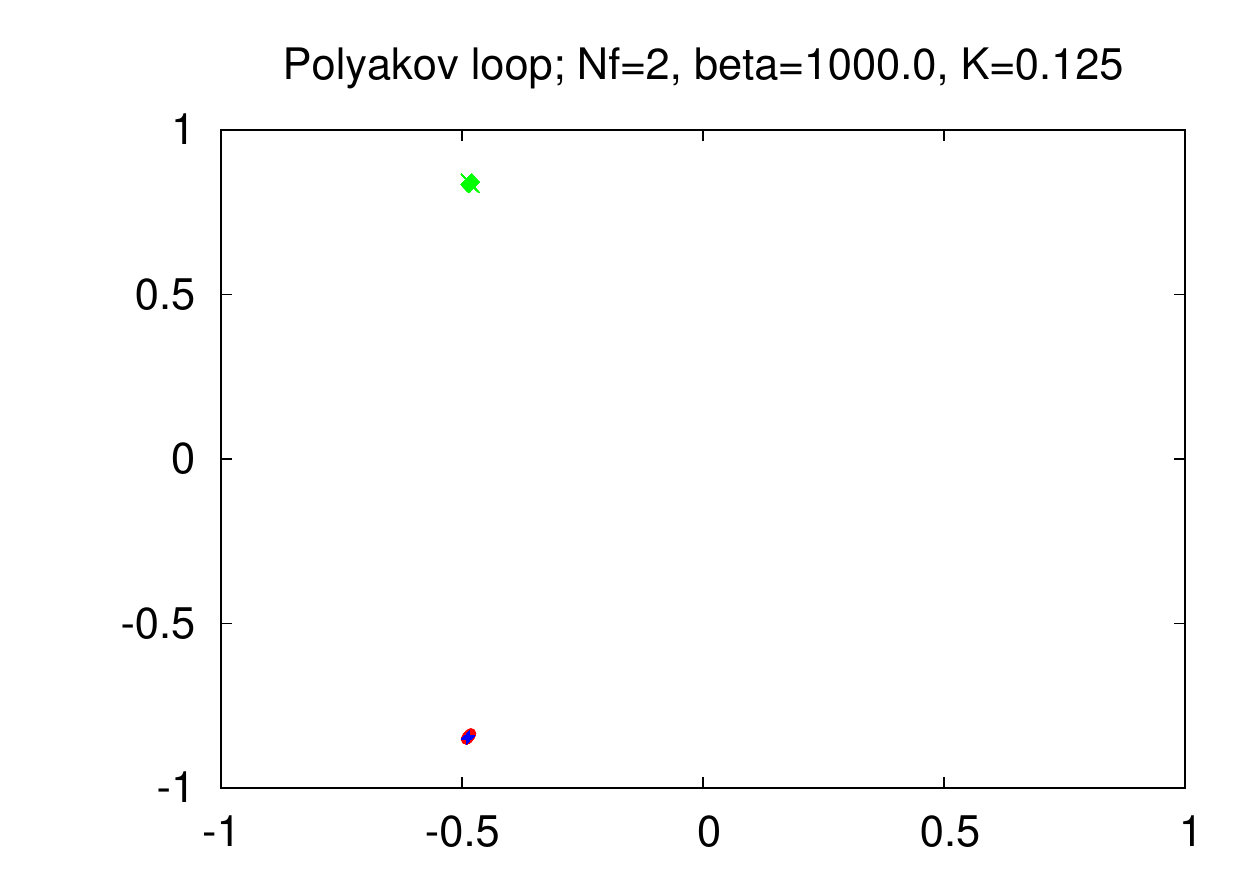}
\hspace{1cm}
\includegraphics [width=6.7cm]{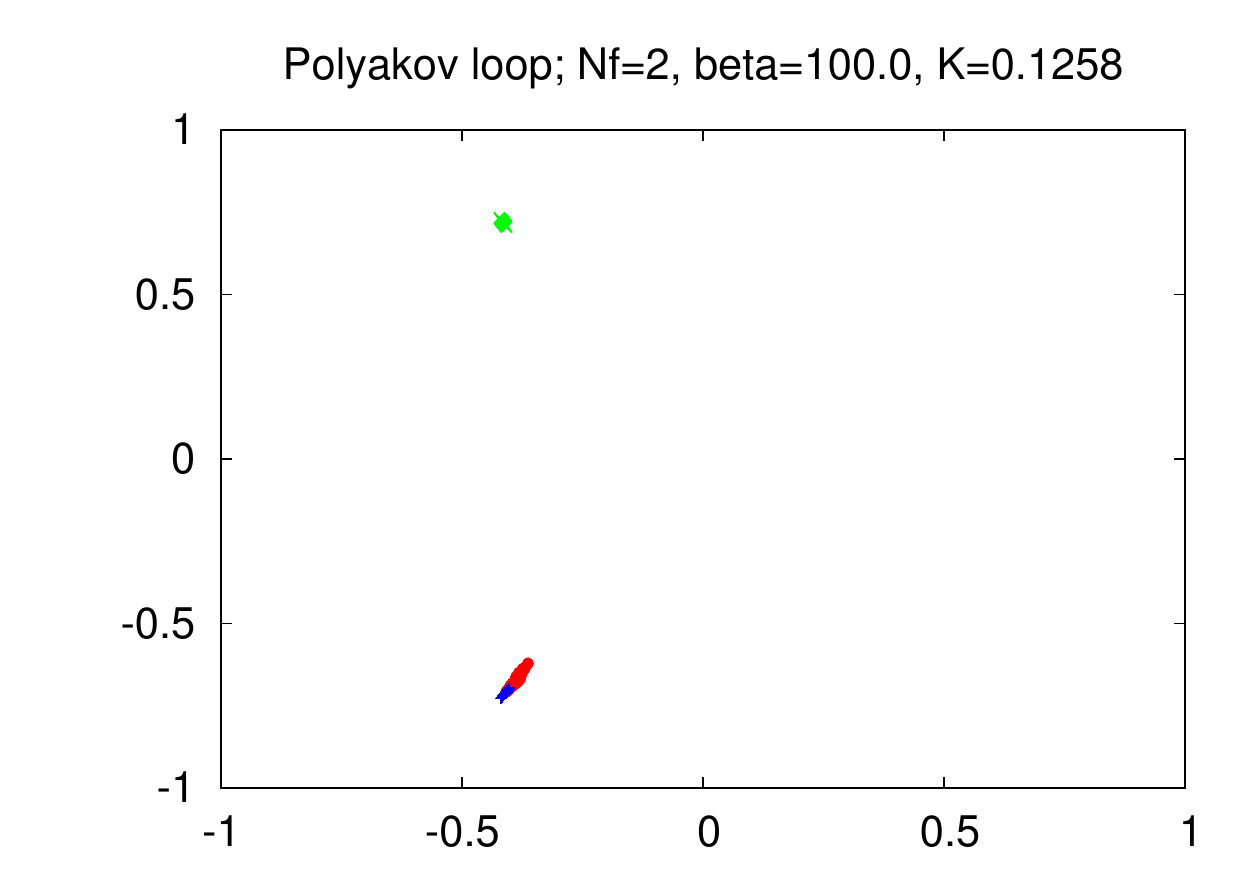}
\includegraphics [width=6.7cm]{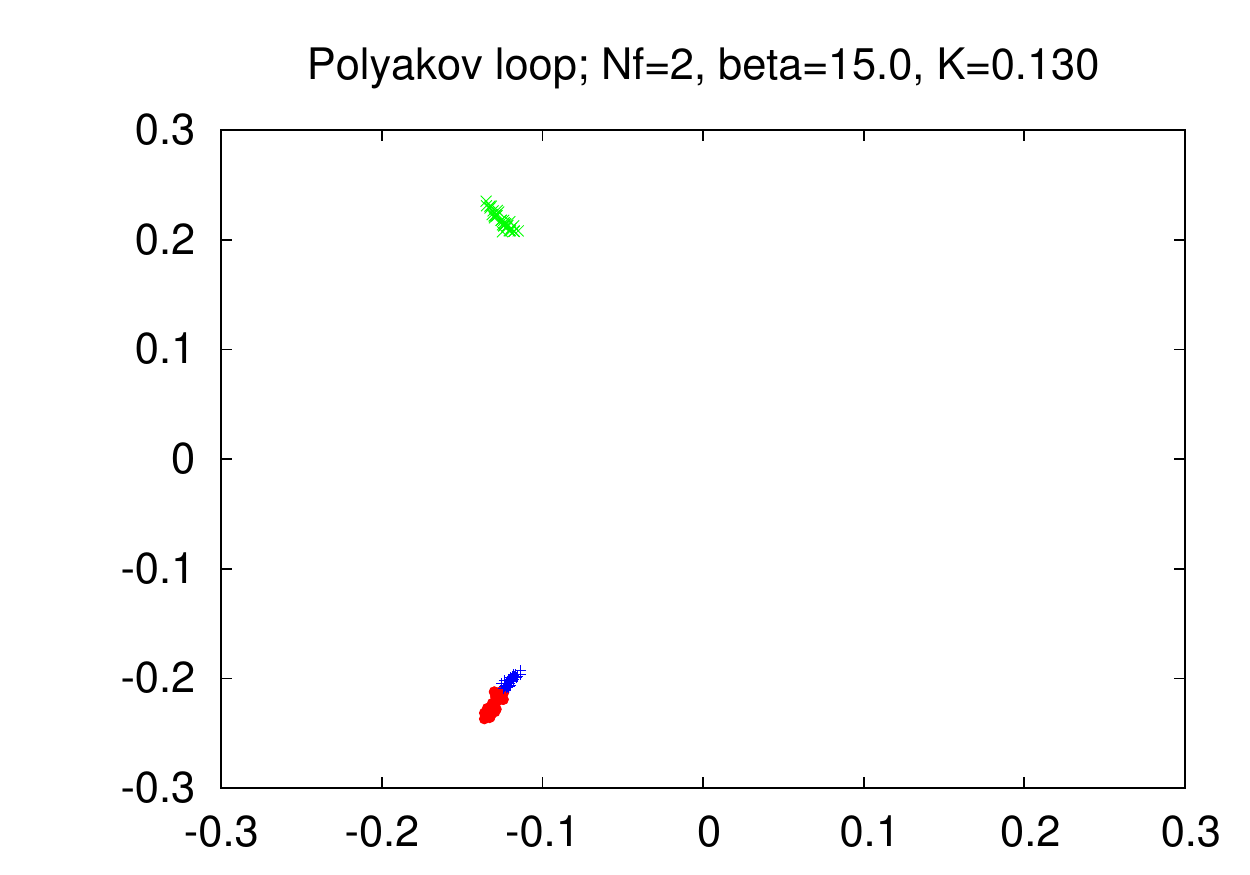}
\hspace{1cm}
\includegraphics [width=6.7cm]{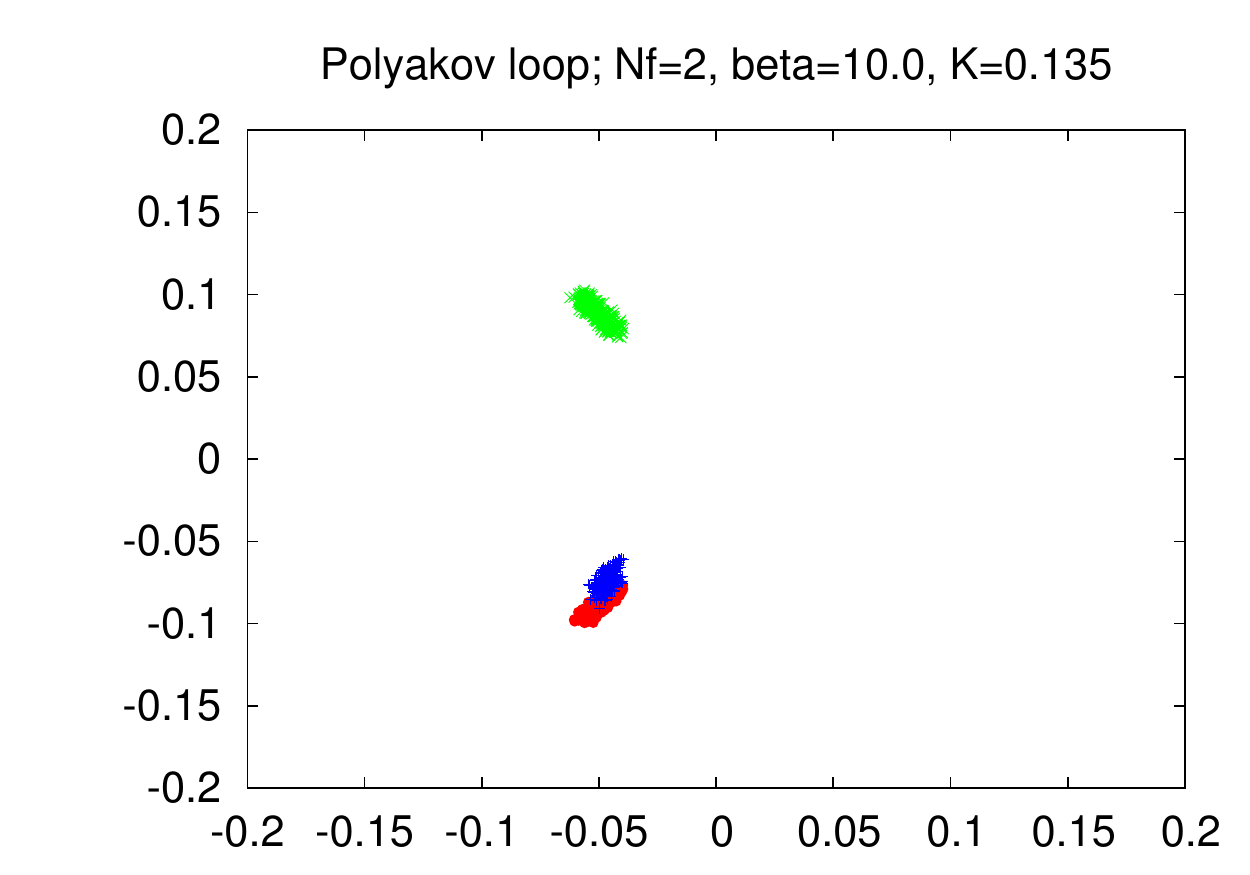}
\caption{(color online)    The scattered plots of Polyakov loops in the $x$, $y$ and $z$ directions overlaid; 
         $N_f=2$ at $\beta=1000.0$ and $K=0.125$, $\beta = 100.0$ and $K=0.1258$, $\beta = 15.0$ and $K=0.130$
         and $\beta = 10.0$ and $K=0.135$.}
\label{Polyakovbeta10-1000}
\end{figure*}

\begin{figure*}[thb]
\includegraphics [width=6.7cm]{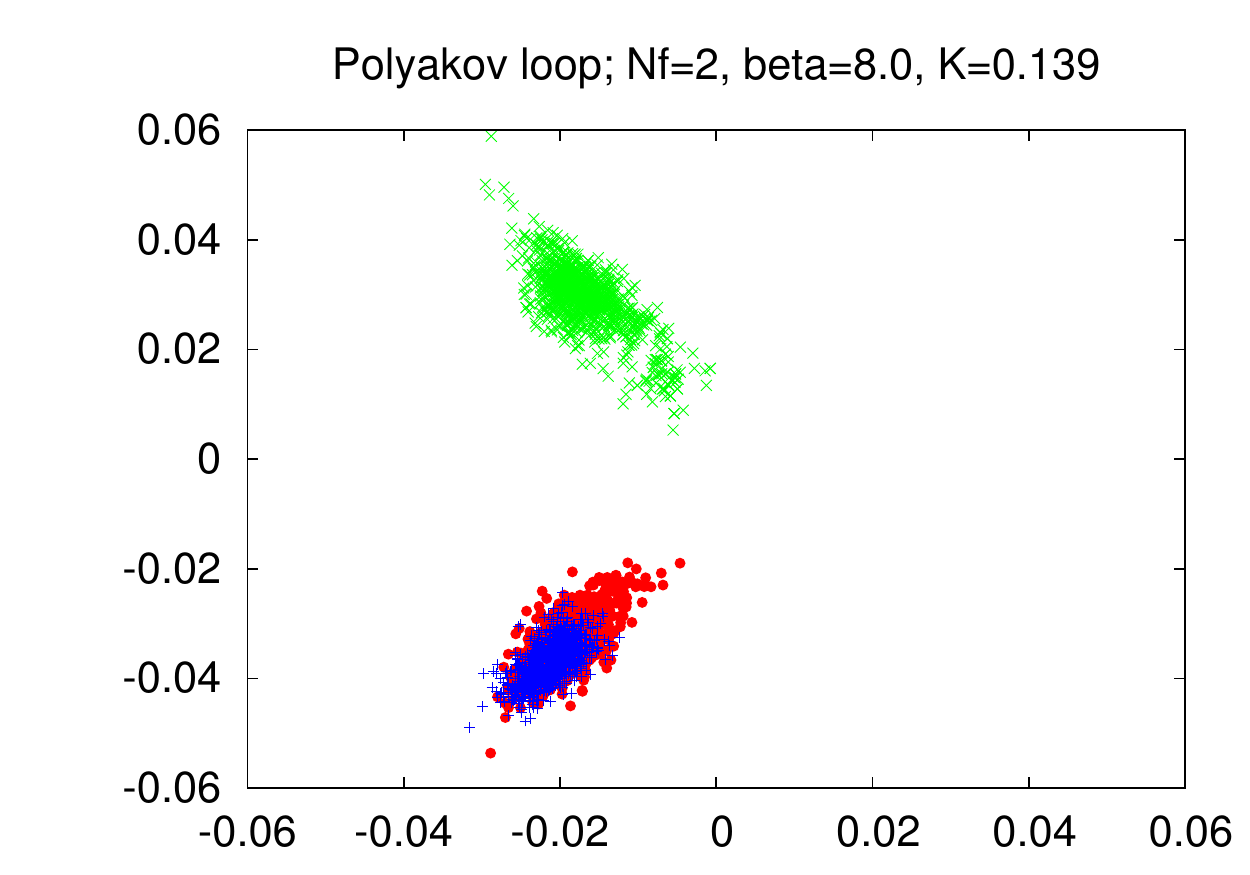}
\hspace{1cm}
\includegraphics [width=6.7cm]{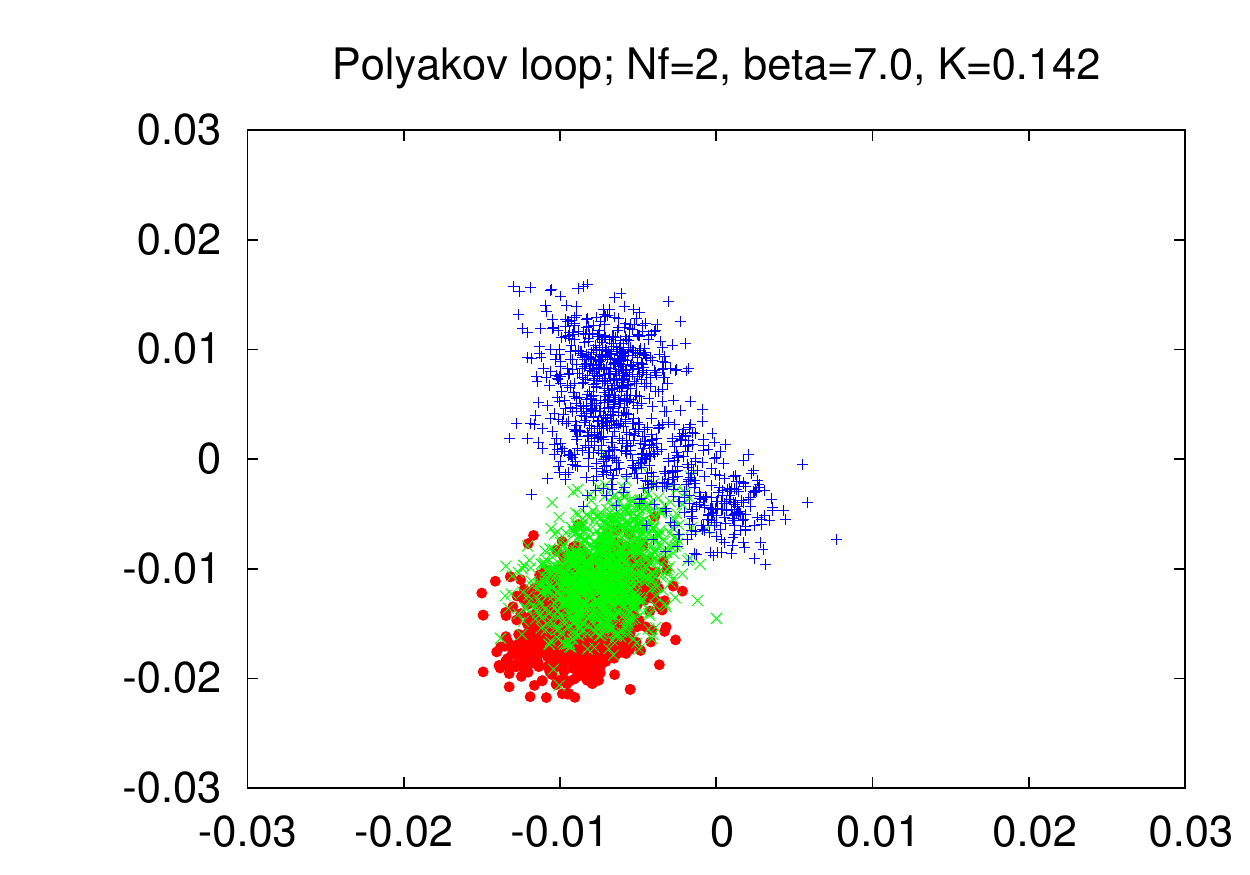}
\includegraphics [width=6.7cm]{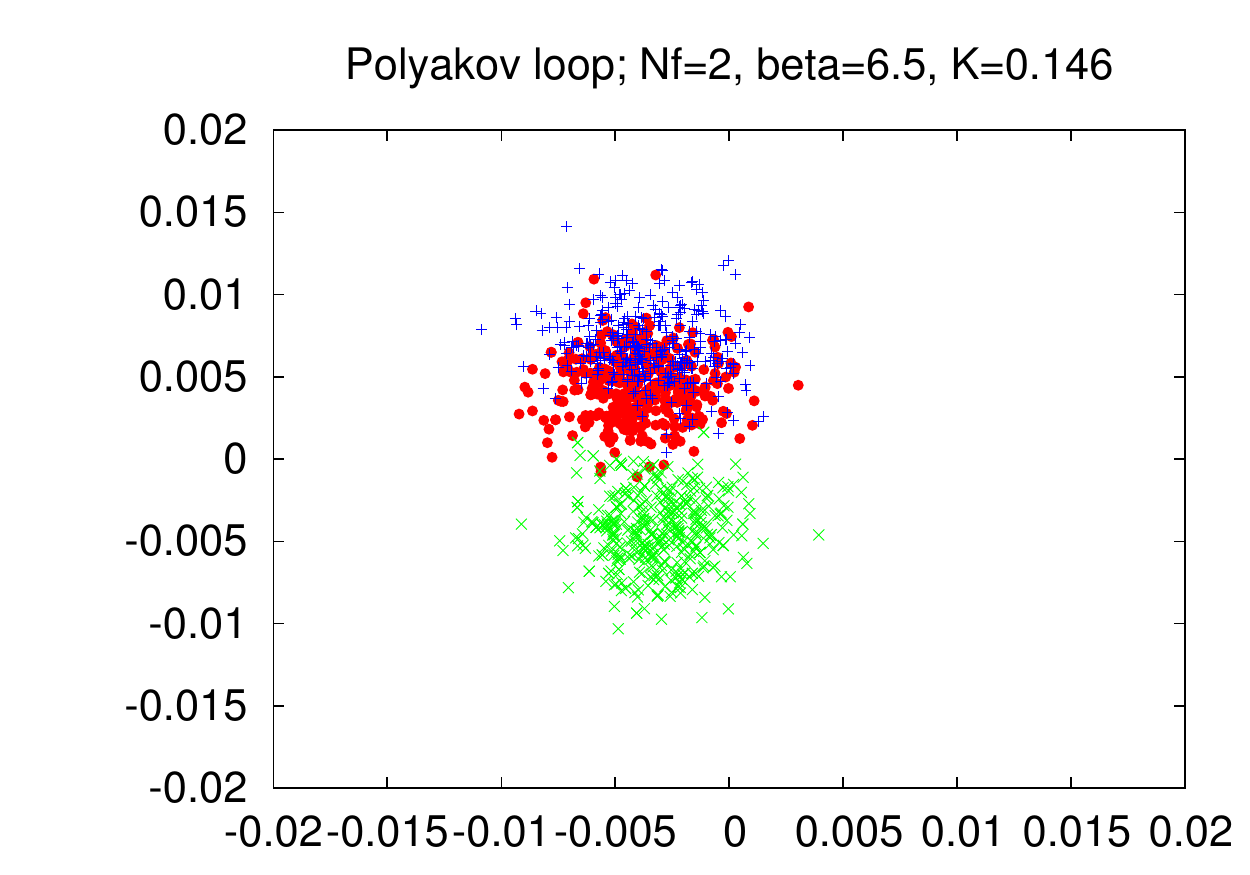}
\hspace{1cm}
\includegraphics [width=6.7cm]{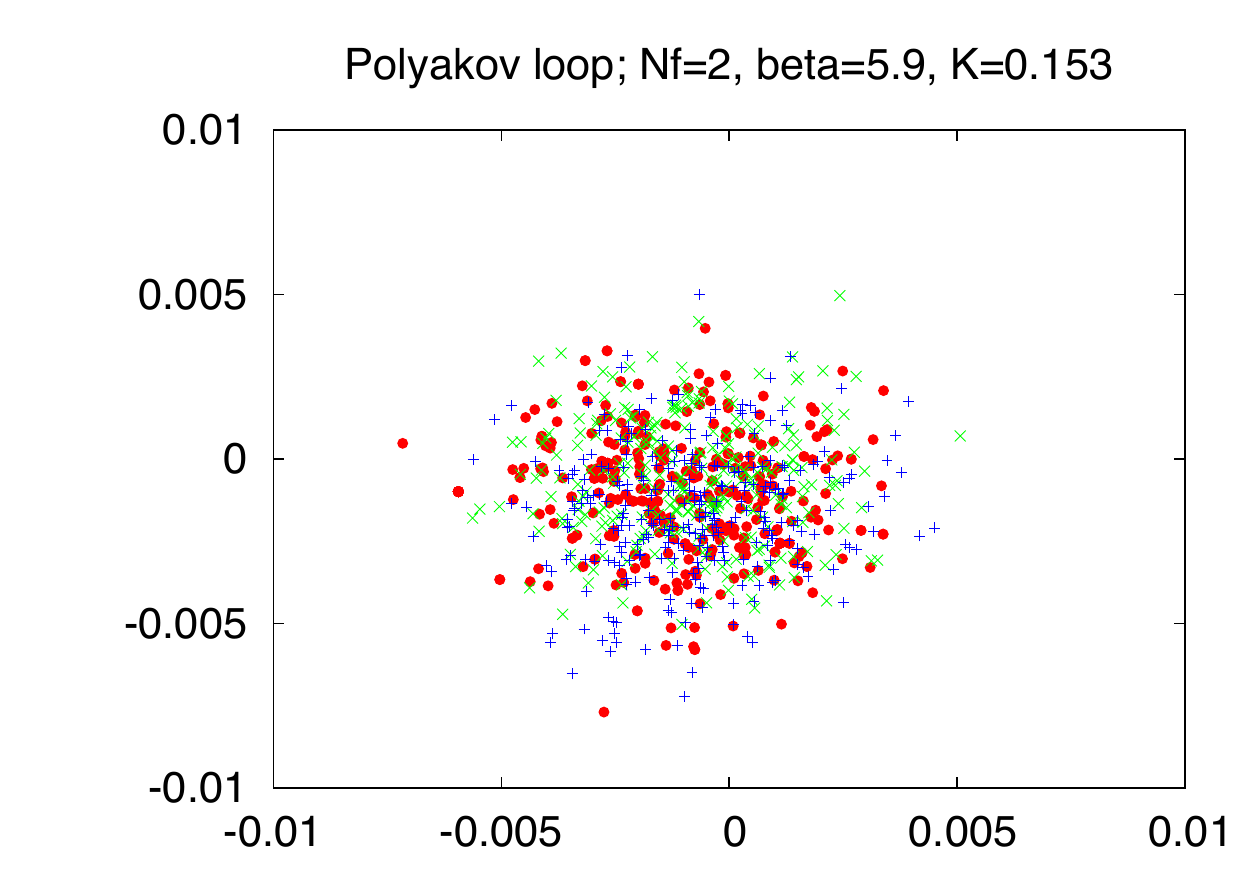}
\caption{(color online)   The scattered plots of Polyakov loops in the $x$, $y$ and $z$ directions overlaid; 
         $N_f=2$ at $\beta=8.0$ and $K=0.139$, $\beta = 7.0$ and $K=0.142$, $\beta = 6.5$ and $K=0.146$
         and $\beta = 5.9$ and $K=0.153$.}
\label{Polyakovbeta5.9-8.0}
\end{figure*}

Denoting the Polyakov loop in the $x, y$ and $z$ directions by $P_x = \frac{1}{3}\mathrm{Tr} U_x$, $P_y=\frac{1}{3}\mathrm{Tr} U_y$, and $P_z=\frac{1}{3}\mathrm{Tr} U_z$, respectively, and writing 
$$(P_x, P_y, P_z)= |P| \exp{(2\pi  i((P_x, P_y, P_z)))}$$
where $|P|$ is the absolute value and $((P_x, P_y, P_z))$ is the arguments in units of $2\pi$ of Polyakov loops,
it turns out that the local extremum of the one-loop energy  is given by elements of the $Z(3)$ center:
$((P_x, P_y, P_z))=$
$$(0, 0, 0), $$
$$(0, 0, \pm 1/3),$$
$$(0, \pm 1/3, \pm 1/3),$$
$$(\pm 1/3, \pm 1/3, \pm 1/3)$$
with $|P| =1.0 $.
They are $3^3$ fold and become degenerate in the quench limit $m_q \to \infty$. This is expected because without the matter the theory must be symmetric under the center of the gauge group.
In the above the order of $P_x, P_y, P_z$ are cyclic. 

From now on we present the mean value of the argument of the state without $\pm$ symbol for simplicity.
We use $(0, 0, 0), (0, 0, 1/3), (0, 1/3, 1/3)$ and $(1/3, 1/3, 1/3)$ 
to denote the phase of the mean values of the Polyakov loops in units of $2\pi$,
without mentioning the absolute value. 
When the state is in a confining region, the mean value of the Polyakov loop as a complex number
is zero, and therefore we denote the state by $(*, *, *)$. 

The effective energy depends on the $Z(3)$ value, and we clearly see from Fig.~\ref{effective potential} and Fig.~\ref{contour} that
the 8-fold states $(1/3,1/3, 1/3)$ are the  lowest energy state
in periodic boundary conditions in the one-loop approximation. 
The (0, 0, 0) state is locally unstable when $m_q$ is light, whereas
it becomes locally stable as the $m_q$ becomes heavy; $m_q=0.15\sim 0.25$.
We can also confirm that the (0, 0, 0) state is unstable at $m=0.0$, but stable at $m=1.0$ from Fig.~\ref{effective potential} and Fig.~\ref{contour} of the effective potential.

In Ref.~\cite{itou1212},
a similar result is obtained in the case of the twisted boundary conditions in the $x$ and $y$ directions:
Assuming a priori the lowest state is represented by the $Z(3)$ center in the $z$ and $t$ directions,
the vacuum takes the non-trivial center in the fermion one-loop approximation.

We have several remarks of the vacuum structure. First of all, in the one-loop approximation, the global vacuum structure does not depend on the number of fermions very much while the shape of the potential does depend on the number of fermions.
For example, the shape of the potential for $N_f=2$ and $m_q=0.0$ is similar to that for $N_f=16$.
 Secondly, the discussion here is done in the zero-temperature limit 
and the phase structure at finite temperature does not necessarily follow the vacuum structure here. 
Finally, we expect that the strong interaction does change the structure of the vacuum 
as we see that the Polyakov loop behaves very differently in the confining region from that in the deconfining region.
At the same time, we will also see that in the perturbative regime in the deconfining region, the one-loop vacuum structure discussed here more or less survives.

\subsection{The vacuum of Conformal QCD}
\subsubsection{$N_f=16$}
We compare the vacuum of $N_f=16$ obtained by simulations at $\beta=11.5$ with $K=0.1315$  ($m_q=0.055$)
with the vacuum in the one-loop approximation obtained above.
The history of the Polyakov loop at $\beta=11.5$ is shown in Fig.~\ref{nf16_poly}.
The argument is very stable during one thousand trajectories, taking the value of $\pm 2/3\pi$.
However the magnitude is about $\sim 0.2$, clearly smaller than $1.0$.
This implies that the vacuum of $N_f=16$ at $\beta=11.5$ is close to the vacuum in the one-loop approximation, 
but is not well described by the perturbation theory.

\subsubsection{Smaller $N_f$}
We show  the Polyakov loop in the cases we have observed the Yukawa-type decay
in Fig.(\ref{Polyakov bf-7-16}) for $N_f=7$, 8, 12 and 16 in conformal QCD .

As $N_f$ decreases,
the magnitude decreases and the fluctuation of magnitude and argument increases.
Therefore the transition among the vacua often occurs in the cases $N_f=8$ and $7$.
However, the mean value of the arguments are $(1/3,1/3, 1/3)$ in units of $2\pi$.

\subsection{The vacuum of High Temperature QCD; $N_f=2$}
\subsubsection{The $\beta=10.0 \sim 1000.0$}
In Fig.(\ref{Polyakovbeta10-1000})
we show  the Polyakov loop 
for $N_f=2$; at $ \beta= 10.0, 15.0, 100.0, 1000.0$.
As the temperature increases, the argument is more stable during simulations, taking the value of $\pm 2/3\pi$.
The magnitude increase as the temperature increases, taking the value $\sim 0.1, 0.2, 0.82$ and $0.96$, respectively.
The approach to the magnitude unity is very slow.
That is, at $T/T_c\simeq10^2$ or $10^5$, $|P|\simeq 0.1\sim0.2$.

\subsubsection{Smaller $\beta$}

In Fig.(\ref{Polyakovbeta5.9-8.0}) 
we show  the Polyakov loop at lower temperatures in the cases we have observed the Yukawa-type decay
for $N_f=2$; at $ \beta=6.5, 7.0$ and $8.0$.

We clearly see that as $\beta$ decreases in High Temperature QCD,
the magnitude decreases and the fluctuation of magnitude and argument increases. Therefore the transition among the vacua often occurs.
Nevertheless the mean value of the arguments are $(1/3,1/3, 1/3)$ down to $\beta=6.5$.

However at $\beta=5.9$ which is smaller than the chiral transition point, the mean value of the Polyakov loop vanishes as a complex number, which shows that the state is in the confining region.
%\red{Refer to the phase structure}

\section{Temporal propagators in various vacua}
\subsection{ Free Wilson fermion and $\beta=100.0$}
In the limit $T/T_c \rightarrow \infty$ it is natural to consider the quark pair becomes a free quark pair.
Therefore we calculate the propagator of the PS channel using  the free Wilson quark propagator
in the vacuum: in all the four species of vacua,
shown in the previous section.

\clearpage

\begin{figure*}[htb]
\includegraphics[width=6.7cm]{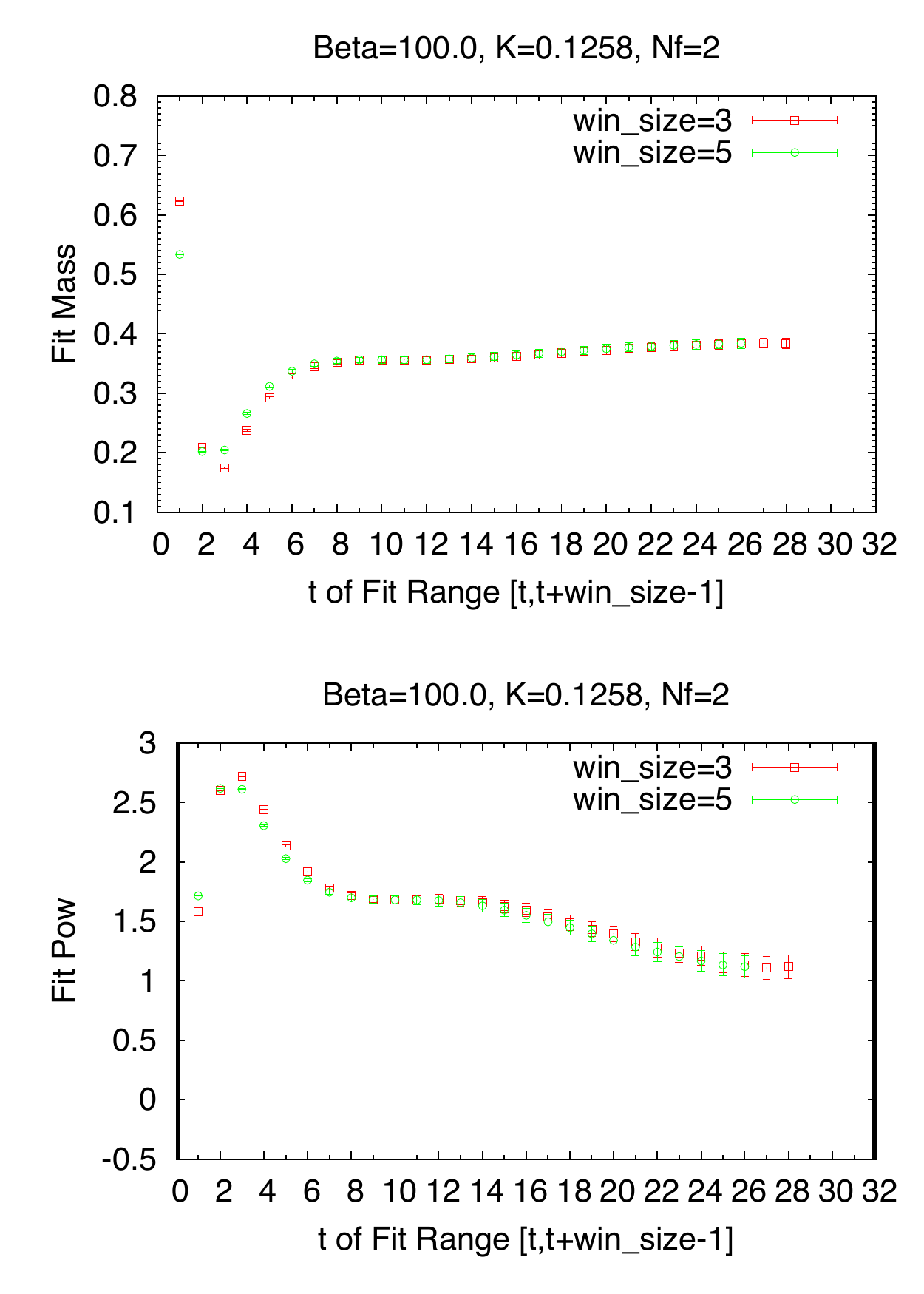}
\hspace{1cm}
   \includegraphics[width=6.7cm]{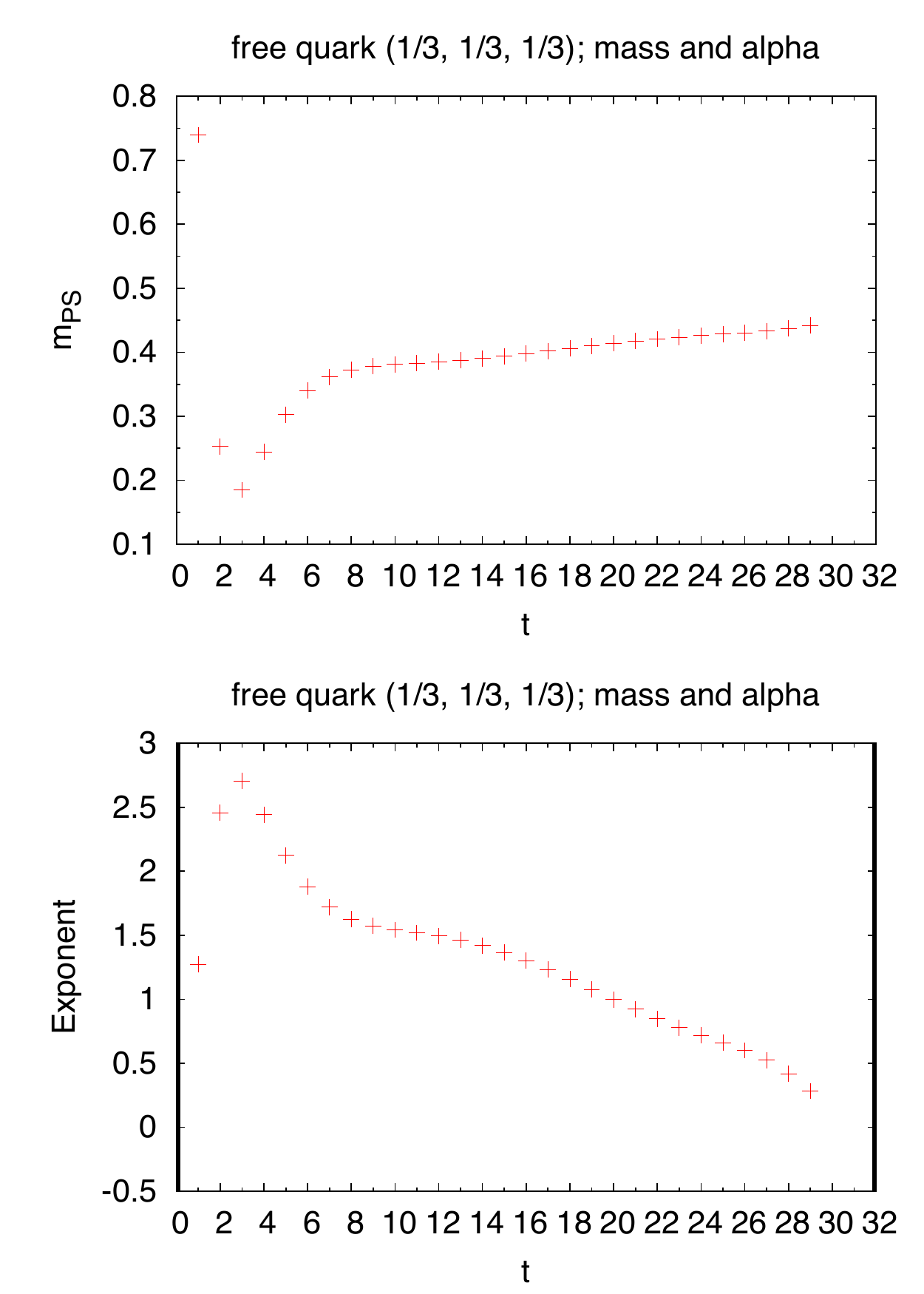}
\caption{(color online)  The local mass $m(t)$ and the local exponent $\alpha(t)$ at $\beta=100.0$ and $K=0.1258$ (left)  and a free particle (1/3,1/3,1/3) with $m_q=0.01$(right).}
\label{(1/3,1/3,1/3) free-beta100}
\end{figure*}

\begin{figure*}[thb]
\includegraphics [width=18cm]{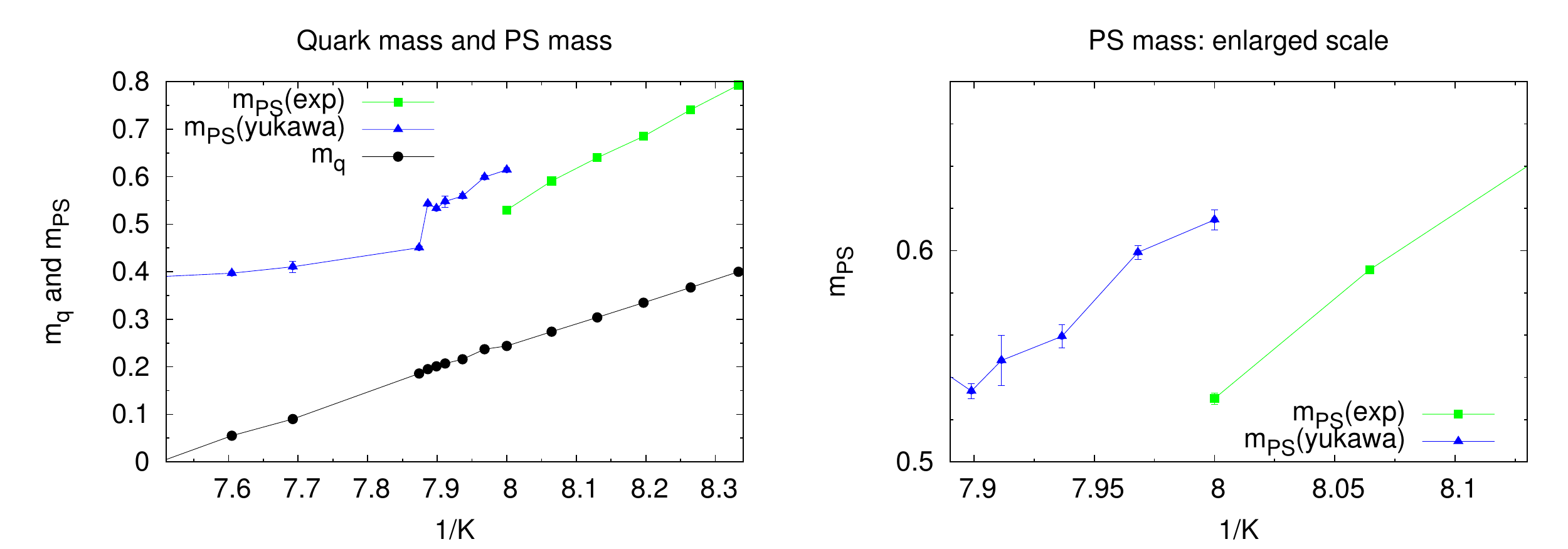}
\caption{(color online)  $m_q$ and $m_{PS}$ (or $\tilde{m}_{PS}$) vs. $1/K$ in the case $N_f=16$ for the range $0.130 \le K \le 0.1472$.
The transition region is enlarged on the right panel.}
\label{nf16_whole}
\end{figure*}

\begin{figure*}[htb]
\includegraphics [width=8cm]{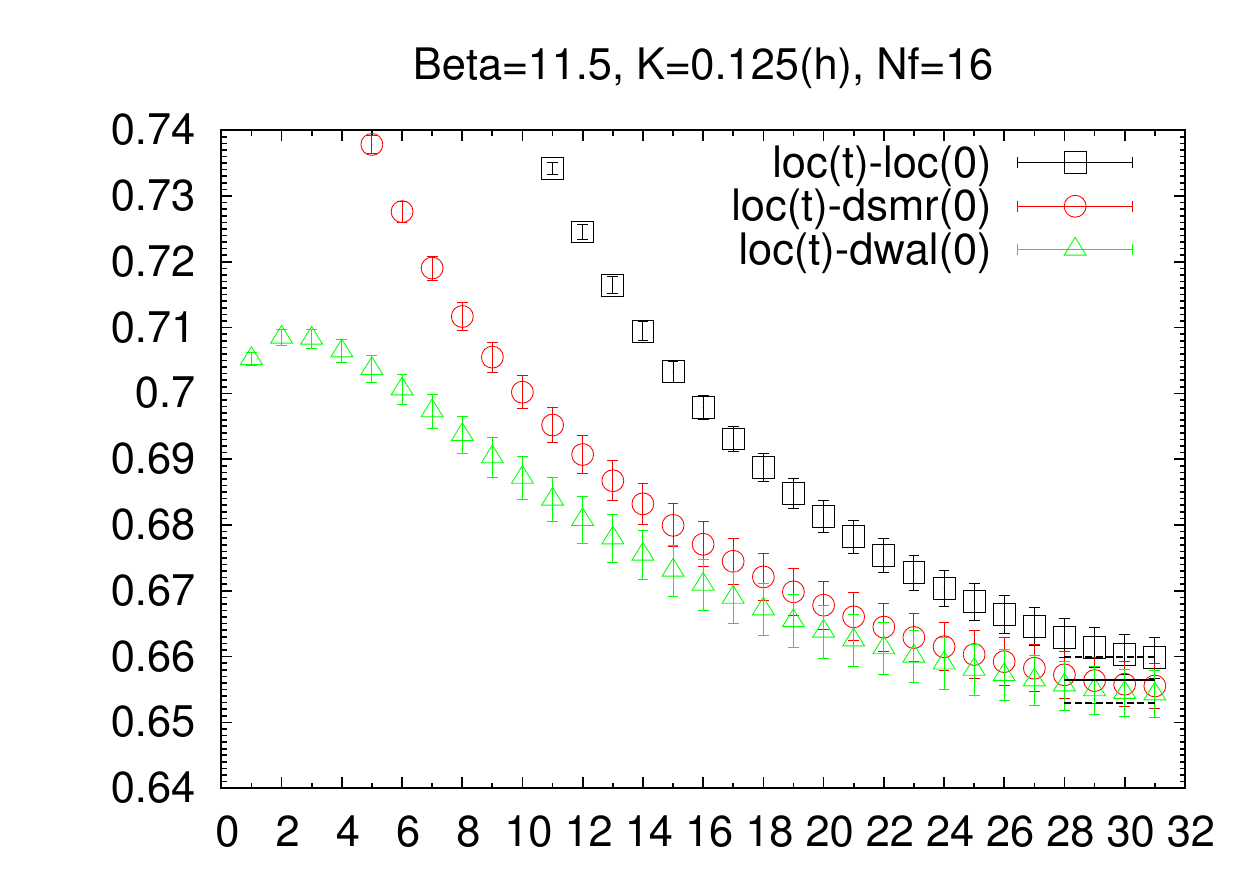}
%  \hspace{1cm}
\includegraphics [width=8cm]{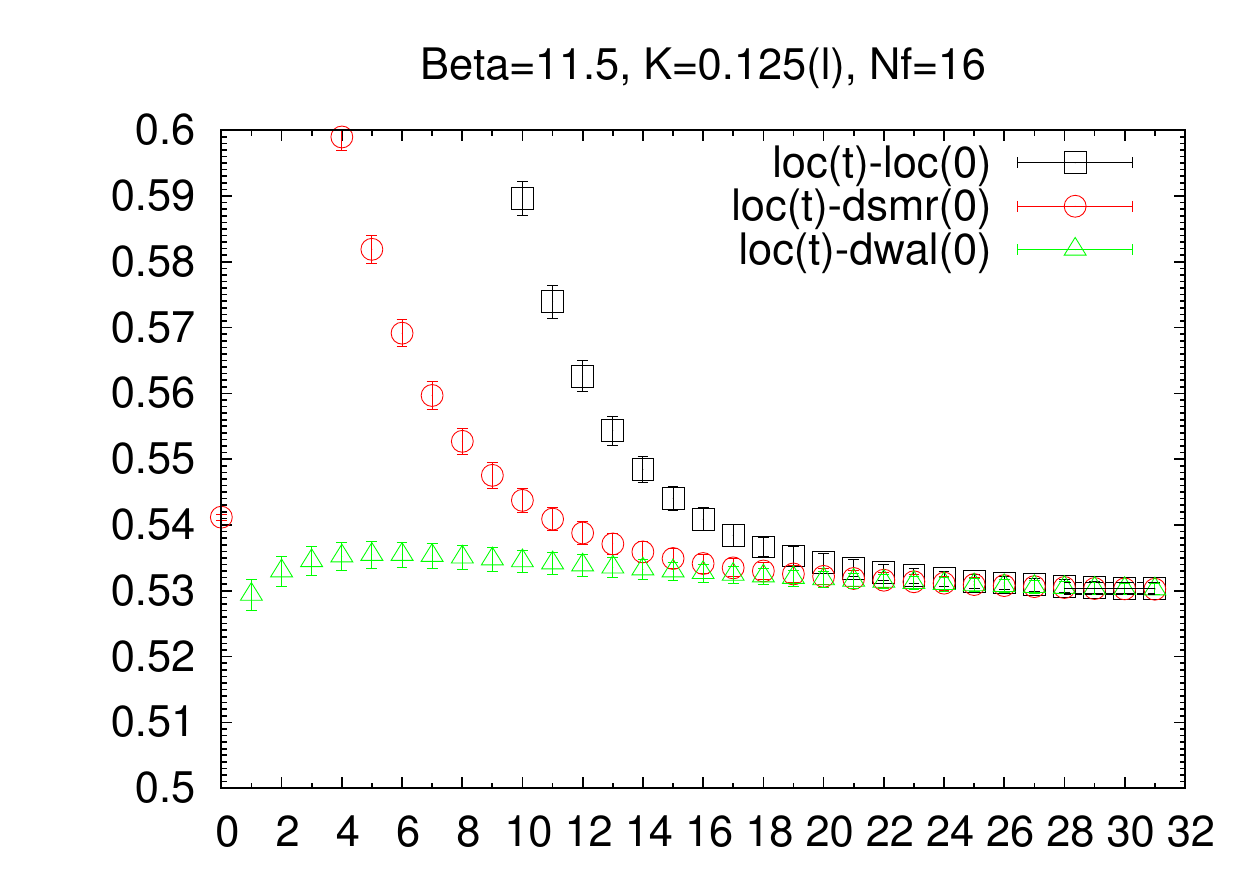}
\caption{(color online) The effective mass: both for $N_f=16$ at $\beta=11.5$ and $K=0.125$: (left) from larger $K$ and (right) from smaller $K$;
See the main text for the three types of sources.}
\label{nf16_effm}
\end{figure*}

\begin{figure*}[htb]
\includegraphics [width=7.5cm]{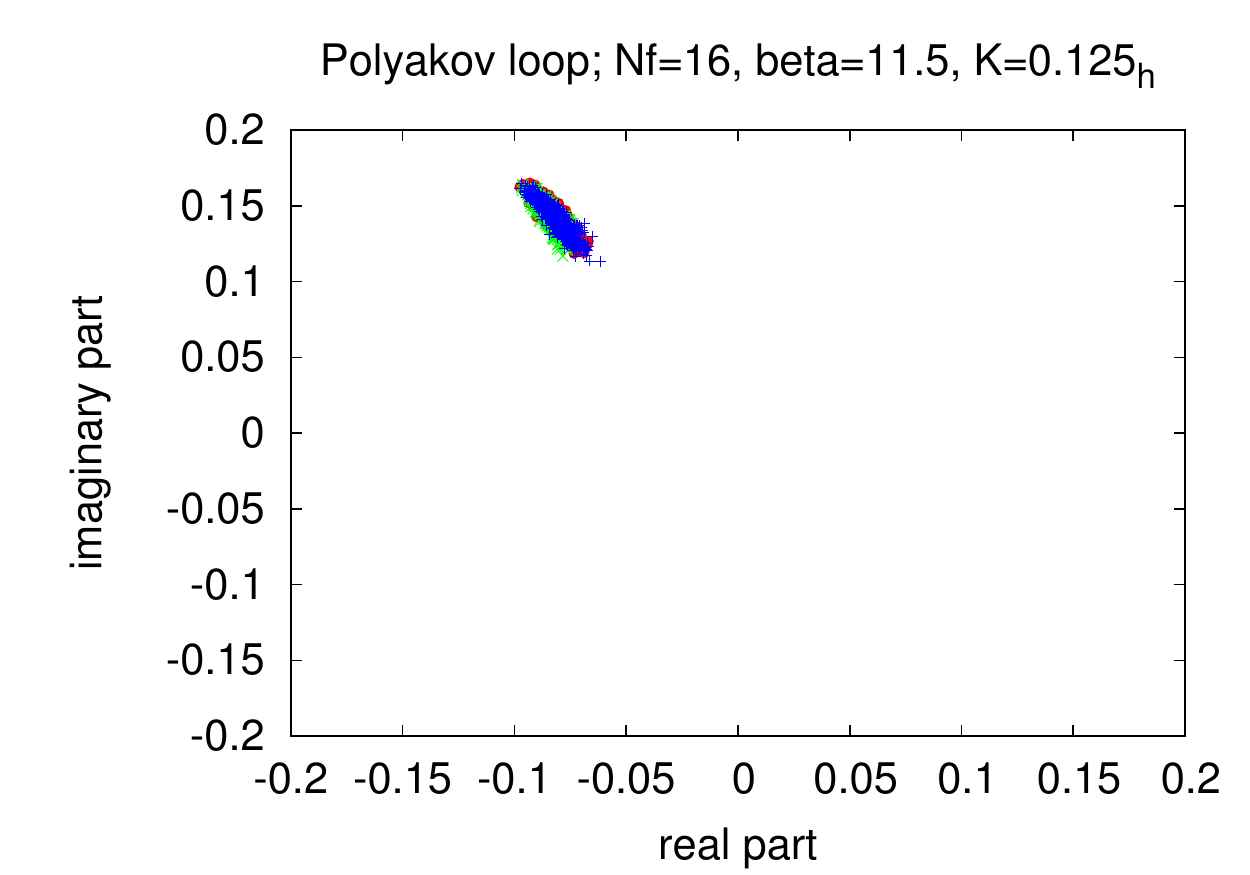}
 \hspace{1cm}
\includegraphics [width=7.5cm]{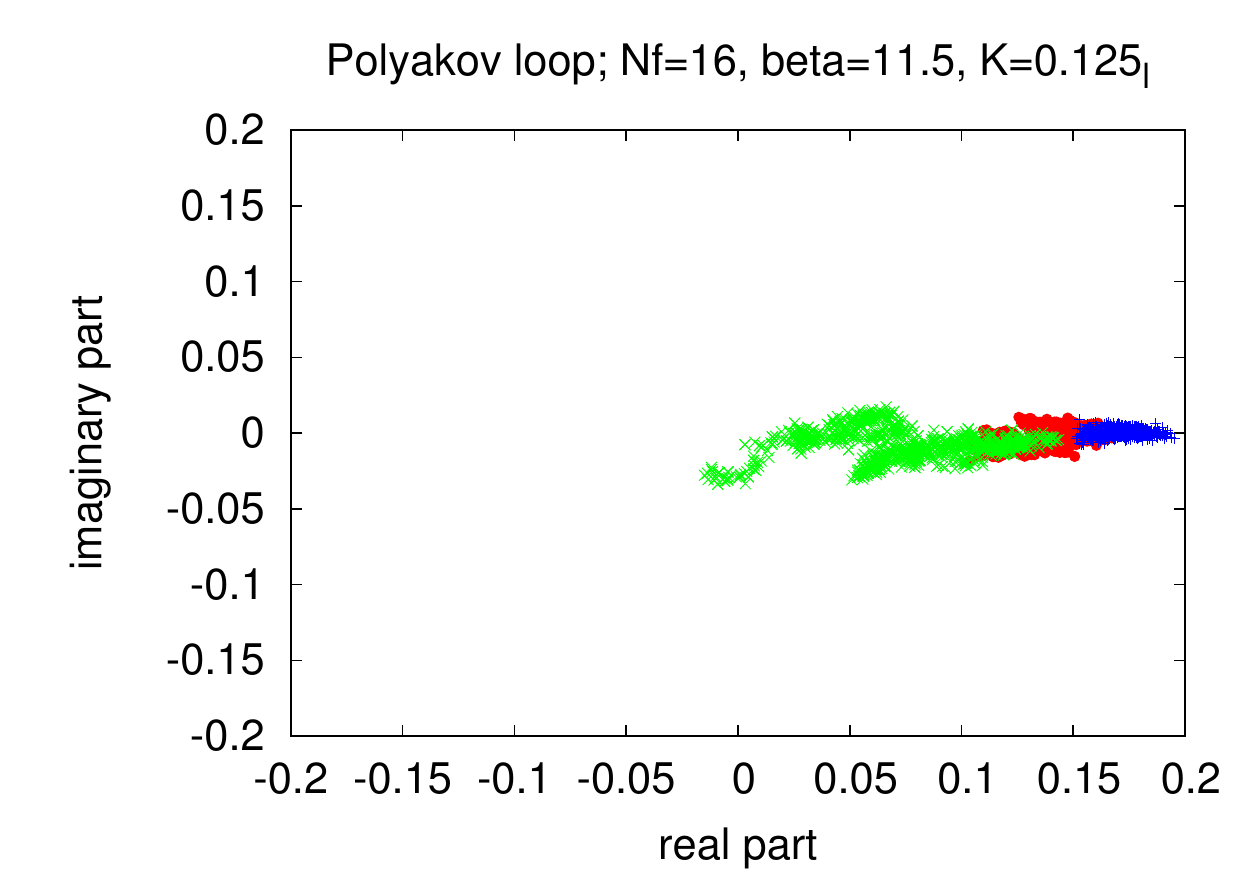}
\caption{(color online) The scattered plots of Polyakov loops in the $x$, $y$ and $z$ directions overlaid; 
        both for $N_f=16$ at $\beta=11.5$ and $K=0.125$: (left) from larger $K$ and (right) from smaller $K$.}
\label{nf16_comp}
\end{figure*}

\begin{figure*}[htb]
\includegraphics [width=18cm]{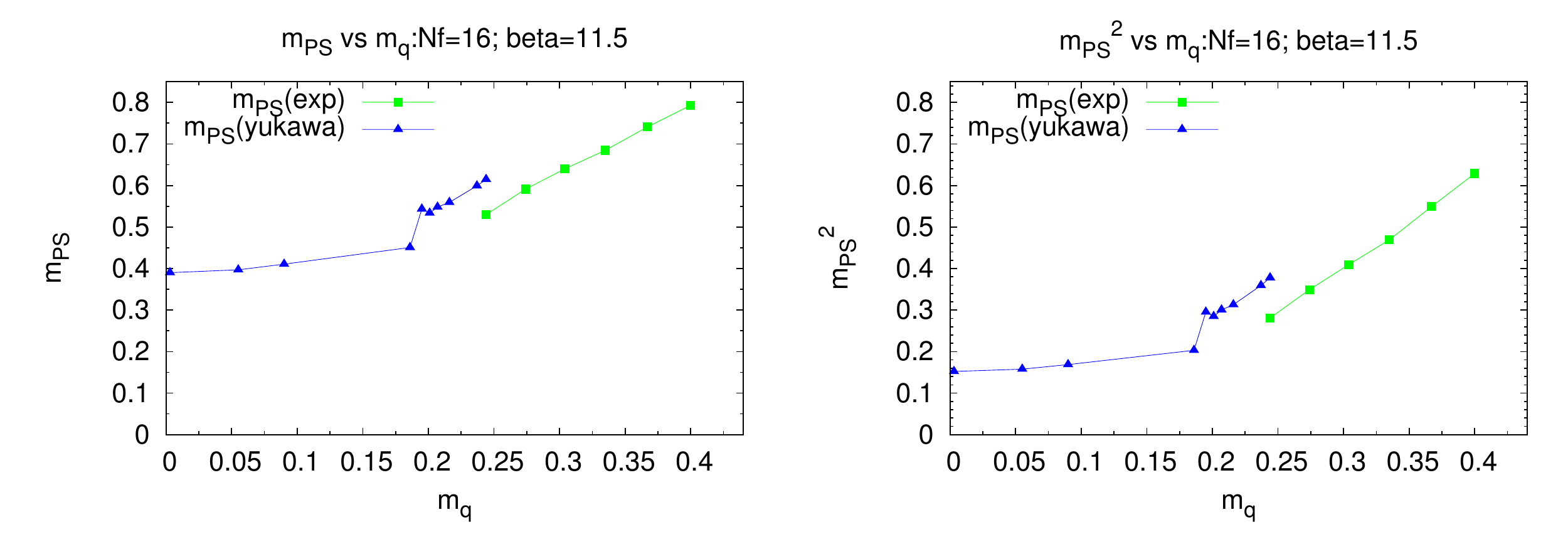}
\caption{(color online) The $m_{PS}$ (or $\tilde{m}_{PS}$) vs. $m_q$ for $N_f=16$: (left) linear $m_{PS}$ and (right) squared $m_{PS}$.}
\label{nf16_mass}
\end{figure*}

\begin{figure*}[htb]
   \includegraphics[width=7.5cm]{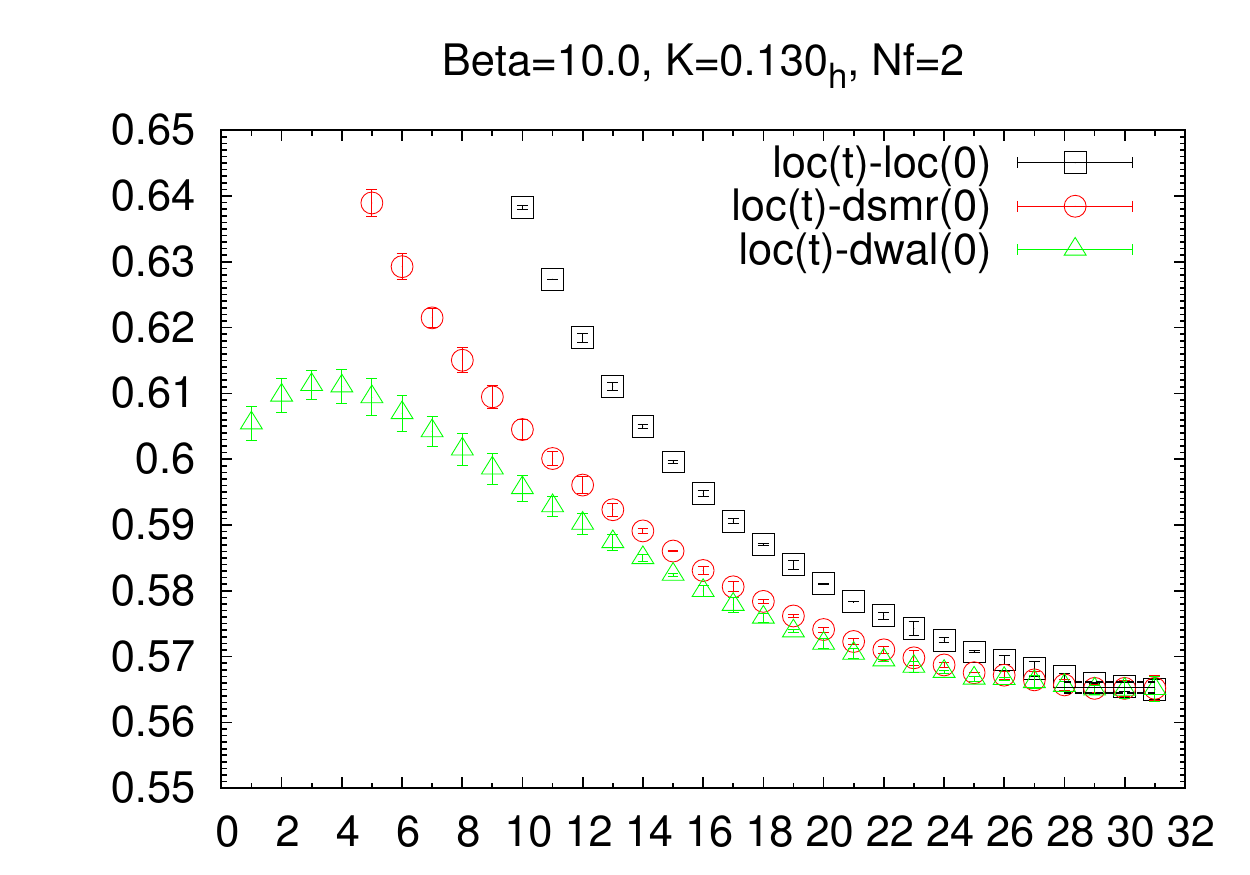}
    \hspace{1cm}
      \includegraphics[width=7.5cm]{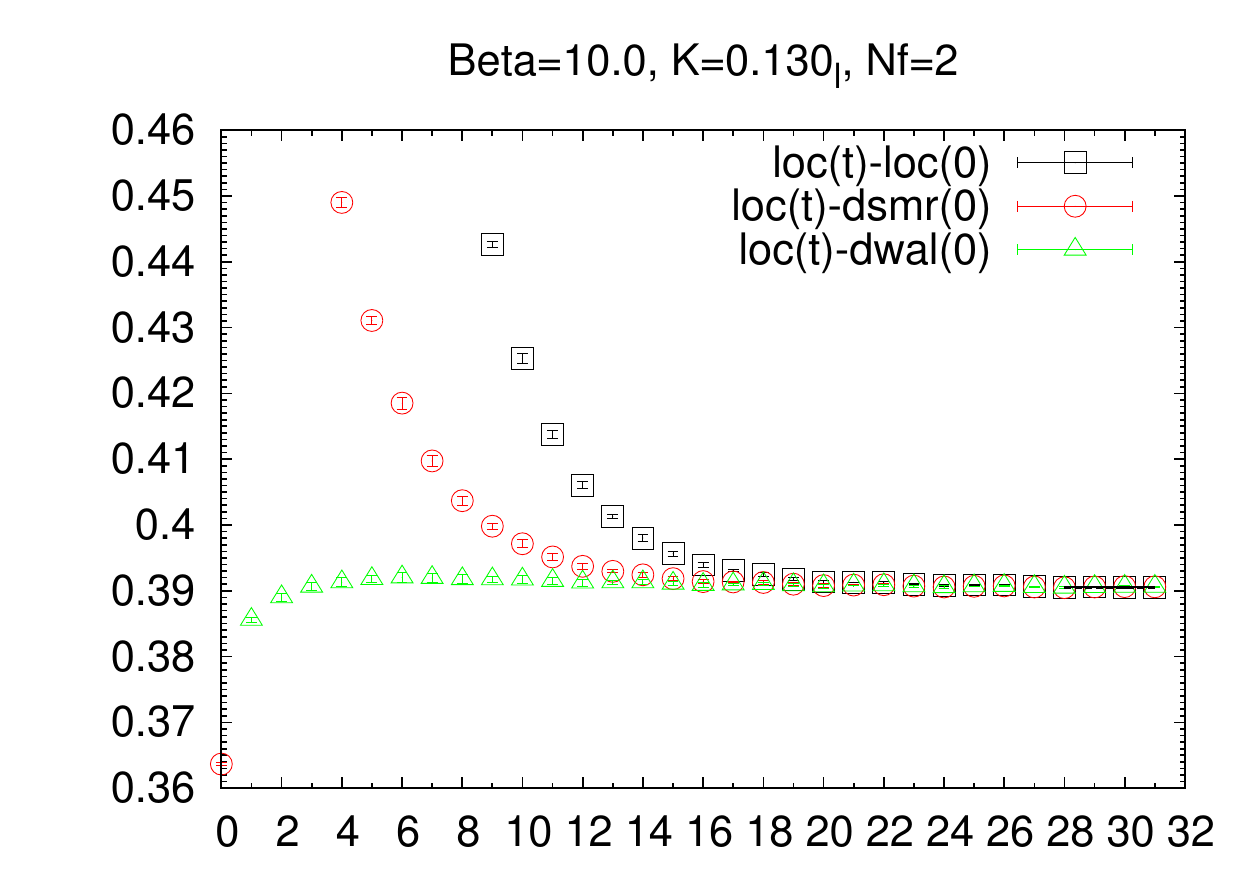}
       \caption{(color online) The effective mass both for $N_f=2$ at $\beta=10.0$ and $K=0.130$: (left) from larger $K$ and (right) from smaller $K$.}
               \label{effm_k130}
\end{figure*}

\begin{figure*}[thb]
   \includegraphics[width=7.5cm]{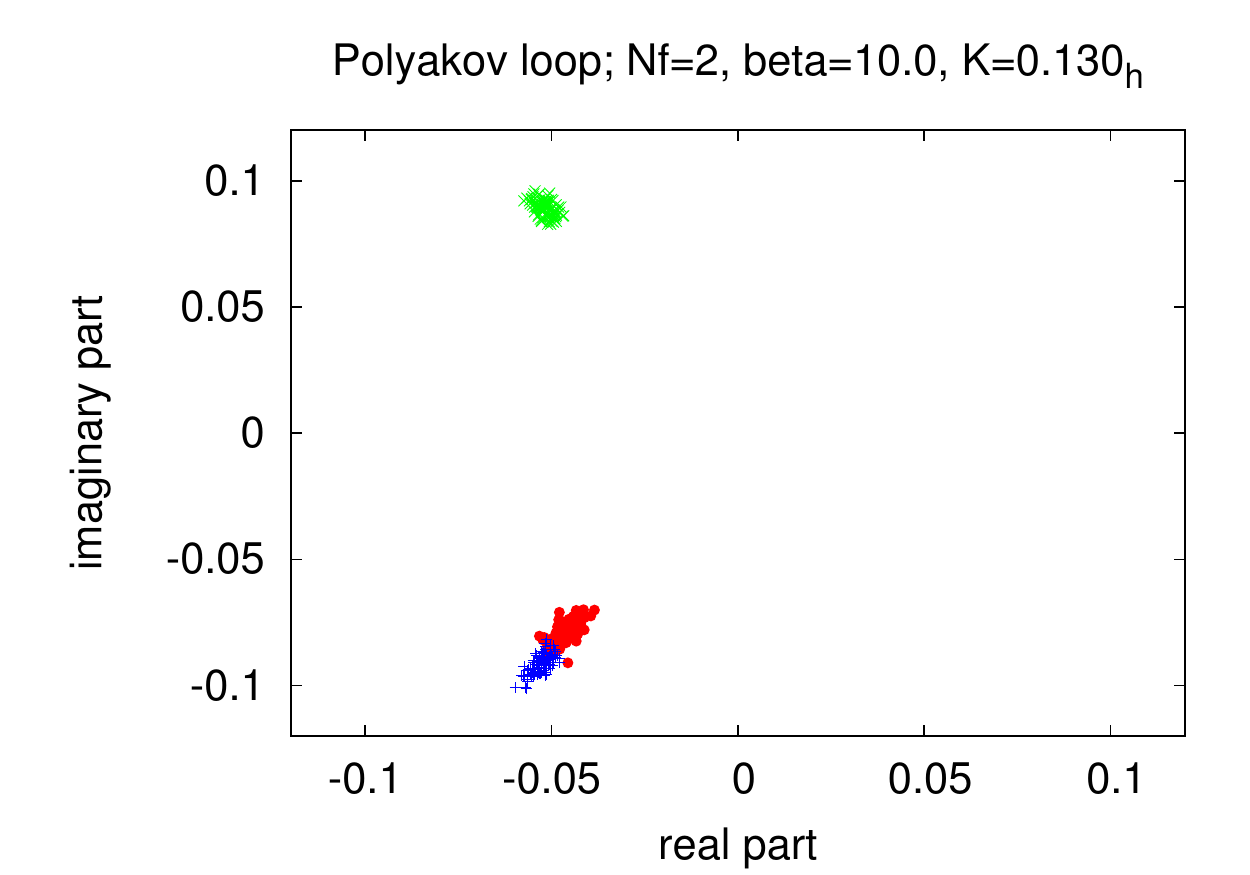}
    \hspace{1cm}
            \includegraphics[width=7.5cm]{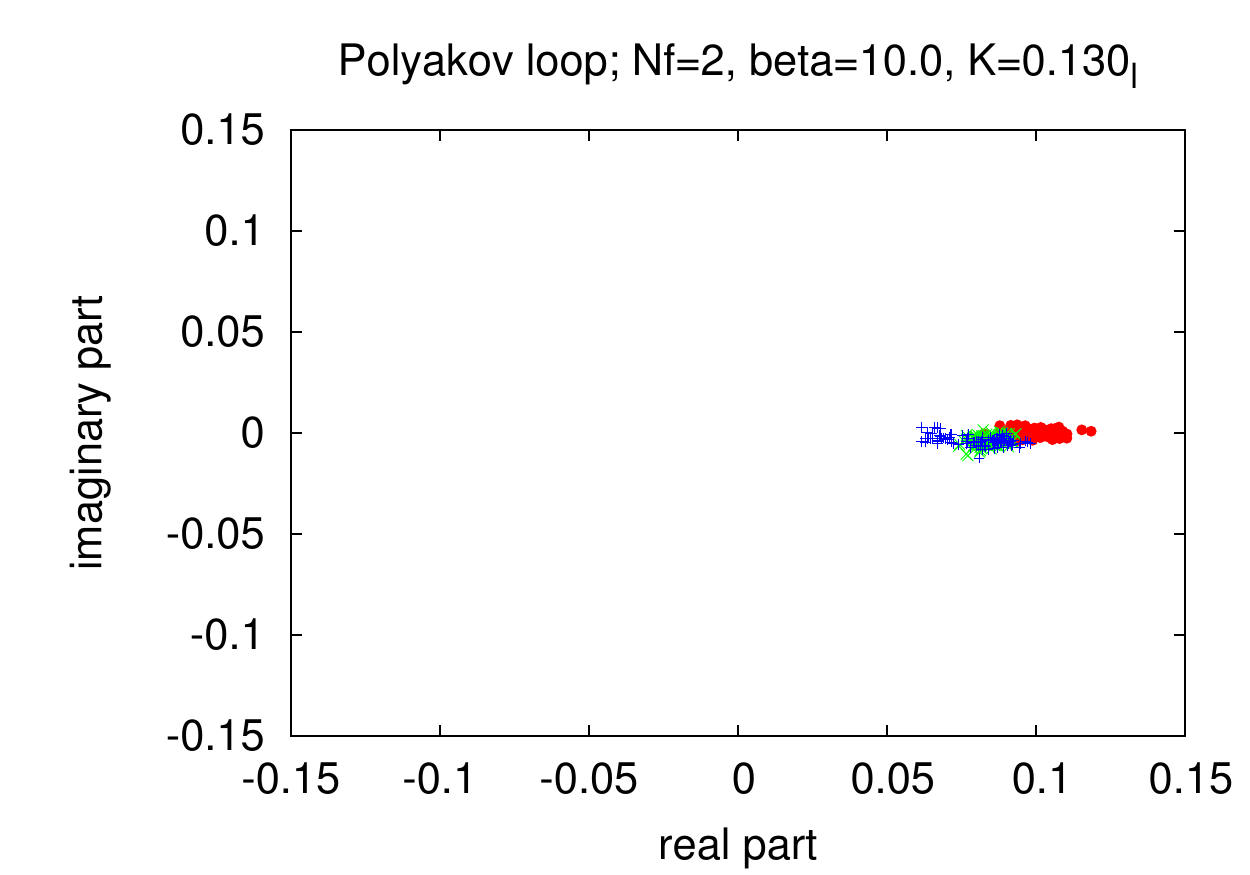}
                                       \caption{(color online) The scattered plots of Polyakov loops in the $x$, $y$ and $z$ directions overlaid; 
        both for $N_f=2$ at $\beta=10.0$ and $K=0.130$: (left) from larger $K$ and (right) from smaller $K$.}
                                             \label{complex_k130}
\end{figure*}

\begin{figure*}[thb]
\includegraphics [width=18cm]{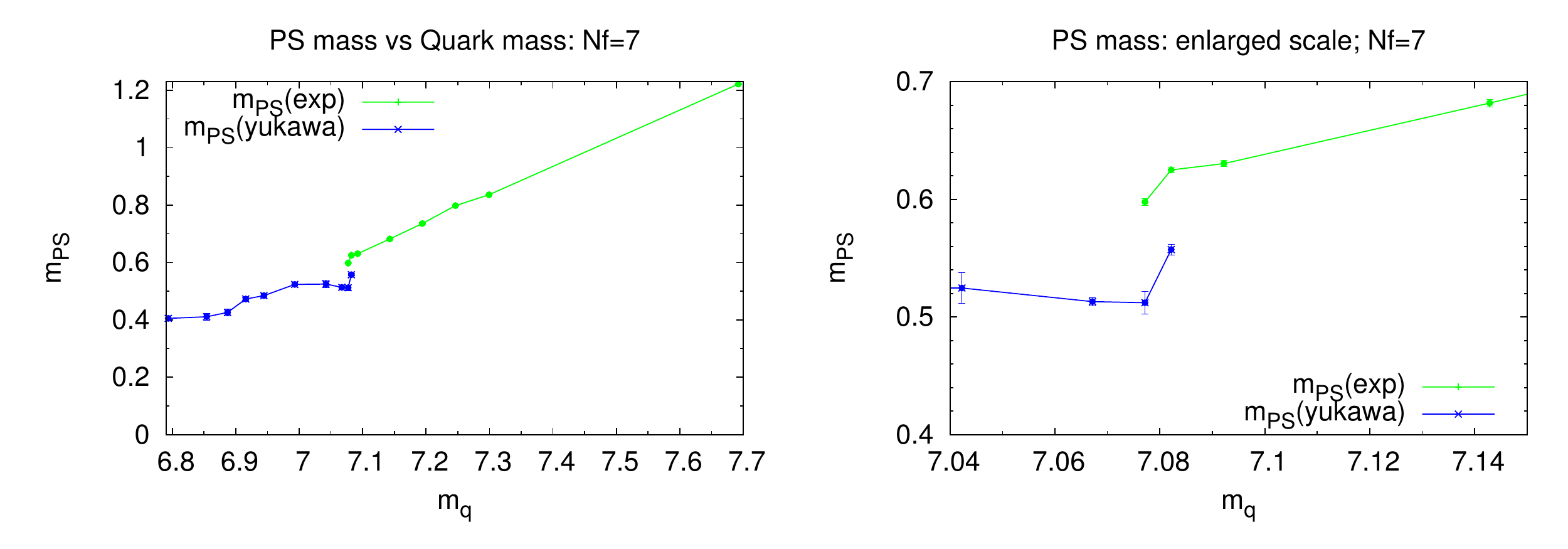}
\caption{(color online) $m_q$ and $m_{PS}$ (or $\tilde{m}_{PS}$) vs. $1/K$ in the case $N_f=7$ for the range $0.130 \le K \le 0.1472$.
The transition region is enlarged on the right panel.}
\label{nf7_mass}
\end{figure*}

\begin{figure*}[htb]
\includegraphics [width=7.5cm]{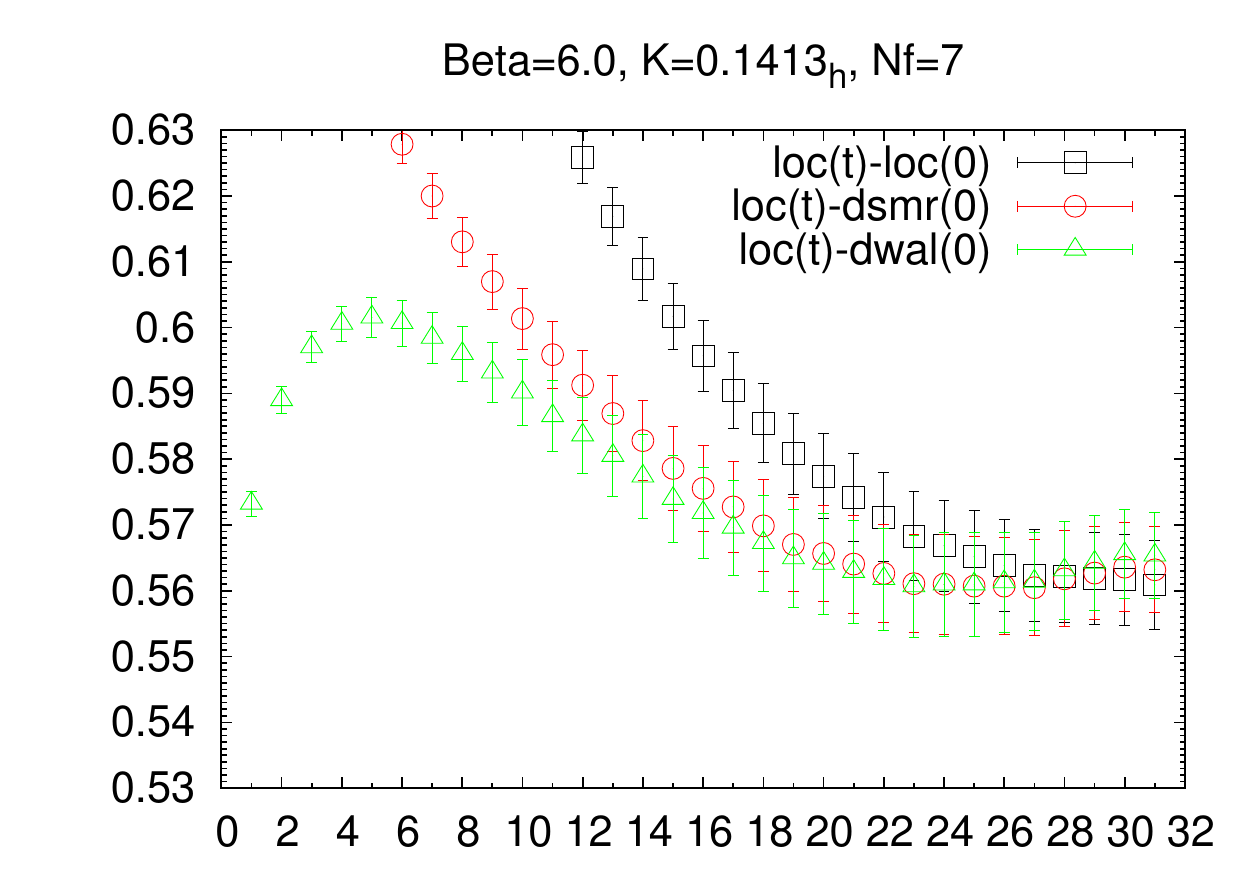}
  \hspace{1cm}
\includegraphics [width=7.5cm]{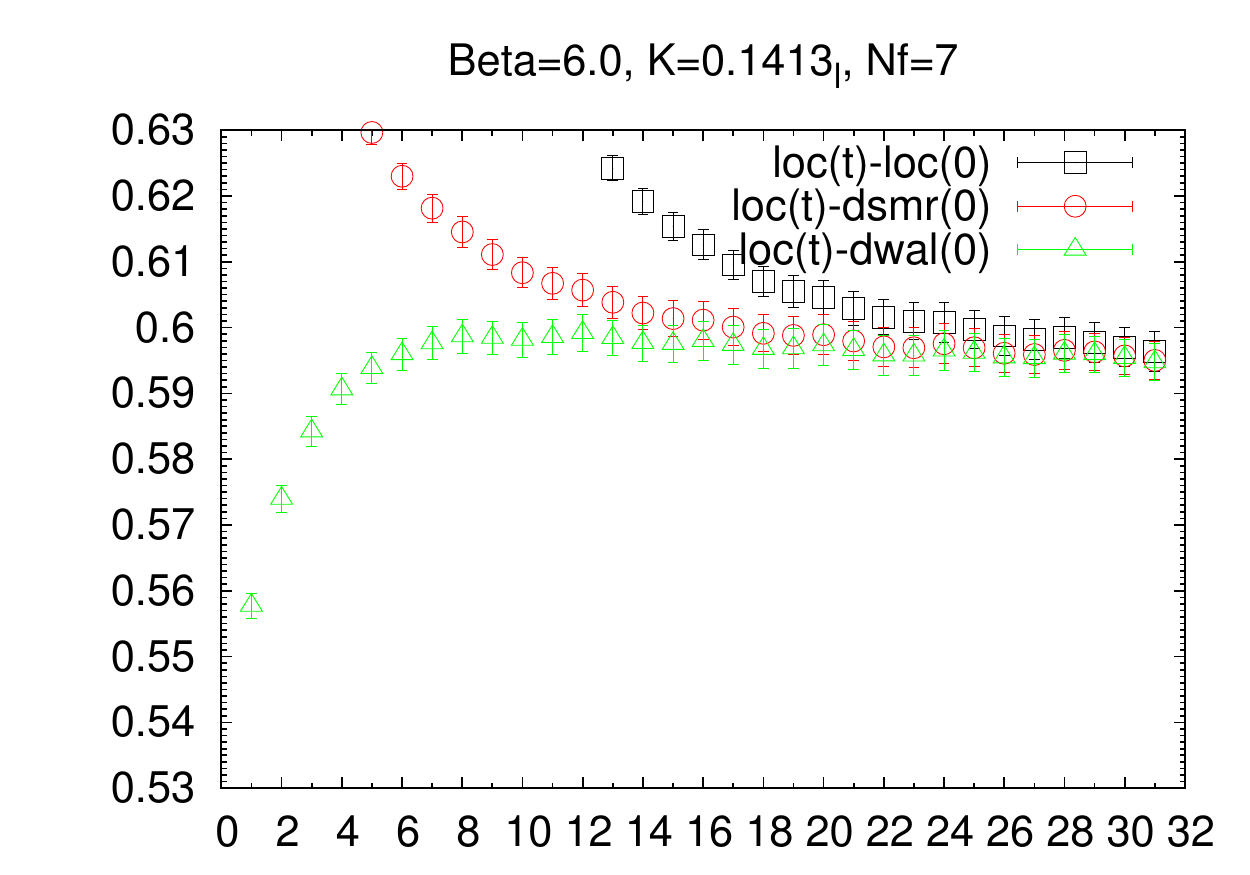}
\caption{(color online) The effective mass both for $N_f=7$ at $\beta=6.0$ and $K=0.1413$: (left) from larger $K$ and (right) from smaller $K$.}
\label{nf7_effm}
\end{figure*}

\begin{figure*}[htb]
\includegraphics [width=18cm]{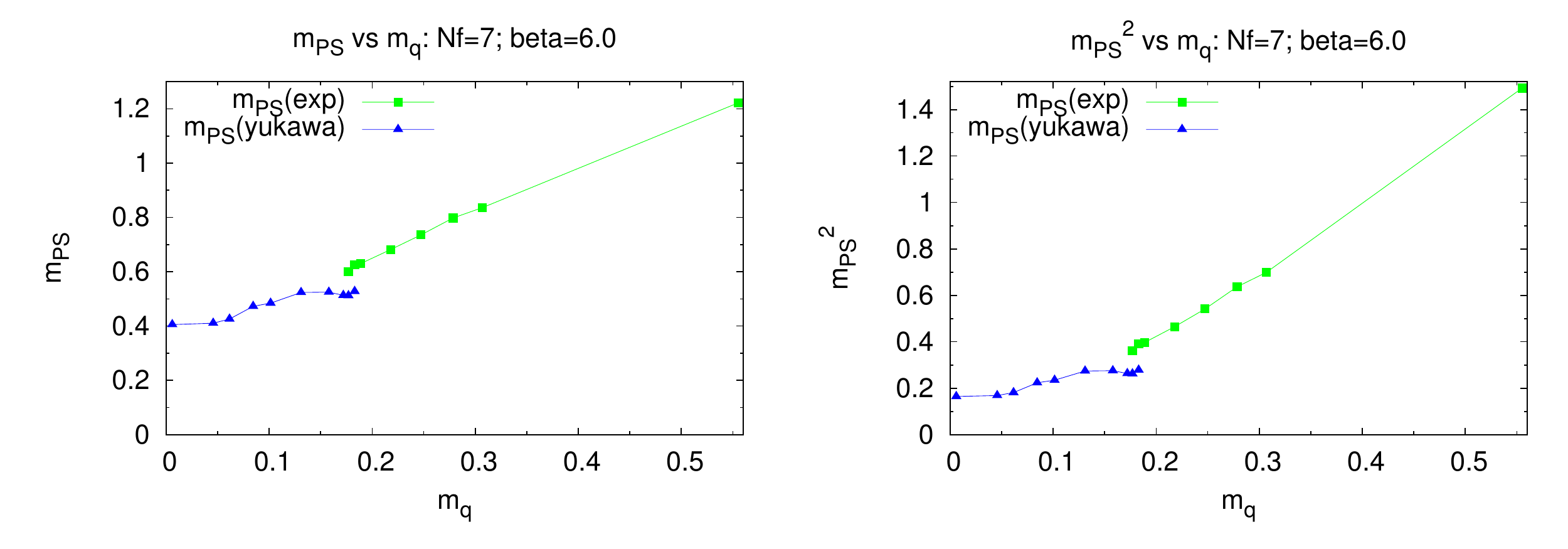}
\caption{(color online) The $m_{PS}$ (or $\tilde{m}_{PS}$) vs. $m_q$ for $N_f=7$:(left) linear $m_{PS}$ and (right) squared $m_{PS}$.}
\label{nf7_mass2}
\end{figure*}

%\red{ how generally hold the statement that the lowest energy state is  $(\exp{(\pm 2\pi/3}), \exp{(\pm 2\pi/3}), \exp{(\pm 2\pi/3})$ 

%\subsection{Polaykov loops and confinement}
%\blue{While the expectation value of the Polyakov loop determines the non-trivial vacuum structure of the perturbative QCD on the lattice, it also serves as the order parameter for the confinement...}

\begin{figure*}[thb]
\includegraphics [width=7.5cm]{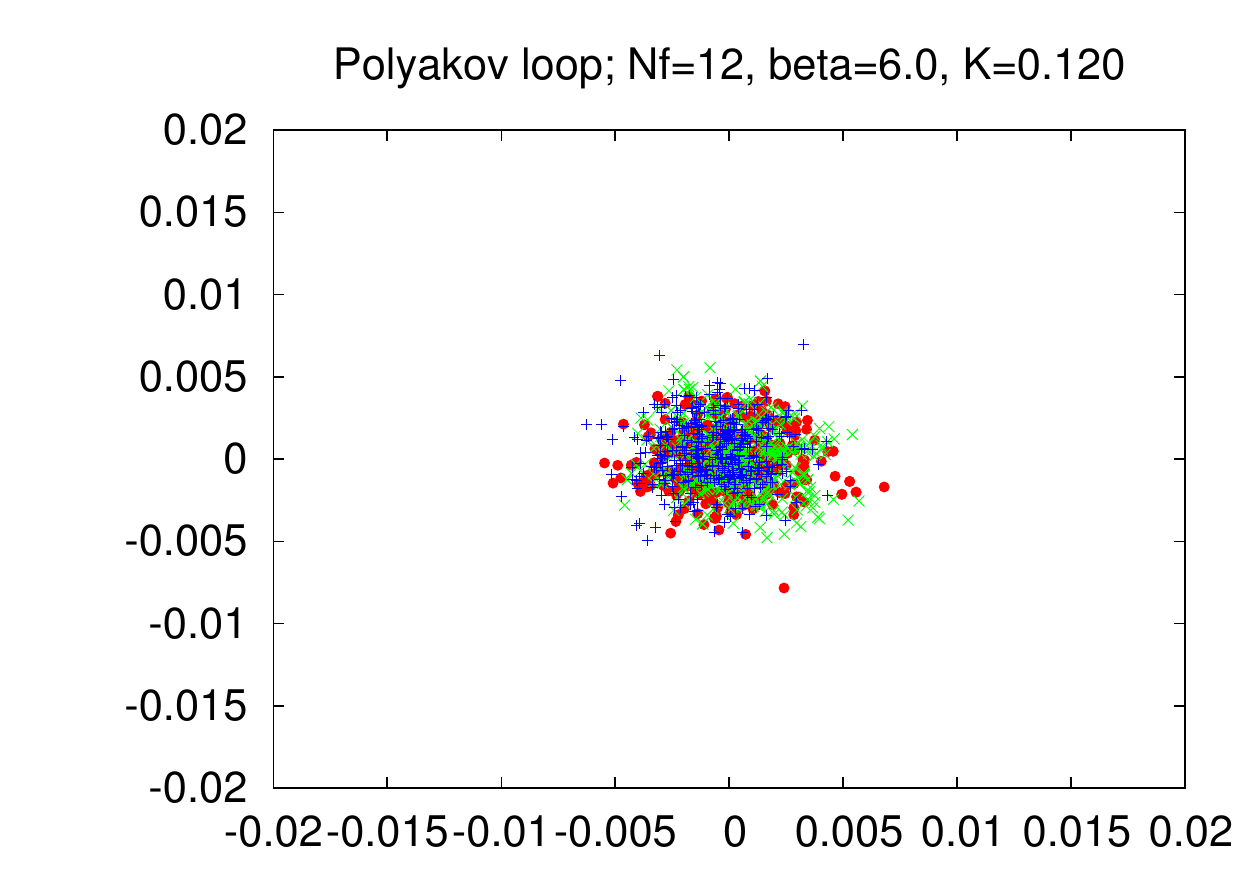}
  \hspace{1cm}
\includegraphics [width=7.5cm]{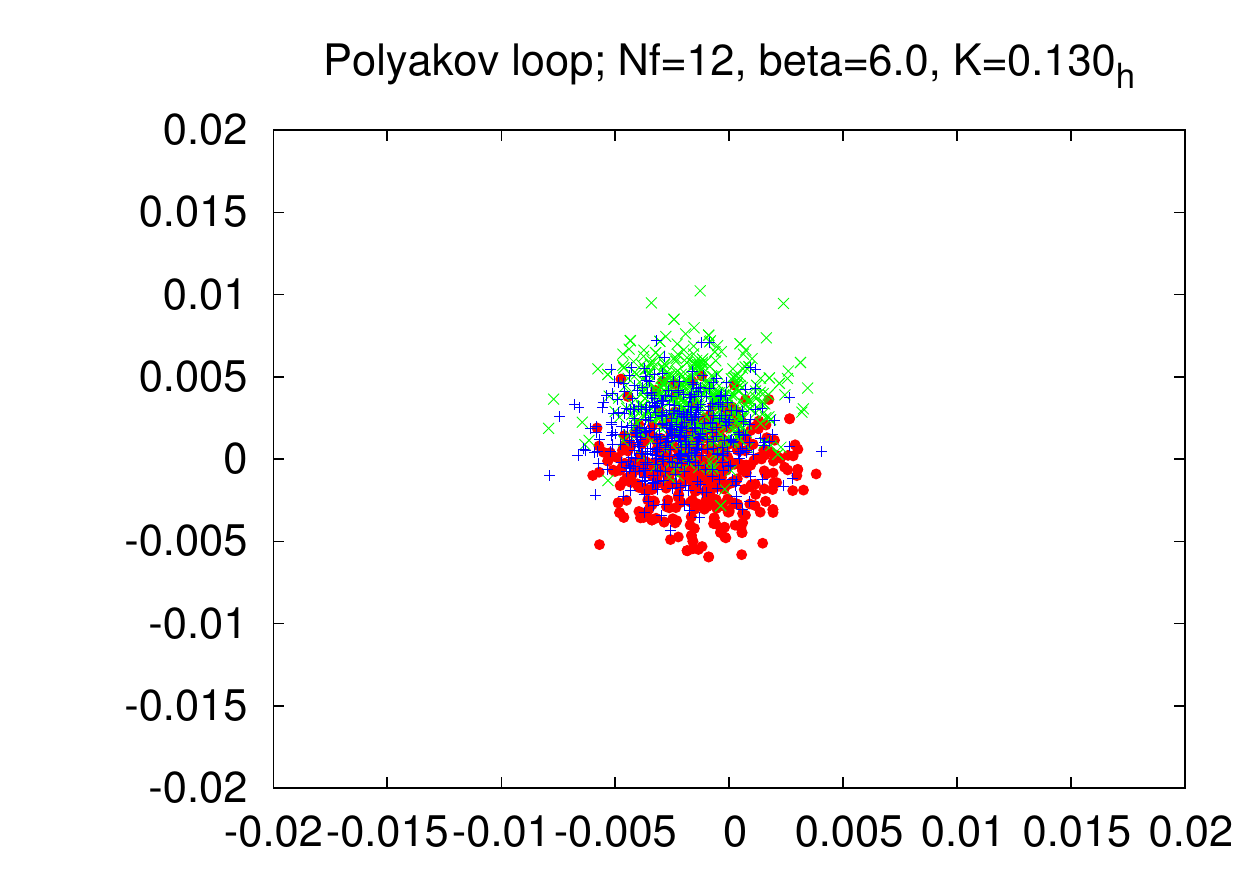}
\caption{(color online) The scattered plots of Polyakov loops in the $x$, $y$ and $z$ directions overlaid; 
        both for $N_f=12$ at $\beta=6.0$: (left) $K=0.120$ and (right) $K=0.130$.}
\label{complex_nf12_beta6.0}
\end{figure*}

\begin{figure*}[htb]
\includegraphics [width=7.5cm]{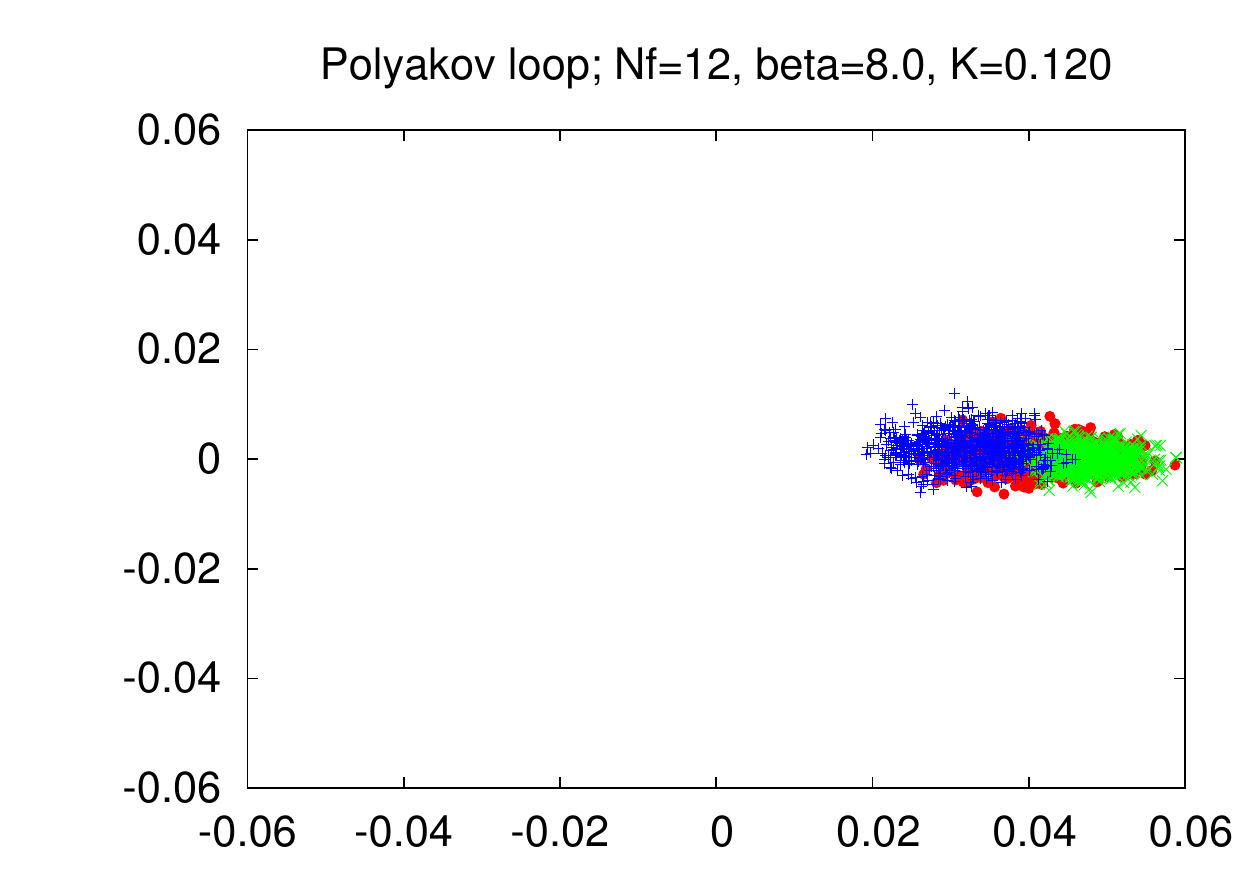}
  \hspace{1cm}
\includegraphics [width=7.5cm]{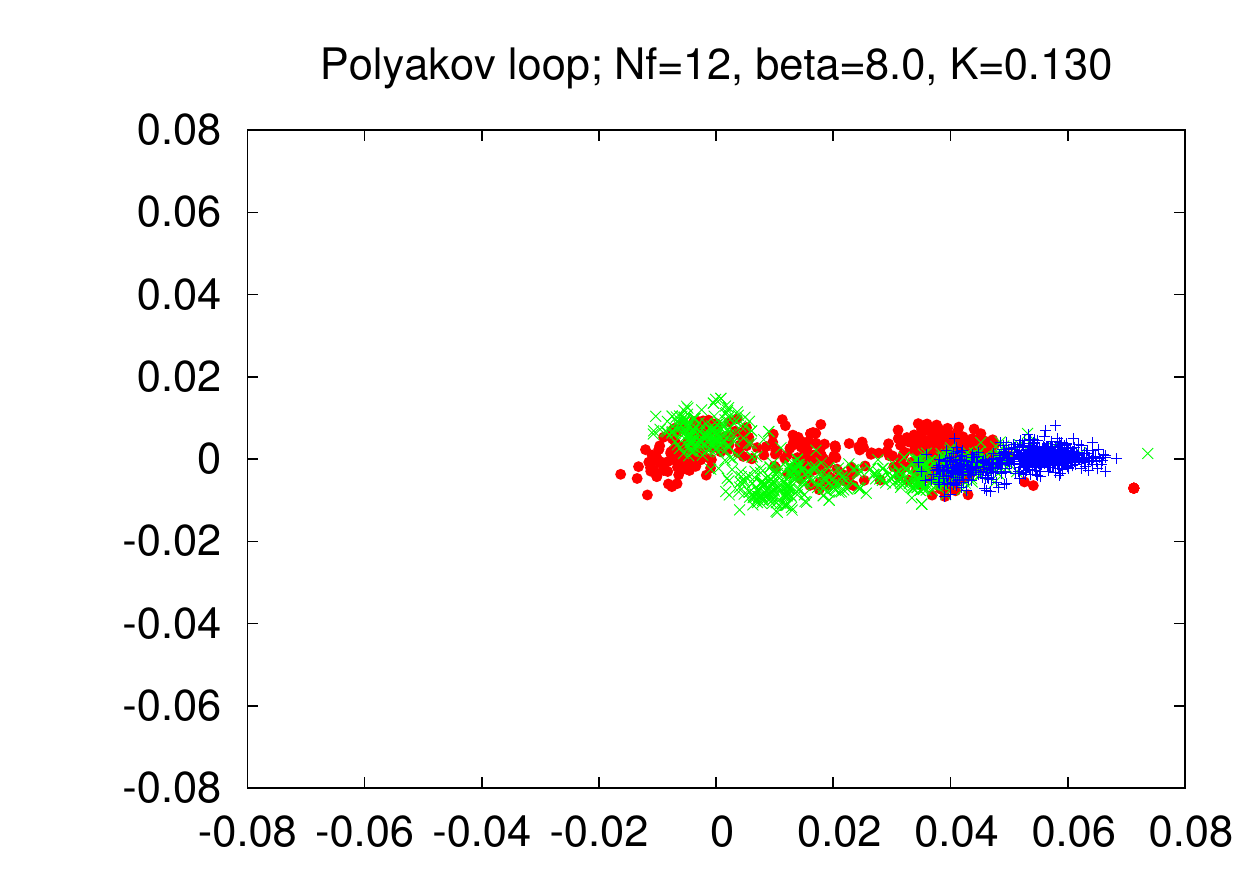}
\caption{(color online) The scattered plots of Polyakov loops in the $x$, $y$ and $z$ directions overlaid; 
        both for $N_f=12$ at $\beta=8.0$: (left) $K=0.120$ and (right) $K=0.130$.}
\label{complex_nf12_beta8.0}
\end{figure*}

\begin{figure*}[htb]
\includegraphics [width=18cm]{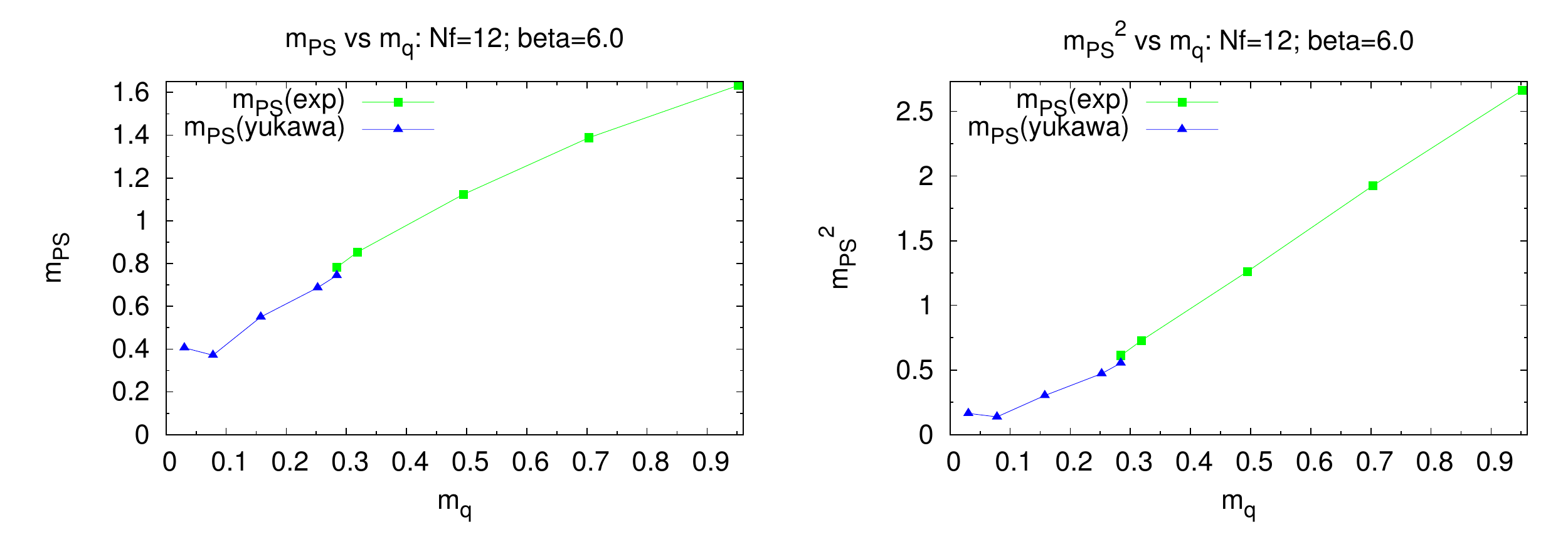}
\caption{(color online) The $m_{PS}$ (or $\tilde{m}_{PS}$) vs. $m_q$ for $N_f=12$ and $\beta=6.0$: left; linear $m_{PS}$ and right; squared $m_{PS}$.}
\label{nf12_beta6.0_mass}
\end{figure*}

\begin{figure*}[htb]
\includegraphics [width=18cm]{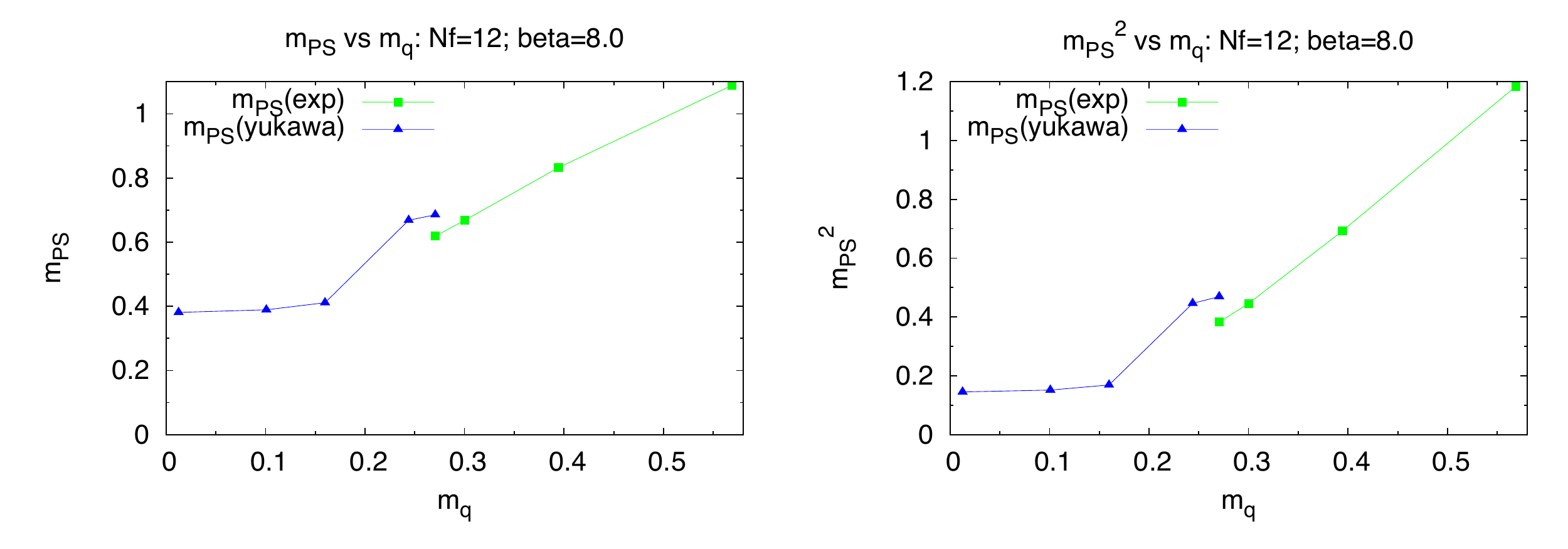}
\caption{(color online) The $m_{PS}$ (or $\tilde{m}_{PS}$) vs. $m_q$ for $N_f=12$ and $\beta=8.0$: left; linear $m_{PS}$ and right; squared $m_{PS}$.}
\label{nf12_beta8.0_mass}
\end{figure*}

\begin{figure*}[thb]
\includegraphics [width=7.5cm]{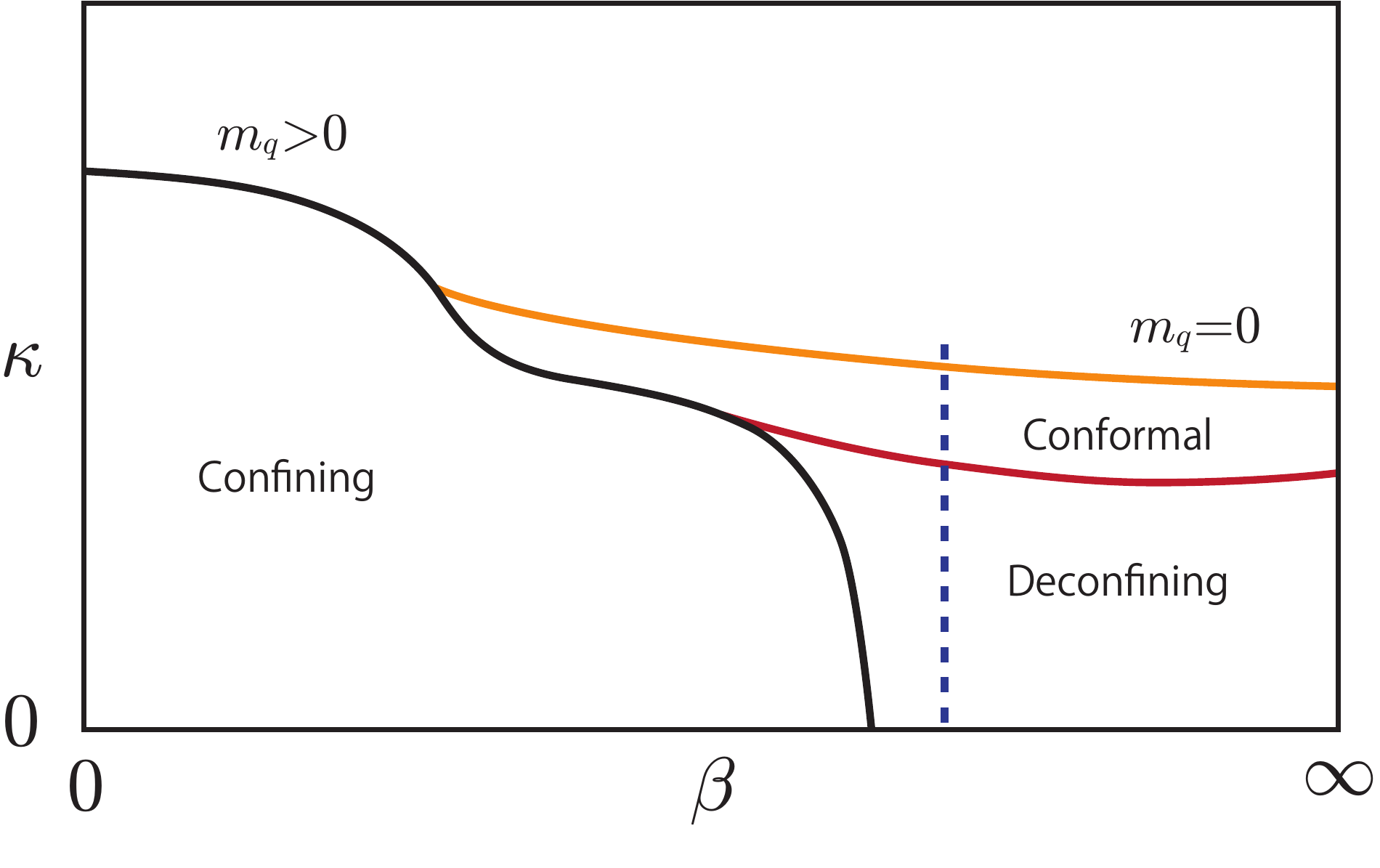}
 \hspace{1cm}
 \includegraphics [width=7.5cm]{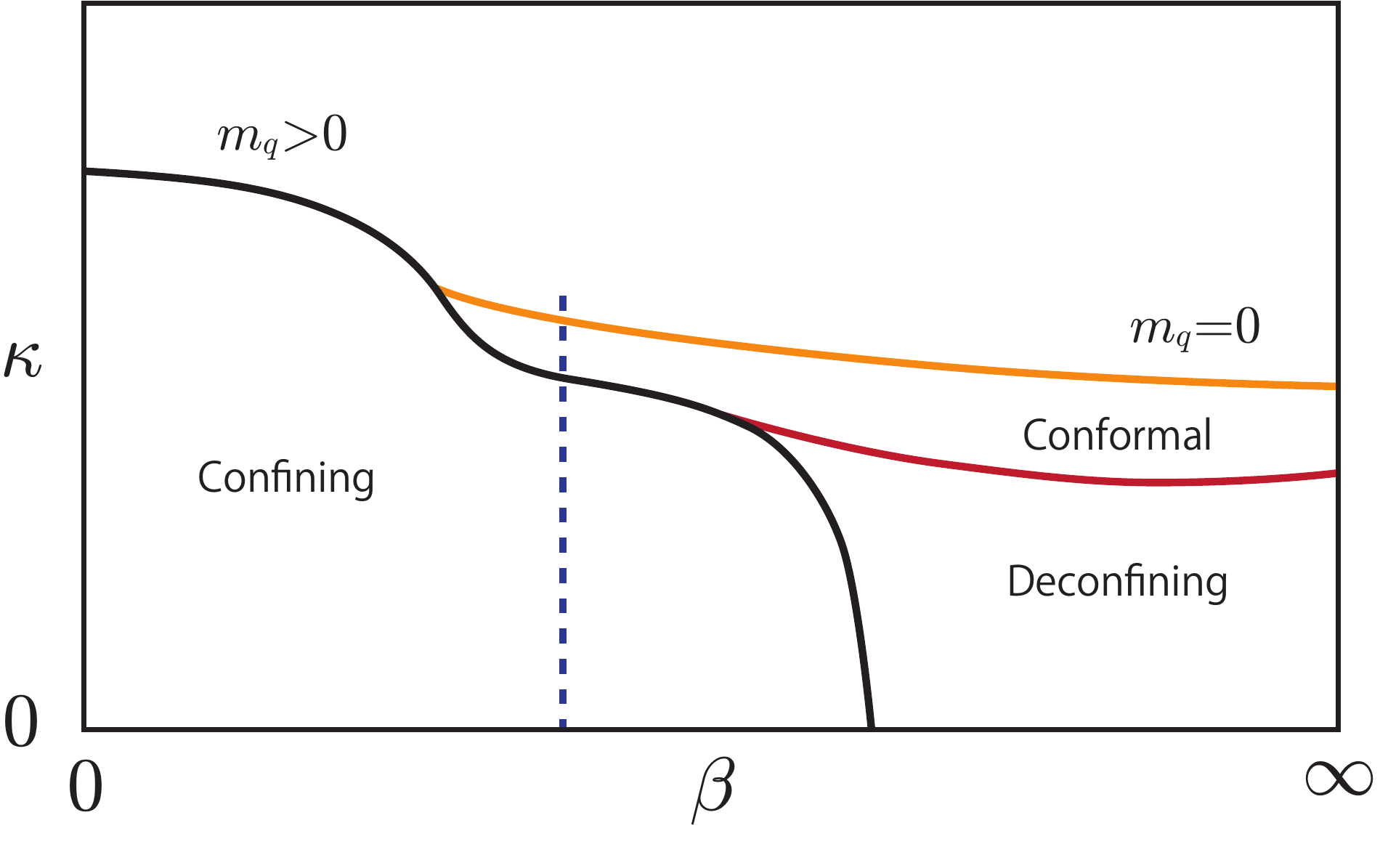}
\caption{(color online) The phase diagram on a finite lattice for $N_f^c \le N_f \le 16$: the solid line toward the quench QCD $K=0$ is
the boundary between the deconfining and confining region. The dashed line represents the simulation line. The massless line hits the bulk transition point at finite $\beta$.}
\label{conformal_region_1}
\end{figure*}

\begin{figure}[htb]
\includegraphics [width=7.5cm]{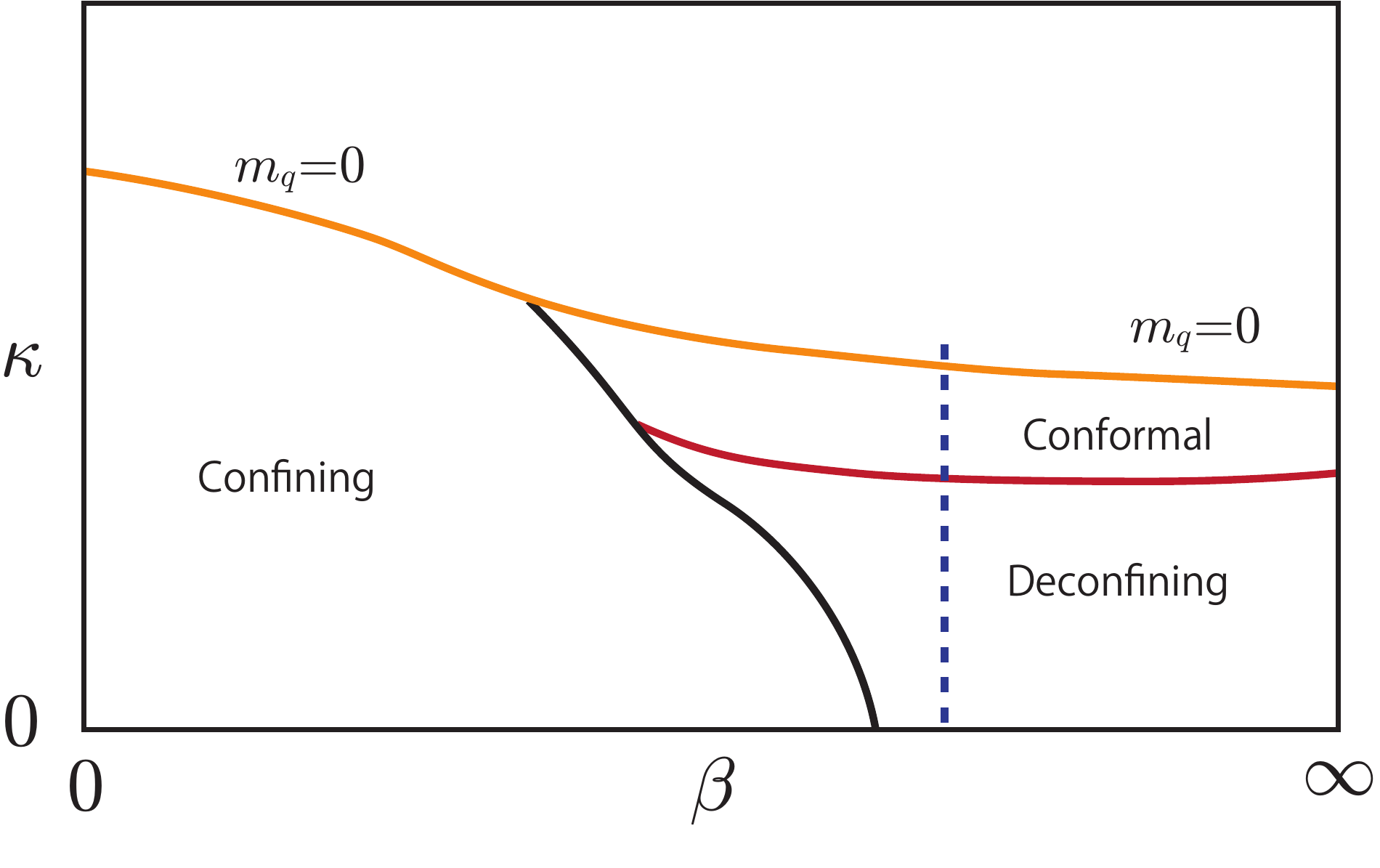}
 \caption{(color online) The phase diagram on a finite lattice for $1 \le N_f^c \le -1$:  the solid line toward the quench QCD $K=0$ is
the boundary between the deconfining and confining region. The dashed line represents the simulation line.  The massless line runs through from  $\beta=\infty$ to $\beta=0$.}
\label{conformal_region_3}
\end{figure}

\begin{figure}[thb]
   \includegraphics[width=7.5cm]{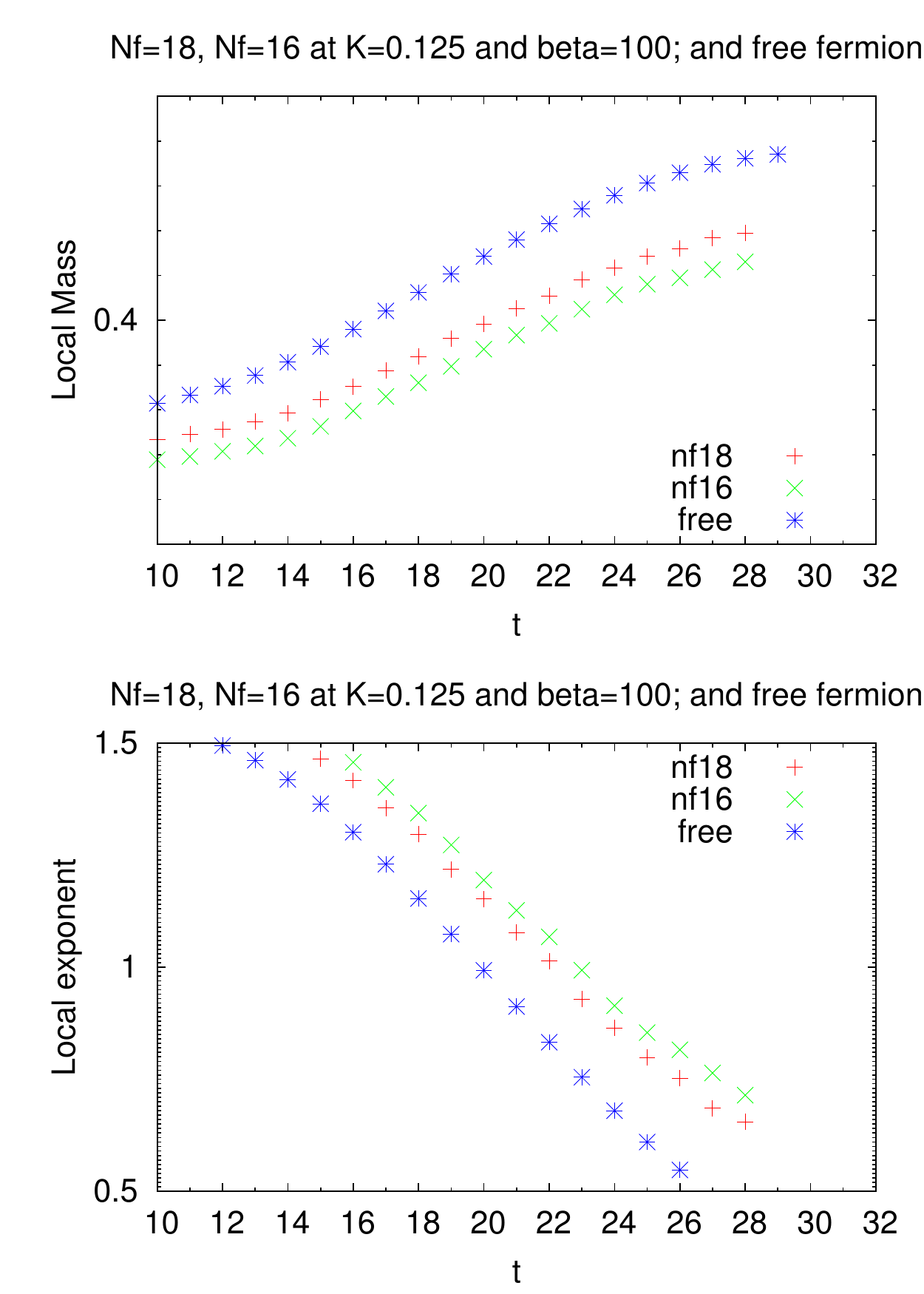}
               \caption{(color online) The local mass $m(t)$ and the exponent $\alpha(t)$ in the large $t$ region for $N_f=16, 18$ and the free case.}
            \label{nf188-nf16-free}
\end{figure}

We also calculate the meson propagators for $N_f=2$, at $\beta=100.0$ and $K=0.1258$,
which correspond to $T\sim 10^{56} T_c$ and $m_q=0.015$

We have performed simulations first at a small value for the temperature $T/T_c\simeq 2$
and gradually increased the temperature. 
At small values of temperatures the transition among the four vacua often occurs,
because the barriers among them are low, and therefore the lowest energy state is chosen during the simulations. 
Then gradually increase the temperature up to  $\beta=100.0$.
In this way the state becomes $(1/3, 1/3, 1/3)$.

Alternatively,  choosing a configuration in the quenched QCD as the initial state and varying the simulation parameters in several ways, we are also able to obtain the state
at $\beta=100.0$ and $K=0.1258$, with % the Polyakov loops 
$(0, 1/3, 1/3)$ and $(0, 0, 1/3)$.
However, we are unable to obtain the state with $(0, 0, 0) .$
This is consistent with the analysis of the vacuum energy that the state $(0, 0, 0)$
is locally unstable, when $m_q\le 0.15$.

Let us show on the right side of Figs.(\ref{(1/3,1/3,1/3) free-beta100}), ( (\ref{(1/3,1/3, 0) free-beta100}), (\ref{(1/3, 0, 0) free-beta100}), and  (\ref{(0, 0, 0) free-beta100}) in Appendix F),
the $m(t)$ and $\alpha(t)$ of the free fermion state with $m_q=0.01$ calculated in the four species of the vacuum,
together with the results of Monte Carlo calculations at $\beta=100.0$ with $K=0.1258$ which
corresponds to $m_q=0.0015$ on the left side.
The  figures corresponding  to  the state $ ( 0, 0,0) $ at $\beta=100.0$ with $K=0.1258$
are missing due to the reason given above.

The similarities between the free Wilson quark states and the $\beta=100.0$ states are excellent.
However. if one closely looks at the both, one notices the exponents $\alpha(t)$ at large $t$  
in the case of $\beta=100.0$ are systematically larger than those of the free Wilson cases.
We will discuss this point later.

The values of the mass  determined by $m(t)$ at $t \rightarrow 31$ are in good tendency  with
the value estimated by the lowest Matsubara frequency:

For $(1/3, 1/3, 1/3)$;\\
\, \, \, \, $m_{PS} =0.384(8)$ vs.$\sqrt{3} \, 4\pi / 3L = 0.45345$.

For $(0, 1/3, 1/3)$,\\
 $m_{PS} =0.328(3)$ vs. $ \sqrt{2}\,  4\, \pi / 3L = 0.37024$.

For $(0, 0, 1/3)$,\\
$m_{PS} =0.226(3)$ vs. $4\, \pi / 3L = 0.2618$.

The differences between the free cases and the simulation results will be discussed below.

\subsection{$\beta=10.0, 15.0, 100.0, 1000.0$}
In order to
understand the difference for $\alpha(t)$ and $m(t)$ 
 at large $t$ between the free case and the state at $\beta=100.0$, 
we calculate the PS propagator in the vacuum mentioned above,
at $\beta=10.0, 15.0, 100.0, 1000.0$.

The results of the $\alpha(t)$
are shown in  Fig. (\ref{b1000.0}).
The $\alpha(t)$'s at  four values of $\beta$ seem almost the same each other at first glance.

However, we know already that
the magnitudes $|P|$ are $0.1, 0.2, 0.82, 0.96$ respectively, for
$\beta=10.0, 15.0, 100.0, 1000.0$.

We see that as the temperature increases up to
$\beta=1000.0$ which corresponds formally to $T/T_c\simeq 10^{598}$,
the exponent $\alpha(t)$ at large $t$ becomes smaller and approaches to the free case.

This implies that even at 
$\beta=1000.0$, non-perturbative effects still work to reduce the magnitude of the Polyakov loop  and the temporal propagators differ slightly from the free case.
Therefore the perturbation around the vacuum may not give a quantitatively good result at $T/T_c=10^2 \sim 10^5$.

In a similar way, the discrepancy of the observed value of the mass and the one estimated by the lowest Matsubara frequency with the twisted boundary condition of the free quarks tends to be resolved at higher temperature. At $\beta=1000.0$ with $K=0.125$, we obtain $m_{PS} = 0.4254(16)$, which is significantly closer to the prediction $m_{PS} = 0.45345$ than at $\beta = 100.0$  with $m_{PS} = 0.384(8)$.

We think that this is closely related with the slow approach of the free energy to the Stefan-Boltzmann ideal gas limit. To conclude that we must perform simulations toward the thermodynamical limit.

\section{Conformal region and Vacuum structure}
\label{sec:conformal_region}

The ``conformal region" is defined by Eq.~(\ref{critical mass}),
$$
m_H   \leq c \,  \Lambda_{\mathrm{IR}},
$$
where the propagator $G(t)$ behaves at large $t$ as
a power-law corrected Yukawa-type decaying form (Eq.(~\ref{yukawa type}))
$$
G_H(t) = \tilde{c}_H\\ \frac {\exp(-\tilde{m}_Ht)}{t^{\alpha_H}},
$$
instead of the exponential decaying form (Eq.(\ref{exp})) observed in the ``confining region" and ''deconfining region''.

We have shown examples of the Yukawa-type decays in many cases;  $N_f=7$, 8, 12 and 16 in conformal QCD and
$N_f=2$; $ \beta=6.5, 7.0, 8.0, 10.0, 100.0, 1000.0$ in High Temperature QCD.

The RG argument implies the boundary is a first order transition.
Indeed, in the previous paper~\cite{coll1} we have shown in the case $N_f=7$ the existence of two states at the same parameters and the transition is first order.

In this section we intensively investigate the ``conformal region" in several cases;  $N_f=16$, $N_f=12$, $N_f=7$ in Conformal QCD and
$N_f=2; \beta=10.0$ in High Temperature QCD.
Doing so, we will clarify that the vacuum of  the ``conformal region" is  $(1/3, 1/3, 1/3)$
and outside the boundary is either the $(0, 0, 0)$ vacuum for the ``deconfining region" or  the $(*, *, *)$ vacuum for the ``confining region".
Therefore the transition across the boundary is the transition between different vacua and we argue that the transition must be
first order being consistent with the gap observed in other physical quantities.

We also make a cautious remark that in order to investigate conformal  properties such as the anomalous mass dimension from the spectrum one must be inside the conformal region. Otherwise one may obtain either the deconfining behavior or the confining behavior depending on the $\beta$ and lattice size, irrespective of the conformal behavior inside the conformal region.
We will show examples in the case $N_f=12$.

In relation to the first order phase transition, the search of the gap in spectrum is crucial. To systematically address the question, we carefully perform simulations in the following way.
We first simulate at small quark mass (large $K$) 
where the propagator $G(t)$ behaves at large $t$ as
a power-law corrected Yukawa-type decaying form.
Then we simulate at a smaller $K$ (larger quark mass) using the
configuration at the larger $K$. We gradually decrease $K$. 
When the step size of $K$ is small enough one will find a gap in the PS mass
at some $K$. Further we decrease $K$.  After reaching some $K$,
we then increase $K$ in the opposite direction to the above.
Then we find a gap at the same (or similar) $K$ to the one in the process of decreasing $K$.

When the step size is large, one may miss the gap,
since each vacuum is quasi-stable. In particular, at large $\beta$, large $N_f$ and large quark mass,
one may obtain the result of a quasi-stable state.
We will give such examples also in the case $N_f=12$.

\subsection{$N_f=16$}
The results  in the $N_f=16$ case for $m_q$ and $m_{PS}$ (or $\tilde{m}_{PS}$) are shown in Fig.~\ref{nf16_whole}.
 We note that the quark mass $m_q$ denoted by filled circles on the left panel is excellently proportional to $1/K$ in the whole region from $0.00$ to $0.4$. 
 
For the propagators of the PS meson,
we observe a clear transition from the exponentially decaying form to the power-law corrected Yukawa-type decaying form
at $K =0.125$.
The transition region is enlarged on the right panel in Fig.~\ref{nf16_whole}.

The values of $m_{PS}$ and  $\tilde{m}_{PS}$ are different  in the limit $K = 0.125$ from smaller $K$,
(denoted by $K=0.125_l$)  and from larger $K$, (denoted by$K=0.125_h$).
The effective mass plot at $K=0.125_h$  and $K=0.125_l$ on the right and left panels of Fig.~\ref{nf16_effm} respectively,
clearly show not only quite a large difference of the effective mass at large $t$
but also completely different decaying behavior.

The scattered plots of the Polyakov loops in the complex plane at $K=0.125_h$ and $K=0.125_l$ are shown
on the right and left panels of Fig.~\ref{nf16_comp}, respectively.
In the conformal region the arguments are $\pm 2/3\pi$ and the magnitudes are $|P|\simeq 0.18$,
while outside the conformal region the arguments are $0$ and the magnitudes are $\sim 0.05 \sim 0.2$: It is characteristic in the deconfining region.

The vacuum of all the other states in the conformal region are  $(1/3, 1/3, 1/3)$,
while all the states outside the conformal region are $(0, 0,0)$.

The $m_{PS}^2$ and $m_{PS}$ are plotted as a function of $m_q$ in Fig.\ref{nf16_mass} on the right and left panels, respectively.
The $m_{PS}$ plotted linearly as a function of $m_q$ outside the conformal region can be smoothly extrapolated to the $m_{PS}=0$ point
in the limit $m_q=0$, compared with the $m_{PS}^2$  extrapolation. This behavior is as expected in the deconfining region.

Thus we conclude that
 the conformal region is the $(1/3, 1/3, 1/3)$ vacuum and the deconfining region is the (0, 0, 0) vacuum,
 and the transition across the boundary is a first order transition between different vacua.

From the analysis we have made, we are able to draw the phase structure as shown on the left panel of Fig.~\ref{conformal_region_1}.
The transition in the quench limit on the lattice $16^3 \times 64$ is estimated to be about $\beta=6.7$.
This value is independent from $N_f$.
Therefore the $\beta=11.5$ is in the deconfining region at least for heavy quarks.
As the quark mass decreases the line of $\beta=11.5$ hits the boundary between the deconfining region and the conformal region.

The $m_q$ dependence of $\tilde{m}_{PS}$ inside the conformal region 
is rather complicated around $m_q\simeq 0.2$.
Apparently the small $\tilde{m}_{PS}$ region suffers from finite size effects.
To verify the scaling relation \cite{miransky}, \cite{DelDebbio:2010ze},\cite{DelDebbio:2013qta}
for $\tilde{m}_{PS}$ as a function of $m_q$ we have to control finite size effects.
 
The transition occurs at $m_{PS} \simeq 0.539$ with $m_q\simeq 0.244$ ($K\simeq 0.125$)
from which we estimate $c\simeq1.94$
with our working definition of $\Lambda_{\mathrm{IR}}=2\, \pi (N^3 \times N_t)^{-1/4}$.

\subsection{$N_f=7$}
We have observed in Ref.~\cite{coll2}  a transition from the exponentially decaying form to the power-law corrected Yukawa-type decaying form around $K =0.1413$: 
the results of new simulations measuring the Polyakov loop on the fly  are shown in Fig.~\ref{nf7_mass},
which are the same as the previous results within errors.
The effective mass plots for $K=0.1413_h$  and $K=0.1413_l$
shown  on the right and left panels of Fig.~\ref{nf7_effm}, 
clearly indicate a gap of the mass cross the transition.

The vacuum structure in terms of the Polyakov loops becomes less clear as $N_f$ decreases as shown in Sec. VIII.
First we show the scattered plot of the Polyakov loops in the complex plane for two states in Fig.~\ref{complex_nf7}:
One is an example state outside the conformal region, $K=0.1400$, and the other in the conformal region,
$K=0.1459$.
The state $K=0.1459$ on the left panel indicates the state is $(1/3, 1/3, 1/3)$,
while  the state $K=0.1400$ on the right panel implies the state is in the confining state $(*,*,*)$.

As these two states approach to the boundary $K=1.413$ from the both sides, the difference between the two states becomes less clear as shown in Fig..~\ref{complex_2_nf7}.
%The arguments for the state within the conformal region are $(1.03(21), 1.04(20), 1.24(11)\times2\pi/3$,
%while for the state outside the conformal region $(1.10(10), 1.12(10), 1.12(12)\times2\pi/3$.
%\red{error estimates should be checked}
%The differences between the two are not so clear as in the case $N_f=16$ case.

%$(Px, Py, Pz) \\= (-0.00157(25), 0.00235(55); \\-0.00232(13), -0.00310(35);\\ -0.00191(11), 0.00113(43)$
%$(Px, Py, Pz) \\= ( -0.00175(34), 0.00175(22); \\-0.00150(21), -0.00155(13);\\ -0.00254(28), -0.00251(50)$

As discussed in Sec.II C, the order parameters, the chiral condensate in  the massless limit
and the Polyakov loop in the time direction in the quenched limit
are not well defined far from the massless limit and the quenched limit, respectively.
The $Z(3)$ center values are good indicators for the structure of the vacuum, when $\beta$ is large or $N_f$ is large.
However, they are not good indicators when $N_f$ becomes small in Conformal QCD or small $\beta$ in High Temperature QCD, due to non-perturbative effects. This is similar to the behavior of the the chiral condensate and the Polyakov loop in the time direction,
discussed above.

Irrespective of the existence of a good indicator in terms of the $Z(3)$ center values, the transition across the boundary is a strong first order transition, manifesting it in terms of the temporal propagators.

All the states outside the conformal region are confining region, from the Polyakov loop analysis.
In accordance with this,
the $m_{PS}$ quadric plotted as a function of $m_q$ in Fig.\ref{nf7_mass2} on the right  panel
can be smoothly extrapolated to the $m_{PS}=0$ point in the limit $m_q=0$,
compared with the linear plot on the left panel.

From the analysis we have made, we also are able to draw the phase structure for $N_f=7$, as on the right panel of Fig.~\ref{conformal_region_1}.
The point $\beta=6.0$ is in the confining region.

The $m_q$ dependence of $\tilde{m}_{PS}$ is rather complicated around $m_q\simeq 0.2$.
Apparently the small $\tilde{m}_{PS}$ region suffers from finite size effects.
To verify the scaling relation \cite{miransky},\cite{DelDebbio:2010ze},\cite{DelDebbio:2013qta}
for $\tilde{m}_{PS}$ as a function of  $m_q$ we have to control finite size effects.
We have more to say about the finite size scaling in section \ref{finite}.
 
The transition occurs at $m_{PS} \simeq 0.601$ with $m_q\simeq 0.216$ ($K\simeq 0.1413$)
from which we estimate $c\simeq 2.16$
with our working definition of $\Lambda_{\mathrm{IR}}=2\, \pi (N^3 \times N_t)^{-1/4}$.

\subsection{$N_f=12$}\label{12}
The $m_q$ dependences of $m_{PS}$ outside the conformal region are different in the cases $N_f=16$ and $N_f=7$:
in the $N_f=16$ case the linear $m_{PS}$ is proportional to $m_q$ which is the relation expected in the deconfining (chiral symmetric) region, while in the case of $N_f=7$ the square of $m_{PS}$ is proportional to $m_q$ which is the relation expected in the confining (chiral symmetry broken) region.
This difference is not originated from the difference of $N_f$,  but from the $\beta$ value.
To make this point clear, we make simulations for $N_f=12$ at $\beta=6.0$ and $\beta=8.0$.

%The results  for $m_q$ and $m_{PS}$ (or $\tilde{m}_{PS}$) are shown in FIG.~\ref{nf12_whole}.
%The effective mass plots at the transitions are given in Fig. \ref{nf12_beta6.0_effm} and ~\ref{nf12_beta8.0_effm}.
%The gap is small at $\beta=6.0$: This is accidental, we think, the mass value $m_{PS}$ at the transition is smaller %in the conformal region for $N_f=7$, while for $N_f=16$ the opposite. Therefore at some point the both values 
%may become comparative.

We show two typical examples of scattered plot of the Polyakov loops in each case: 
 Fig.\ref{complex_nf12_beta6.0} shows the examples in the case $\beta=6.0$.
 They are consistent with that the states are  $(*, *, *)$, which is  characteristic in the confining region.
On the other hand, the plots of Fig.~\ref{complex_nf12_beta8.0} are for $\beta=8.0$.
The Polyakov loops are on the real axis. This implies that
the states are $(0, 0,0)$, which is characteristic in the deconfining region.

The $m_{PS}^2$ and $m_{PS}$ are plotted as a function of  $m_q$ for $\beta=6.0$ and $\beta=8.0$,
respectively, in Fig.\ref{nf12_beta6.0_mass} and Fig.\ref{nf12_beta8.0_mass}.
The $m_{PS}$ quadrically plotted as a function of $m_q$  at $\beta=6.0$
can be more smoothly extrapolated to the $m_{PS}=0$ point in the limit $m_q=0$,
compared with the linear plot on the left panel.
On the other hand, at $\beta=8.0$,
the $m_{PS}$ linearly plotted in $m_q$
can be more smoothly extrapolated to the $m_{PS}=0$ point in the limit $m_q=0$,
compared with the quadric plot on the right panel.

From all results
we are able to draw the phase structure for $N_f=12$:  on the right panel of Fig.~\ref{conformal_region_1}
for $\beta=6.0$ (confining region), while on the left panel for $\beta=8.0$ (deconfining region).

Here is a cautious remark.
The $m_q$ dependence of $m_{PS}$ outside the conformal region is determined by the lattice size and the beta.
It is irrelevant to the conformal behavior.
In order to obtain conformal properties, one should be inside the conformal region.

Another comment concerns quasi-stable states.
Doing the simulation decreasing $K$ in a small step taking a state $(1/3, 1/3, 1/3)$ as the initial state,
one find a transition at some $K$ to a state $(0, 0, 0)$ or $(*, *, *)$. This implies that the potential energy 
of the states $(0, 0, 0)$ or $(*, *, *)$ is smaller than that of the state $(1/3, 1/3, 1/3)$ for 
$K \le K^c$.
It should be stressed that this is different from
the perturbation theory where the state $(1/3, 1/3, 1/3)$ is the lowest state for all $K$.
Because of this fact the first order transition occurs.

As mentioned above, each vacuum is quasi-stable, in particular, at large $\beta$, large $N_f$ and large quark mass,
Instead of the process in small steps of $K$, when one jumps from $K > K^c$ to some smaller $K$,
the state may stay at a state $(1/3, 1/3, 1/3)$.
We have checked in the $N_f=12$ case that at $\beta=6.0$, $m_q^c=0.284$ ($K^c=0.136$) and the allowed region for the next step from a state $(1/3, 1/3, 1/3)$ is $0.285 \le m_q \le 0.704$. On the other hand, at $\beta=8.0$, 
$m_q^c=0.271$ ($K^c=0.129$) and the allowed region is $0.271 \le m_q \le 0.394$.
If we would take a next step wider than this allowed region, the state will be a state $(1/3, 1/3, 1/3)$.
The state is quasi-stable for order of one thousand trajectories.
The allowed region becomes smaller for increasing $\beta$ as expected.

Furthermore if one would not systematically decrease or increase $K$, but would use a bisection-like method to choose $K$,
one would obtain a complicated phase structure with quasi-stable vacua.

\subsection{$N_f=2$ at $\beta=10.0$}
Now we investigate a case in High Temperature QCD.
It was shown in subsection \ref{long_distance}
for $N_f=2$ at $\beta=10.0$, the propagator behaves at large $t$ 
exponential decay  at  $K=0.125$ ($m_q=0.30$), while Yukawa-type decay
at $K=0.135$ ($m_q=0.028$).

Here we further investigate the state at $K=0.130$.
We observe two states;
one   at $K=0.130_h$ continued from $K=0.135$ and the other at $K=0.130_l$ from $K=0.125$.
Effective mass plots are shown in Fig.~\ref{effm_k130},
and the scattered plots of the Polyakov loops in the complex plane in Figs.\ref{complex_k130}. 
Thus, we see the difference between the two sates is exactly the same as in the $n_f=16$ case:
the conformal region can be identified with the vacuum $(1/3, 1/3, 1/3)$,
while the deconfining region with $(0, 0,0)$.
The transition across the boundary is a first order transition between different vacua in this case
also.

We are able to draw the phase structure as shown in Fig.~\ref{conformal_region_3},
which is similar to the left panel of Fig.~\ref{conformal_region_1}.
However, in Fig.~\ref{conformal_region_1} there is no $m_q=0$ in the confining phase, which is quite different
from  Fig.~\ref{conformal_region_3}.

\subsection{Phase structure on a lattice}
From the above detailed analyses for $N_f=16$, $N_f=12$, $N_f=7$ and $N_f=2; \beta=10.0$,
we make one of main conclusions in this article, announced in subsection~\ref{sec:phase},  that the phase structures on a finite lattice are as shown in Fig.~\ref{phase diagram finite lattice}. Thus, there exists on a finite lattice the conformal region in addition to the confining region and the deconfining region
both in Conformal QCD and High Temperature QCD.

%\red{continuum limit}

\subsection{The vector channel}
We have mainly discussed the temporal propagators in the ${PS}$ channel so far.
We have also measured the propagator in the vector channel, the results being listed in the tables.
In general, the signal is worse in the vector channel, in particular, at the very small quark mass.
This is common to the usual QCD.

In the confining region, the pion mass in  the chiral limit, satisfies $m_\pi^2 \sim m_q$ as
a softly broken Goldstone particle. Therefore $m_V$ deviates from $m_\pi$ and takes a non-zero value in the chiral limit. in the case $N_f=2$, we are able to take a very small quark mass $m_q=0.0332(1)$
at $\beta = 5.9$ which is lower than the chiral transition at $\beta=6.0$.
We see clearly the vector meson deviates from the pseudo-scalar meson; $m_\pi=0.328(6)$ and $m_V=0.449(6)$. 
At $\beta=6.0$ for $N_f=12$ and $N_f=7$, there is a tendency that $m_V$ slightly larger than $m_\pi$. However, the quark mass is heavy to conclude the deviation.

In the deconfining region at $\beta=11.5$ for $N_f=16$, at $\beta=6.0$ for $N_f=12$, and at $\beta \geq 6.5$ for $N_f=2$, $m_\pi$ agrees with $m_V$ within errors.
This is consistent with that the chiral symmetry is not spontaneously broken.
However the quark mass is relatively heavy and we have to measure the scalar meson to conclude the chiral symmetry is conserved.

In the conformal region we measure both the mass $\tilde{m}$ and the exponent of power modified Yukawa type decay.
From the tables we see the $\tilde{m}_V$  agrees with $\tilde{m}_\pi$ with almost one standard deviation,
while $\alpha_V$ is in general systematically larger than $\alpha_{PS}$ albeit errors are large.
Theoretically it is possible that the anomalous dimension in the vector channel is different form the $PS$ cannel.
To conclude the difference we need much more high statistics data.

\begin{table*}
\caption{Numerical results with anti-periodic and periodic boundary conditions in spatial directions 
for $N_f=7$ at $\beta=6.0$: the meanings of the columns are the same as $N_f=16$ except for the second column.}
\begin{tabular}{lrrllllllll}
\hline
\hline
 & \multicolumn{4}{c}{$N_f=7$}   & \multicolumn{2}{c}{$\beta=6.0$} &&& \\
\hline
$K$ & s& $N_{tra}$&  plaq & $m_q$ & $m_\pi$ & $m_V$ & $\tilde{m_{\pi}}$ &$\alpha_\pi$ &$\tilde{m_{V}}$ & $\alpha_V$\\
\hline
0.1446 &pbc & 1000 &   0.634723(22) &   0.0842(5) &-&-& 0.4726(46)&    0.46(6)  &  0.4834(47) &  0.54(3)\\
0.1446 &apbc  &1000&    0.634656(33) &   0.0880(3) &    0.5462(41) &    0.5690(49)&-&-&-&-\\
0.1459 &pbc   & 1000 &   0.637062(17) &   0.0450(2) &  - & -& 0.4106(117) & 0.80(14) &  0.4131(135) & 1.01(18)\\
0.1459 &apbc &  1000 &   0.637104(21)  &  0.0479(1) &  0.5479(19) &  0.5690(32) &-&-&-&-\\
\hline
\end{tabular}
\end{table*}

\subsection{$N_f=18$}
It is believed, when $N_f \ge 17$, the theory is a free theory in the continuum limit,
since the point $g_0=0$ and $m_0=0$ is an IR fixed point in this case.
The simulations are performed both at heavy quarks; $K=0.100$ ($m_q=1.00$)
and light quarks; $K=0.125$ ($m_q=0.027$).
The result for heavy quarks is in complete agreement with the free case with $m_q=1.00$
as shown in Fig.(\ref{nf18}).
The result for light quarks is also in good agreement with the free case. However,
the behavior in the IR limit are slightly different. In Fig.\ref{nf188-nf16-free},
the local mass $m(t)$ and the local exponent $\alpha(t)$ enlarged at large $t$ for $N_f=18$ and $N_f=16$ (at $K=0.125$ and $\beta=10..0$) together with the free fermion $m_q=0.01$ are shown.
The $N_f=18$ case is closer to the free case, but there is still a gap between the $N_f=18$ case and the free case.
Although it would be intriguing to take the continuum limit of the $N_f=18$  case, since the direction of the RG flow is opposite
to the $N_f \le 16$, it is out of the scope of this article. The results obtained so far are consistent with  the common lore that the theory must become a free theory in the continuum limit.

\subsection{Finite size scaling}
\label{finite}

Since our discussions  on the conformal field theory with an IR cut-off crucially depend on the existence
of the finite lattice, in particular at zero temperature, it would be important  to ask if our theoretical as well as numerical results
are consistent with the finite size scaling argument based on the renormalization group analysis.
We would like to stress that the finite size scaling is based on very general properties
of the renormalization group, and it is applicable to any vacua irrespective of if the 
theory is in the conformal region or in the confining region. 
Some of the implicit assumptions, 
however, may be valid only when the theory stays
for a sufficiently long time close to the fixed point.
Such an assumption does not necessarily hold in the confining region
as we will discuss in the following.

Let us consider the conformal QCD in a finite box with the size $L$.
The simple scaling argument (see e.g. \cite{DelDebbio:2009fd}\cite{DelDebbio:2010hx}\cite{DelDebbio:2010ze}\cite{Bursa:2010xn}\cite{DelDebbio:2010jy}\cite{Bursa:2011ru} and reference therein for the 
argument as well as attempts for its verification in numerical simulations) tells that the any dimensionful quantity
such as $m_{PS}$ (or $\tilde{m}_{PS}$) is a function of the scaling variable
$x = L^{1+\gamma^*} m_q$ as
\begin{align}
L  \cdot m_{PS} = f(x) \ .
\end{align}
In the thermodynamics limit $x \to \infty$, 
it was claimed \cite{DelDebbio:2010ze} that the scaling function should behave as $f(x) \sim x^{\frac{1}{1+\gamma^*}}$
under the assumption that the RG flow stays for a sufficiently long time close to the 
fixed point. Note that this assumption is not valid if we take $m_q \to \infty$ with
a fixed $L$, so the naive limit in the confining region may not be used to determine
the mass anomalous dimension. This scaling relation also assumes that the continuum limit $a\to 0$ is implicitly taken.
Indeed, we have seen that $m_{q}$ dependence on $m_{PS}$ in the confining 
region in the large $m_q$ limit (for any $N_f$) does not satisfy this scaling behavior with non-trivial $\gamma^*$
corresponding to the fixed point.
We stress that this is not in contradiction with the underlying IR fixed point in any means.
It is rather due to the fact that the large $m_q$ limit is not ideal to probe the underlying IR fixed point
from the scaling behavior of $f(x)$ with the fixed lattice size. 

On the other hand, our numerical results in the conformal region predict the behavior of the scaling function in the opposite limit
($x \sim 0$):
\begin{align}
f(x) = c_0 + g(x) \ ,
\end{align}
where $g(0) = 0$. The power ansatz $g(x) \sim x^\alpha$ is typically employed in the literature e.g. \cite{Fodor:2012ty}\cite{Fodor:2012et}.
The function $f(x)$ from our numerical simulations can be read from Fig. 22, 27, 30 and 31, where we have also discussed some qualitative features of the shape of $f(x)$ in the main text.
One remark here is that the contribution from the constant $c_0$ dominates in the conformal region (as also
noted in \cite{Fodor:2012ty}\cite{Fodor:2012et} for small $x$  region in $N_f =12$). A theoretical explanation of this constant is given by studying the effective potential for the Polyakov loops, where we have shown that when $\beta \to \infty$, the constant $c_0$ is determined by the non-trivial condensation of the Polyakov loops. 
In addition, the presence of the first order transition tells that $f(x)$ is discontinuous as a function of $x$. 
Indeed, we have shown that the value of $c_0$ is affected by the choice of the vacuum. 
Apart from this, our results have no discrepancy with the existing finite size scaling argument as they should because the 
finite size scaling is just a consequence of the renormalization group with the fixed point.
Note that the crucial assumption that the RG flow stays close to the fixed point is much more
reliable in the conformal region than in the confining region.
It remains an open problem to determine the large $x$ behavior within the conformal region
to see the scaling behavior of $f(x)$.

While our results give a prediction of the scaling function $f(x)$ for small $x$ as long as 
our assumption that the RG flow stays for a sufficiently long time close to the 
fixed point is valid, we have not directly checked the $L$ (in)dependence of the scaling function 
numerically except for the trivial limit of $\beta \to \infty$ because our numerical
analysis is done with the fixed lattice size. It would be interesting to 
study the finite size scaling function  $f(x)$ systematically under the 
change of the lattice size to see if further evidence for the fixed point may be obtained.

As we have already mentioned at the end of section \ref{contlimit}, the continuum limit is subtle in the conformal QCD. We would like to propose to take the limit while keeping the condition
\begin{align}
L \cdot m_{PS} < c 
\end{align}
in order to be within the conformal region as we have demonstrated for a fixed $L$. 
This condition is equivalent to evaluating $f(x)$ below the discontinuity.
If we would like to take the limit in the confining region, we have to pay extra care to ensure that the RG flow stays close to the fixed point for a sufficiently long time. It is, however, beyond our scope of this paper to take the continuum limit, and we leave it for the future study.

\section{Boundary conditions and the structure of the vacuum}
We have shown that 
in the case of periodic boundary conditions in spatial directions for fermions,
the lowest effective energy state (the vacuum) in the one-loop approximation are the 8-fold states $(1/3, 1/3, 1/3)$.
On the other hand,
the $(0, 0,0)$ state is locally unstable when $m_q$ is light, whereas
it becomes locally stable as the $m_q$ becomes heavy; $m_q=0.15\sim 0.25$.

The lowest  energy state depends on the boundary conditions.
We discuss the other relevant cases here.

In the case of  anti-periodic boundary conditions in spatial directions,
we are able to show  that
the $(0, 0,0)$ state is the  lowest energy state by performing the one-loop computation of the vacuum energy at zero temperature, as in the case of periodic boundary conditions. %Figures????

We compare, in the cases of anti-periodic boundary conditions in spatial directions,
analytic results for free Wilson fermions with $m_q=0.01$ on the left panel of 
Fig.~\ref{apbc_free_beta10.0}
with the results of simulations shown on the right panel
for $N_f=2$ at $\beta=100.0$ with $K=0.125\, (m_q=0.03)$.
Both of them show the characteristic pattern for the vacuum $(0, 0,0)$ and they are in good agreement with each other.
However, when one closely looks at the details, one notices the exponent of the free case is smaller and the mass is
larger than those of $N_f=2$ case  at large $t$.	
The situation is similar to the periodic case. The difference is due to non-perturbative effects.

The value of the mass $m(t)$ at large $t$  should be compared with the lowest Matsubara frequency 
$2 \sqrt{3}\, \pi / L= 0.680115$, which is larger than that of the periodic boundary conditions with twisted vacuum $2 \sqrt{3} \, 2\pi/(3L) = 0.45345$.
The result for the $N_f=2$ case $m(t)$ at large $t$ is $m(t)=0.634(2)$, which is close to 0.680.

We make short runs of simulations for other cases such as $N_f=7, 8, 12, 16$, and $N_f=2$ at $\beta=6.5, 7.0,  8.0$.
The most characteristic shared feature for them is that $m(t)$ becomes larger than those for periodic boundary conditions.
Therefore it is more difficult to investigate the conformal properties.

We perform a long run for  $N_f=7$ at $K=0.1446$ and $0.1459$.
We obtain $m_{PS}=0.5462(41)$  at $K=0.1446$ and $m_{PS}=0.5479(19)$  at $K=0.1459$,  with the fit range $[28:31]$. 
%For periodic boundary conditions, although there are no plateau, if we made the fit with the fit range $[28:31]$, 
%they are $0.4886(31)$ and $0.4380(74)$, respectively.
We do not see the power-law corrected Yukawa-type decay, since we think, $m_{PS}=0.5462(41)$ and $0.5479(19)$ are larger than the critical mass.

When anti-periodic and periodic boundary conditions are mixed, the Polyakov loop in the
 lowest energy state takes either $\exp{(\pm i2\pi/3)}$ or $1$ depending on the boundary conditions
 in that direction.

In the limit $L\rightarrow \infty$, physical quantities will not depend on boundary conditions.
Therefore it is natural to conjecture that the true vacuum in the limit $L\rightarrow \infty$ is a
 weighted superposition of  27-fold local minima with four different  species.

When $\beta$ is large on a finite lattice with a medium size, the transition between two different local minima is hard to occur.
However, as the lattice size is  increased,
the barrier between the vacua is decreased as $O(1/L)$.
Therefore, although it takes time, it will eventually reach an equilibrium state.

In order to obtain a physical quantity in the continuum limit, we first perform simulations on a large lattice 
for a long runs in such a way that the transition among different vacua occurs with a non-negligible probability. We repeat the same computation by changing the lattice size. Finally we fit the data with a constant plus a $1/L$ term and extract the physical quantity in the continuum limit.

Ideally we would like to repeat a similar procedure with anti-periodic boundary conditions and get the physical quantity
in the continuum limit. Then we should be able to check the result does not depend on the boundary conditions.
Probably this requires a lots of CPU times.

We need certainly more works to investigate the vacuum
and conformal properties  in the limit $L\rightarrow \infty$.

\section{unparticle models}
\label{sec:unparticle}
In order to understand the relation between the power exponent obtained in the Yukawa-type power decaying form of the propagators and the mass anomalous dimensions, we need a concrete theoretical model that realizes the effects of an IR cutoff in (strongly coupled) conformal field theories.
For this purpose, let us discuss
a meson unparticle model, which is motivated by the soft-wall model in AdS/CFT correspondence \cite{Cacciapaglia:2008ns}.
(For details see Appendix C).
We regard the unparticle models as effective descriptions of the conformal field theory with an IR cutoff in the continuum limit.

 The soft-wall model predicts the form of the propagator in the momentum space as
\begin{equation}
\langle O(p) O(-p) \rangle = \frac{1}{(p^2+m^2)^{1-\alpha}} \ .
\label{meson_unparticle}
\end{equation}
The spectrum in the momentum representation has a cut instead of a pole.
As we see in Appendix C, this ansatz of the propagator explains the power-law corrected Yukawa-type form of the propagator in position space.

When $m\, t \ll 1,$ the mass anomalous dimension and the power is related by
$\alpha(t)=3-2 \, \gamma^{*}$. This is model independent and universal.
On the other hand, when $m\, t \gg 1,$ the computation in Appendix C shows that
$\alpha(t)=2-\gamma^{*}$ for $t \gg \Lambda_{\mathrm{CFT}}^{-1}$. Here $\Lambda_{\mathrm{CFT}}$ is the scale under which the coupling constant does not effectively run. When $\beta_0$ is sufficiently close to the fixed point value, it is very close to the UV cutoff.

A possible scenario for $T/T_c\gg 1$ is to treat $\bar{\psi}\gamma_5 \psi(x)$ as the non-bound state of unfermions.
The soft-wall model predicts the form of the propagator with scale dimension $\Delta_f$ in the momentum space as
\begin{equation}
\langle \Psi(p) \bar{\Psi}(-p) \rangle = (p^\mu \gamma_\mu + m) \frac{1}{(p^2+m^2)^{\frac{5}{2}-\Delta_f}} \ .
\label{cut model}
\end{equation}

As shown in Appendix C, one can compute the power corrections for the meson operators out of the fermion unparticle model.
When $m\, t \ll 1,$
$\alpha(t)=3-2 \, \gamma^{*}$.
On the other hand, when $m\, t \gg 1,$
$\alpha(t)=1.5-\gamma^{*}$ for $t \gg \Lambda_{\mathrm{CFT}}^{-1}.$ 

\begin{figure*}[thb]
\includegraphics [width=15cm]{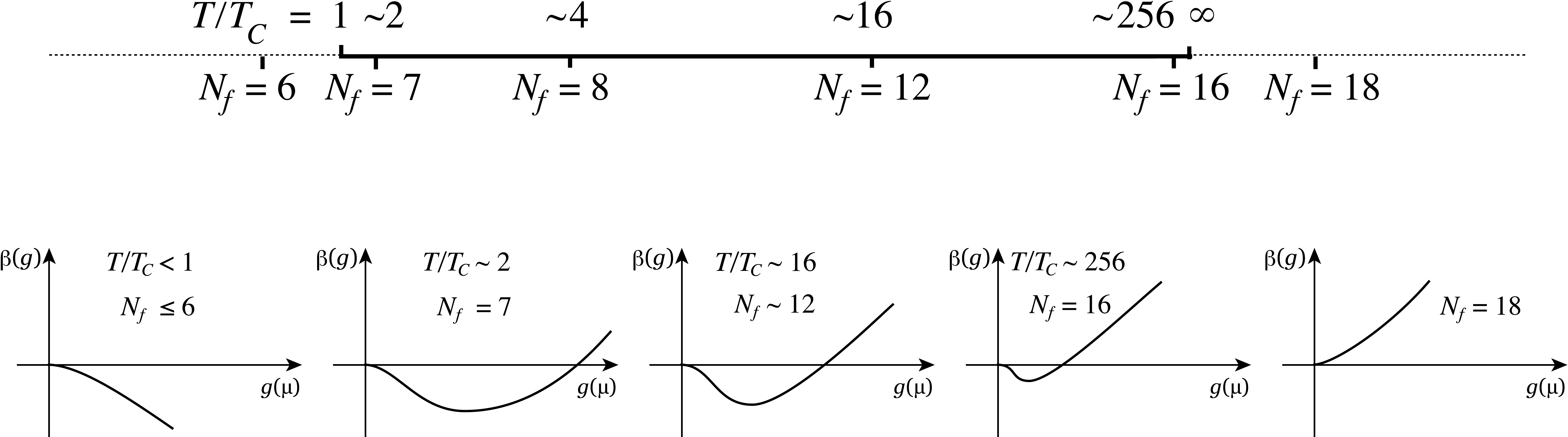}
\caption{(color online) Correspondence between Conformal QCD and High Temperature QCD in terms of the beta function.
The horizontal line on the top represents the correspondence between the number of flavor $N_f$ and the temperature $T/T_c$. }
\label{betaf2}
\end{figure*}

\section{ Correspondence between Conformal QCD and High Temperature QCD }
As a highlight of our discussions on the ``conformal field theories with an IR cutoff", we propose the direct correspondence between Conformal QCD and High Temperature QCD in the conformal region. This will enable us to understand the boundary of the conformal region and the computation of the mass anomalous dimension.

\subsection{Similarity of the beta function}
We first observe the similarity of the beta functions
on the $N_f$ dependence of the Conformal QCD and the $T/T_c$ dependence of High Temperature QCD as shown in Fig.(\ref{betaf2}):
When $N_f=N_f^{c}$ and $T/T_c \sim 1$, the beta function changes the sign at large $g$,
as $N_f$ and $T/T_c$ increase, the point of  the sign change moves toward smaller $g$.
When $N_f=16$ and $T/T_c \gg 1$ it changes sign at very small $g$.

This fact is only suggestive for the similarity on the dynamics of the two sets of conformal theories with an IR cutoff.
We will argue that the similarity is more than that.

 \begin{figure*}[thb]
\includegraphics [width=6.7cm]{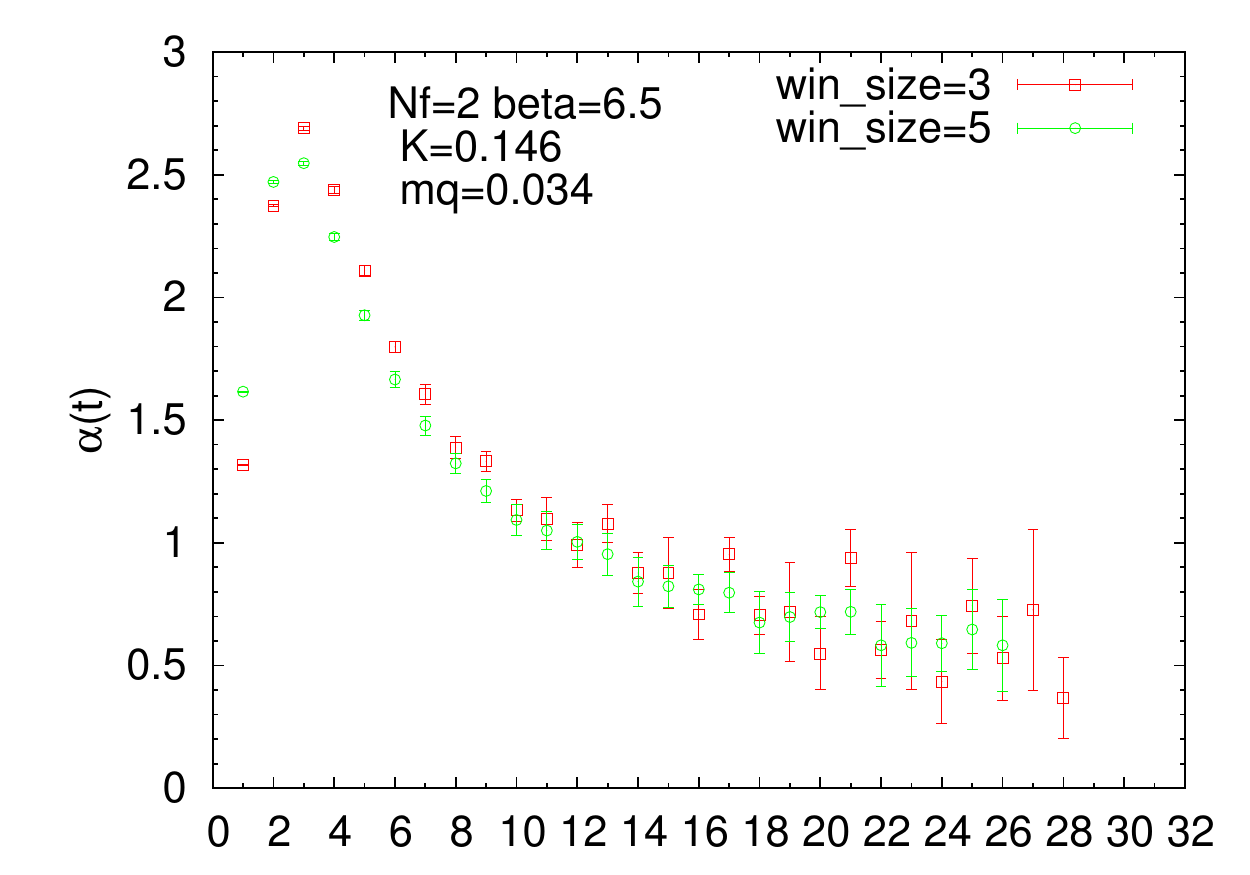}
\hspace{1cm}
\includegraphics [width=6.7cm]{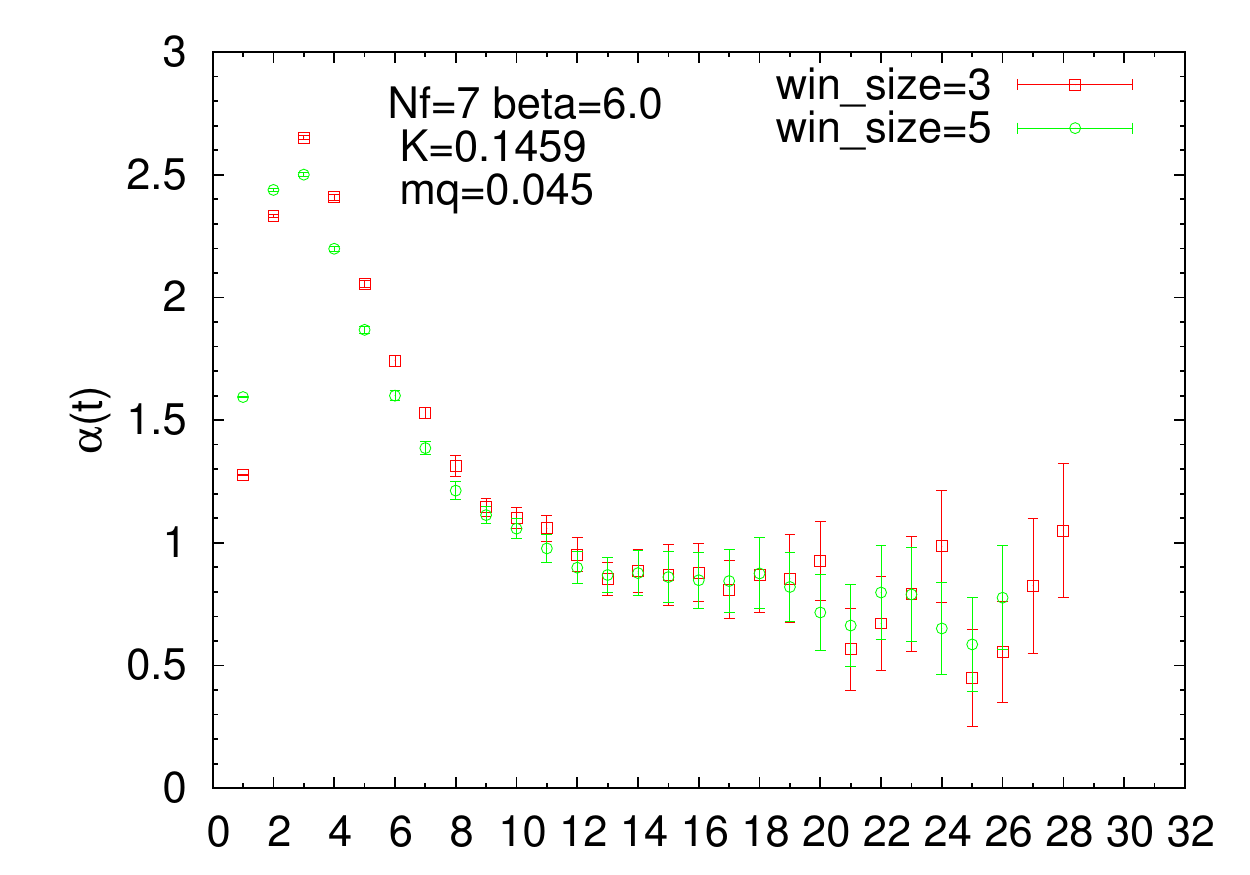}
%\vspace{1cm}
\includegraphics [width=6.7cm]{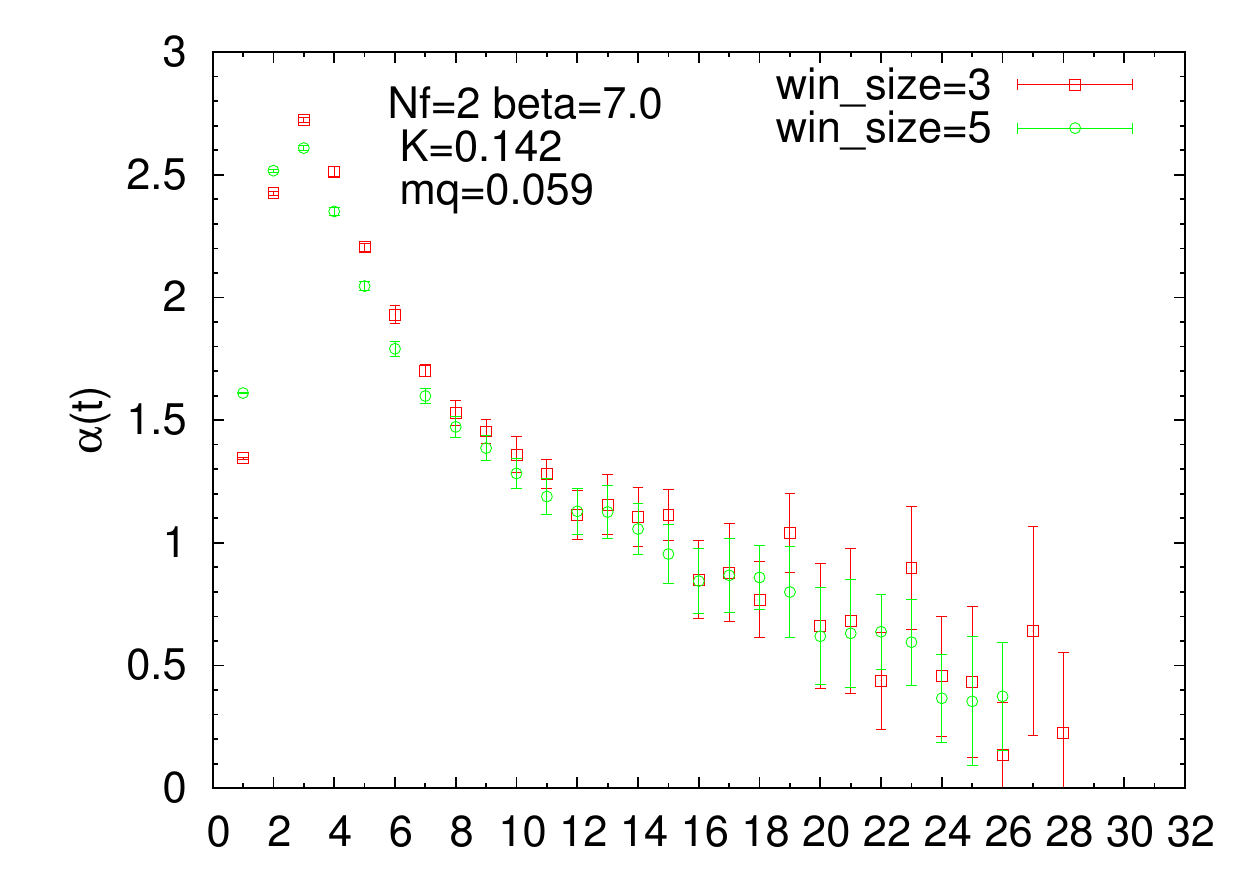}
\hspace{1cm}
\includegraphics [width=6.7cm]{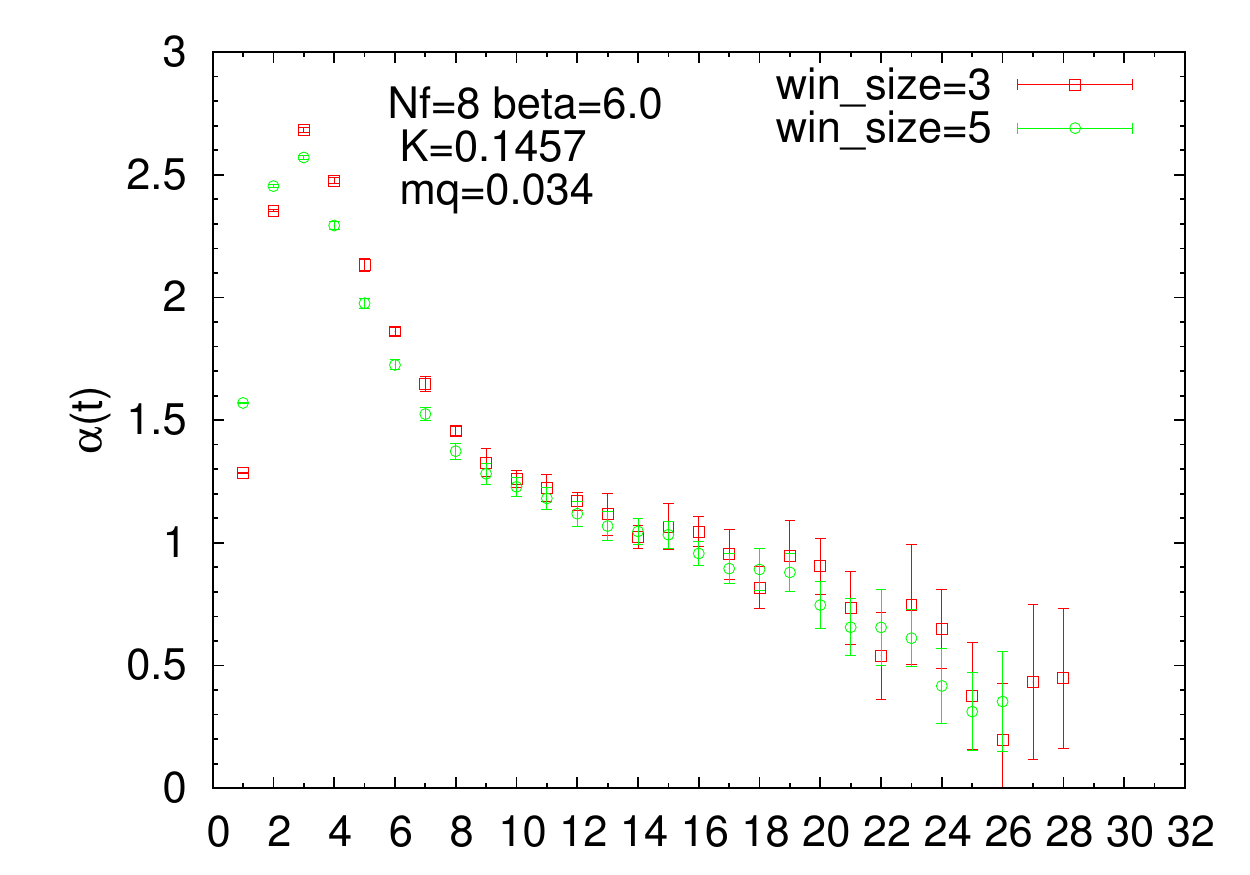}
%\vspace{1cm}
\includegraphics [width=6.7cm]{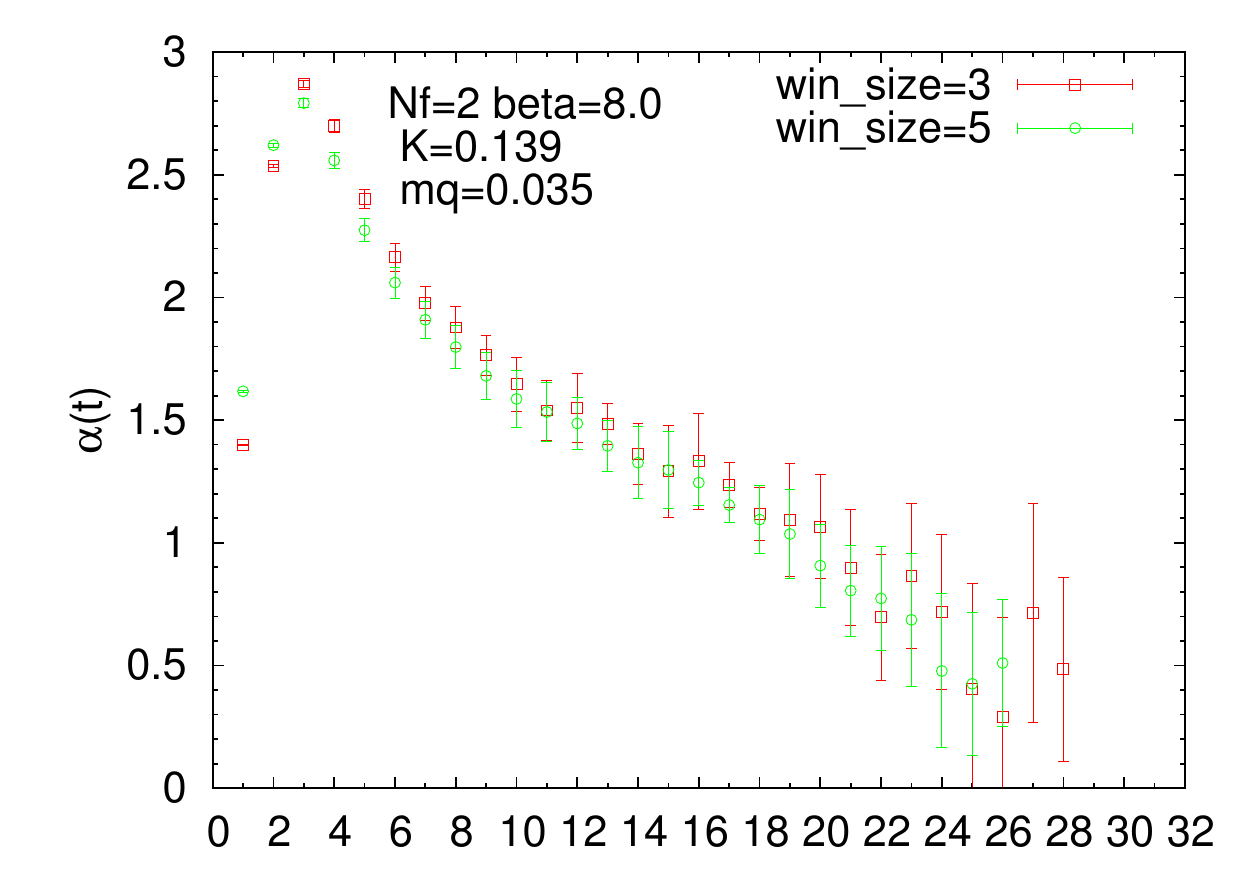}
\hspace{1cm}
\includegraphics [width=6.7cm]{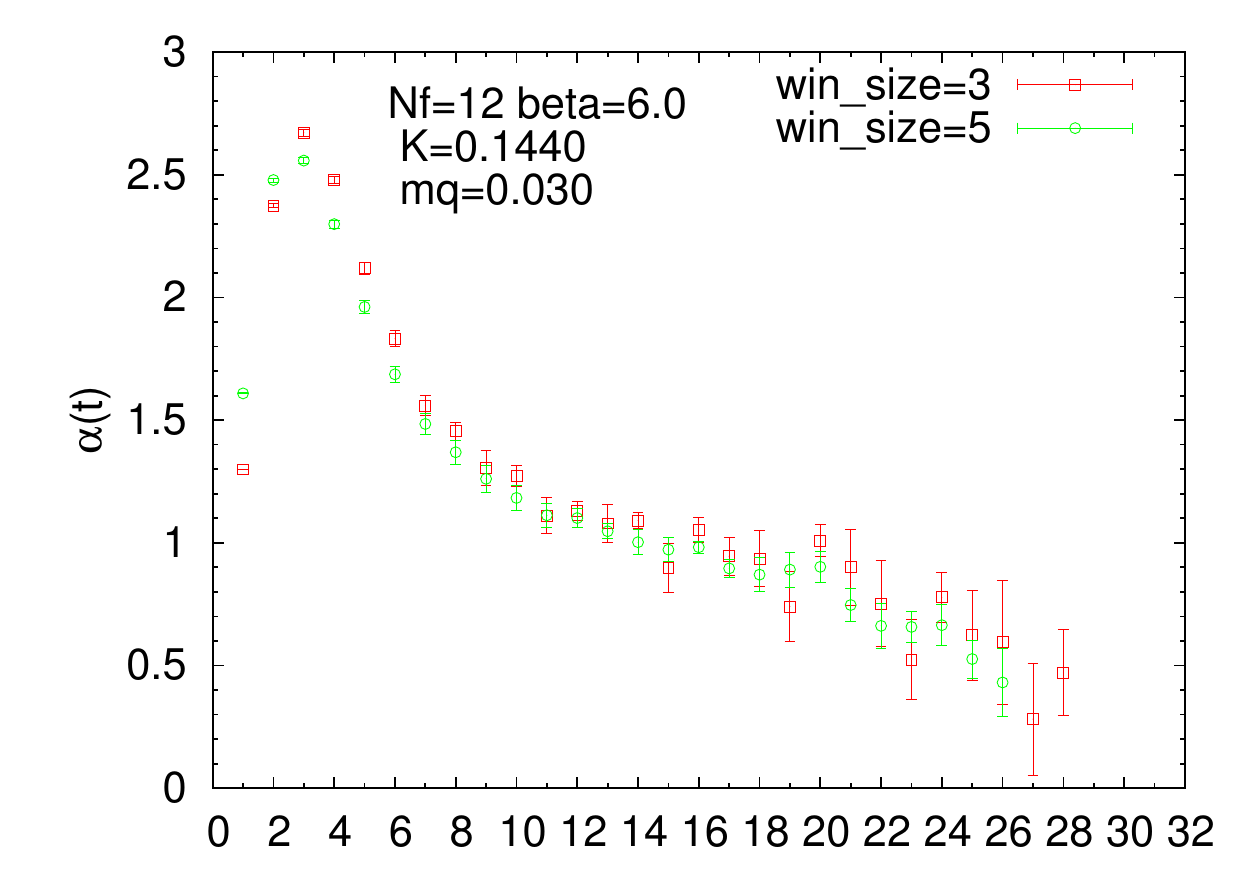}
\includegraphics [width=6.7cm]{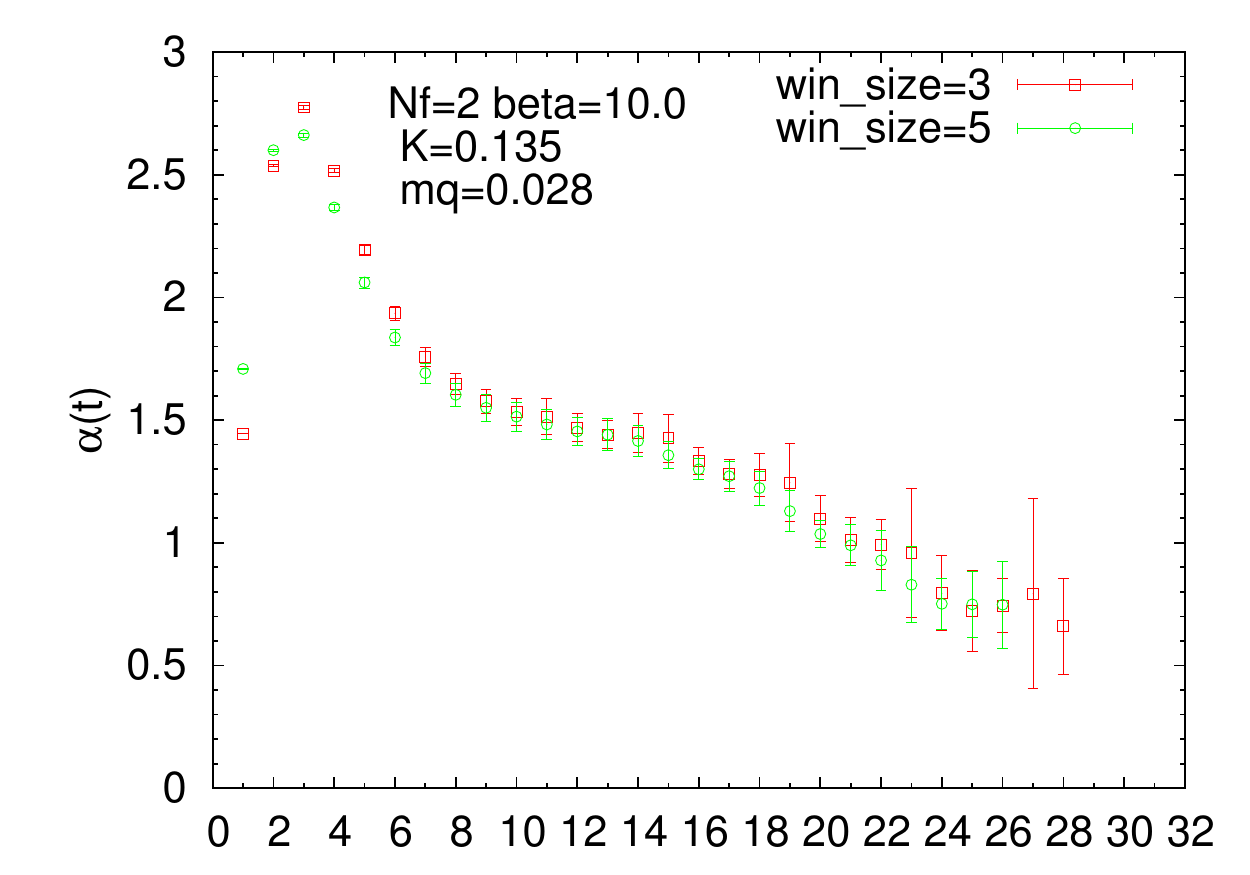}
\hspace{1cm}
\includegraphics [width=6.7cm]{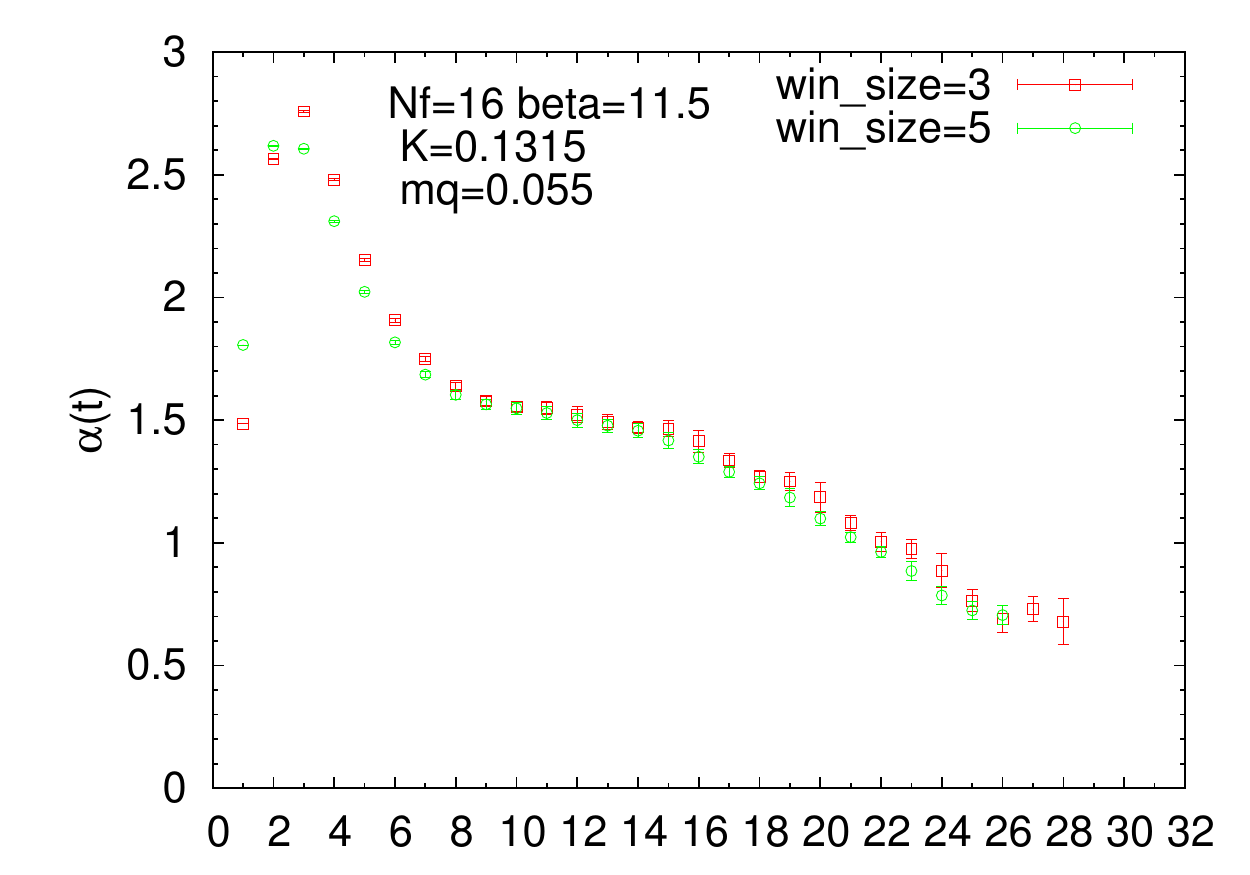}
\caption{(color online) 
The correspondence of the local exponent $\alpha(t)$ for High Temperature QCD (left) and 
for Conformal QCD (right).}
\label{correspondence of exponent}
\end{figure*}

\subsection{Correspondence on the $t$ Dependence of Propagators}

Now let us compare  the exponent $\alpha(t)$ and the local mass $m(t)$
between Conformal QCD and High Temperature QCD.
The form of  $\alpha(t)$ changes with the temperature $T/T_c$ in High Temperature QCD or the number of flavor $N_f$ in Conformal QCD.
On the other hand the $t$ dependence of $m(t)$ does not depend very much on them.

We show the two sets of $\alpha(t)$
side by side in Figs.~\ref{correspondence of exponent},
the Conformal QCD data on the right panel
and the High Temperature QCD data on the left panel in order to compare them directly.
We take a quark mass which is not close to the boundaries of the conformal region in order to avoid boundary effects. We also check in several cases the quark mass dependence is weak in the case the quark mass is well 
between the boundaries. We do not make a fine-tuning to state the correspondence.

We observe the correspondence on the $t$ dependence of $\alpha(t)$ between the two sets of data is excellent with the following each pair:
$T\sim 2 T_c$ and  $N_f=7$;
$T\sim 4 T_c$ and  $N_f=8$;
$T\sim 16 T_c$ and  $N_f=12$;
$T\sim 256T_c$ and  $N_f=16$.

Thus we plot schematically the correspondence between Conformal QCD and High Temperature QCD
as in Fig.(\ref{betaf2}). 
The correspondence is a powerful tool to investigate the properties of conformal theories.
The $T/T_c$ is a continuous variable, while the $N_f$ is a discrete variable.
Therefore we are able to use the information in High Temperature QCD to understand the properties of Conformal QCD.

On the other hand, for Conformal QCD we are able to extend the region to $N_f\ge17$
out of the conformal window. This extension is useful to investigate 
the limiting behavior of High Temperature QCD in the limit $T/T_c\rightarrow \infty$,
since there is no state for $T/T_c > \infty$.

The correspondence we propose can be supported by
the following RG argument. When the quark mass
$m_q$ is sufficiently small, the RG equation
for our propagator is governed by just one
number, the mass anomalous dimension $\gamma^*$, so whenever the mass
anomalous dimension is the same, they must satisfy the same equation (e.g.~\cite{DelDebbio:2010ze}),
which gives a heuristic support how the correspondence works.
The implicit assumption here is that they are in the same vacuum. Due to this reason, the correspondence does not necessarily work outside of the 
conformal region because the confining vacuum and the deconfining vacuum are different.

%See also the discussion in Sec. X1.???

While this agreement of the RG equation is a
necessary kinematical condition for the correspondence, we would like
to stress that our correspondence tells us more than that.
The agreement of the detailed form of the propagator (even with the
finite size corrections) suggests the underlying universal dynamics
beyond what conformal invariance dictates. 

In order to make the underlying dynamics clear, in addition to the vacuum structure, we need a comparison with the unparticle models we proposed in section~\ref{sec:unparticle}.

\subsection{$N_f=7$ and $T/T_c \simeq 2$}

We note that both in the $N_f=7$ case of Conformal QCD and 
at $T\sim 2 T_c$  in High Temperature QCD, 
we have 
a plateau in the $\alpha(t)$ at large $t$ ($15\,  \le t \le \, 31$).

This behavior of  the propagator at large $t$ implies 
 Eq.(\ref{meson_unparticle})  in the momentum representation.
Thus  in the both cases the IR behavior of the state is well described by the meson unparticle model.

The value of $\alpha(t)$ at plateau($t=15\sim 31$) is $0.8(1)$ for $K=0.1452$ 
and $K=0.1459$ in the $N_f=7$ case.
We have taken, to avoid boundary effects,
the quark masses middle among those within the conformal region.

Applying the formula $\alpha(t)=2 -\gamma^*$, we have
$\gamma^* = 1.2(1).$ 
Although this value  should be refined in the future by taking the continuum limit,
this value implies the anomalous mass dimension is of order unity.

\subsection{$N_f=16$ and $T/T_c \simeq 256$}
As discussed in Sec.~\ref{sec:structure of vacuum}, the vacuum of $N_f=16$ at $\beta=11.5$ (and $N_f=2$ at $T/T_c\simeq100.0$) is close to the twisted $Z(3)$ vacuum,
but is different in the magnitude of the Polyakov loop taking $|P|\simeq 0.2$.
It is tempting to identify the corresponding effective theory as the fermion unparticle model in the continuum limit. Smallness of the deviation from the free fermion certainly suggests that the model cannot be meson unparticle.

It is important to note that the unparticle models are effective descriptions and do not directly encode the vacuum structure nor boundary conditions. It is an interesting question if the fermion unparticle model with twisted boundary condition might explain this difference. 
However, it turns out to be hard to resolve the difference by the fermion unparticle model within the lattice size we have studied. 

Comparing the previous subsection with this subsection, we note
the effect of the finite lattice size is smaller for the meson unparticle models due to the point-like nature of the bound states, so we expect that the meson unparticle models near $N_f = N_f^c$ are more trustable in comparison with the lattice simulation.

\section{Two sets of conformal theories with an IR cutoff}
Now we have  two sets of conformal theories with an IR cutoff:\\
(1) $7 \le N_f\le 16$ in Conformal QCD\\
and \\
(2) $1\le T/T_c \le \infty$ in High Temperature QCD.\\

We have verified on a finite lattice $16^3\times 64$ that 
the two sets satisfy the properties of conformal theories with an IR cutoff.

We have pointed out from our theoretical analysis based on the RG flow and our
numerical simulations that there is a precise correspondence between
Conformal QCD and High Temperature QCD.
The correspondence between the two sets is realized between 
a continuous parameter $T/T_c$ and a discrete parameter $N_f$
as depicted in Fig.~\ref{betaf2}:\\
$T\sim 2 T_c$ and  $N_f=7$;\\
$T\sim 4 T_c$ and  $N_f=8$;\\
$T\sim 16 T_c$ and  $N_f=12$;\\
$T\sim 64T_c$ and  $N_f=16$:\\
the one boundary is
close to meson states and the other is close to free quark states.

Now, 
we have systematic understanding of  High Temperature QCD for $1\le T/T_c \le \infty$.
In the limit $T \rightarrow T_c$, the ground state becomes a meson state.
As the $T/T_c$ increases, a meson becomes a meson unparticle.
The meson unparticle with $\alpha$ 
in Eq.~(\ref{meson_unparticle}) smoothly changes the state with $\alpha$.
When $N_f=2$, the transition at $T=T_c$ is second order (or weak first order) and
therefore it is expected that the $\alpha$ changes smoothly at $T \rightarrow T_c$.
As $T \rightarrow T_c$, $\alpha \rightarrow 0$.
Note $\alpha=0$ means $\gamma^*=2$, from the formula $\alpha=2 -\gamma^*$.
(See Appendix~\ref{detail-unparticle} for a debate on the critical value for confinement.)
We will check in the future the smoothness for other cases $N_f=3 \sim 6$.

The state gradually changes following $T/T_c$ through non-meson unparticle states
toward a free fermion anti-fermion pair state in the twisted  $Z(3)$ vacuum. Here, of course, $\gamma^{*}=0.0$.

Therefore at $T/T_c\rightarrow \infty$, $\gamma^{*}=0.0$ and 
at $T/T_c=1$, $\gamma^{*}=2.0$.
It is natural to assume that $\gamma^{*}$ is a monotonous function of $T/T_c$.
We may regard the set of  High Temperature QCD  for $1\le T/T_c \le \infty$ as a conformal window.
The window  from $T/T_c =1$ to $T/T_c\rightarrow \infty$ is complete in the sense that it covers 
 $0.0 \le \gamma^{*} \le 2.0$.

Corresponding to this viewpoint, we also have reached systematic understanding of the range of conformal windows  $7 \le N_f\le 16$ in Conformal QCD.
Similarly a meson at $N_f=6$ becomes an unparticle at $N_f=7$, changes through non-meson unparticle states
and finally reaches close to a free fermion  state at $N_f=16$. When $N_f\ge 17$, it is a free quark state due to the loss of the asymptotic freedom.

It is natural to assume that the mass anomalous dimension $\gamma^{*}$ takes a monotonously increasing discrete value  between $0.0$ and $2.0$ from $N_f=16$ and $N_f=7$.

Since High Temperature QCD covers  $0.0 \le \gamma^{*} \le 2.0$ and Conformal QCD takes
discrete values of $\gamma^*$  between $0.0$ and $2.0$,
the correspondence is realized between a continuous parameter $T/T_c$ and a discrete parameter $N_f$.
This is the precise origin of the correspondence between the two
 observed in the local-analysis of propagators.

The plateau at $15 \le t  \le 31$ in $\alpha(t)$ for $T \sim 2\, T_c$ disappears,
as the temperature increases to $T \sim 4\, T_c$.
Translating this fact into Conformal QCD is that the plateau in $\alpha(t)$ at $15 \le t \le 31$
observed  as  the IR behavior of $N_f=7$ disappears for $N_f=8$.

We stress that the IR behavior of the $N_f=7$ reported in this paper is numerically verified independently  from the assumption of the conformal window.
However, solely from this fact we are not able to conclude that $N_f=7$ is within the conformal window.
It implies
the $N_f=7$ is either in the conformal window or in High Temperature QCD.
If the $N_f=7$ would be in High Temperature QCD, there should be a chiral phase transition point and there should be
a confining region below the critical point $\beta^c$.
Nevertheless,  since there is no confining region on the massless line for actions composed of the Wilson fermion action and
any type of gauge actions, it is unlikely that $N_f=7$ belongs to High Temperature QCD.
This is the same logic as in Ref.~\cite{iwa2004}.

Another viewpoint is following: if the case $N_f=7$ were outside of the conformal window, it would imply that 
there is no corresponding state to  $T \sim 2\, T_c$, which is closest state to a meson in our proposed correspondence.
Logically this possibility cannot be excluded.
However, we believe that the physical picture for the case where $N_f=7$ is within the conformal window is more appealing.

Thus our analyses presented in this article is consistent with our conjecture $N_f^{c}=7$.
We would like to conclude the conjecture $N_f^{c}=7$ by investigating directly the existence of the IR point in the future study.

\section{Summary and Discussion}
%Introducing a concept `'conformal theories with an IR cutoff",
Motivated by RG argument, we theoretically conjectured that Conformal QCD with an IR cutoff  and  High Temperature QCD 
show the common feature as the ``conformal theories with an IR cutoff'':
In the ``conformal region'' where the quark mass is smaller than the critical value,
a propagator $G(t)$ of a meson behaves at large $t$ as
a power-law corrected Yukawa-type decaying form (Eq.(\ref{yukawa type})) instead of the exponential decaying form observed in the ``confining region" and ``deconfining region'':
$$
G_H(t) = \tilde{c}_H\\ \frac {\exp(-\tilde{m}_Ht)}{t^{\alpha_H}}.
$$

We note that the behavior Eq.(\ref{yukawa type}) is proposed based on the AdS/CFT correspondence with a softwall cutoff in the literature \cite{Cacciapaglia:2008ns}.
 The meson propagator in the momentum space has a cut instead of a pole: 
$G_H(p)= 1/(p^2+\tilde{m}_H^2)^{1-\alpha_H}$.
%\end{equation}
The propagator in the position space (after space integration)
takes the form Eq.(\ref{yukawa type}) in the limit  $t \, \tilde{m}_H \gg 1$.

In the continuum limit  with $L= \infty$ (i.e. $\Lambda_{\mathrm{IR}} = 0$), the propagator on the massless quark line takes the form (Eq.~(\ref{massless})):
$$ G_H(t) = \tilde{c} \, \frac {1}{t^{\alpha_H}}.$$
%consistent with $\tilde{m}_H=0$ limit of Eq. (\ref{yukawa type}).
%since   there is no physical quantities with physical dimensions.
If we take the coupling constant $g_0=g^{*}$ at the UV cutoff, $\alpha_H$ takes a constant value, and the RG equation demands (Eq.(\ref{anoma})):
$$ \alpha_H=3 - 2 \gamma^{*},$$
for the pseudo-scalar (PS) channel
with $\gamma^{*}$ being the anomalous mass dimension $\gamma$ at $g=g^{*}$.
The theory is scale invariant (and shown to be conformal
invariant within perturbation theory \cite{Polchinski:1987dy}.
See also e.g. \cite{Nakayama:2013is} and references therein from AdS/CFT approach).
When  $0 \le g_0 <  g^*$, $\alpha_H$ depends slowly on $t$ as a solution of the  RG equation. In the IR limit $t\to \infty$, we must retain $\alpha_H(t) \to 3-2\gamma^{*}$.

Clarifying the vacuum structure and properties of temporal propagators
in QCD with $N_f$ flavors in fundamental representation,
we have verified numerically on a lattice $16^3\times 64$ the following:
The conformal region exists together with the confining region and the deconfining region
in the phase structure parametrized by $\beta$ and $K$
both in Conformal QCD and in High Temperature QCD.
The structure of the vacuum of the conformal region is 
characterized by the Polyakov loops in spatial directions,
and the vacuum is the nontrivial $Z(3)$ twisted vacuum modified by non-perturbative effects.
On the other hand, the vacua of the confining region and the deconfining region are
the vacuum characterized by the zero expectation values  and the untwisted vacuum, respectively.

We find the transition from the conformal region to the deconfining region or the confining region
is  a transition between different vacua in our finite lattice simulations, 
and therefore we conjecture that the transition is a first order transition
both in Conformal QCD and in High Temperature QCD.
However, we do not exclude the possibility that the phase transition becomes weaker or the 
discontinuities vanish as crossovers in the continuum/thermodynamic limit, whose confirmation will need further studies.
 
The results for the existence of the conformal region mean that when  the quark mass decreases from a heavy mass with fixed $\beta$, there is a first
order transition at the critical quark mass $m_q^c$ and after the critical mass there is no singular point up to
the zero mass. In particular, the hadronic mass smoothly change from very small quark mass to the zero quark mass.
This is realized indeed in the fact that
Eq.(\ref{massless}) is consistent with $\tilde{m}_H=0$ limit of Eq. (\ref{yukawa type}).
Conversely we can say that the smoothness from massive quark to the $m_q=0$ limit and Eq.(\ref{massless}) at $m_q=0$
requires the behavior Eq.(\ref{yukawa type}) or similar one. The exponential form is expected from the view point of physics.
The form also corresponds to a cut in the momentum representation, as mentioned above.
Furthermore, all numerical results are beautifully fitted with Eq.(\ref{yukawa type}).
From all these we conclude that the ansatz based on the RG argument is to the point.

It should be noted that 
when $\Lambda_{\mathrm{IR}}$ is finite, even at the massless quark $m_H$ is not zero in general. Therefore the propagator 
behaves as eq(\ref{yukawa type}) at large $t$.
In the weakly coupled region (i.e. high temperature or $N_f \sim 16$), we have shown that the main source of the $m_H$ is due to the non-trivial Polyakov loop condensate with some non-perturbative contributions.

We argue from our theoretical analysis based on the RG flow and our
numerical simulations that there is a precise correspondence between
 Conformal QCD and High Temperature QCD in the temporal propagators 
%and in the phase structure 
under the change of the parameters $N_f$ and $T/T_c$
with the same anomalous mass dimension.

Thereby we clarify the global structure of conformal theories with an IR cutoff on the finite lattice. 
The conformal window  from $T/T_c =1$ to $T/T_c\rightarrow \infty$ is complete in the sense that it covers 
 $0.0 \le \gamma^{*} \le 2.0$,
while the conformal window from $N_f=16$ to $N_f=7$  takes a discrete value of $\gamma^{*}$ between $0.0$ and $2.0$. The one boundary of the two sets is
close to meson states and the other is close to free quark states.
This observation turns out to be very useful to reveal the characteristics of each theory.

In particular, we find the correspondence between Conformal QCD with
$N_f = 7$ and High Temperature QCD with $N_f=2$ at $T\sim 2\, T_c$ being in
close relation to a meson unparticle model.
From this we estimate the anomalous mass dimension $\gamma^* = 1.2 (1)$ for $N_f=7$.
We also show that the asymptotic state in the limit $T/T_c \rightarrow \infty$
 is a free quark state in the $Z(3)$ twisted vacuum.
The approach to a free quark state is very slow; even at $T/T_c \sim 10^5$,
the state is affected by non-perturbative effects.

We have verified our conjectures on the finite lattice $16^3\times64$.
Since our conjectures are based on the general RG argument, we expect that the conjectures are also satisfied on a
larger lattice. This will be a future research.

For now, let us theoretically speculate what will happen in the continuum limit of the Conformal QCD and the High Temperature QCD, separately.
In the case of Conformal QCD, the vacuum structure of the conformal region is 
the nontrivial $Z(3)$ twisted vacuum modified by non-perturbative effects, as far as the lattice size
is not very large and the periodic boundary conditions are imposed for fermions in spatial directions. 
As lattice size $N$ increases, the transition to other vacua occurs since the energy difference decreases as $1/N$.
Finally in the limit $N\rightarrow \infty$, we are able to obtain physical quantities in the continuum theory.

The phase diagrams that we expect in the continuum limit are shown in Figs.\ref{phase diagram infinity lattice}
(left: for $\Lambda_{\mathrm{IR}}=0$) and (right: for $\Lambda_{\mathrm{IR}}=$ finite).
The shaded strong coupling region for small quark masses does not exist in the $\beta - m_q$ plane {\cite{iwa2012}}.
When the phase structure is described in terms of $\beta - K$, the corresponding phase belongs to a region for Wilson doubles. Therefore when it is mapped to the $\beta - m_q$ planes, the region corresponding to the shaded one does not exist.

In High Temperature QCD, 
our conjecture is applicable to any QCD with compact space, even for the case $L < 1/T$. 
However, the thermodynamical limit where $L=\infty$ is most relevant to our Universe.

In the thermodynamical limit at finite temperature, the exponential type decay
Eq.~(\ref{exp}) and the power-law corrected Yukawa-type decay Eq.~(\ref{yukawa type}) are valid only approximately due to the finiteness of the $t$ range.
In order to obtain physical quantities in the thermodynamical limit,				
a more rigorous way would be to make the spectral decomposition of $G_H(t)$ by using e.g.		
the maximal entropy method (MEM)\cite{hatsuda2004}.

Let us consider what can be conjectured in the thermodynamical limit.
The existence of an IR fixed point is deduced from general argument.
In addition, the temperature plays as an IR cutoff.
Thus we safely conjecture the conformal behavior in the conformal region in the continuum limit.
However, as mentioned earlier, the vacuum structure is not necessarily 
the nontrivial $Z(3)$ twisted vacuum.

As one application of the conformal field theory with an IR cutoff in the thermodynamical limit, we have recently pointed out that 
the hyper-scaling relation of physical  observables may modify the existing argument about the order of the chiral phase transition in the $N_f=2$ case. We recapitulate our argument in Appendix E.

We also believe that the very slow approach to a free quark state in the limit $T/T_c\rightarrow \infty$
on the $16^3\times 64$ lattice is closely connected with the slow approach of the free energy to the
Stefan-Boltzmann ideal gas limit. We would like to investigate the case where the aspect ratio 
$N_t/N$ is small like $N_t/N=1/4$ to conclude it in the future.

Based on the global structure of conformal theories with an IR cutoff on the finite lattice we have established in this article,
we would like to investigate further the global structure of conformal theories in the continuum limit.
In parallel,
we would like to confront the nature from the viewpoint of the conformal theories. In particular, we believe in the scenario in which the model for the beyond standard model and the thermodynamics in the early Universe are described by the conformal theories (with an IR cutoff).
In addition to these phenomenological applications, it is of our utmost priority to unambiguously establish the lower critical flavor number for the conformal window.

\section*{Acknowledgments}
We would like to express our gratitude to T. Yanagida for making a chance to start this collaboration.
We would also like to thank 
S. Aoki, H. Fukaya, E. Itou, K. Kanaya, T Hatsuda, Y. Taniguchi, A. Ukawa and N. Yamada
for useful discussion.

The calculations were performed on 
Hitachi SR16000 at KEK under its Large-Scale Simulation Program
and HA-PACS computer at CCS, University of Tsukuba
under HA-PACS Project for advanced interdisciplinary computational sciences by exa-scale computing technology.

The work by Yu.~Nakayama is supported by the World Premier International Research Center Initiative (WPI Initiative), MEXT, Japan.

%\clearpage

\appendix
\section{Our previous works}
\label{our-previous-works}
We started our projects at an early stage. In 1992~\cite{iwa1992} we pointed out that when $N_f\ge 7$ with Wilson fermions, in the strong coupling limit 
there is a bulk transition when the quark mass is decreased from the large value and there is no massless state in the confining region.

To investigate the continuum limit of the theory we then started the analysis of the phase structure.
In particular, in order to understand the phase for the region which corresponds to the small quark mass,
we made the analysis of the phase structure for very large $N_f$ up to 300.
When $N_f$ is very large, there is no bulk transition and the massless quark line from $\beta = \infty$ smoothly moves to
the point at $\beta=0$. 
Decreasing $N_f$ gradually,
we conclude that the region which corresponds to the small quark mass corresponds to the region of doublers, that is, $K\ge1/8$ at $\beta=\infty$~\cite{lat93, report97}. 

We further applied a MCRG method to investigate the RG flow(the  last in \cite{lat93})
 in the $N_f=12$ case. However we noticed the well-known subtlety of MCRG method:
Without very precise calculations, it depends on the observable to match and the number of RG transformation.
Furthermore,  because of the fact that the massless line for $m_q$ hits a bulk transition
around $\beta=4.0$, it is difficult to determine the location of the IR fixed point.
We only stated that the lower limit of the IR fixed point is $\beta \le 5.0$.

In 2004, we published the results obtained so far in~\cite{iwa2004}.
The salient facts we found are the following:
In the case $7  \le N_f \le 16$,
the massless line originating from the UV fixed point hits a bulk transition at finite $\beta$. The massless line belongs to a deconfining region all through the line. In contrast, in the confining region at the strong coupling region there is no massless line. 
Thus this confining region is irrelevant to the continuum theory.

On the other hand, in the case $N_f \le 6$, there is a chiral transition on the massless quark line originating from the UV fixed point~\cite{iwa96}. In the strong coupling region $\beta < \beta^c$ the massless line exists in the confining region. As the lattice size increases the confining region enlarges and finally the confining region occupies the phase space as far as the coupling constant is kept larger than the chiral transition value.

From this analysis we conjectured that the conformal window  is $7 \le N_F \le 16$ for the $SU(3)$ and similarly $3 \le N_F \le 10$ for the $SU(2)$.

\begin{figure*}[htb]%		
\includegraphics[width=5cm]{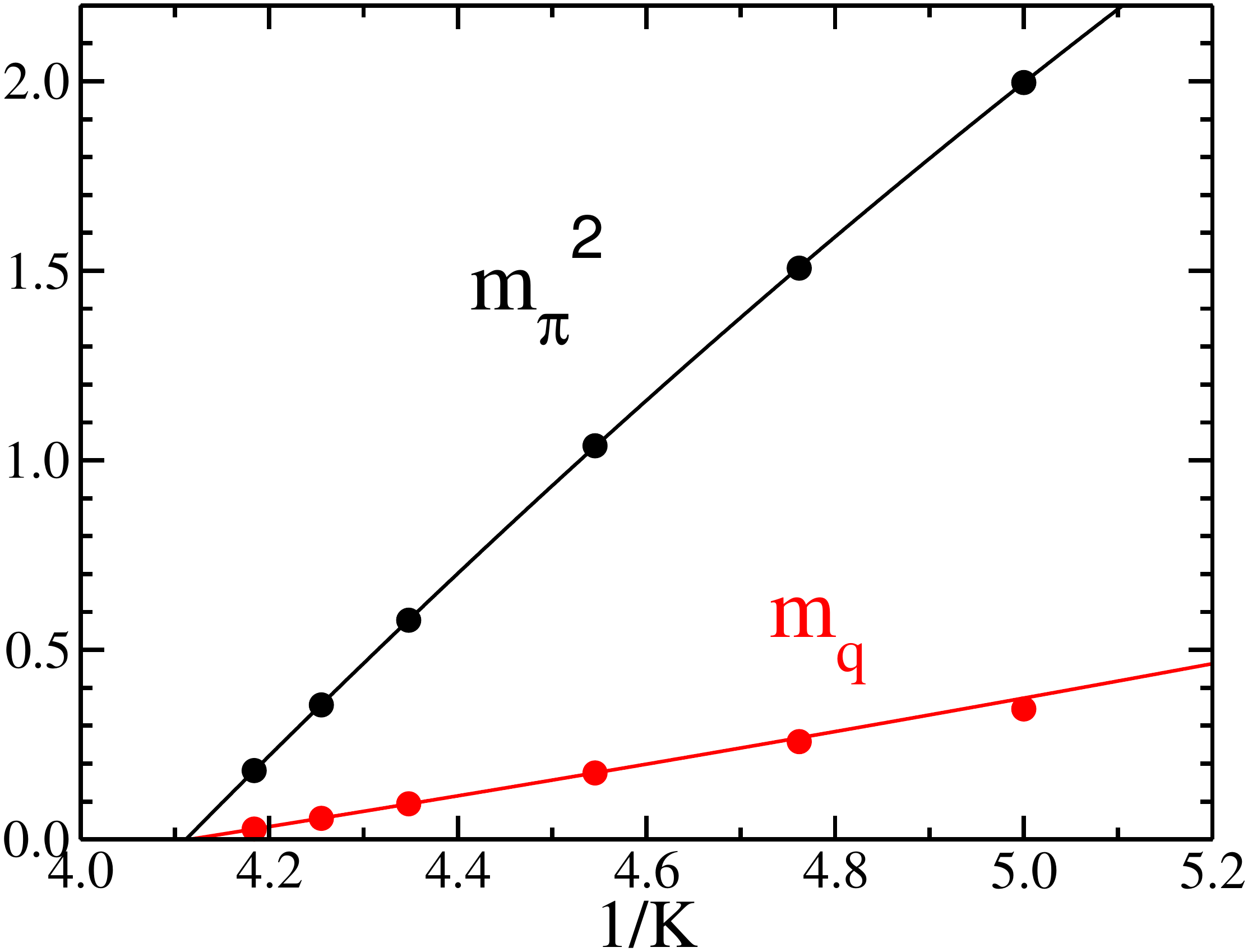}
\hspace{0.5cm}
\includegraphics[width=5cm]{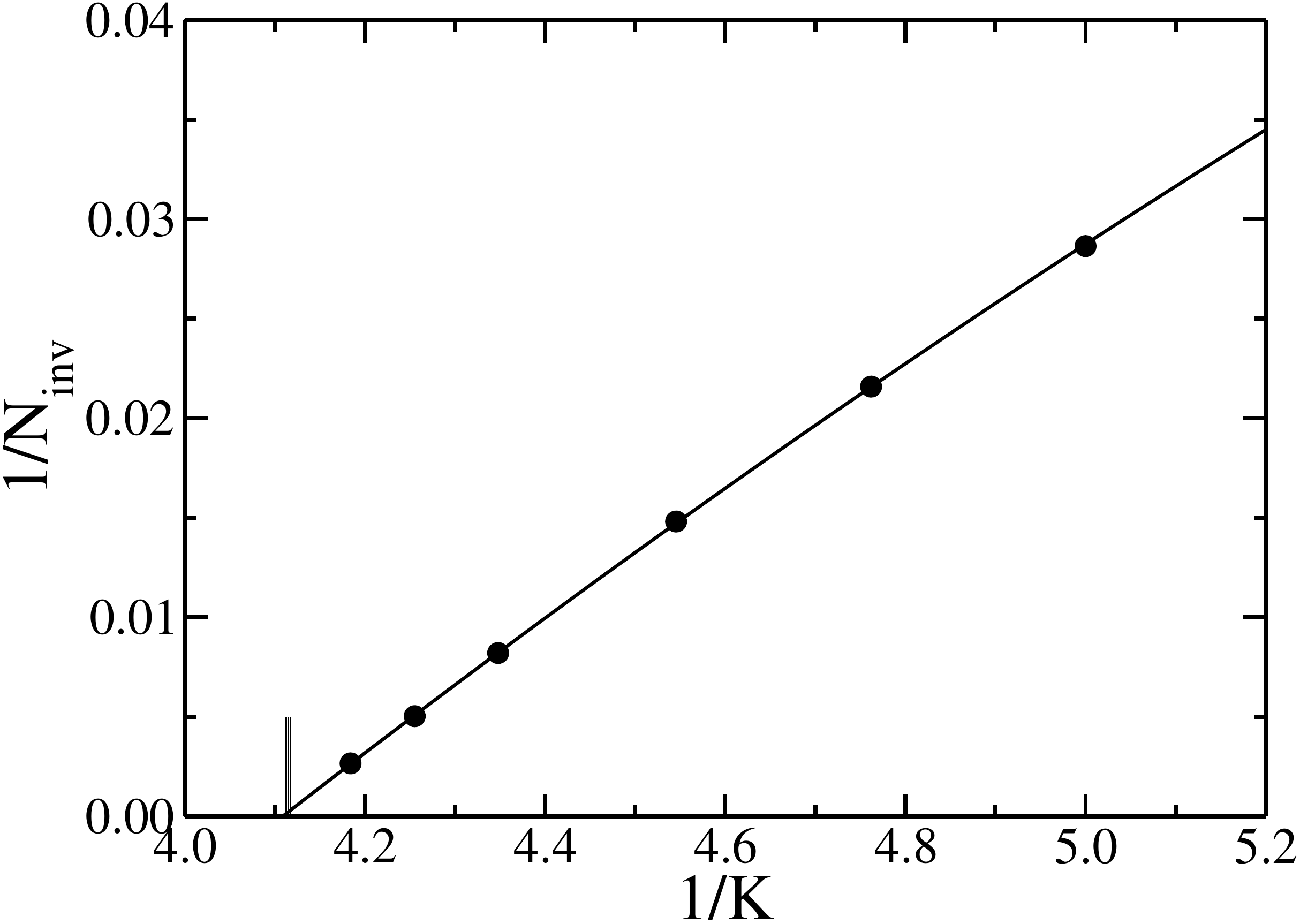}
\hspace{0.5cm}
\includegraphics[width=5cm]{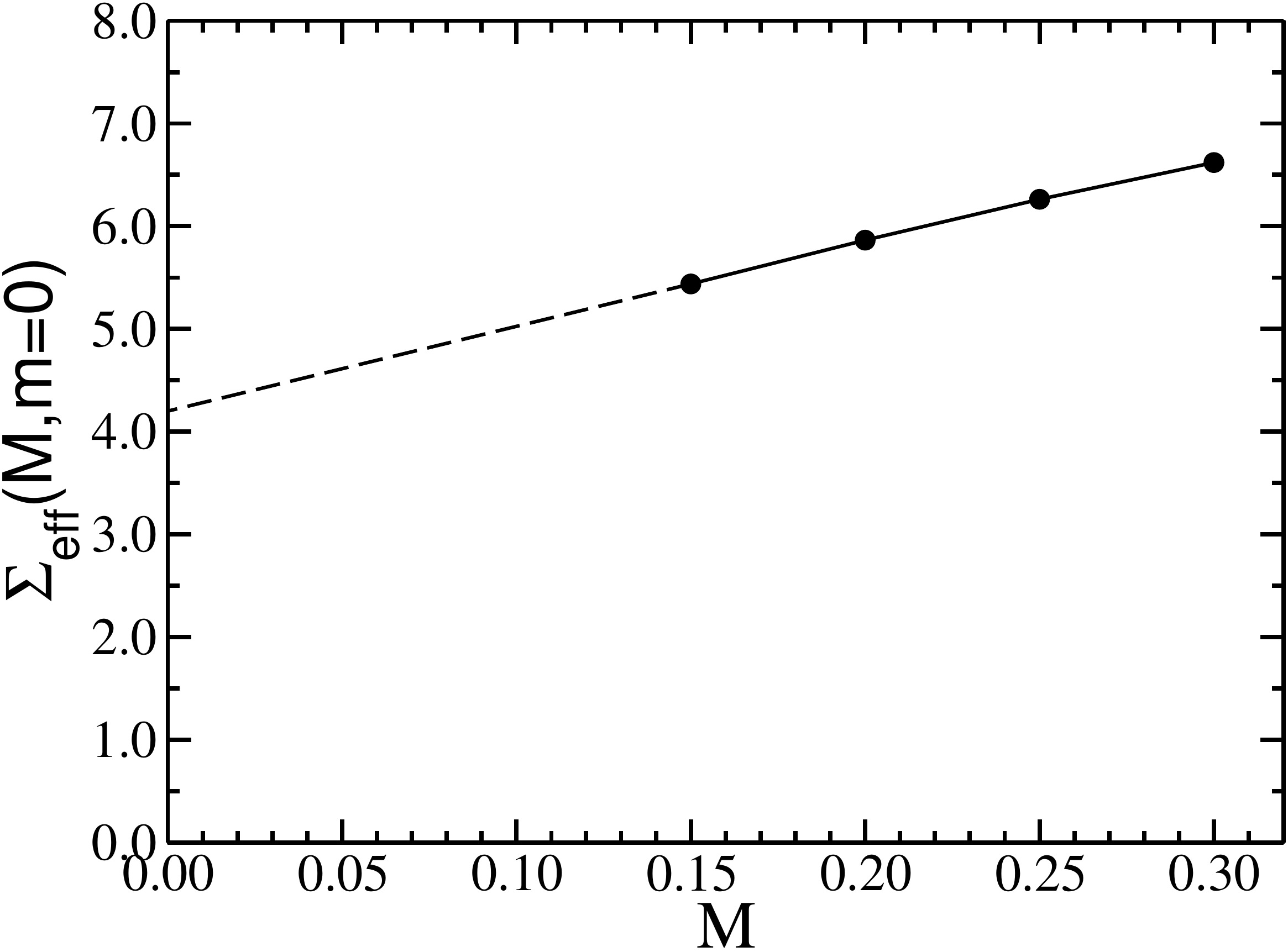}
\caption{(color online) (left) $m_\pi^2$ and $m_q$ vs. $1/K$ 
and their chiral extrapolations.
(center) $1/N_{\rm inv}$ vs. $1/K$. Vertical line
indicates $1/K_c(m_q)$.
(right)Effective chiral condensate at $m=0$. Solid and dashed lines
are guide for eye.
}
\label{fig:N6B0mpmq}
\end{figure*}

We make a side remark.
For $N_f \le 6$ at strong coupling constant  the system is in the confining region even in the chiral limit.
In particular, at $\beta=0$ in the chiral limit the pion mass should vanishes within order $a$ correction.
We had assumed that the chiral point at $\beta=0$ is $K = 0.25$, which is the quenched value.
In fact the  chiral point  $K_c$  decreases as $N_f$ increases, as pointed out in Ref.~\cite{nagai-a}.
However, as stated clearly on the fourth page (right column; 17th line) in  Ref.~\cite{nagai-a} as
\begin{itemize}
\item
 this fact is sufficient to leave the conclusions of Ref. [1] intact.
\end{itemize}
Here Ref.[1] corresponds to Ref.\cite{iwa2004}.
That is, the conclusions of Ref.\cite{iwa2004} that there is a bulk transition and there is no massless state in the strong region for $7 \le N_F \le 16$
are intact.
All the results and statements in Ref.~\cite{nagai-a} are consistent with our results. We just have to refine the results with the correct value for $K_c$.

Actually  new simulations at $\beta=0$
for $N_f=6$ QCD with Wilson fermions have been carried out, in order to 
reinforce our conclusion \cite{iwa1992}\cite{iwa2004} that 
the chiral limit  is in the confining region with
the chiral symmetry spontaneously broken.
The results are given below.

Please do not confuse the un-refereed conference report~\cite{nagai-b} with the refereed paper Ref.~\cite{nagai-a}.
In the un-refereed paper there are some statements which are difficult for us to understand such as:
The $N_f=6$ case in $SU(3)$ gauge theory with Wilson fermions is not in the confinement region because the pion mass at $K_c=0.25$ is not zero. However, we believe this is logically incorrect.
Of course, one must measure the pion mass at the correct $K=0.243(2)$.
We have updated the calculation at $\beta=0$ for $N_f=6$ with the corrected value.  The result clearly shows that 
the $N_f=6$ case in $SU(3)$ gauge theory with Wilson fermions is in the confining region, as usually expected.
 
They said that they do not support the claim that the critical flavor of the conformal window is 7,
since we only made simulations on the lattice with $N_t=4$. 
This is not correct. In fact, we made simulations on lattice with $N_t=4, 6, 8$ and $18$, as written in the paper~\cite{iwa2004}.

\subsubsection{Results at $\beta=0$ for $N_f=6$ QCD}
\label{beta0-nf6}
The simulations are performed as similarly as possible to those 
in Ref.~\cite{nagai-a}.
Configurations are generated on $  8^3\times 16$ lattices with periodic boundary
conditions for both gauge and fermion fields in all directions (different from the boundary conditions in the other parts of this article) for $K=$ 0.2, 0.21, 0.22, 0.23, 0.235 and 0.239.
We employ the HMC algorithm. The run parameters are chosen in such a way the acceptance is about $70\%$

Chiral extrapolations of $m_\pi^2$ and $m_q$ are made with
quadratic polynomial functions of $1/K$ (Fig.~\ref{fig:N6B0mpmq}).
Fits to $m_\pi^2$ and $m_q$ with lightest 4
data points reproduce data well with reasonable $\chi^2/{\rm dof}$
of 0.20 and 0.84, respectively.
We find that $m_\pi^2$ and $m_q$ vanish at almost the same $K_c$:
\begin{equation}
 K_c(m_\pi^2) = 0.24326(7),\hspace{0.5cm}
 K_c(m_q)   =   0.24301(13).
\end{equation}
This strongly supports that the chiral symmetry is
spontaneously broken in the critical limit.

In addition, we measure the number of 
iterations $N_{\rm inv}$ of the solver (BiCGStab-L2)
necessary to invert the Dirac operator.
As $K$ increases, $N_{\rm inv}$ diverges toward $K_c$.
as shown in Fig.~\ref{fig:N6B0mpmq}.
This implies that zero eigen-values appear in the Dirac operator 
around $K_c$. We also have tried to simulate at $K=K_c(m_\pi^2)$
 and found that the BiCGStab solver fails to converge.
These observations are consistent with that the system is in the
confining region at $K_c$.

We also estimate the chiral condensate using the Banks--Casher 
relation~\cite{ref:BanksCasher}.
Following Ref.~\cite{ref:GiustiLuscher}, we first calculate
the effective chiral condensate
\begin{equation}
\Sigma_{\rm eff}(M,m) =\frac{\pi}{2V}
\sqrt{1-\left(\frac{m}{M}\right)^2} 
\frac{\partial \nu(M,m)}{\partial M},
\end{equation}
where $\nu(M,m)$ is the average number of eigen-modes of
the Hermitian Dirac operator $\gamma_5 D(m)$ with eigenvalues
$\alpha$ in the range $-M < \alpha < M$, $m=(1/K-1/K_c)/2$
and $V$ is the lattice volume.
We extrapolate $\Sigma_{\rm eff}(M,m)$ to $m=0$ and then to
$M=0$. 
As Fig.~\ref{fig:N6B0mpmq} shows, the chiral condensate
takes a finite value in the limits of $m=0$ and $M=0$.
This implies that chiral symmetry is spontaneously broken.

\section{The running coupling constant, the beta function and the trace anomaly at finite temperature}

Let us consider the case where the renormalized quark mass is zero.
Then the renormalized coupling constant is the only relevant variable in the theory. 
A running coupling constant $g(\mu; T)$  at temperature $T$ can be defined as in the case of $T=0$.
The following discussion can be applied to any definition of the running coupling constant $g(\mu; T)$. 
Several ways to define the running coupling constant $g(\mu; T)$ are proposed in the 
 literatures (see e.g. Ref. \cite{karsch}). For example, in \cite{karsch} 
a running coupling constant $g(r; T)$ is defined in terms of the  quark anti-quark free energy (Eq.(8) in 
\cite{karsch}), where $r$ the distance between the static quark and anti-quark, plays the running scale.
An alternative way is the Wilson MCRG method to investigate the running coupling, fixing the temperature $T=1/N_t\, a$ at the block transformation.

In the UV regime, since the theory is asymptotically free, 
the running coupling constant at finite $T$ can be expressed as a power series of the running coupling constant at $T=0$ as long as  $g$ is small~\cite{step_scaling}. 
% the following relation 
%\begin{equation}g^2(\mu;T)= g^2 (\mu; T=0) +c g^4 (\mu; T=0),\end{equation}
%with $c$ constant, is satisfied. The constant $c$ depends on both of the schemes of $g (\mu; T=0)$ and $g(\mu; T)$ in addition to $N_f.$
The leading term is universal in the limit $g \rightarrow 0$.

However, in the IR region, $g(\mu; T)$ is quite different from $g(\mu;T=0)$, since the IR cutoff $\Lambda_{\mathrm{IR}}$ in the time direction is $T$, 
while the IR cutoff is zero at zero temperature.
%the IR cutoff $\Lambda_{IR}$ in  the $T=0$ case,being $0.$ 
Furthermore, 
when $T/T_c  > 1$, where the quark is not confined, the running coupling constant $g(\mu; T)$ cannot be arbitrarily large. This means that there is an IR fixed point with non-trivial zero of the beta  function when $T/T_c  > 1$.
This is the key observation in this article.

As long as  $T < T_c,$ the beta function is negative all through $g$.
As the temperature is increased further, the form of the beta function will change
as in Fig.\ref{beta_gT}:  (left) When $T > T_c$ but $T \sim T_c$, the beta function changes the sign from negative to positive at large $g$;  As temperature increases the fixed point moves toward smaller $g$; (right) When $T \gg T_c$ it changes the sign at small $g$.

Numerical results of the running coupling constant $g(r; T)$ shown in Fig. 2 in \cite{karsch} are consistent with
the above statement: the running coupling constant $g(r; T)$ increases as $r$ increases up to some value and does not further increases more than that, and the maximum value decrease as $T/T_c$ increases.

To avoid a possible confusion about the implication of vanishing of the beta function at finite temperature we have just introduced, we recall the relation between the trace anomaly of energy momentum tensor and the beta function with massless quarks: 
%\begin{align}
$$
\langle T^{\mu}_{\ \mu} \rangle|_T = \mathcal{B}(g^{-2}(\mu))  \langle \mathrm{Tr}(F_{\mu\nu} (\mu))^2 \rangle|_T \ , %\label{equiv}
$$
%\end{align}
where $\mathcal{B}(g^{-2}(\mu))$ is the zero temperature beta function evaluated at $g = g(\mu)$, and 
$\langle \mathrm{Tr}(F_{\mu\nu} (\mu))^2 \rangle|_T$ is the field strength squared at temperature $T$ renormalized at scale $\mu$. 
%
%
%$T^{\mu}_{\ \mu} = \beta[g(\mu;T=0)] \mathrm{Tr}F_{\mu\nu} F^{\mu\nu}$.  
%This anomaly relation is the operator identity and the coefficient is the beta function at zero temperature.
%Therefore when we take the expectation values of operators even at finite temperature,
%the coefficient is the beta function of the gauge coupling constant at zero temperature.

The derivation of  the relation is a standard method which is probably well-known. For the reader who is not familiar with it,
a simple note is attached as a subsection below.

In Lorentz invariant zero-temperature field theories, the vanishing
beta function means that the theory is scale invariant. In general, scale invariance and conformal invariance are two distinct concepts \cite{Polchinski:1987dy} because the requirement of scale invariance alone is weaker.
In (perturbative) QCD, we can further show from Eq. \eqref{equiv} that the trace anomaly vanishes and the theory is conformal invariant in the chiral limit $m_q = 0$ (see e.g. \cite{Nakayama:2013is} for a review). 
In our situation, however, we claim that the beta
function at finite temperatures vanishes,
 which does not  imply vanishing of the trace  of the energy-momentum tensor.
 Thus the vanishing beta function at $T>T_c$ does not contradict with the non-vanishing of
the difference of energy density and three times the pressure.

\subsection{note on the trace anomaly at finite temperature}

The trace of the energy-momentum tensor  $\epsilon -3p$ is given by
\begin{align}
\langle T^{\mu}_{\ \mu} \rangle|_T = \epsilon - 3p &= -T^5 \frac{\partial}{\partial T} (T^{-4} f) \ , \label{trace}
\end{align}
where $f$ is the free energy density given by $f = - \frac{T}{Z} \log Z $.

In massless QCD, the free energy density is given by
\begin{align}
f = T^4 \bar{f} (T,\Lambda_0, g_0) \ ,
\end{align}
where $\bar{f}$ is dimensionless. $\Lambda_0$ is cutoff and $g_0$ is the bare QCD coupling constant. 
(In lattice QCD, $g_0$ appears in the action, and $\Lambda_0$ and $T$ are defined implicitly through
$N_t /\Lambda_0 =T$, where $N_t$ is the lattice size in the $t$ direction.)
Since $\bar{f}$ is dimensionless, the dependence on $T$ is only through $T/\Lambda_0$, and we have the trivial identity
\begin{align}
\Lambda_0 \frac{\partial}{\partial \Lambda_0} \bar{f} = - T \frac{\partial}{\partial T} \bar{f} \ . \label{trivial}
\end{align}

On the other hand, renormalizability of QCD means that the cutoff $\Lambda_0$ and the bare coupling constant $g_0$ must be correlated so that the QCD scale $\Lambda$ is fixed (irrespective of the temperature). This is governed by the RG equation 
\begin{align}
\Lambda_0 \frac{\partial}{\partial \Lambda_0} \bar{f} = \mathcal{B}(g_0^{-2}) \frac{\partial}{\partial g^{-2}_0} \bar{f} \ , \label{renormalization}
\end{align}
where $\mathcal{B}(g_0^{-2})$ is the zero-temperature QCD beta function at cutoff scale. This is nothing but the statement that $\bar{f}$ is a function of $T/\Lambda$.  For this to hold, $\mathcal{B}(g_0^{-2})$ must be the zero-temperature QCD beta function.

Combining (\ref{trivial}) and (\ref{renormalization}), we can rewrite the thermodynamic trace identity (\ref{trace}) in QCD as
\begin{align}
\langle T^{\mu}_{\ \mu} \rangle|_T = \epsilon - 3p &=  -T^5 \frac{\partial}{\partial T} (T^{-4} f) \cr
&= T^4 \mathcal{B}(g_0^{-2})  \frac{\partial}{\partial g^{-2}_0} \bar{f}  \ .
\end{align}

At this point, we recall the (bare) Schwinger action principle
\begin{align}
\frac{\partial}{\partial g^{-2}_0} \bar{f} &= -\frac{1}{T^3 V} \frac{\partial}{\partial g^{-2}_0} \log Z \cr
& = \frac{1}{T^3 V} \langle \int d^4x \mathrm{Tr}(F^0_{\mu\nu})^2 \rangle|_T \cr
& \sim  T^{-4} \langle \mathrm{Tr}(F^0_{\mu\nu})^2 \rangle|_T \ ,
\end{align}
where $\mathrm{Tr}(F^0_{\mu\nu})^2$ is the bare  field strength squared defined at the cutoff scale. In the third line, the translational invariance was assumed.
Thus, indeed, we arrive at the anomalous trace identity
\begin{align}
\langle T^{\mu}_{\ \mu} \rangle|_T = \mathcal{B}(g_0^{-2})  \langle \mathrm{Tr}(F^0_{\mu\nu})^2 \rangle|_T \ . \label{anot}
\end{align}
We again emphasize that $\mathcal{B}(g_0^{-2})$ is the zero-temperature beta function 
%
%because the anomalous trace relation $T^{\mu}_{\ \mu} = \beta(g_0^{-2})  \mathrm{Tr}(F^0_{\mu\nu})^2$
% is the operator identity. However since we know that the energy-momentum tensor is not renormalized in (perturbative) QCD, 
 
 The right hand side of (\ref{anot}) is RG invariant. Therefore we may change the renormalization scale as we wish. 
\begin{align}
\langle T^{\mu}_{\ \mu} \rangle|_T = \mathcal{B}(g^{-2}(\mu))  \langle \mathrm{Tr}(F_{\mu\nu} (\mu))^2 \rangle|_T \ , \label{equiv}
\end{align}
where $\mathcal{B}(g^{-2}(\mu))$ is the zero temperature beta function evaluated at $g = g(\mu)$, and $\langle \mathrm{Tr}(F_{\mu\nu} (\mu))^2 \rangle|_T$ is thermal expectation value of the field strength  squared at temperature $T$,  renormalized at scale $\mu$. 
In particular, we may put $\mu=T$ in (\ref{equiv}). However, the $\mathcal{B}(g^{-2}(\mu=T))$ is different from 
the beta-function at fixed temperature $T$, $\mathcal{B}(g^{-2}(\mu ; T))$ we have defined in this appendix and used for the fixed point for the High Temperature QCD.

\section{Massive unparticle correlator}
\label{detail-unparticle}
In general, there is no universal way to construct the temporal propagators of mass deformed conformal field theories. Some particular proposals are made in the literature of unparticles. They are motivated by the soft-wall model in AdS/CFT correspondence \cite{Cacciapaglia:2008ns}.

\subsection{Mass deformed scalar unparticle correlator}
Let us discuss the mass deformed scalar unparticle correlator $\langle O(x) O(0)\rangle$ with scale dimension $\Delta$. The soft-wall model predicts the form in the momentum space as
\begin{align}
\langle O(p) O(-p) \rangle = \frac{1}{(p^2+m^2)^{2-\Delta}} \ .
\end{align}
In position space, the Fourier transform gives (up to constant)
\begin{align}
\langle O(x) O(0) \rangle = \frac{K_{\Delta}(m|x|)}{|x|^{\Delta}} \ ,
\end{align}
where $K_{\Delta}(z)$ is the modified Bessel function.

We would like to study the temporal propagator $\int d^3\vec{x} \langle O(x) O(0)\rangle$. In $mt \ll 1$ limit, we can approximate $K_\Delta(m|x|) \sim \frac{1}{(m|x|)^\Delta}$, so the integral gives
\begin{align}
\int d^3 \vec{x} \langle O(x) O(0) &\rangle \sim \int d^3 \vec{x} \frac{1}{(\sqrt{t^2 + \vec{x}^2})^{2\Delta}} \cr
&\sim \frac{1}{t^{-3+2\Delta}} \ .
\end{align}
In terms of the anomalous dimension $\Delta = 3-\gamma_m$, we have $\sim \frac{1}{t^{3-2\gamma_m}}$.
%For instance, free massless meson operator $\bar{\psi}\gamma_5\psi(x)$ has $\Delta = 3$ so that $\sim \frac{1}{t^3}$ as expected. 

On the other hand, in the other extreme limit $m t \gg 1$, one can approximate $K_\Delta(m|x|) \sim \frac{e^{-m|x|}}{\sqrt{m|x|}}$, and integrate over $\vec{x}$ by Gaussian integral by expanding $\sqrt{t^2 + \vec{x}^2} \sim t + \frac{\vec{x}^2}{2t}$. The result is
\begin{align}
\int d^3 \vec{x} \langle O(x) O(0) \rangle &\sim \int d^3 \vec{x} \frac{e^{-m\sqrt{t^2 + \vec{x}^2}}}{(\sqrt{t^2 + \vec{x}^2})^{\Delta + \frac{1}{2}}} \cr 
%&\sim \int d^3 \vec{x} \frac{e^{-m(t+\frac{\vec{x}^2}{2t})}}{t^{\Delta + \frac{%1}{2}}} \cr 
&\sim \frac{e^{-mt}}{t^{\Delta-1}} \ .\end{align}
%Note that the assumption is that $\bar{\psi}\gamma_5 \psi(x)$ creates bound state scalar much like in the confined picture. 
With the anomalous dimension $\Delta = 3-\gamma_m$, we have
$\frac{e^{-mt}}{t^{2-\gamma_m}}$. For a free scalar (= confined free hadrons), $\Delta = 1$, so the temporal propagator is $\sim e^{-{mt}}$ with no power as expected. 

There is a debate whether $\gamma_m = 1$ \cite{Ryttov:2007cx} or $\gamma_m=2$ \cite{Kaplan:2009kr} would be the critical value for confinement. The CFT unitarity argument suggests $\gamma_m = 2$ and it is realized here in the naive application of AdS/CFT. On the other hand, ``conformality lost" scenario cited above suggests $\gamma_m = 1$. It is possible that AdS/CFT accommodates the latter possibility because when $\Delta <2$, we observe the ambiguities in the boundary condition in the soft-wall model.

\subsection{Mass deformed unfermion correlator}
Another plausible scenario is to treat $\bar{\psi}\gamma_5 \psi(x)$ as the non-bound state of unfermions. We will see that it has a different $mt \gg 1$ asymptotic.

Let us discuss the mass deformed unfermion correlator $\langle \Psi(x) \bar{\Psi}(0)\rangle$ with scale dimension $\Delta_f$. The soft-wall model predicts the form in the momentum space as
\begin{align}
\langle \Psi(p) \bar{\Psi}(-p) \rangle = (p^\mu \gamma_\mu + m) \frac{1}{(p^2+m^2)^{\frac{5}{2}-\Delta_f}} \ .
\end{align}
In position space, we have
\begin{align}
\langle \Psi(x) \bar{\Psi}(0) \rangle = (\partial^\mu \gamma_\mu + m) \frac{K_{\Delta_f-\frac{1}{2}}(m|x|)}{|x|^{\Delta_f-\frac{1}{2}}} \ .
\end{align}

We would like to study $\int d^3 \vec{x} \langle \bar{\Psi}\gamma_5 \Psi(x) \bar{\Psi}\gamma_5 \Psi(0) \rangle$. When $ mt \ll 1$, we can neglect mass and obtain
\begin{align}
\int d^3 \vec{x} \langle \bar{\Psi}\gamma_5 \Psi(x) \bar{\Psi}\gamma_5 \Psi(0) \rangle &\sim \int d^3 \vec{x} \frac{1}{(\sqrt{t^2 + \vec{x}^2})^{4\Delta_f}} \cr
&\sim \frac{1}{t^{-3+4\Delta_f}} \ .
\end{align}
For free fermion, we have $\Delta_f = \frac{3}{2}$, and we obtain $\sim \frac{1}{t^3}$. If $\Delta_f = \frac{3}{2} - \frac{\gamma_m}{2}$ (so that $\bar{\Psi}\Psi$ has dimension $ \Delta = 3-\gamma_m$), we obtain $\sim \frac{1}{t^{3-2\gamma_m}}$ as in the scalar unparticle scenario. This is uniquely determined from the  scale invariance.

On the other hand, in the other extreme limit $m t \gg 1$, one can approximate $K_{\Delta_f-\frac{1}{2}}(m|x|) \sim \frac{e^{-m|x|}}{\sqrt{m|x|}}$, and integrate over $\vec{x}$ by Gaussian integral  by expanding $\sqrt{t^2 + \vec{x}^2} \sim t + \frac{\vec{x}^2}{2t}$. The result is
\begin{align}
\int d^3 \vec{x} \langle  \bar{\Psi}\gamma_5 \Psi(x) \bar{\Psi}\gamma_5 \Psi(0)\rangle &\sim \int d^3 \vec{x} \frac{e^{-m\sqrt{t^2 + \vec{x}^2}}}{(\sqrt{t^2 + \vec{x}^2})^{2\Delta_f}} \cr 
%&\sim \int d^3 \vec{x} \frac{e^{-m(t+\frac{\vec{x}^2}{2t})}}{t^{2\Delta_f}} \cr
&\sim \frac{e^{-2mt}}{t^{2\Delta_f-\frac{3}{2}}} \ .
\end{align}
For free fermion, we have $\Delta_f = \frac{3}{2}$, and we obtain $\sim \frac{e^{-2mt}}{t^{\frac{3}{2}}}$ as expected. If $\Delta_f = \frac{3}{2} - \frac{\gamma_m}{2}$ due to the anomalous dimension (so that $\bar{\Psi}\Psi$ has dimension $ \Delta = 3-\gamma_m$), we obtain $\sim \frac{e^{-2mt}}{t^{\frac{3}{2} - \gamma_m}}$. 

%\subsection{Comment}
%We believe that in the strong coupling regime, scalar unparticle is a better de%scription (note that $\gamma_m \to 2$ gives zero power).
%In the weak coupling regime, the unfermion is a better description (directly go%es to free fermion picture) in $\gamma_m \to 0$ limit. 

\section{vacuum}
\subsection{Periodic boundary condition}
In general quantum field theories, the one-loop corrections to the zero-temperature vacuum energy is obtained by the sum over the (tree-level) on-shell energy
\begin{align}
E = \sum_{\mathrm{boson}} \frac{E_B}{2} - \sum_{\mathrm{fermion}} \frac{E_F}{2} \  \label{oneloop}
\end{align}
which is same as computing the one-loop determinant $\pm \mathrm{Tr} \log (D)$ in the path integral formulation. 
In the perturbative QCD at zero temperature on the lattice, tree-level degenerate vacua is characterized by the flat-connection.  
On $\mathbb{T}^3$, the most generic Polyakov loop (in fundamental rep of $SU(3)$) with flat connection would be 
\begin{align}
 U_x &= \exp( i\int A_x dx) =  \mathrm{diag} (e^{i2\pi a_x}, e^{ i2\pi b_x}, e^{i 2\pi c_x}) \cr
 U_y &= \exp( i\int A_y dy)  = \mathrm{diag} (e^{i2\pi a_y}, e^{ i2\pi b_y}, e^{i 2\pi c_y}) \cr
 U_z &= \exp( i\int A_z dz)  = \mathrm{diag} (e^{i2\pi a_z}, e^{ i2\pi b_z}, e^{i 2\pi c_z})  
\end{align}
with $a_i + b_i + c_i \in \mathbb{Z}$ for $i=x,y,z$ up to gauge transformation.
Note that $a_i = b_i = c_i = \frac{1}{3}, \frac{2}{3}$ gives a non-trivial center of the gauge group. Due to the one-loop corrections \eqref{oneloop}, we obtain a non-trivial potential for $(a_i,b_i,c_i)$, which will determine the one-loop  vacua.

For free Wilson fermion, the on-shell energy used in \eqref{oneloop} can be obtained by 
\begin{align}
&k^2[k_x,k_y,k_z]  \cr
&= (\sin^2(k_x) + \sin^2 (k_y) + \sin^2(k_z)) \cr
&m^2[k_x,k_y,k_z]  \cr
&= (m_q + 3-\cos(k_x) - \cos(k_y) - \cos(k_z))^2
\end{align}
where $m_q$ is the quark mass in the action, with the implicit form:
\begin{align}
\cosh(E[k_x,k_y,k_z]) = 1 + \frac{k^2 + m^2}{2(1+m)} \ . \label{energy}
\end{align}
The mode number $k_i$ is determined from the boundary condition for the quarks.

If we do the singular gauge transformation, the Wilson-line can be encoded in the twisted boundary condition for the quark field, which in turn changes momentum quantization in the summation. 
Therefore the one-loop potential is obtained by
\begin{align}
-V_F(a_i,b_i,c_i) = \sum_{n_i = a_i}^{N -1 + a_i} E[2\pi n_x/N,2\pi n_y/N, 2\pi n_z/N] \cr
 + \sum_{n_i = b_i}^{N -1 + b_i} E[2\pi n_x/N,2\pi n_y/N, 2\pi n_z/N] \cr
+ \sum_{n_i = c_i}^{N -1 + c_i}E[2\pi n_x/N,2\pi n_y/N, 2\pi n_z/N] \ .
\end{align}
The summation is taken for $n_i = a_i, a_i + 1, a_i + 2, \cdots$.
In the figure, we subtracted $V(0,0,0)$ since the absolute value is unphysical,
and thereby cancels the singular behavior in the massless quark limit.

%The gauge field contribution can be estimated as follow. For $SU(3)$, the adjoint representation (octet) of the gauge group obtains the shift of momentum in $(a-b), (b-a), (c-a), (a-c), (b-c), (c-b), 0 , 0$. 
%This can be understood as follows. Set $A_{\mu} = A^0_{\mu} + \delta A_\mu$, where $A^0_{\mu}$ is the background field that gives the specified Wilson line. The gauge transformation is given by $A_\mu^0 \to U^\dagger A_\mu^0 U + U^\dagger \partial_\mu U$ and $\delta A_{\mu} \to U^\dagger \delta A_{\mu} U$ (with $\psi \to U \psi$ for fundamental matter). Now, we use the singular gauge transformation to get rid of $A_\mu^0$. Then the matrix $U$ acts as the twisted boundary condition for the fluctuation $\delta A_{\mu}$ (and matter field $\psi$), which transforms as adjoint representation of the gauge group.
% This leads to the above mentioned shift of momentum. The path integral over $\delta A_\mu$ gives the one-loop energy from the one-loop determinant  (it is instructive to see that when $a = b = c = 1/3$, there is no contribution to the potential due to center symmetry).

One can compute the one-loop shift of energy (vacuum energy) of the gauge field by using the similar formula to the above by the momentum shift for the adjoint representation.  For $SU(3)$, the adjoint representation (octet) of the gauge group obtains the shift of momentum in $(a-b), (b-a), (c-a), (a-c), (b-c), (c-b), 0 , 0$:
\begin{align}
+V_B(a_i,b_i,c_i) &= \sum_{n_i = a_i-b_i}^{N -1 + a_i-b_i} E_G[2\pi n_x/N,2\pi n_y/N, 2\pi n_z/N]  \cr
 &+ (\text{7 other shift in the momentum})  \ . 
\end{align}
Here $E_G(\vec{k})$ is determined from the pole of the propagator of the gauge fields:
\begin{align}
\sinh(E_G[k_x,k_y,k_z]/2) \cr
= \sqrt{\sin^2(k_x/2) + \sin^2(k_y/2) + \sin^2(k_z/2)} \ .
\end{align}

Note that the one-loop energy is typically divergent both in IR and UV, but since we are only interested in the energy difference, if we subtract the energy by $V(0,0,0)$, the result is finite.
The total effective energy in terms of $(a_x, a_y)$ is shown in Fig.~\ref{effective potential} and 
the contour of the effective energy is shown in Fig.~\ref{contour}.
The minimums are at  $(a_x=1/3, a_y=1/3)$ and $(a_x=2/3, a_y=2/3)$.

%I used massless limit of \eqref{energy} in the potential computation.

\subsection{Antiperiodic boundary condition}
We could instead use the anti-periodic boundary condition for the quarks. With the above Wilson line introduced, the one-loop potential for quark fields becomes\begin{align}
-V_F(a_i,b_i,c_i) = \sum_{n_i = a_i+1/2}^{N -1 + a_i+1/2}E[2\pi n_x/N,2\pi n_y/N, 2\pi n_z/N] \cr
 + \sum_{n_i = b_i+1/2}^{N -1 + b_i+1/2} E[2\pi n_x/N,2\pi n_y/N, 2\pi n_z/N] \cr
+ \sum_{n_i = c_i+1/2}^{N -1 + c_i+1/2} E[2\pi n_x/N,2\pi n_y/N, 2\pi n_z/N] \ 
\end{align}
The one-loop potential from gauge field does not change.
We realize that $a_i = b_i = c_i = 0$ is the minimum of the total potential.

\section{The order of the chiral phase transition in $N_f=2$ case}
\label{sec:nf2}
Here we discuss an implication of the existence of the IR fixed point in high temperature QCD
for the issue of the order of the chiral phase transition in the $N_f=2$ case.
Our key observation is the existence of an IR fixed point at $T>T_c$.
 We stress that the reasoning for the existence can be justified even in the  thermodynamic limit.

Pisarski and Wilczek \cite{pisarski}
mapped $N_f=2$ QCD at high temperature 
to the three dimensional sigma model and 
pointed out that if $U_A(1)$ symmetry is not recovered at the chiral transition temperature,
the chiral phase transition of QCD in the $N_f=2$ case is 2nd order 
with exponents of the  three dimensional $O(4)$ sigma model.

For the Wilson quarks it was shown
that
the chiral condensate satisfies remarkably  the $O(4)$  scaling relation,
with the RG improved gauge action and the Wilson quark action \cite{iwa1997}  and 
with the same gauge action and the clover-improved Wilson quark action \cite{cppacs2001} (See, for example, Fig.6 in Ref.~\cite{iwa1997}).
It was also shown for staggered quarks the scaling relation is satisfied 
 in the $N_f=2 + 1$ case \cite{staggered}, extending the region from $T/T_c >1$ adopted in
 \cite{iwa1997} and  \cite{cppacs2001}
 to the region including $T/T_c<1$.
 These results imply the transition is second order.

However, recently, it was shown that the expectation value of 
 the chiral susceptibility $\chi_{\pi} -\chi_{\sigma}$ is
 zero~\cite{aoki2012} in thermodynamic limit
 when the $SU(2)$ chiral symmetry is recovered under the assumptions we will discuss below. This is consistent with that the  $U_{A}(1)$ symmetry is recovered, which
implies the transition is 1st order according to \cite{pisarski}. 
Apparently the  two conclusions are in contradiction.

Here we revisit this issue with the new insight of conformal field theories with an IR cutoff.
It is assumed in Ref.~\cite{aoki2012}  that
the vacuum expectation value of mass-independent observable is an analytic function of 
$m_q^2$, if the chiral symmetry is restored.
However, in the conformal region
the propagator of a meson behaves as Eq.(3) and the relation between the $m_H$ and the $m_q$ is given by the hyper-scaling relation~\cite{miransky} \cite{DelDebbio:2010ze}
$$m_H = c \, m_q^{1/(1+\gamma*)},$$
with $\gamma^*$ the anomalous mass dimension.
This anomalous scaling implies $m_H$  is not analytic in terms of $m_q^2$ and  the analyticity assumption does not hold.
%as mentioned as a viable possibility there.
It should be noted that the Ward -Takahashi Identities
in~\cite{aoki2012} are proved in the thermodynamic limit and therefore
the numerical verification of the hyper-scaling in the limit will be decisive.
We stress however that the hyper-scaling is theoretically derived with
the condition of the existence of the IR fixed point and
multiplicative renormalization of $m_q$. We believe that this
violation of the analyticity assumption resolves the apparent
discrepancy as also mentioned in~\cite{aoki2012} as a viable
possibility.

%\clearpage
%\section{Tables}
%\input{data_table}
%\clearpage

\section{figures}

\begin{figure*}[hbt]
\includegraphics [width=7.0cm]{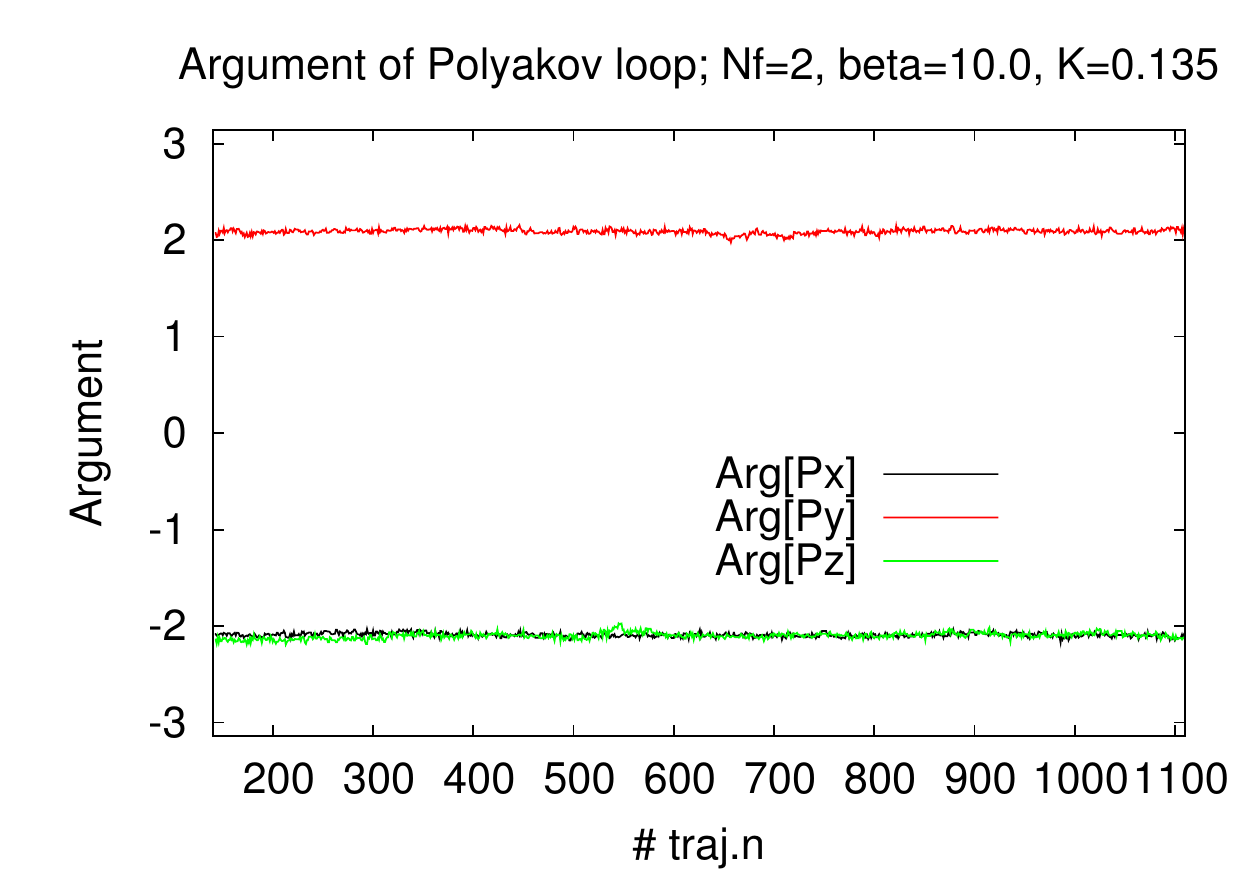}
\hspace{1cm}
\includegraphics [width=7.0cm]{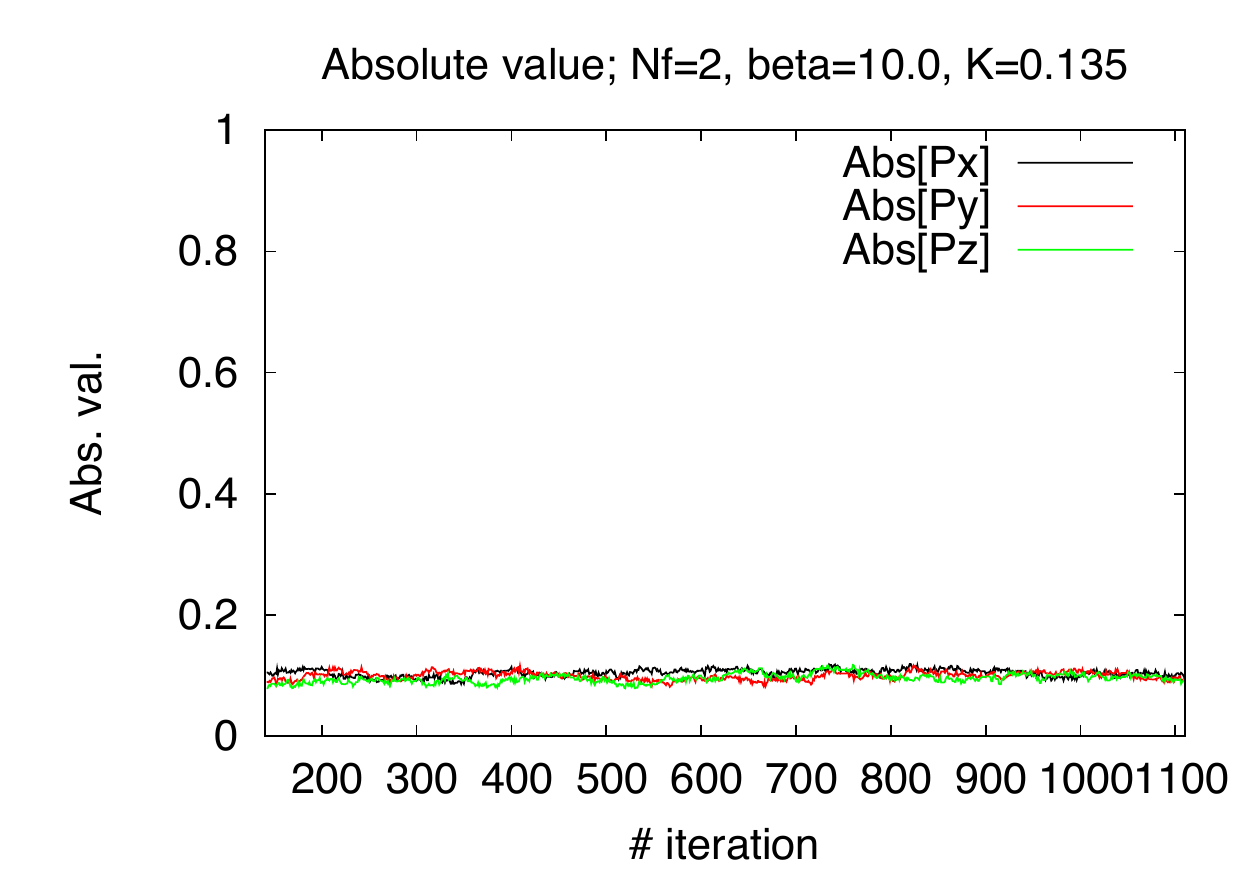}
\includegraphics [width=7.0cm]{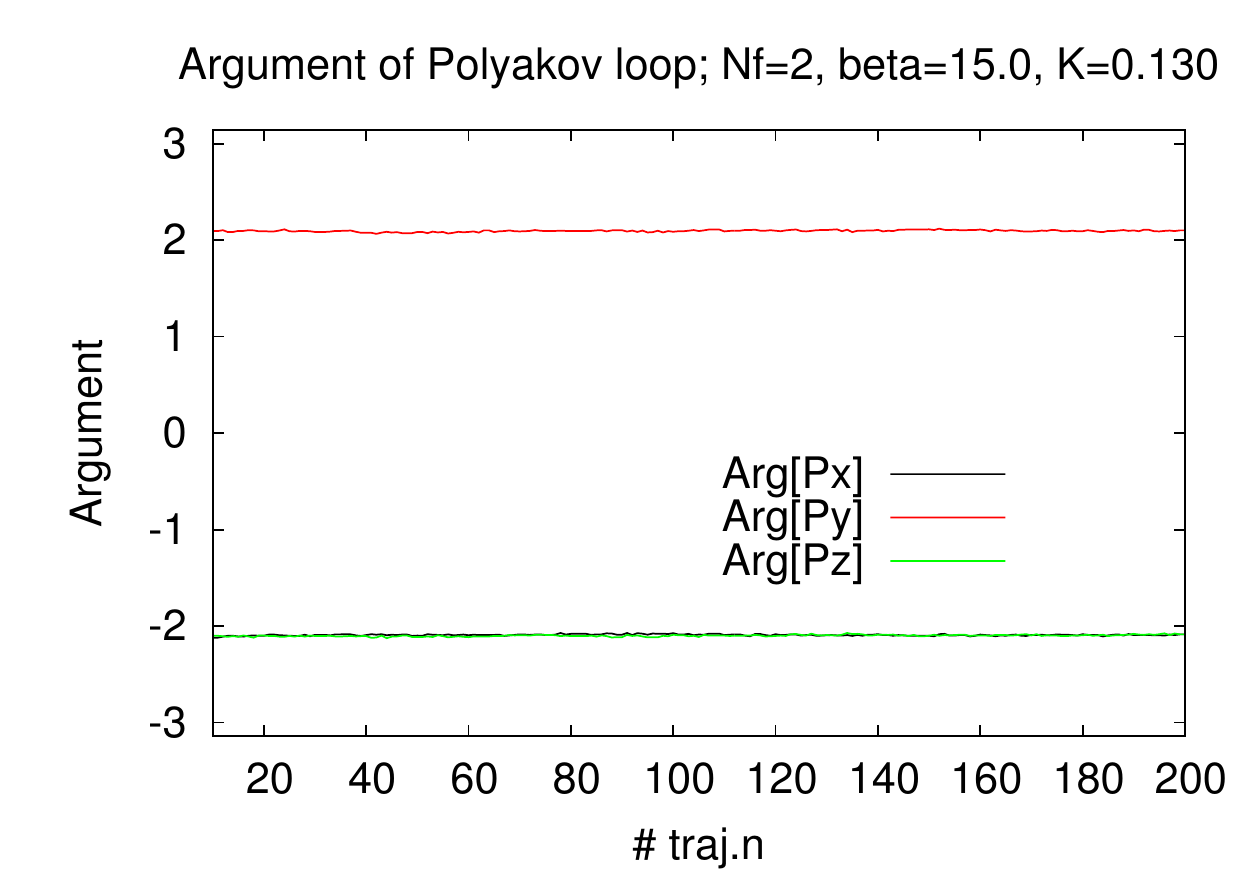}
\hspace{1cm}
\includegraphics [width=7.0cm]{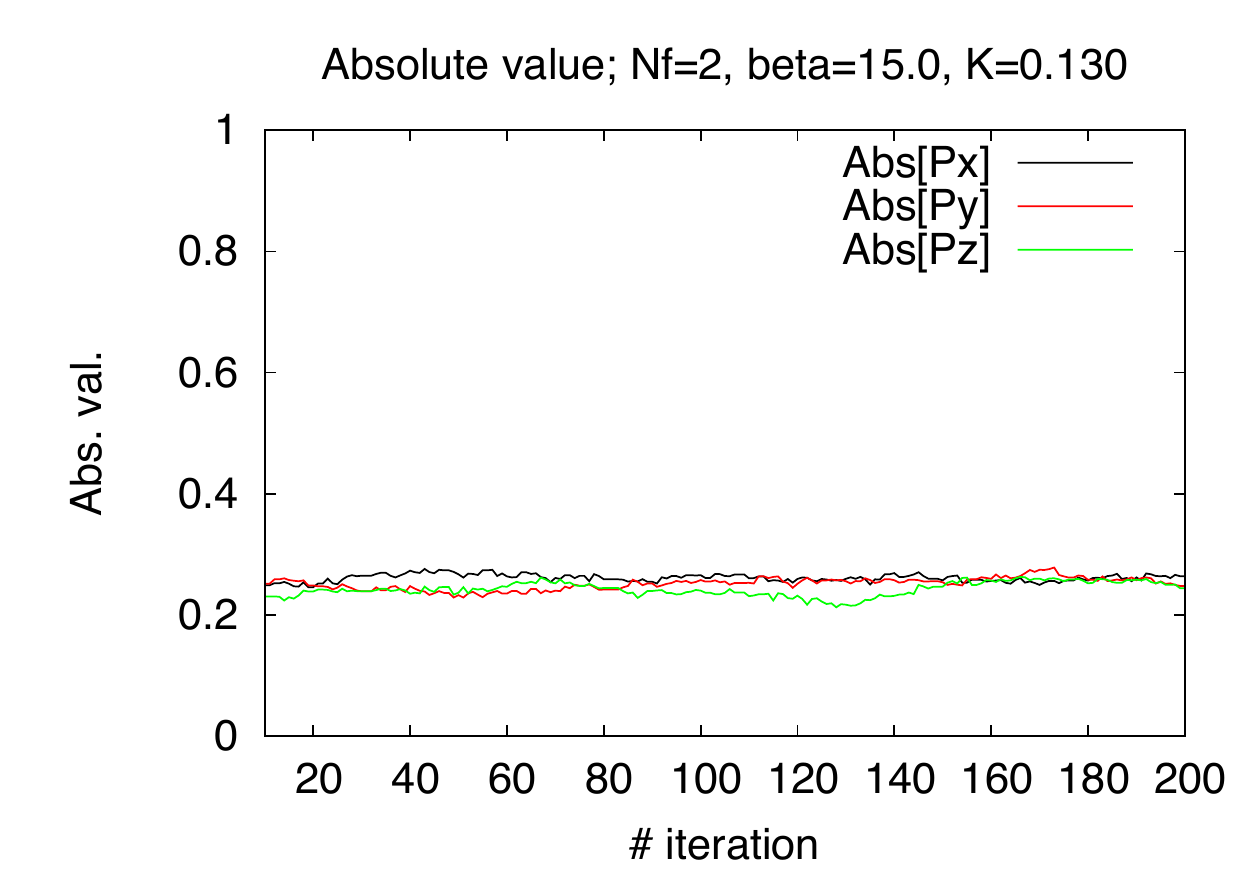}
\includegraphics [width=7.0cm]{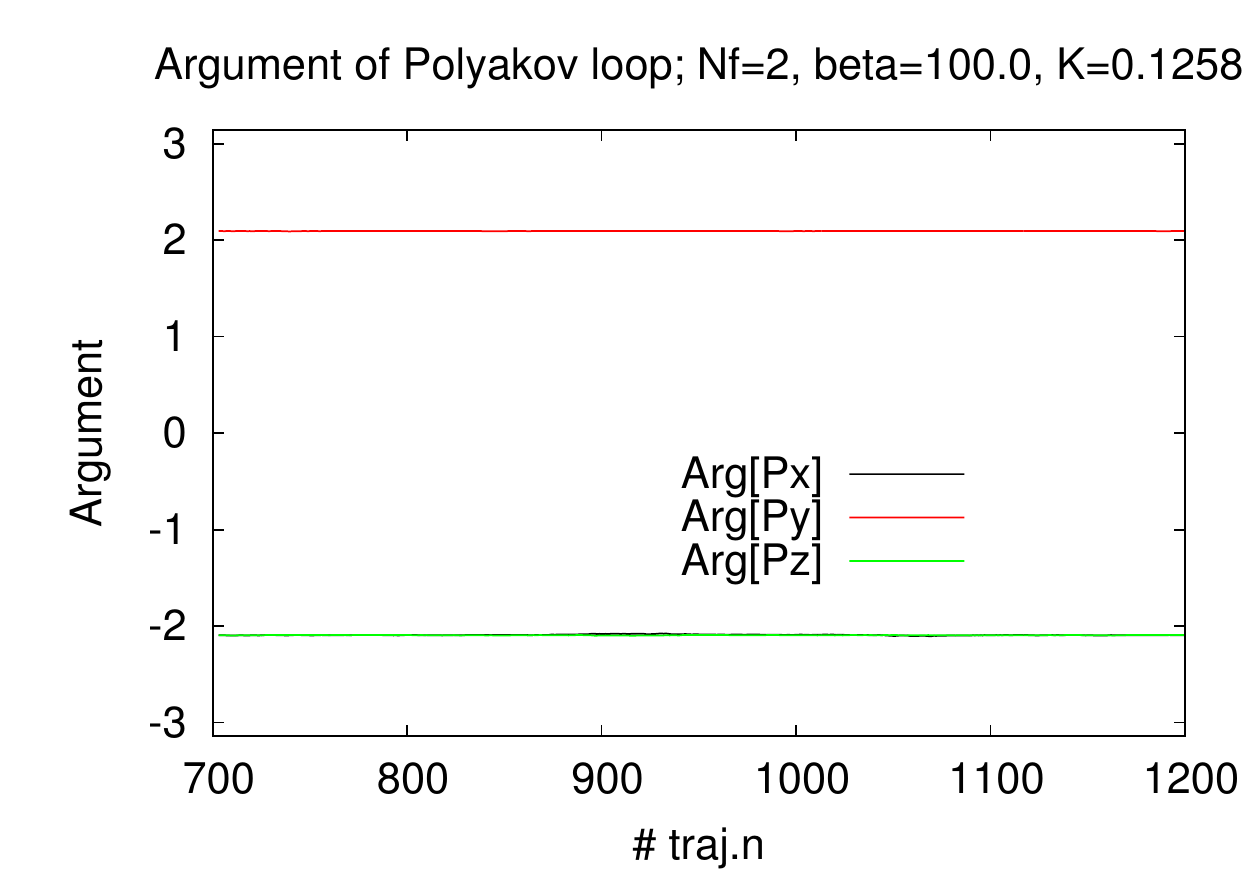}
\hspace{1cm}
\includegraphics [width=7.0cm]{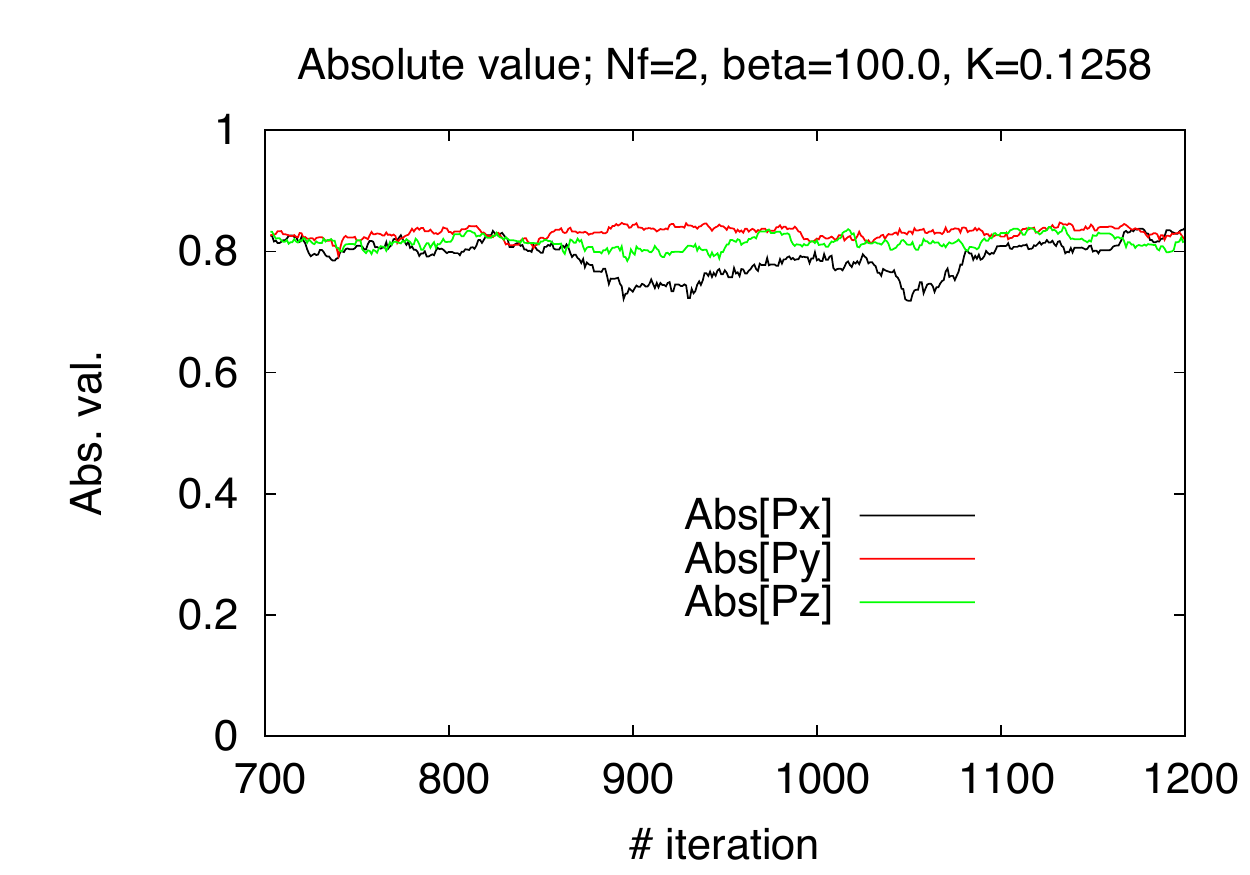}
%\vspace{1cm}
\caption{(color online) The time history of the argument and the absolute value of Polyakov loops for $N_f=2$ at $\beta=10.0$ and $K=0.135$, $\beta=15.0$ and $K=0.130$, and $\beta=100.0$ and $K=0.1258$.}
\label{nf2_poly}
\end{figure*}

\begin{figure*}[thb]
\includegraphics [width=7.0cm]{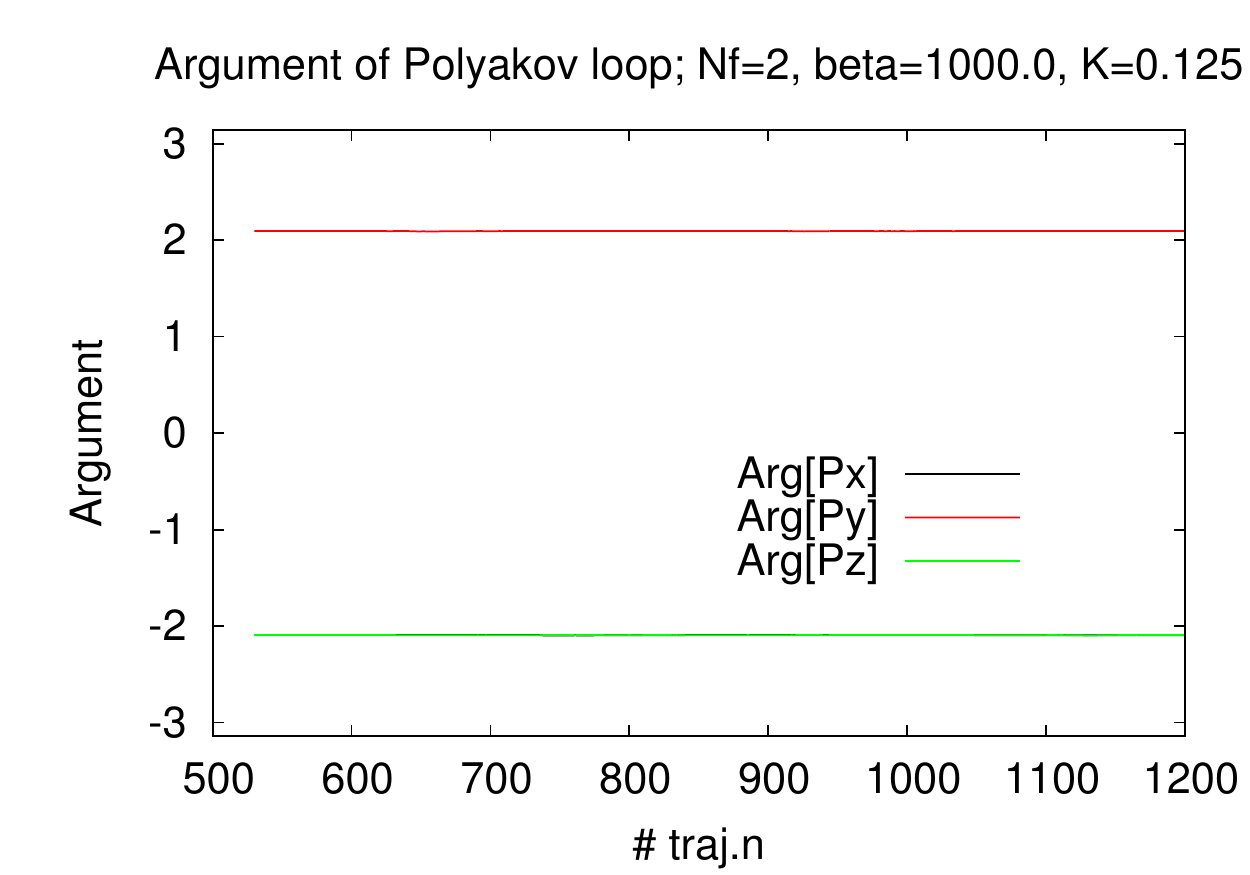}
\hspace{1cm}
\includegraphics [width=7.0cm]{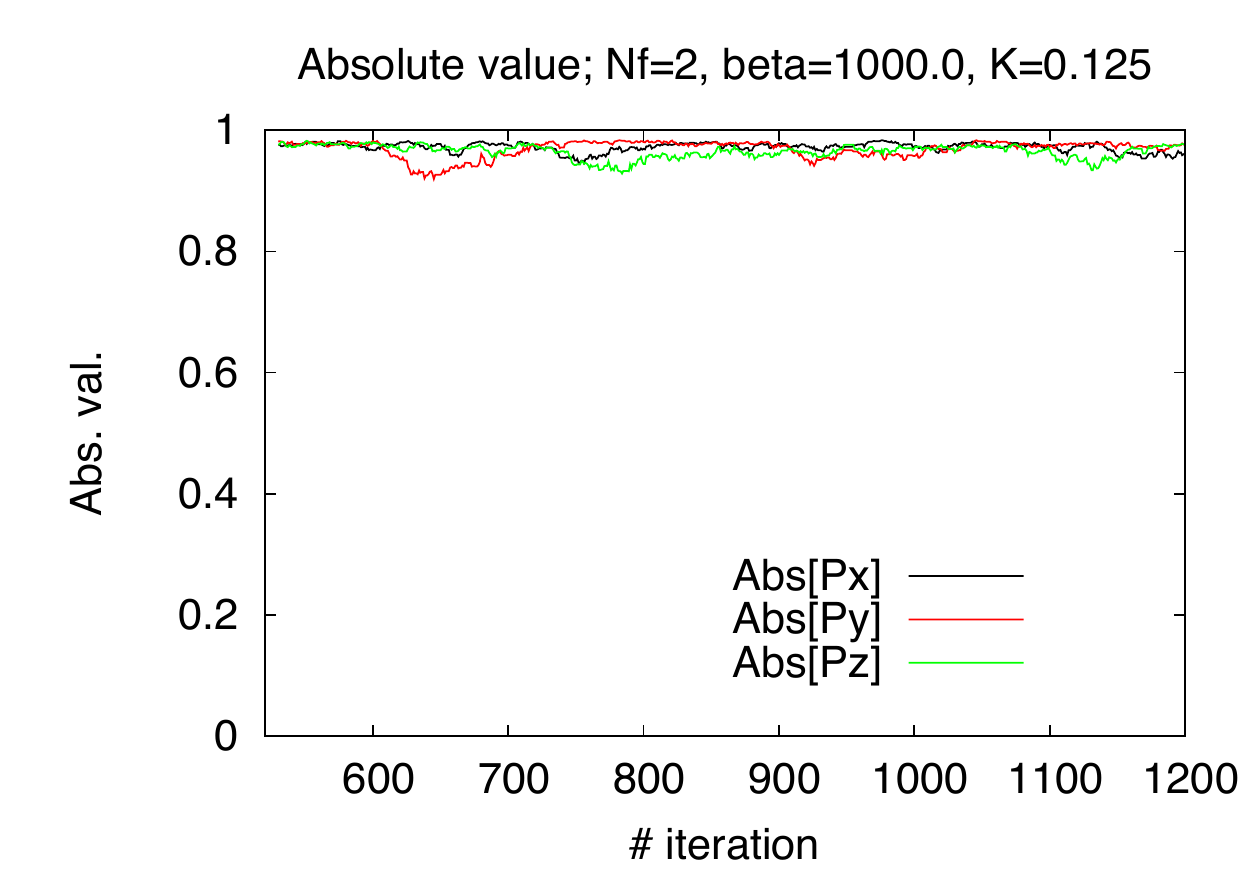}
%\vspace{1cm}
\caption{(color online) The time history of the argument and the absolute value of Polyakov loops for $N_f=2$ at $\beta=1000.0$ and $K=0.125$.}
\label{nf2_b1000}
\end{figure*}

%\section{$K=0.1258$}
%\clearpage

\begin{figure*}[thb]
\includegraphics[width=6.7cm]{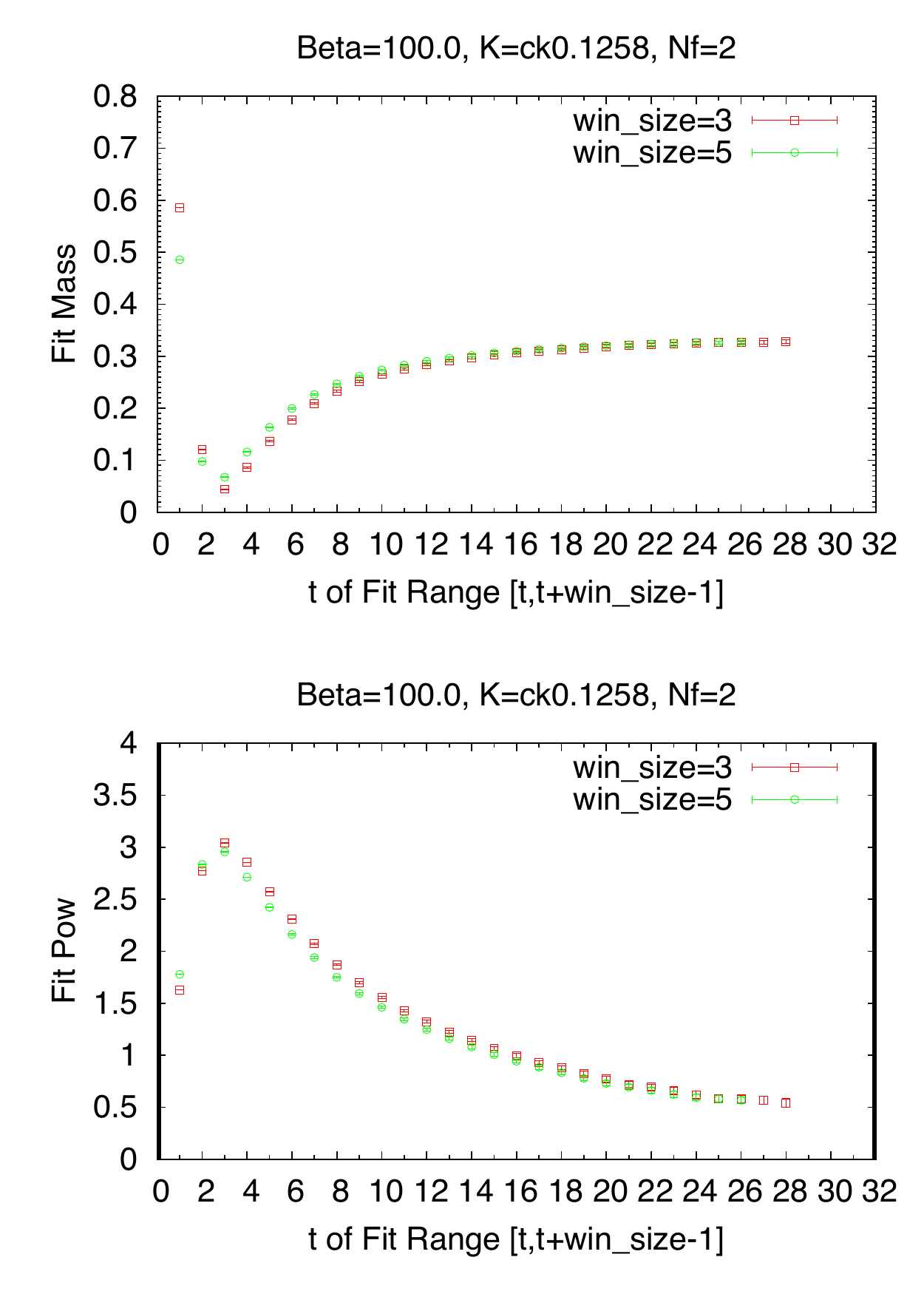}
\hspace{0.5cm}
\includegraphics[width=6.7cm]{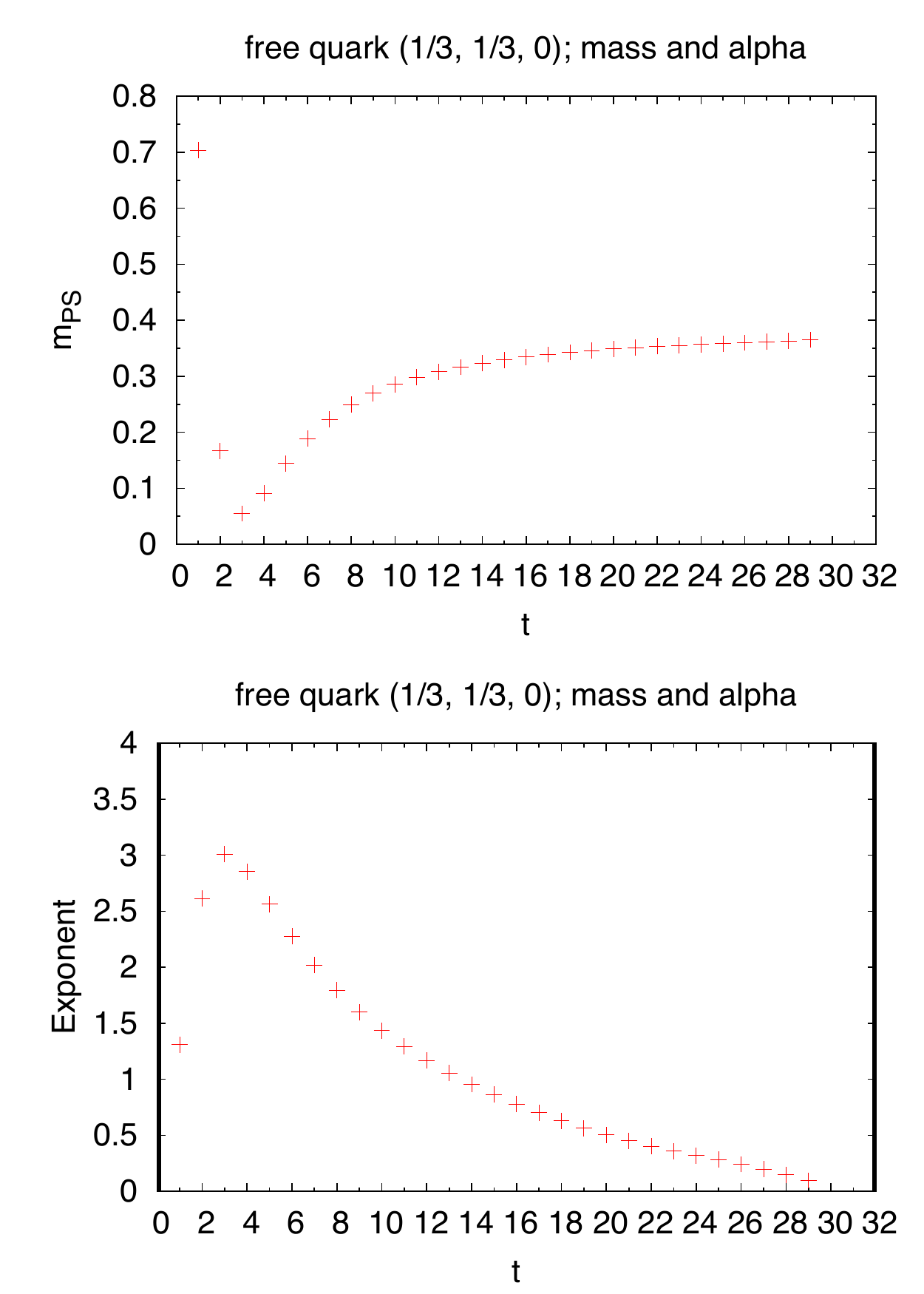}
\caption{(color online) The local mass $m(t)$ and the local exponent $\alpha(t)$ for $N_f=2$ at $\beta=100.0$ and $K=0.1258$ (left) and for a free particle (1/3, 1/3, 0) with $m_q=0.01$ (right)}
\label{(1/3,1/3, 0) free-beta100}
\end{figure*}

\begin{figure*}[thb]
   \includegraphics[width=6.77cm]{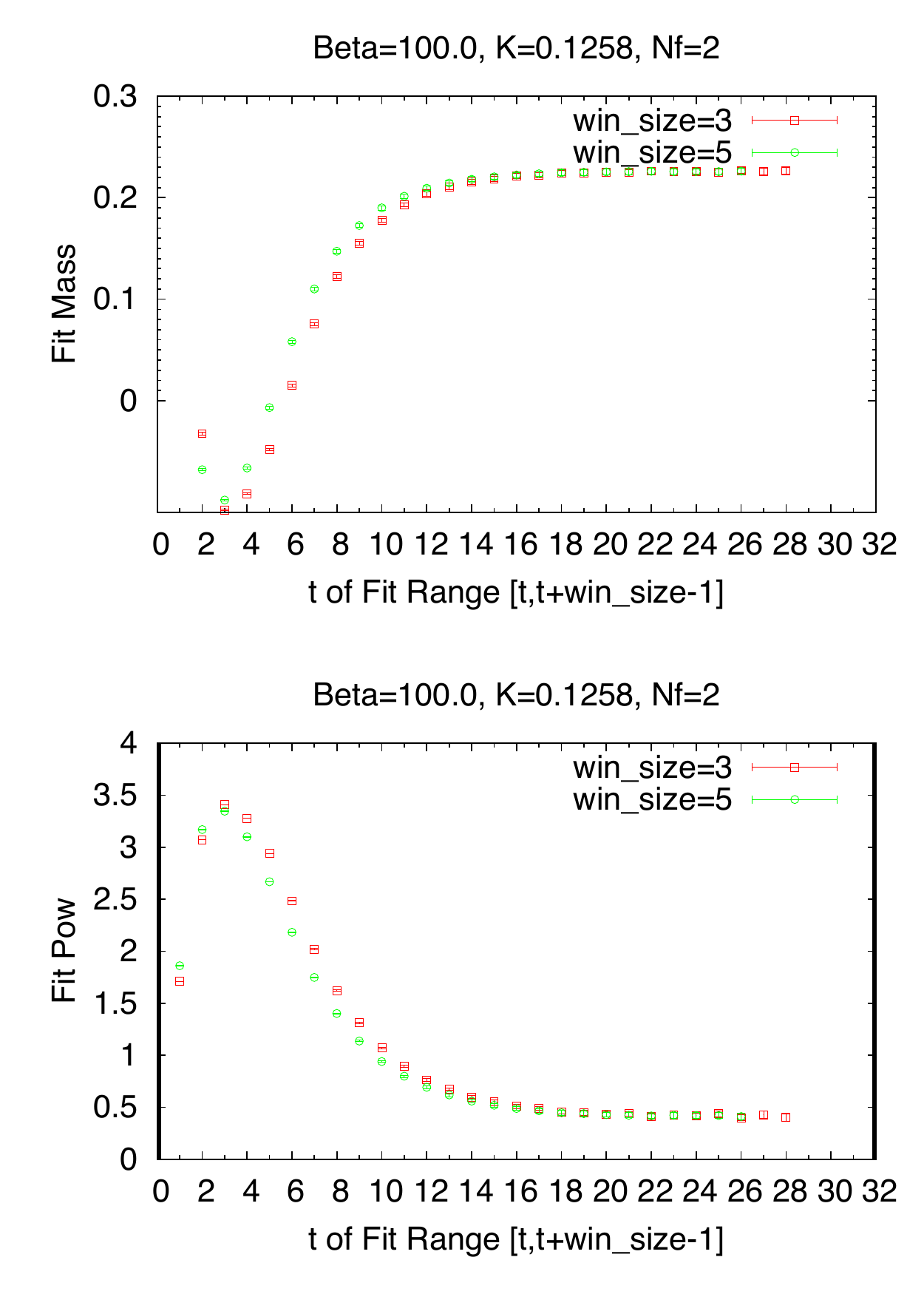}
  \hspace{0.5cm}
   \includegraphics[width=6.7cm]{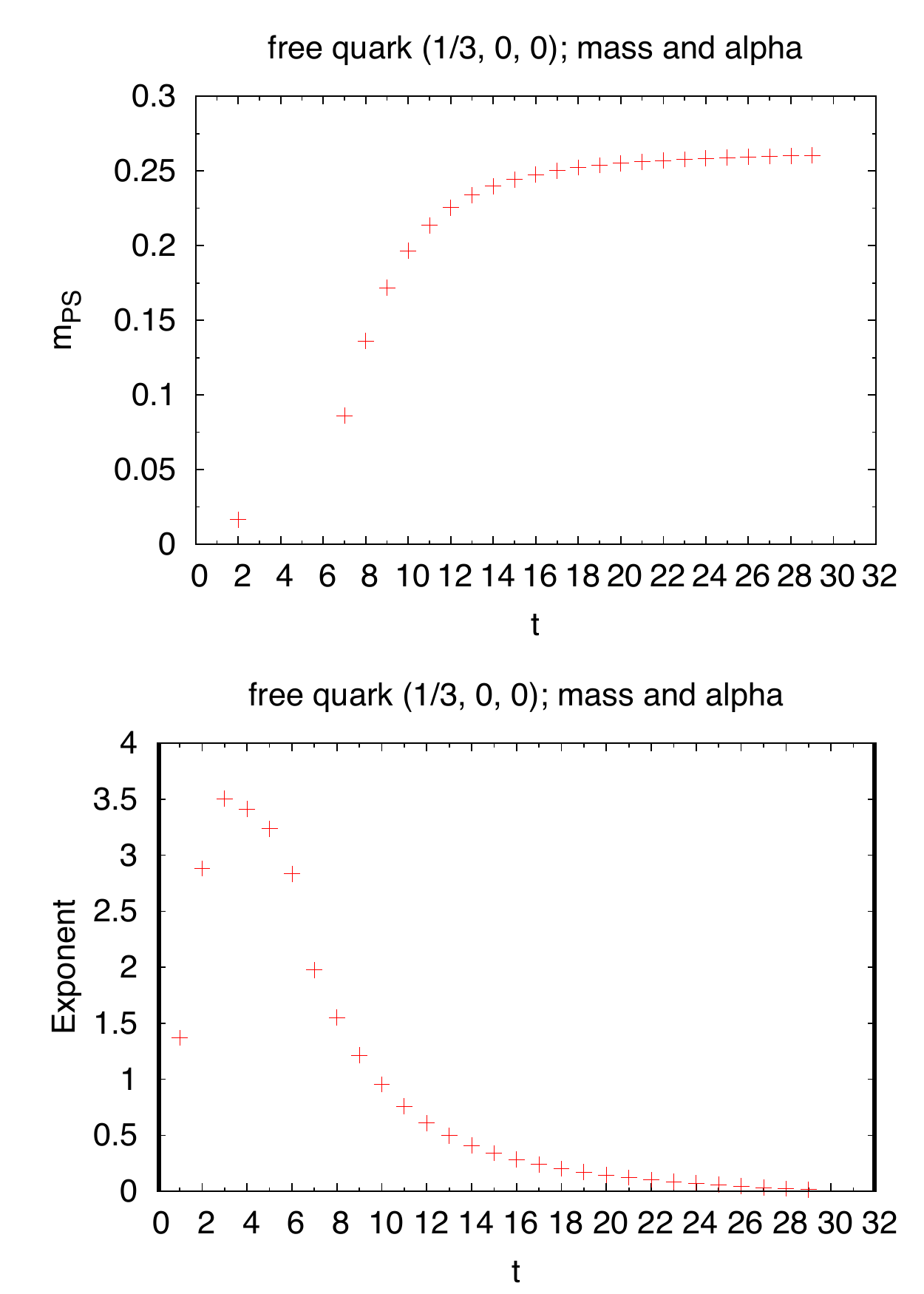}
   \caption{(color online) The local mass $m(t)$ and the local exponent $\alpha(t)$ for $N_f=2$ at $\beta=100.0$ and $K=0.1258$ (left) and  for a free particle (1/3, 0, 0) with $m_q=0.01$ (right)}
   \label{(1/3, 0, 0) free-beta100}
\end{figure*}
  
\begin{figure}[thb]
%\begin{minipage}[t]{0.5\columnwidth}
\hspace{8.5 cm}
%  \begin{center}
   \includegraphics[width=6.77cm]{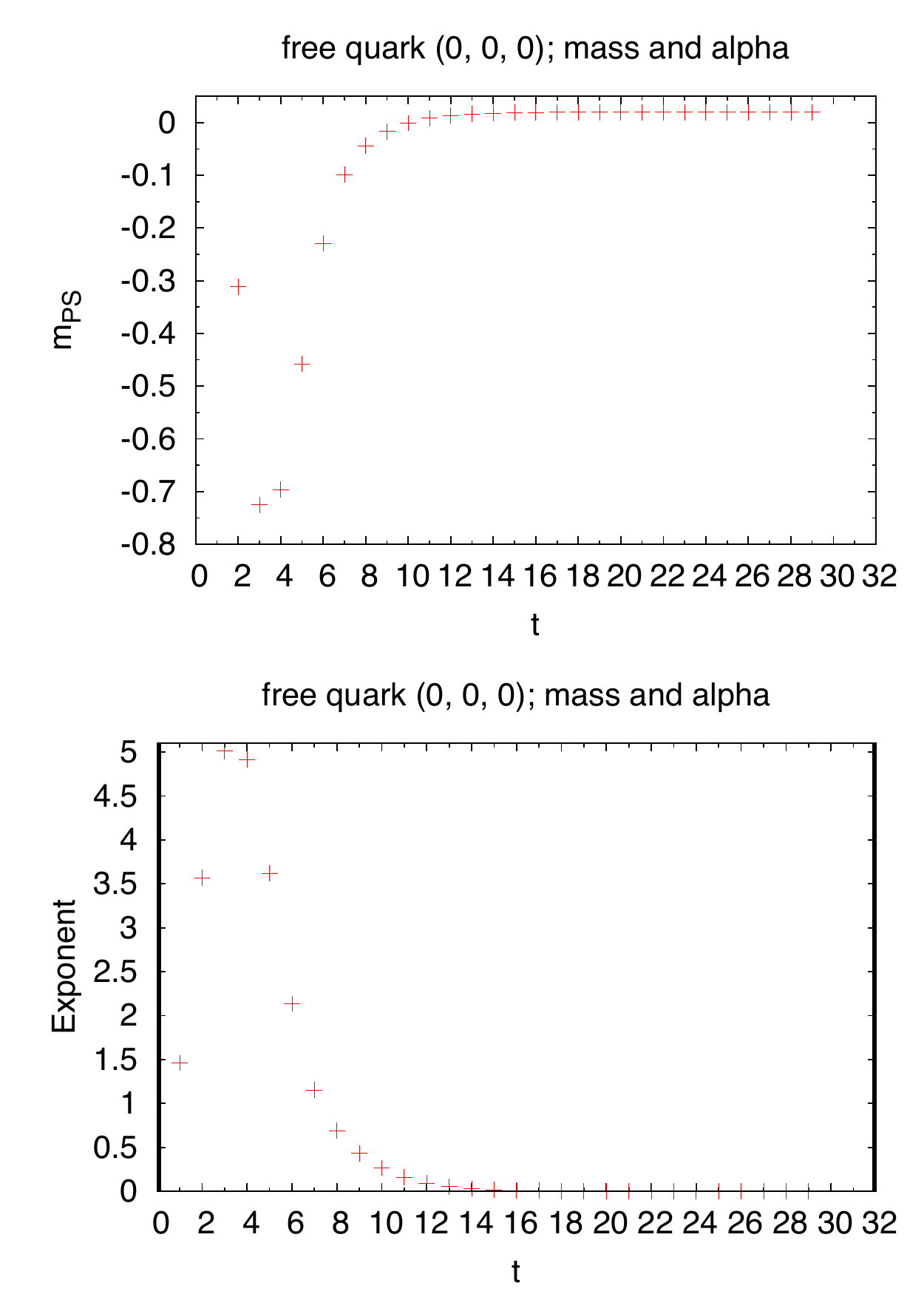}
   %         \end{center}
   %\end{minipage}%
%\begin{minipage}[t]{0.5\columnwidth}
  %\begin{center}
   %\includegraphics[width=7.5cm]{beta100_Fig/Fig_fermiontwist16641.pdf}
   %\end{center}
  % \end{minipage}%
   \caption{(color online) The local mass $m(t)$ and the local exponent $\alpha(t)$ for a free particle (0,0,0)  with $m_q=0.01$}
  \label{(0, 0, 0) free-beta100}
\end{figure}

\begin{figure*}[htb]
\includegraphics [width=7.5cm]{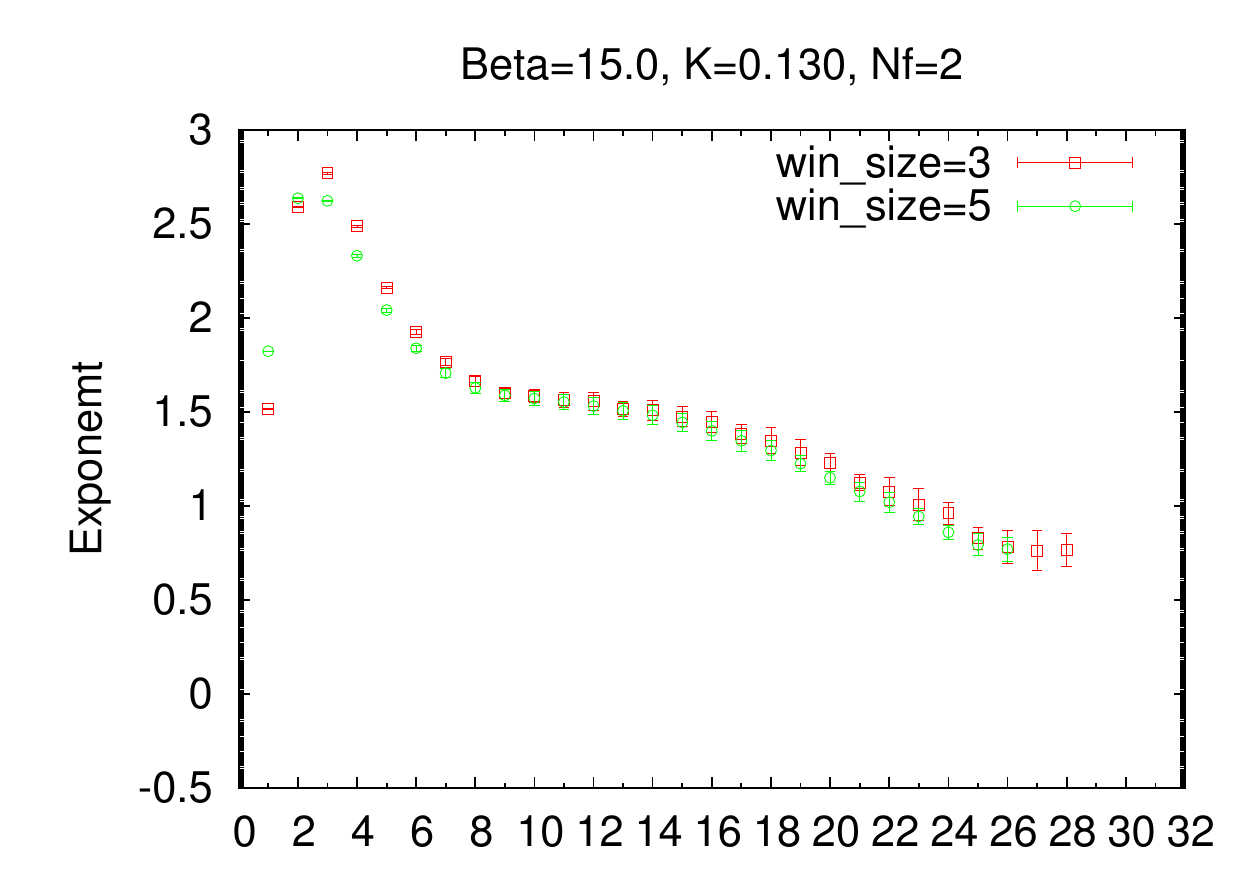}
\hspace{1cm}
\includegraphics [width=7.5cm]{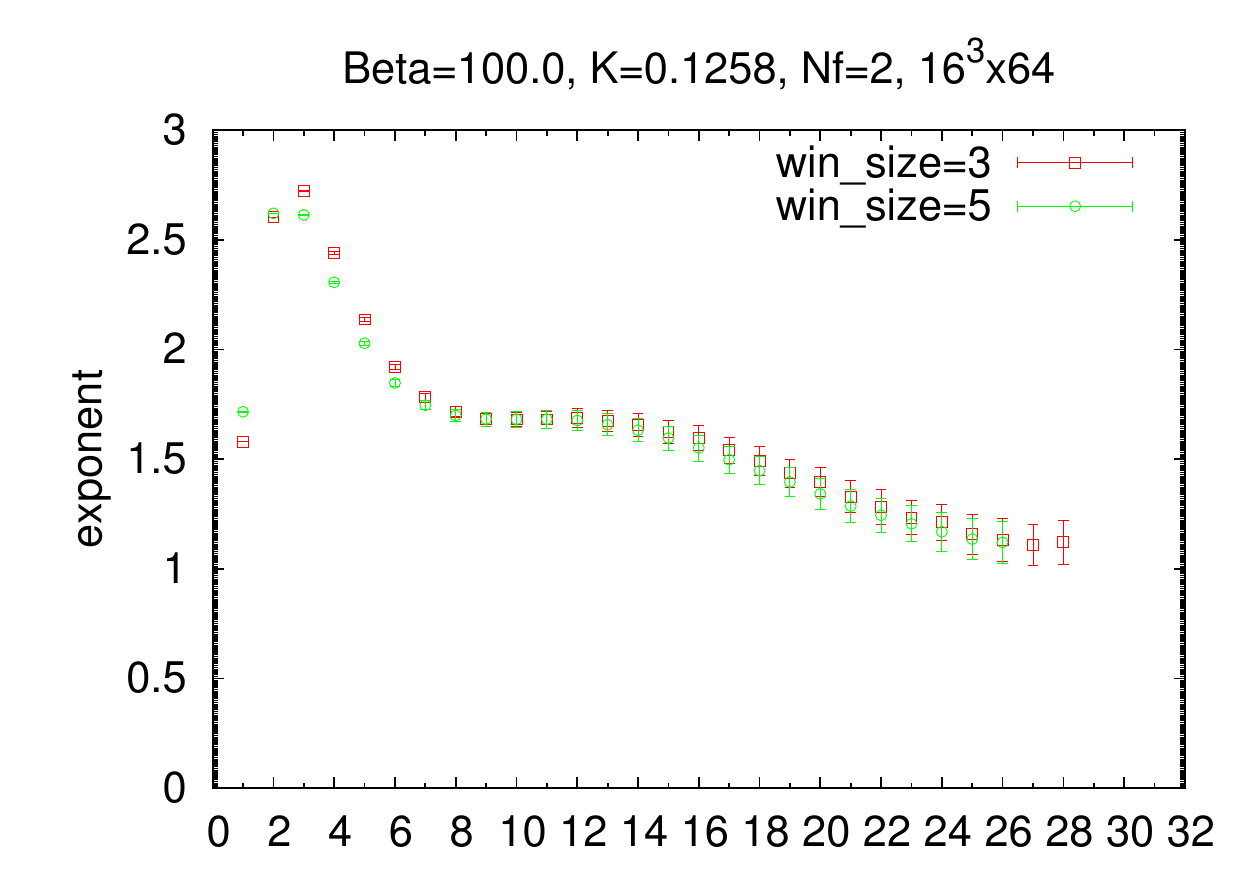}
\vspace{2cm}
\includegraphics [width=7.5cm]{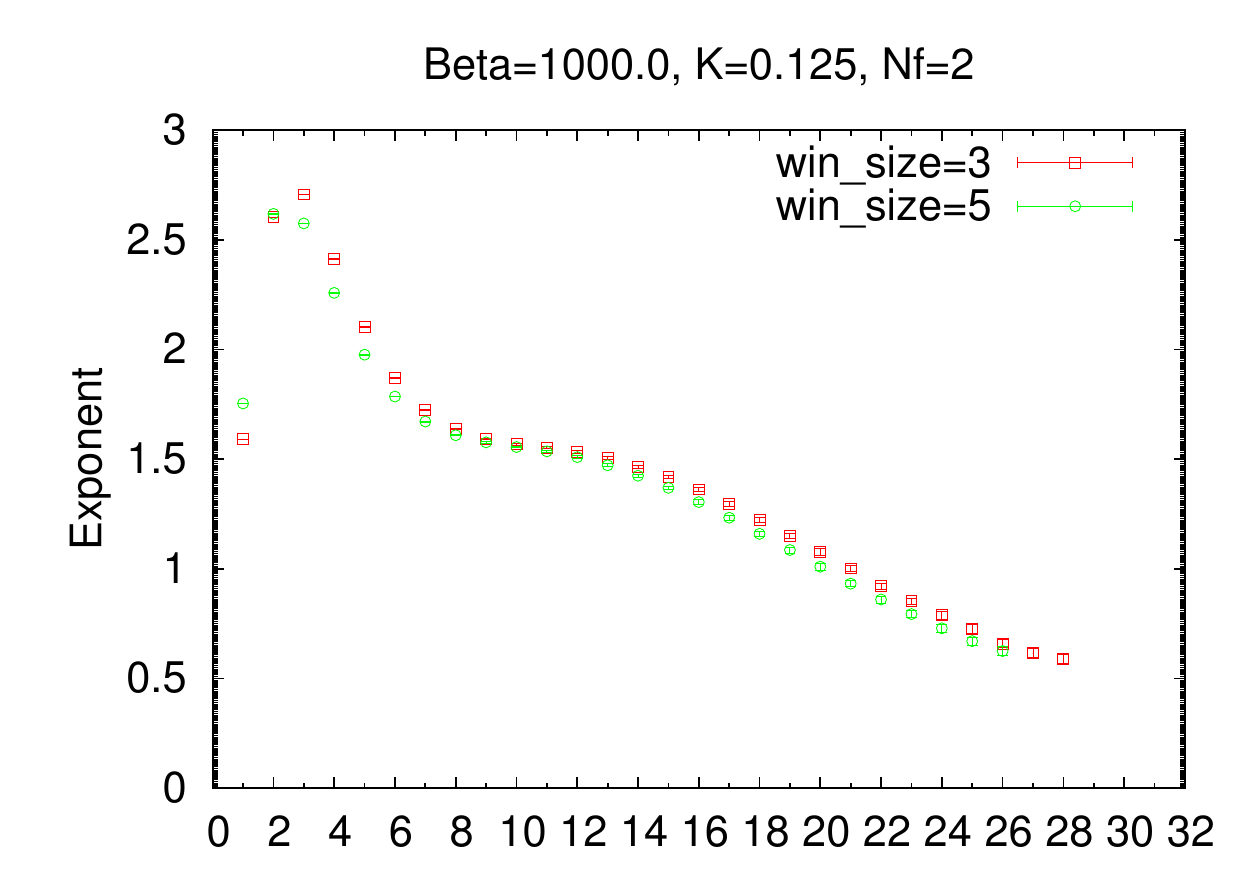}
\hspace{1cm}
\includegraphics [width=7.5cm]{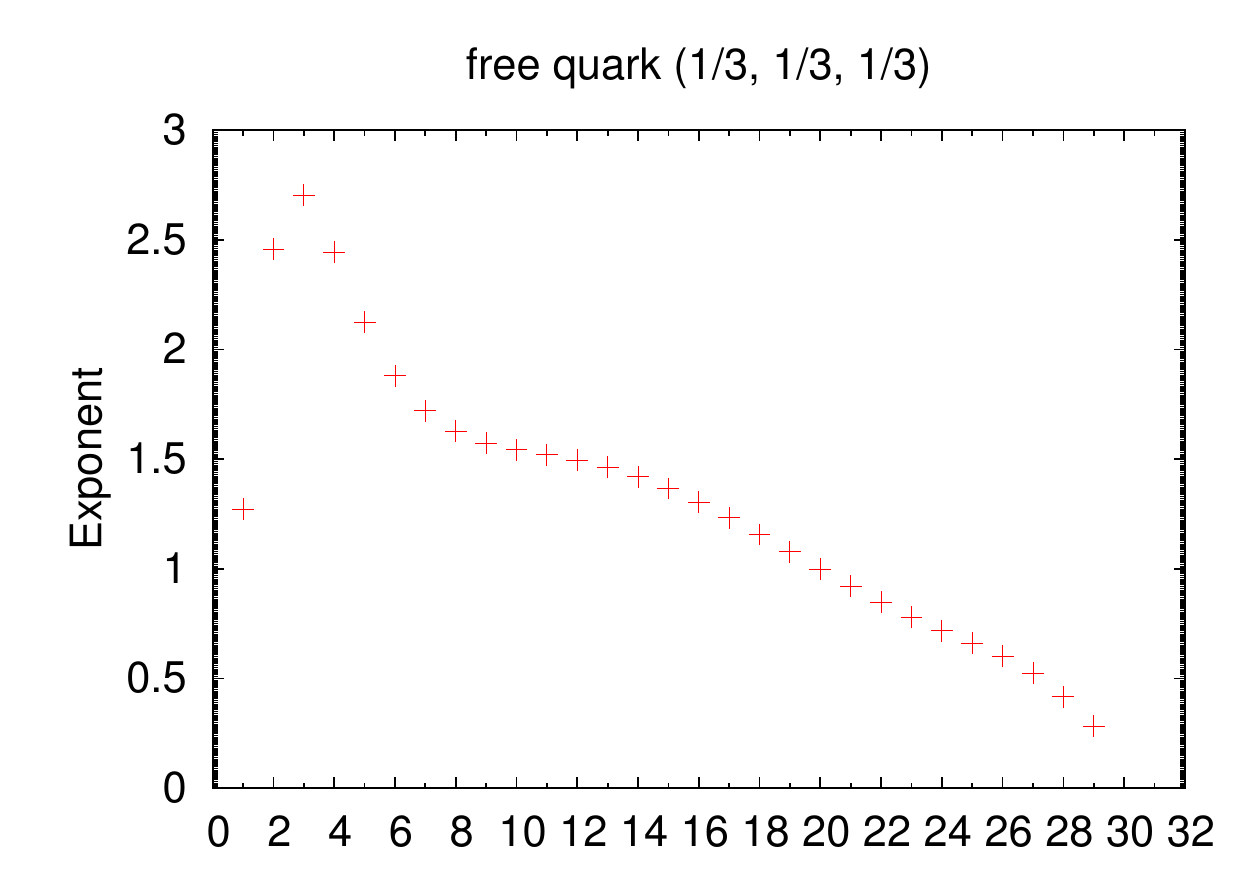}
\caption{(color online) The local exponent $\alpha(t)$:
$N_f=2$ at $\beta = 15.0$ and $K=0.130$, $\beta = 100.0$ and $K=0.1258$, $\beta=1000.0$ and $K=0.125$,   
         and for a free particle (1/3, 1/3, 1/3) with $m_q=0.01$}
\label{b1000.0}
\end{figure*}

%\clearpage

\begin{figure*}[htb]
   \includegraphics[width=7.5cm]{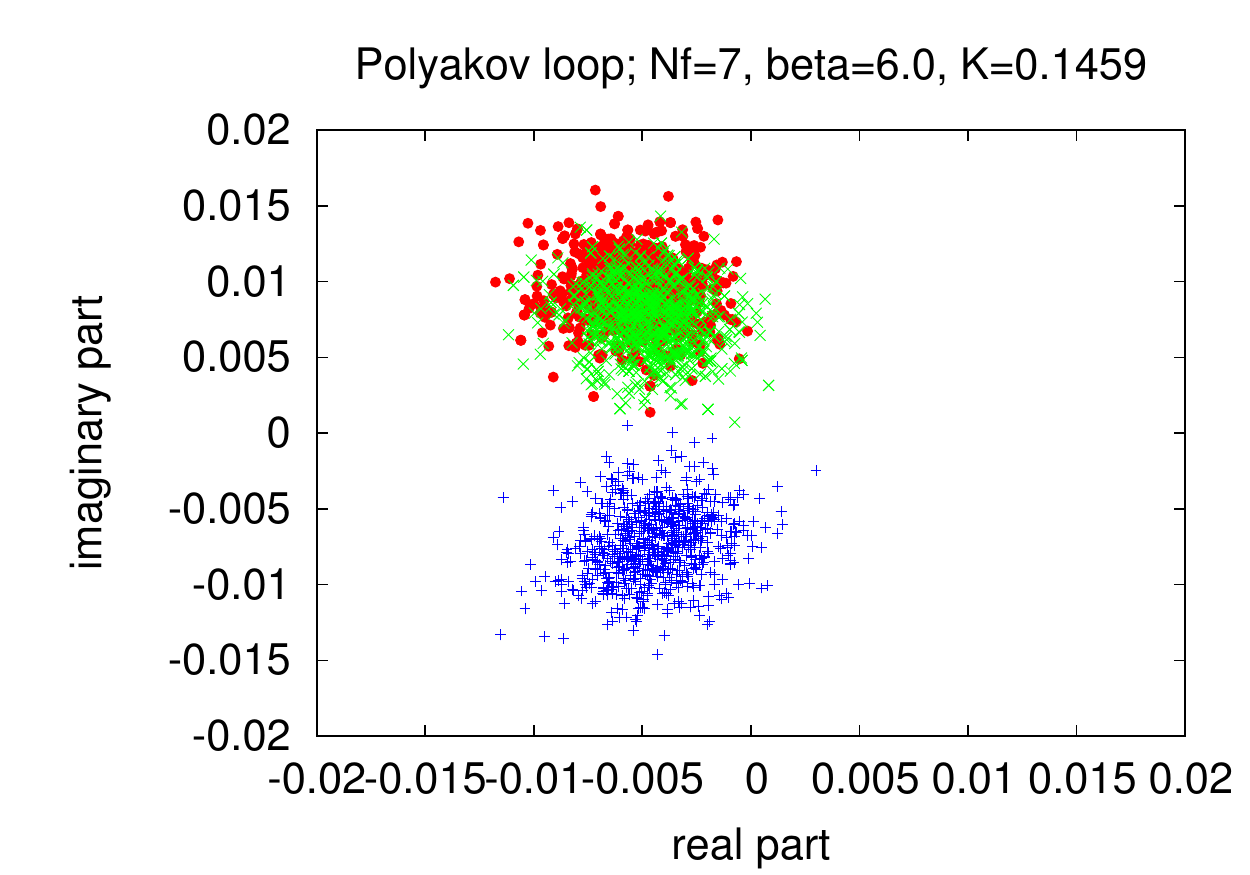}
       \hspace{1cm}
       \includegraphics[width=7.5cm]{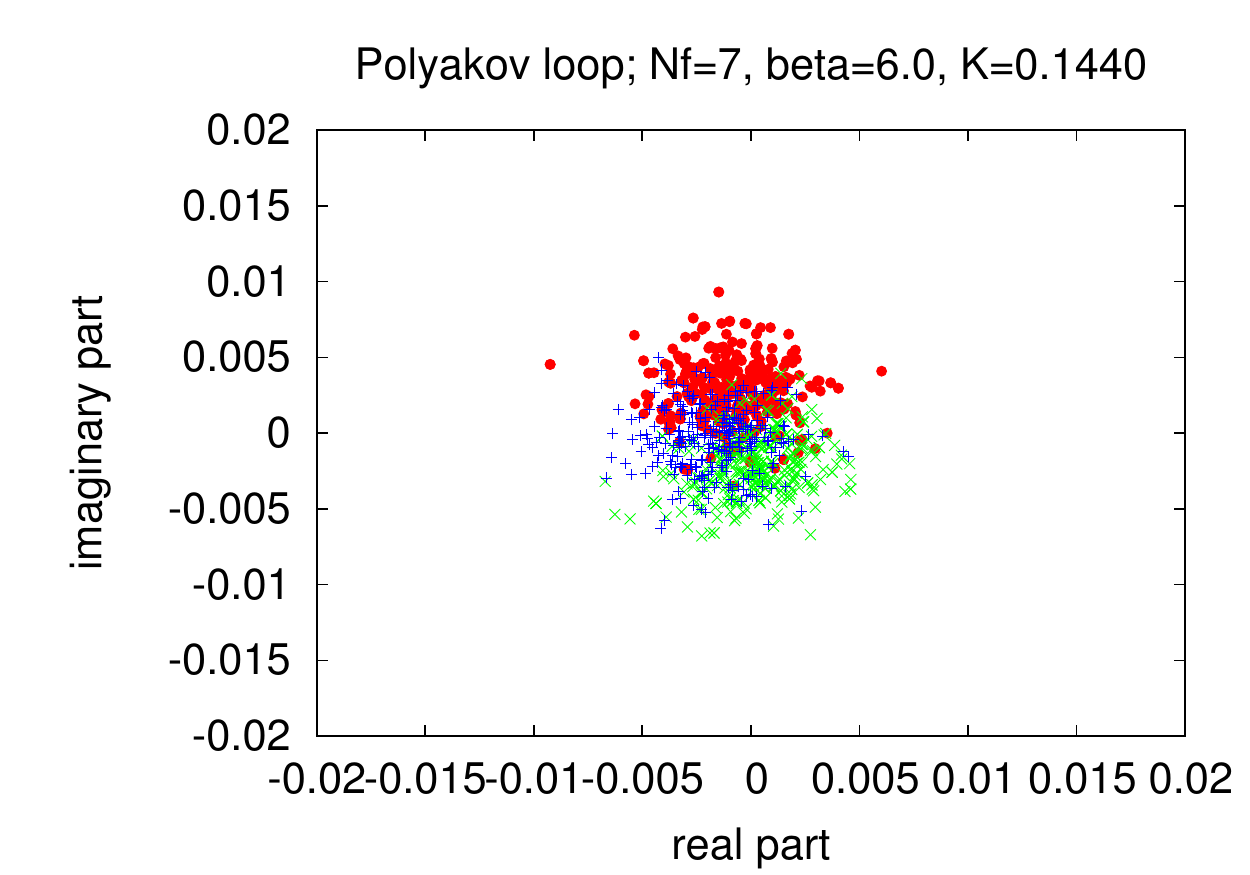}
\caption{(color online) The scattered plots of Polyakov loops in the $x$, $y$ and $z$ directions overlaid; 
        both for $N_f=7$ at $\beta=6.0$; (left) $K=0.1459$: (right) $K=0.1440$.}
\label{complex_nf7}
\end{figure*}
%\clearpage

\begin{figure*}[htb]
   \includegraphics[width=7.5cm]{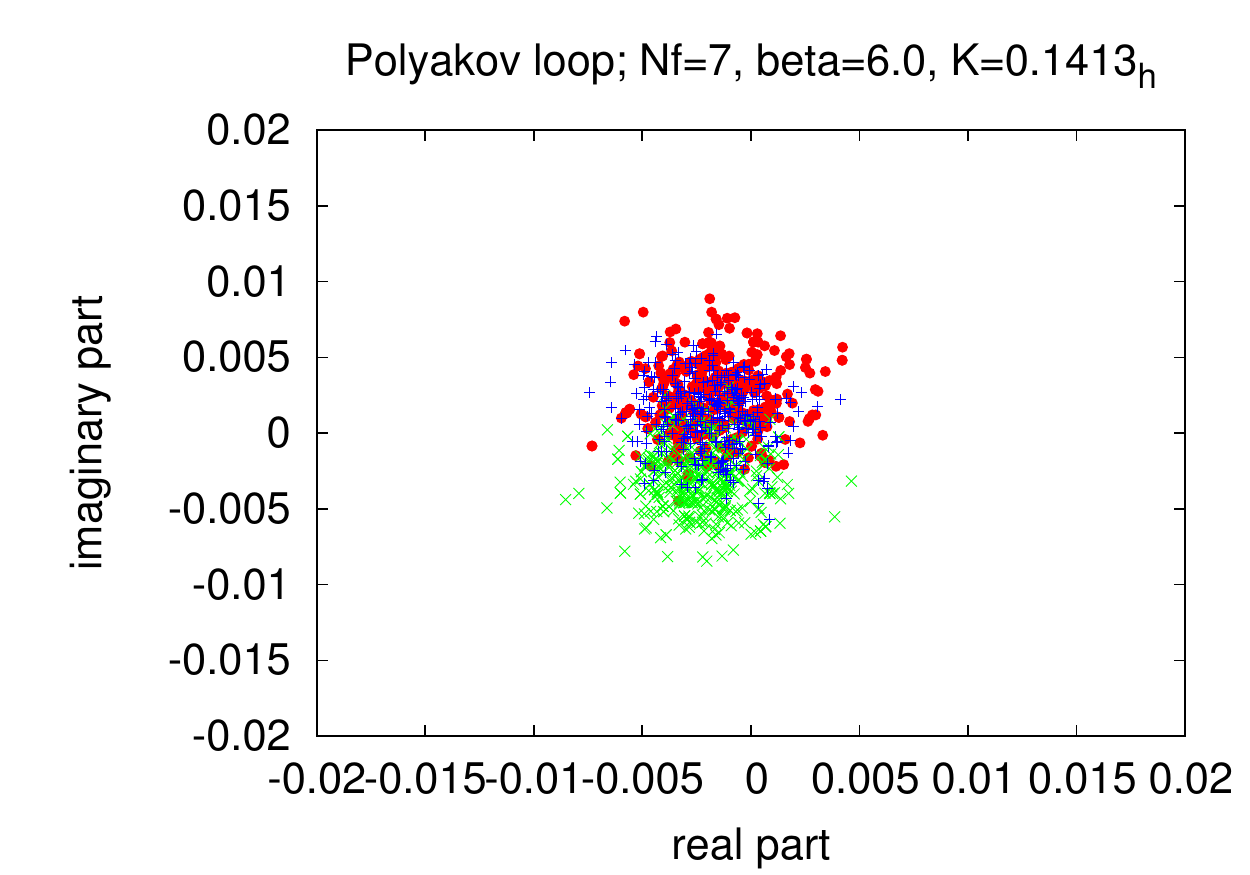}
          \hspace{1cm}
       \includegraphics[width=7.5cm]{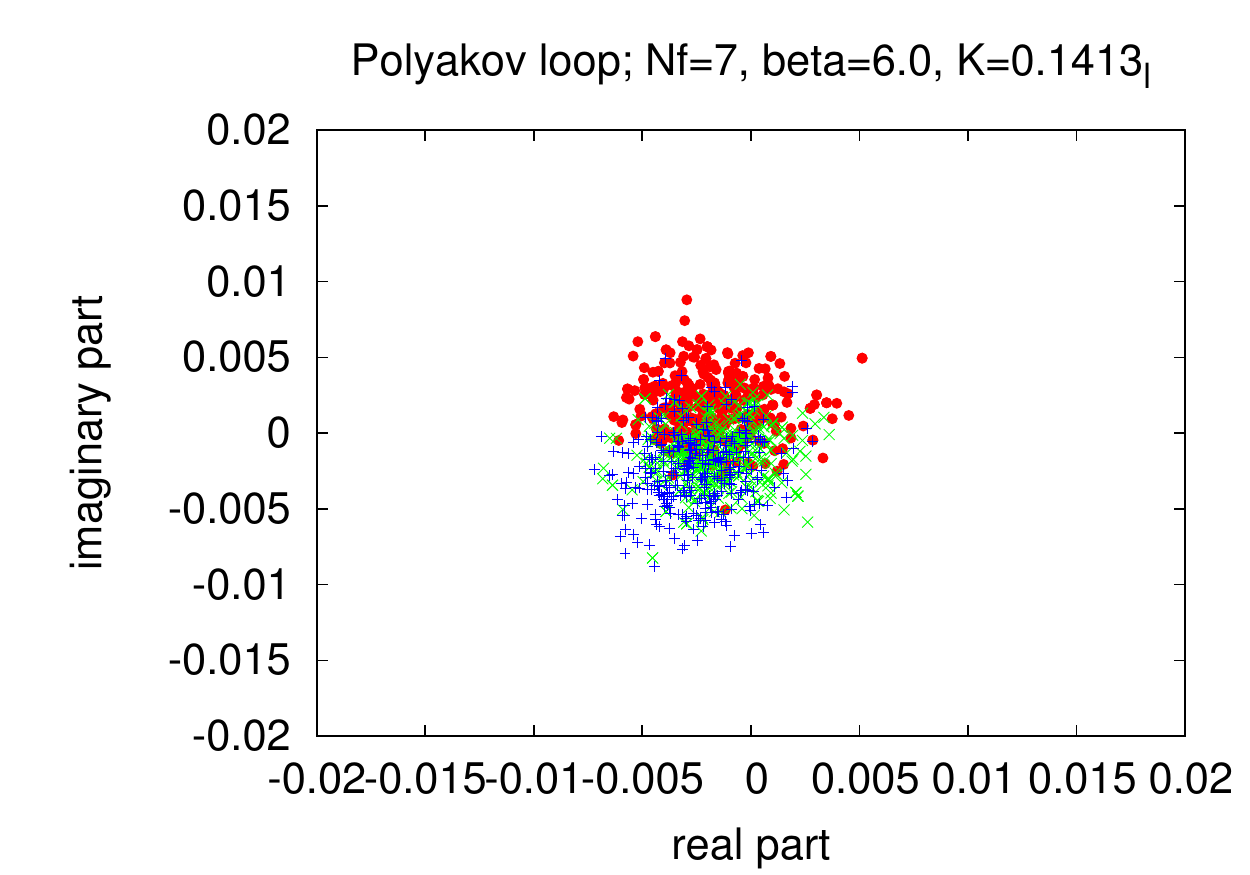}
                      \caption{(color online) The scattered plots of Polyakov loops in the $x$, $y$ and $z$ directions overlaid; 
        both for $N_f=7$ at $\beta=6.0$ and $K=0.1413$. (left) from larger $K$ and (right) from smaller $K$.}
\label{complex_2_nf7}
\end{figure*}

%\clearpage

\begin{figure*}[thb]
\includegraphics [width=7.5cm]{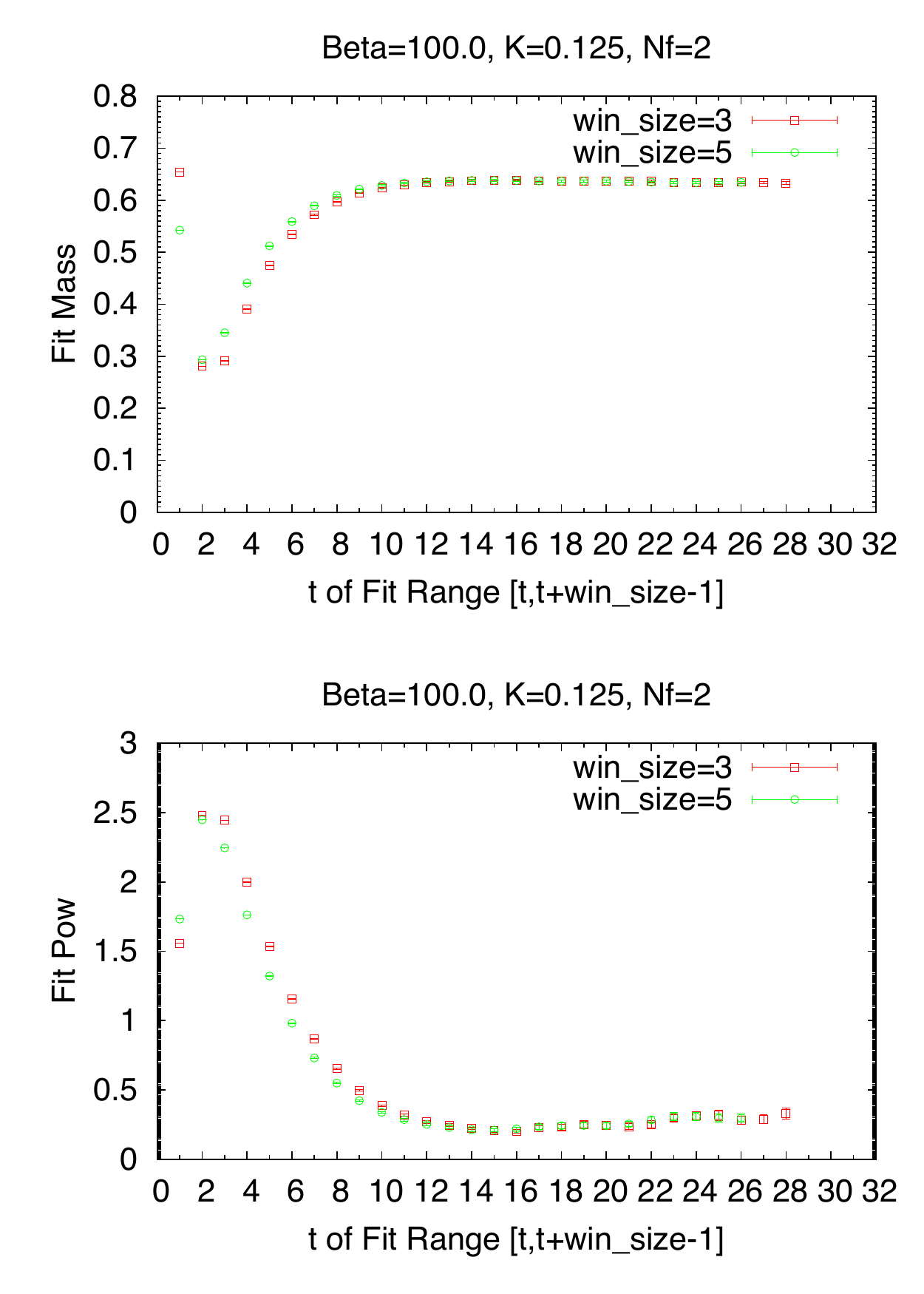}
 \hspace{1cm}
\includegraphics [width=7.5cm]{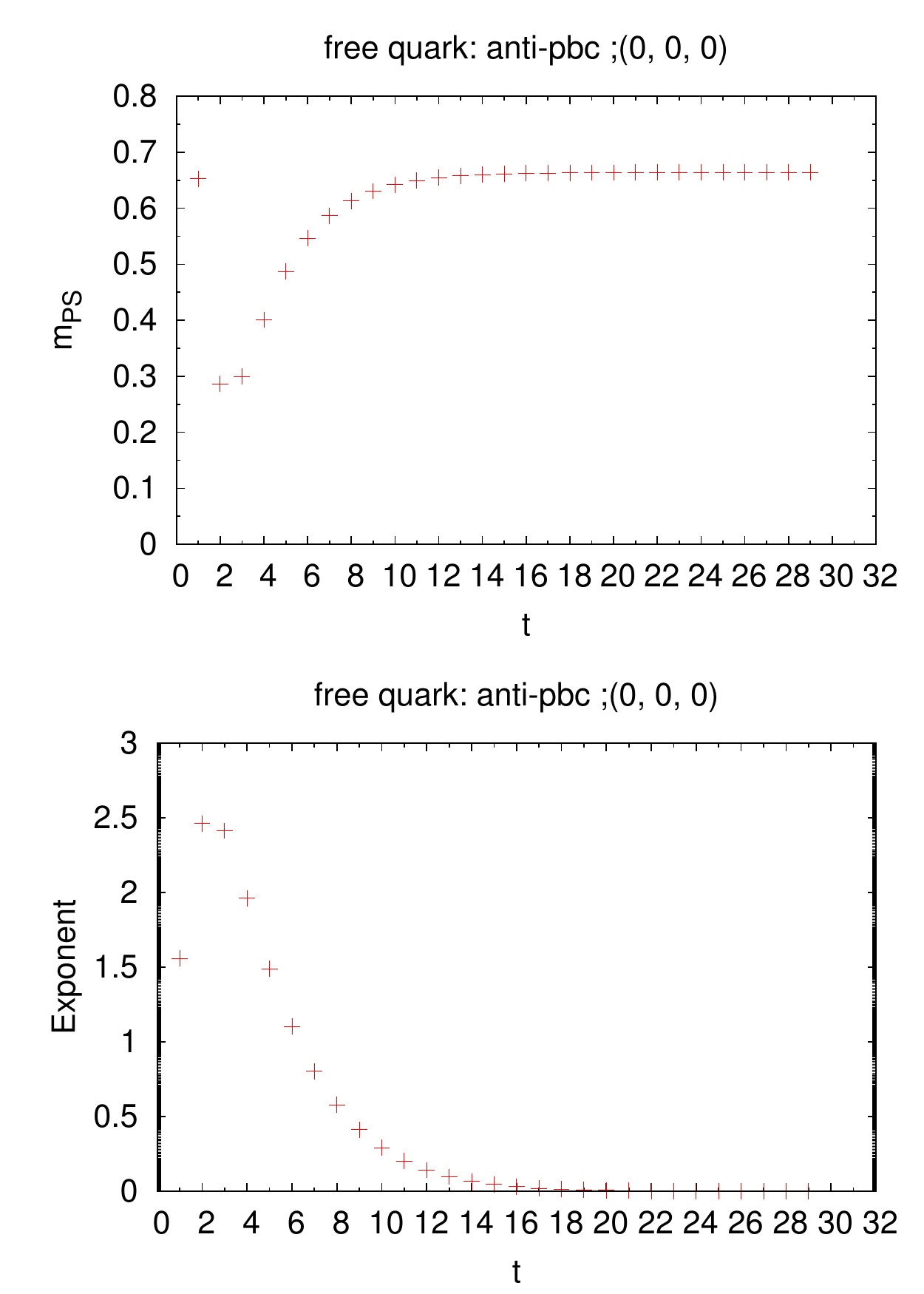}
\caption{(color online) 
The anti-periodic boundary conditions for fermion in the spatial directions;
The local mass $m(t)$ and the local exponent $\alpha(t)$ for a free particle (0,0,0) (right)  with $m_q=0.01$
and for $N_f=2$ at $\beta=100.0$ and $K=0.125$ (left).}
\label{apbc_free_beta10.0}
\end{figure*}

\begin{figure*}[hbt]
   \includegraphics[width=7.3cm]{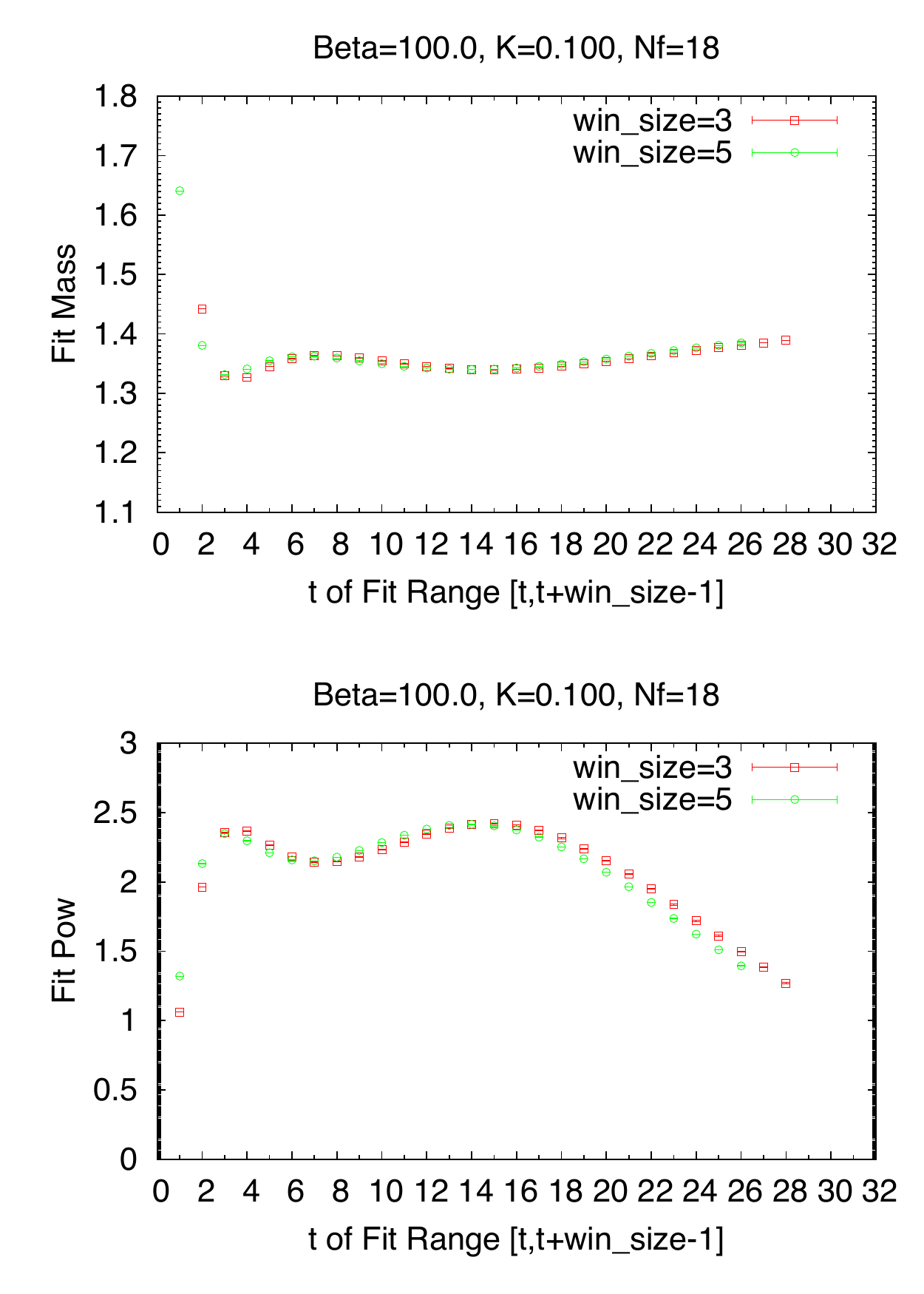}
 \hspace{1cm}
      \includegraphics[width=7.3cm]{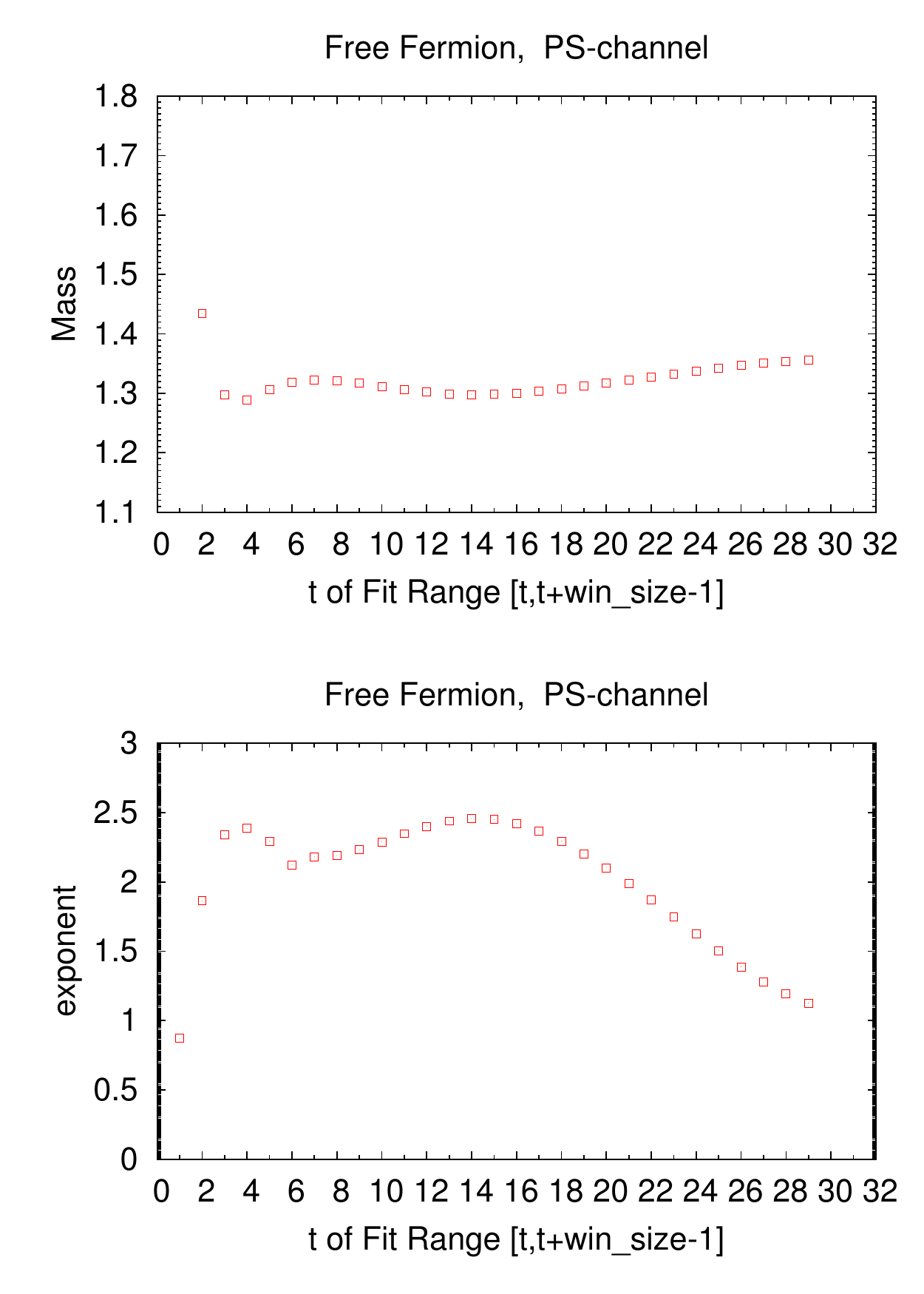}
            \caption{(color online)             
            The local mass $m(t)$ and the local exponent $\alpha(t)$ for $N_f=18$ at $\beta=100.0$ and $K=0.100$ (left) 
                        and for a free particle (0,0,0) with $m_q=1.0$ (right)}
            \label{nf18}
\end{figure*}

\end{document}